\documentclass[useAMS,usenatbib]{mn2e}
\usepackage{graphicx}
\usepackage{amsmath}
\usepackage{grffile}
\usepackage{array}
\usepackage{color}

\newcommand{\be}{\begin{equation}}
\newcommand{\ee}{\end{equation}}

\newcommand{\ifm}[1]{\relax\ifmmode#1\else$\mathsurround=0pt #1$\fi}
\newcommand{\kms}{\ifmmode\,{\rm km}\,{\rm s}^{-1}\else km$\,$s$^{-1}$\fi}

\newcommand{\ltsima}{$\; \buildrel < \over \sim \;$}
\newcommand{\lsim}{\lower.5ex\hbox{\ltsima}}
\newcommand{\gtsima}{$\; \buildrel > \over \sim \;$}
\newcommand{\gsim}{\lower.5ex\hbox{\gtsima}}

\definecolor{green}{rgb}{0,0.5,0}
\definecolor{grey}{rgb}{0.4,0.5,0.7}

\def\M11{M_{11}}
\def\V100{V_{100}}
\def\R1{R_{Mpc}}
\def\T6{T_6}

\begin{document}

\title[]{Cold-mode and hot-mode accretion in galaxy formation: an entropy approach}

\pagerange{\pageref{firstpage}--\pageref{lastpage}} \pubyear{2022}

\author[Tollet et al.]{{\'E}douard~Tollet$^{1,2}$, Andrea~Cattaneo$^2$, Andrea~V.~Macci{\`o}$^{3,4,5}$, Xi~Kang$^6$
\\
\\
$^1$Centre de Recherche Astrophysique de Lyon, 9 avenue Charles André, 69230 Saint-Genis-Laval, France \\
$^2$Observatoire de Paris, LERMA, PSL University, 61 avenue de l'Observatoire, 75014 Paris, France\\
$^3$New York University of Abu Dhabi, P.O. Box 129188, Saadiyat Island, Abu Dhabi, United Arab Emirates\\
$^4$Center for Astro, Particle and Planetary Physics (CAP$^3$), New York University Abu Dhabi\\
$^5$Max-Planck-Institut f\"ur Astronomie, K\"onigstuhl 17, 69117 Heidelberg, Germany \\
$^6$Purple Mountain Observatory, the Partner Group of MPI für Astronomie, 2 West Beijing Road, Nanjing 210008, China}

\date{Accepted for publication in MNRAS 16 June 2022. Received 14 June 2022; in original form 10 Frebruary 2022}

\maketitle

\label{firstpage}


\begin{abstract}

We have analysed two cosmological zoom simulations with $M_{\rm vir}\sim 10^{12}{\rm\,M}_\odot$ from the NIHAO series, both  with and without feedback. 
We show that an entropy criterion based on the equation of  state of the intergalactic medium can successfully separate cold- and hot-mode accretion.
The shock-heated gas has non-negligible turbulent support and cools inefficiently.
In the simulations without feedback, only a small fraction ($\lsim 20$ per cent) of the stellar mass comes from baryons that have been in the hot circumgalactic medium,
although quantitative conclusions should be taken with caution due to our small-number statistics.
With feedback, the fraction is larger because of the reaccretion of gas heated by supernovae, which has lower entropies and  shorter cooling times than the gas heated by accretion shocks. 
We have compared the results of NIHAO to predictions of the {\sc GalICS~2.1} semianalytic model of galaxy formation.
The shock-stability criterion implemented in {\sc GalICS~2.1}  successfully reproduces  the transition from cold- to hot-mode accretion.

\end{abstract} 

\begin{keywords}
{
galaxies: evolution ---
galaxies: formation 
}
\end{keywords}

\section{Introduction}

Galaxies can grow through the accretion of cold filamentary flows
or through the cooling of hot gas (they  can also grow through mergers, but they are not the focus of this article).
Since \citet{keres_etal05} and \citet{dekel_birnboim06} {  linked the galaxy bimodality to these two regimes}
(hereafter, cold and hot accretion, respectively), 
there has been a lot of discussion on their role in the formation and evolution of galaxies,
mainly in the context of cosmological hydrodynamic simulations {  \citep{ocvirk_etal08,keres_etal09,brooks_etal09,vandevoort_schaye12,nelson_etal13,nelson_etal16}, but also in that of semianalytic models of galaxy formation
(hereafter SAMs; \citealp{benson_bower11,cattaneo_etal20}).}

Attempts at separating cold and hot accretion in cosmological
hydrodynamic simulations have normally used a temperature criterion
{  \citep{keres_etal05,keres_etal09,ocvirk_etal08,vandevoort_schaye12}.}
\citet{nelson_etal13}  {  have shown that this approach is sensitive to the assumed temperature threshold and that 
criteria based on the absolute
temperature $T$ give different results  from those based on $T/T_{\rm vir}$}, where $T_{\rm vir}$ is the virial temperature.
In analytic studies \citep{dekel_birnboim06,cattaneo_etal20},  it is
more convenient to use a stability criterion {  based on the ratio between the gravitational compression time $t_{\rm comp}$ and the radiative cooling time $t_{\rm cool}$} (the formation of a hot
atmosphere requires the propagation of a stable shock and thus  $t_{\rm comp}<t_{\rm cool}$ ).

{A third route to identify the shock-heated gas is to use an entropy criterion, since shocks present an entropy jump (in a strong shock, the temperature increase is much more significant than the density increase).}
 \citet{brooks_etal09} explored a combination of an
entropy-jump and a temperature criterion, and
found that this approach could separate the filamentary cold gas from the more spherical hot phase.
{A criterion that is purely based on entropy dispenses  from the need of a temperature threshold altogether, while
precluding the possibility of confusing the hot interstellar medium (ISM) with the hot circumgalactic medium (CGM), since
both have high temperature, but the hot ISM is denser and has thus lower entropy.
{\citet{correa_etal18} explored an entropy criterion and its effect on the shock-heated fraction compared to other selection criteria for the hot phase.

Here, we propose a criterion based on the entropy of the gas to that of the intergalactic medium (IGM). The gas is in the cold mode if its entropy decreases (i.e. if it looses heat) while it is accreted. It is in the hot mode mode if there is a heat gain (i.e. if its entropy increases).}

Our first question is whether an entropy approach and an approach based on $t_{\rm comp}/t_{\rm cool}$
return a consistent picture when applied to a same astrophysical object. 
Is $t_{\rm comp}/t_{\rm cool}$ a good predictor of the shock-heated fraction measured in cosmological 
simulations?

{  Secondly, the proposal that shock heating quenches star formation \citep{dekel_birnboim06,cattaneo_etal06}  assumes that the shock-heated gas will not cool efficiently.  
To what extent is this assumption verified?}

{  In this article, we address these questions with cosmological zoom simulations from a Numerical Investigation of a Hundred Astrophysical Objects
(NIHAO; \citealp{wang_etal15}). We also investigate
whether a SAM based on \citet{dekel_birnboim06}'s shock-stability criterion} ({\sc GalICS~2.1}; \citealp{cattaneo_etal20}) can
account for  the simulations'  key features
(e.g., the time at which the cold filaments disappear and the transition to the hot mode occurs).

Feedback complicates the analysis. 
Galactic winds can disrupt cold filaments through shear instabilities.
Violent collisions  between inflows and outflows can provide additional shock-heating.
In milder interactions, winds can make  radiative cooling more efficient by compressing the filaments.
{  Feedback also increases the cooling rate by enriching the CGM with metals.
The problem is  complicated by the impossibility to replicate the same exact feedback recipes in hydrodynamic simulations and SAMs.
In simulations, shock-heating, supernova (SN)-heating  and shear instabilities occur simultaneously. In SAMs, they are modelled sequentially and independently of one another.
Hence, it is difficult to interpret the results with feedback  if one does not fully understand the case without it. 

However, because of the simulations' computational cost, only two massive spirals have been resimulated without feedback (dwarf galaxies lack hot gas and are thus irrelevant
for this study).
Obviously, two galaxies cannot be used to draw  statistical conclusions on the galaxy population, but neither is that the goal of our article.
Our goal is to test whether our assumptions on the role of $t_{\rm comp}/t_{\rm cool}$ and on the inefficient cooling of the hot CGM  are correct.
If two massive spirals with different accretion histories converge on a same picture,  it is unlikely that a larger sample
will lead to substantially different conclusion unless both galaxies are  atypical in the same manner.
}

The structure of the article is as follows. In Section~2, we present the key properties of the objects 
retained for this study 
(virial mass, stellar mass, growth history) and use temperature maps for
a qualitative analysis   of the mutiphase
structure of the CGM from high to low redshift.
In Section~3, we analyse the {  effective} equation of state of the IGM and use it to separate the hot CGM from the cold gas within haloes. We also
test the assumption that the hot CGM is quasi-hydrostatic. 
In Section~4, we examine the entropy distribution of the baryons within the virial radius and its evolution with redshift, comparing the results without and with SN feedback.
In Section~5, we recall the key assumptions of \citet{cattaneo_etal20}'s shock-heating model and check to what extent its assumptions and predictions agree with what we measure in NIHAO.
Finally, in Section~6, we discuss and summarise the main conclusions of the article.

\section{Simulations}

NIHAO 
is a series of $\sim 90$ zoomed cosmological smoothed-particle-hydrodynamics (SPH)  simulations in a flat LCDM universe with $\Omega_{\rm m}=0.3175$, $\Omega_{\rm b}=0.049$ and $H_0=67.1{\rm\,km\,s}^{-1}$
(\citealp{wang_etal15}; also see \citealp{tollet_etal19}).  About 10 of the initial $100$ 
objects were discarded because of their merging histories, since the NIHAO project focusses on {\it isolated} galaxies.
The NIHAO series include feedback from SNe, but not active galactic nuclei.

{  The virial masses at $z=0$ of the two massive spirals retained for this study, g7.55e11  and g1.12e12, are 
$M_{\rm vir}= 9.29\times 10^{11}{\rm\,M}_\odot$ 
and $M_{\rm vir}=1.12\times 10^{12}{\rm\,M}_\odot$, respectively. For comparison, the virial mass of the Milky Way is $M_{\rm vir}=1.3\pm 0.3\times 10^{12}{\rm\,M}_\odot$ \citep{bland_gerhard16}.}

\begin{figure*}
\begin{center}$
\begin{array}{c}
\includegraphics[width=0.45\hsize]{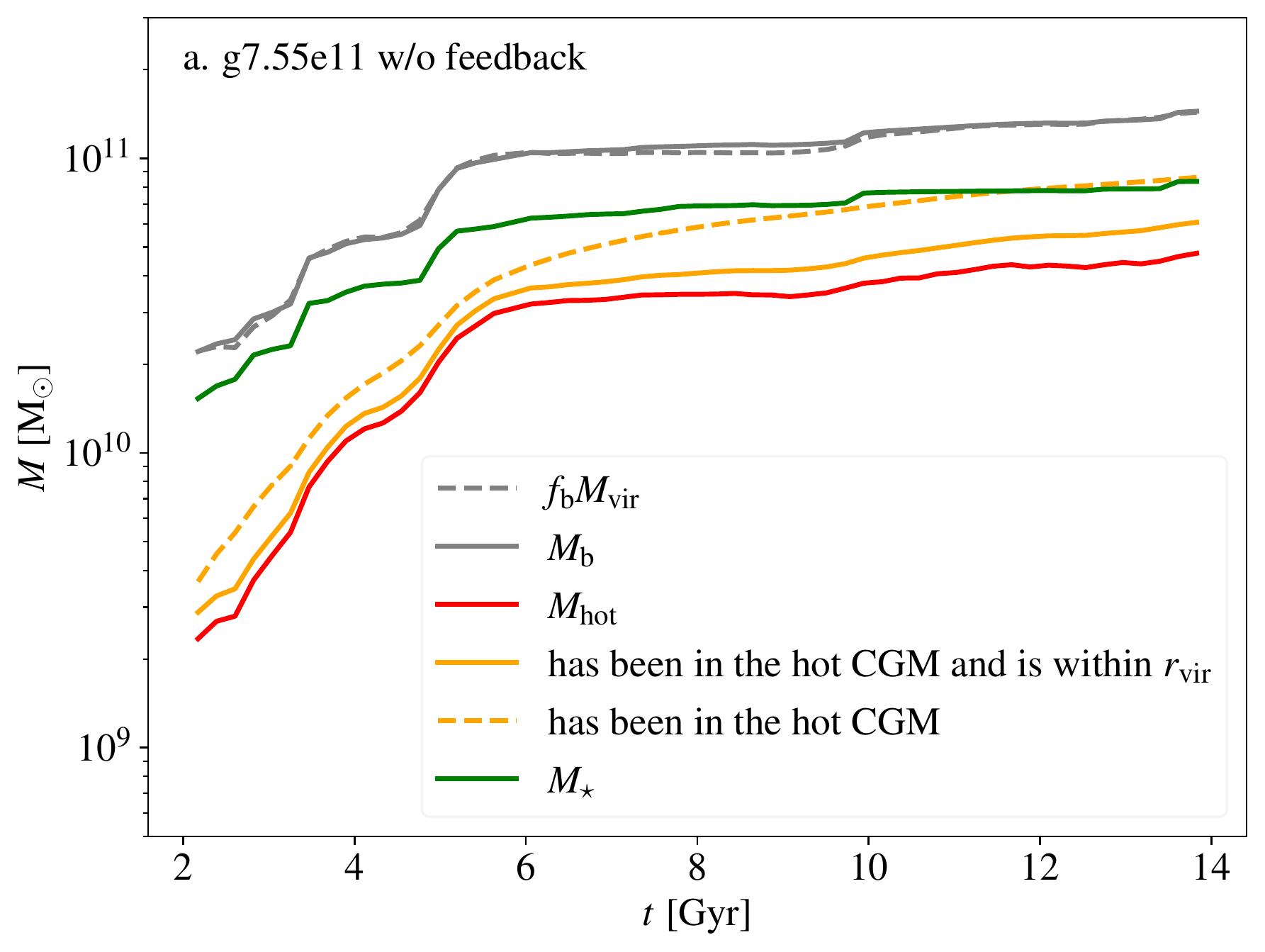} 
\includegraphics[width=0.45\hsize]{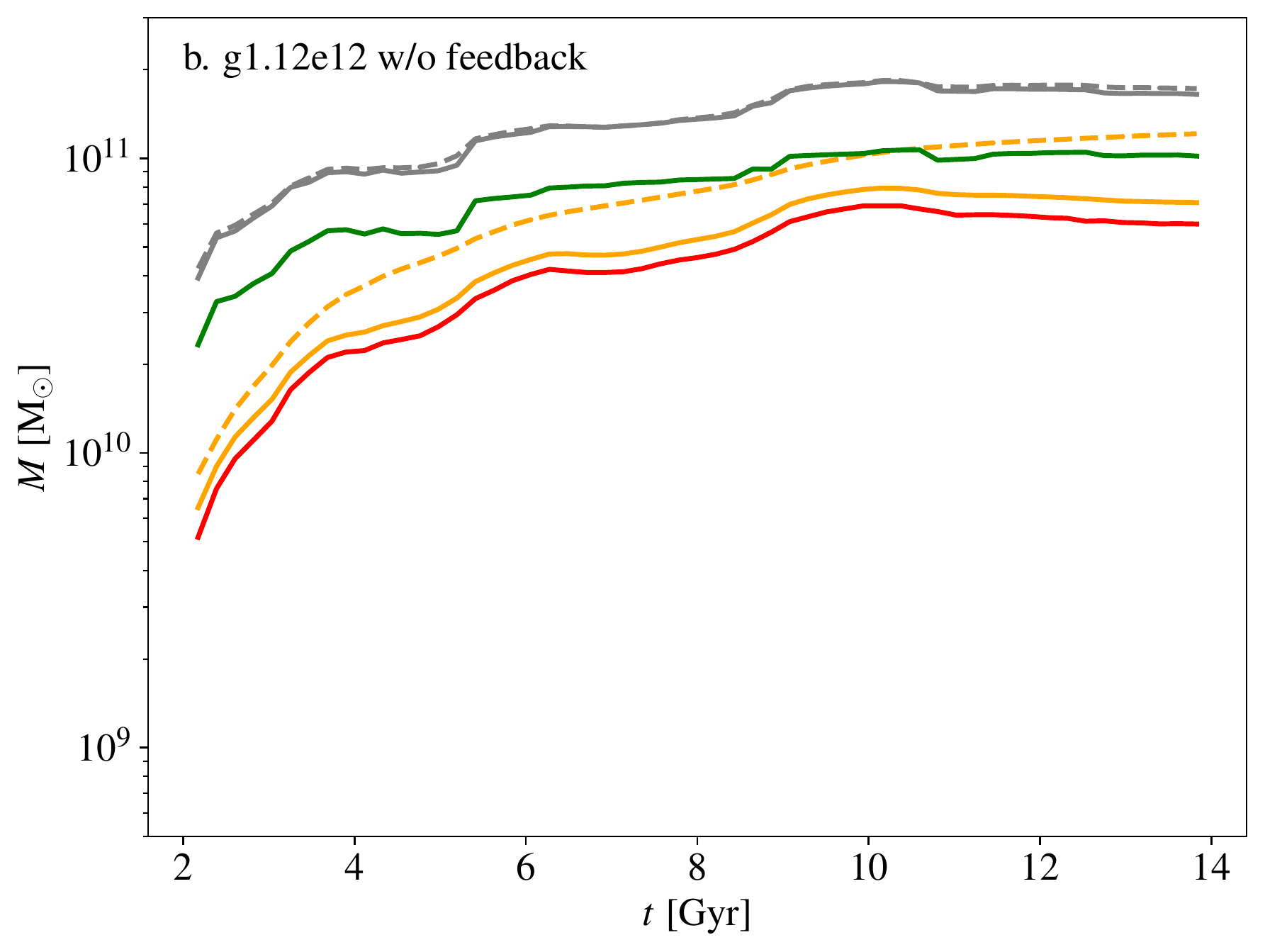} \\
\includegraphics[width=0.45\hsize]{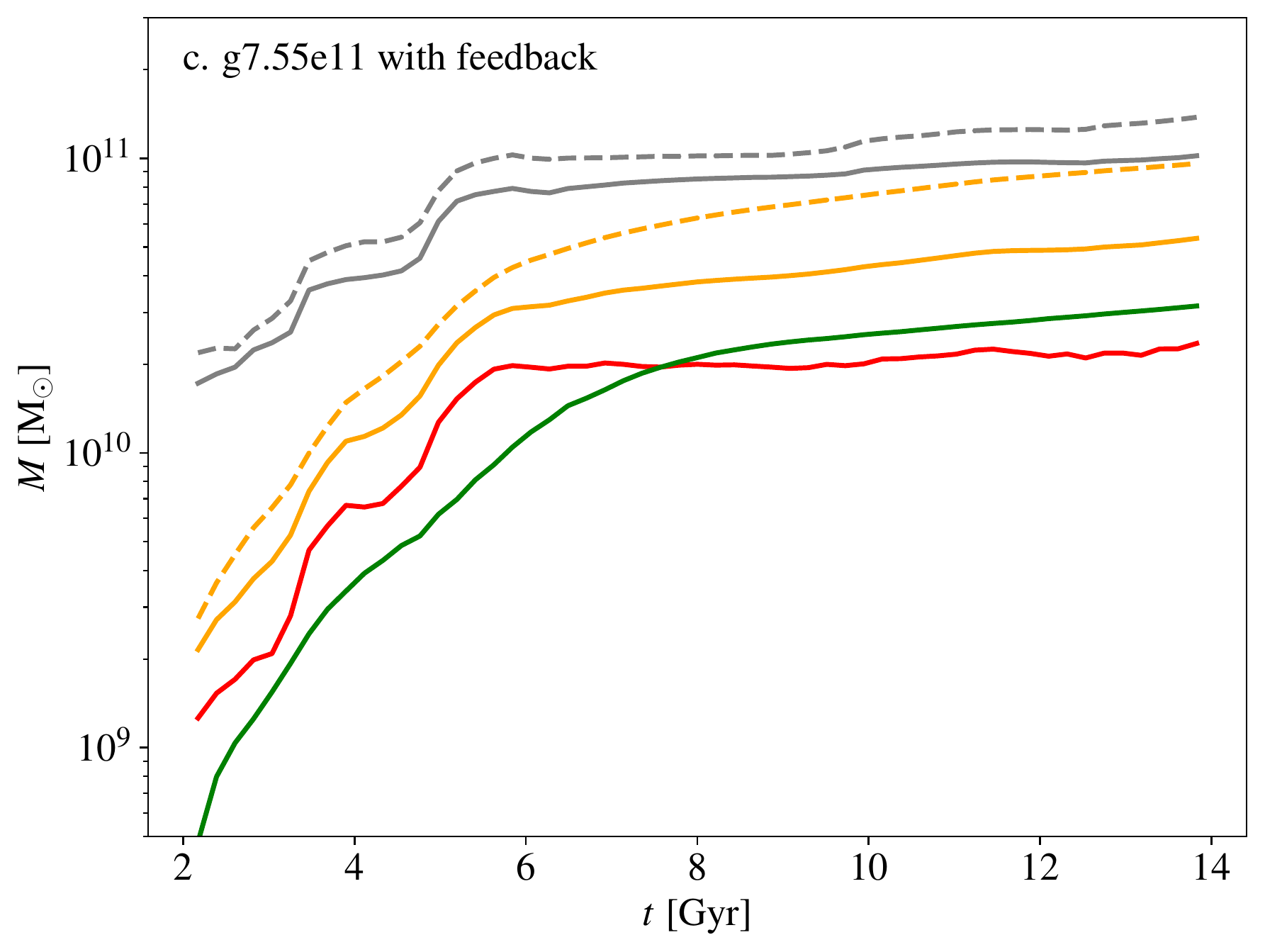}
\includegraphics[width=0.45\hsize]{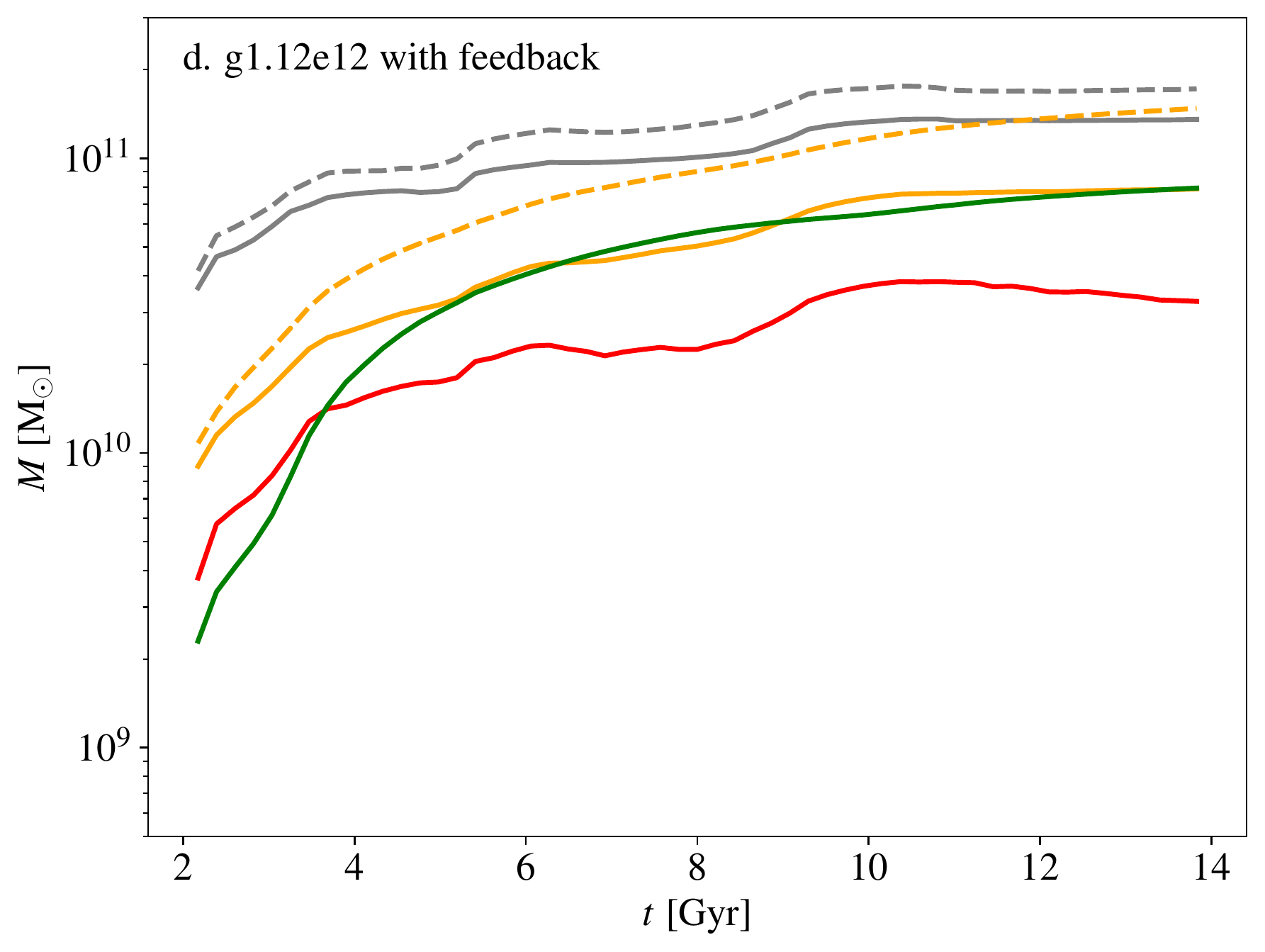}\\
\end{array}$
\end{center}
\caption{The gray curves compare the total baryonic mass $M_{\rm b}$ within the virial radius (solid curves) with the virial mass rescaled by the universal baryon fraction $f_{\rm b}=0.15$ (dashed curves), both
as a function of the cosmic time $t$.
The green curves show the stellar mass $M_\star$ within $r_{\rm vir}$, almost all of which is in the central galaxy.
The red curves  show the mass $M_{\rm hot}$ in the hot CGM.
The orange dashed curves show the total mass of all gas particles that have been in the hot CGM. The orange solid curves exclude the particles that are no longer within $r_{\rm vir}$ at the time $t$.
The panels to the left are for g7.55e11. Those to the right are for g1.12e12. The upper and lower panels show the simulations without and with feedback, respectively.}
\label{massgrowth}
\end{figure*}

\begin{figure*}
\begin{center}$
\begin{array}{cccc}
  \includegraphics[width=0.25\hsize]{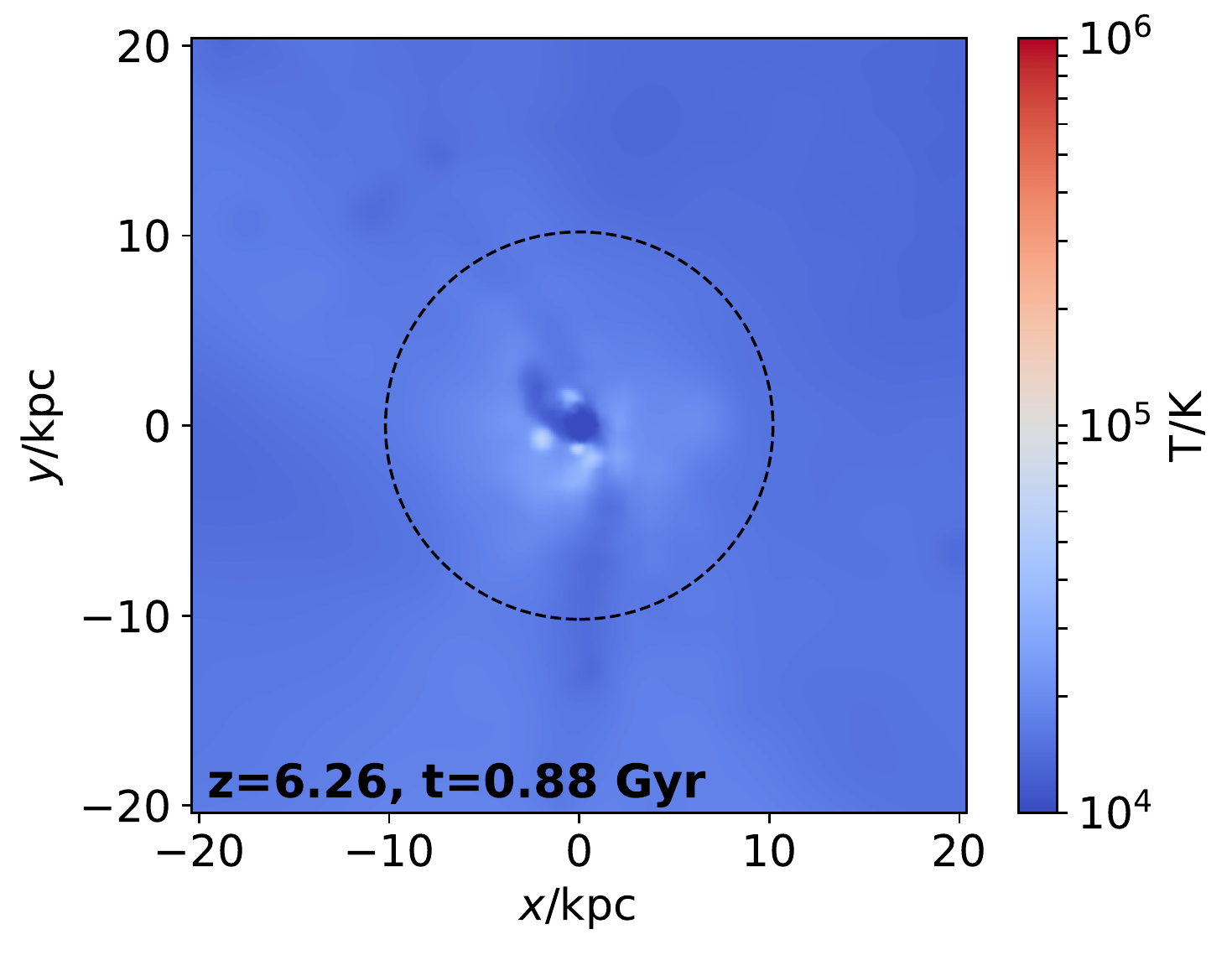} 
  \includegraphics[width=0.25\hsize]{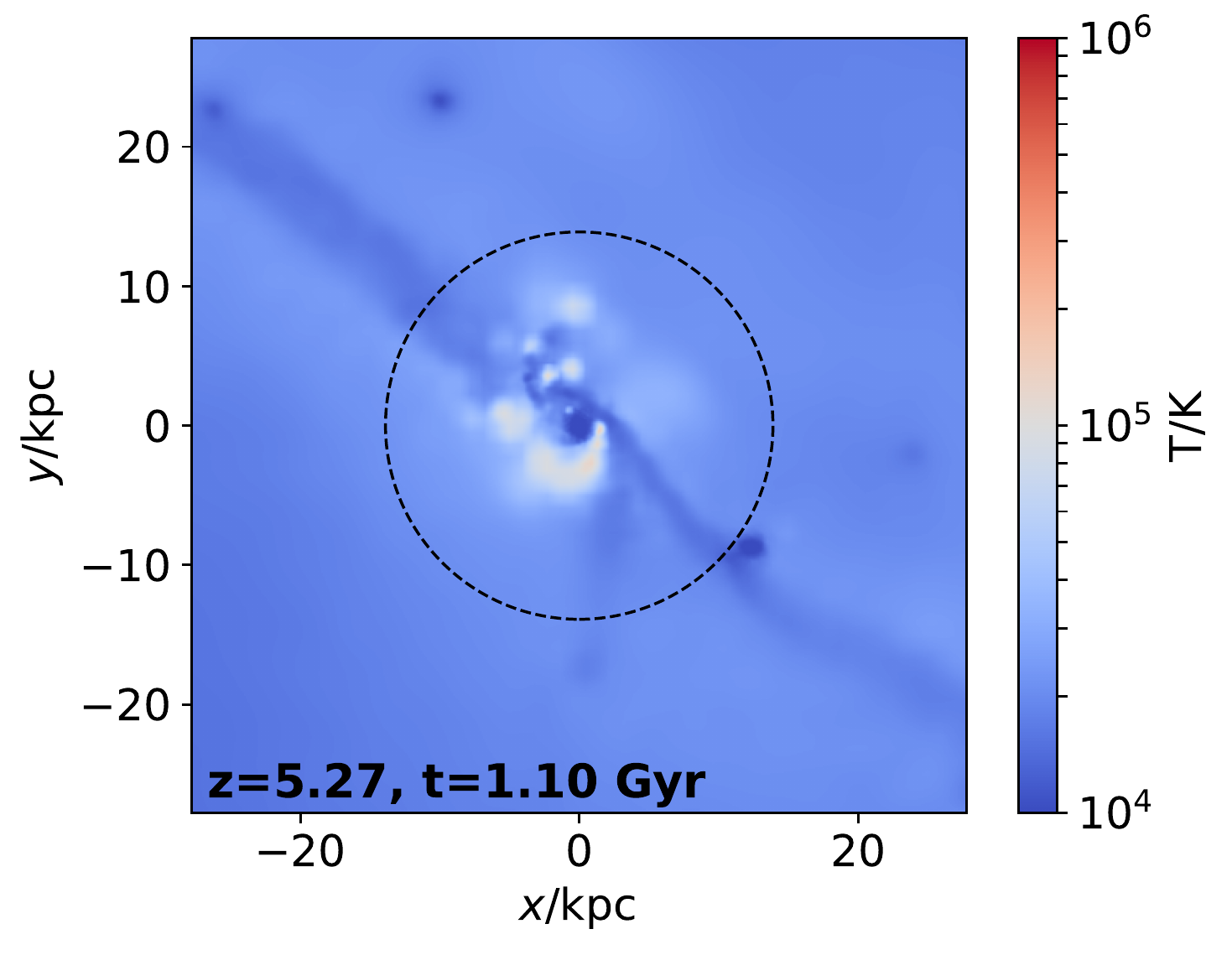}
   \includegraphics[width=0.25\hsize]{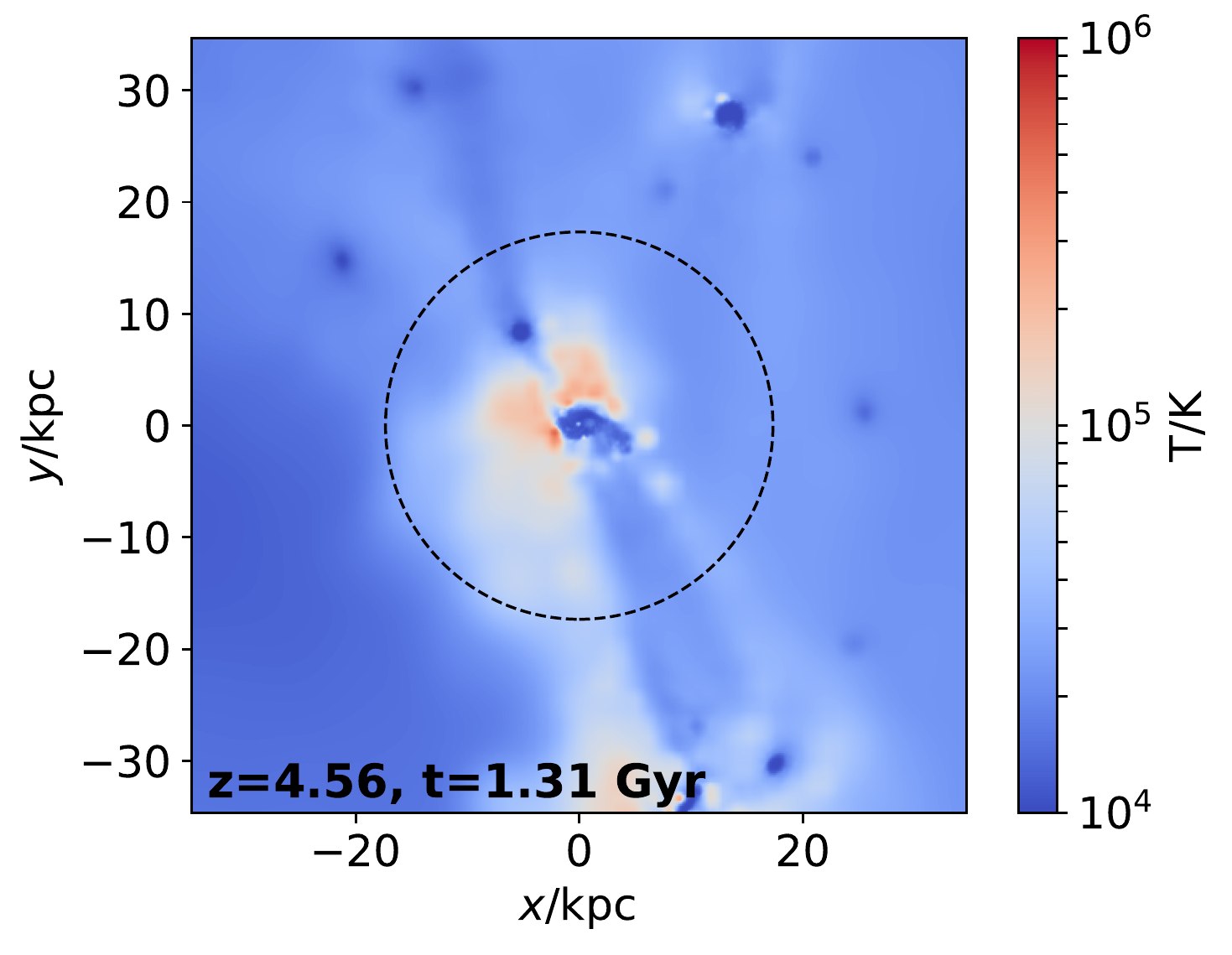}
  \includegraphics[width=0.25\hsize]{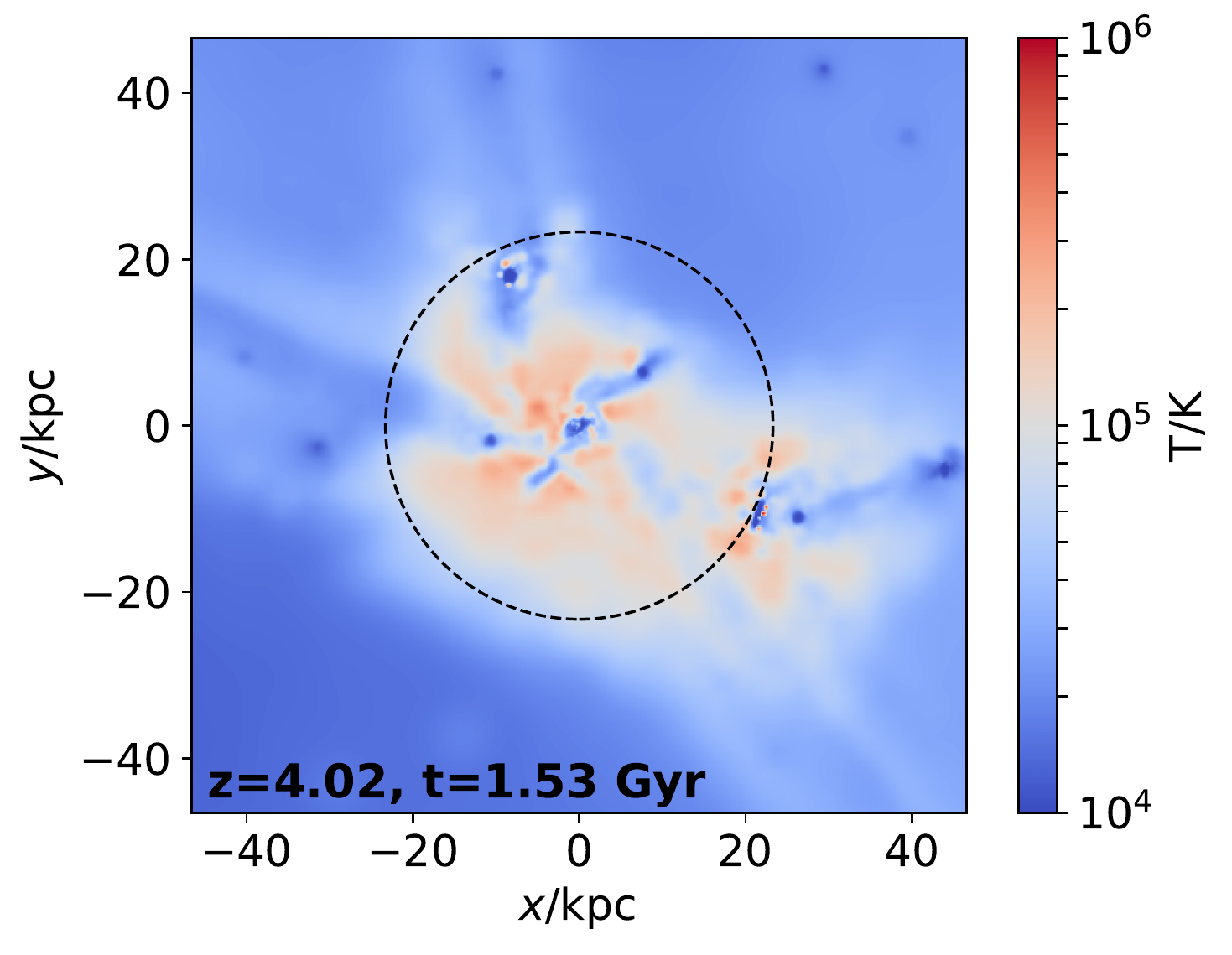}\\
  \includegraphics[width=0.25\hsize]{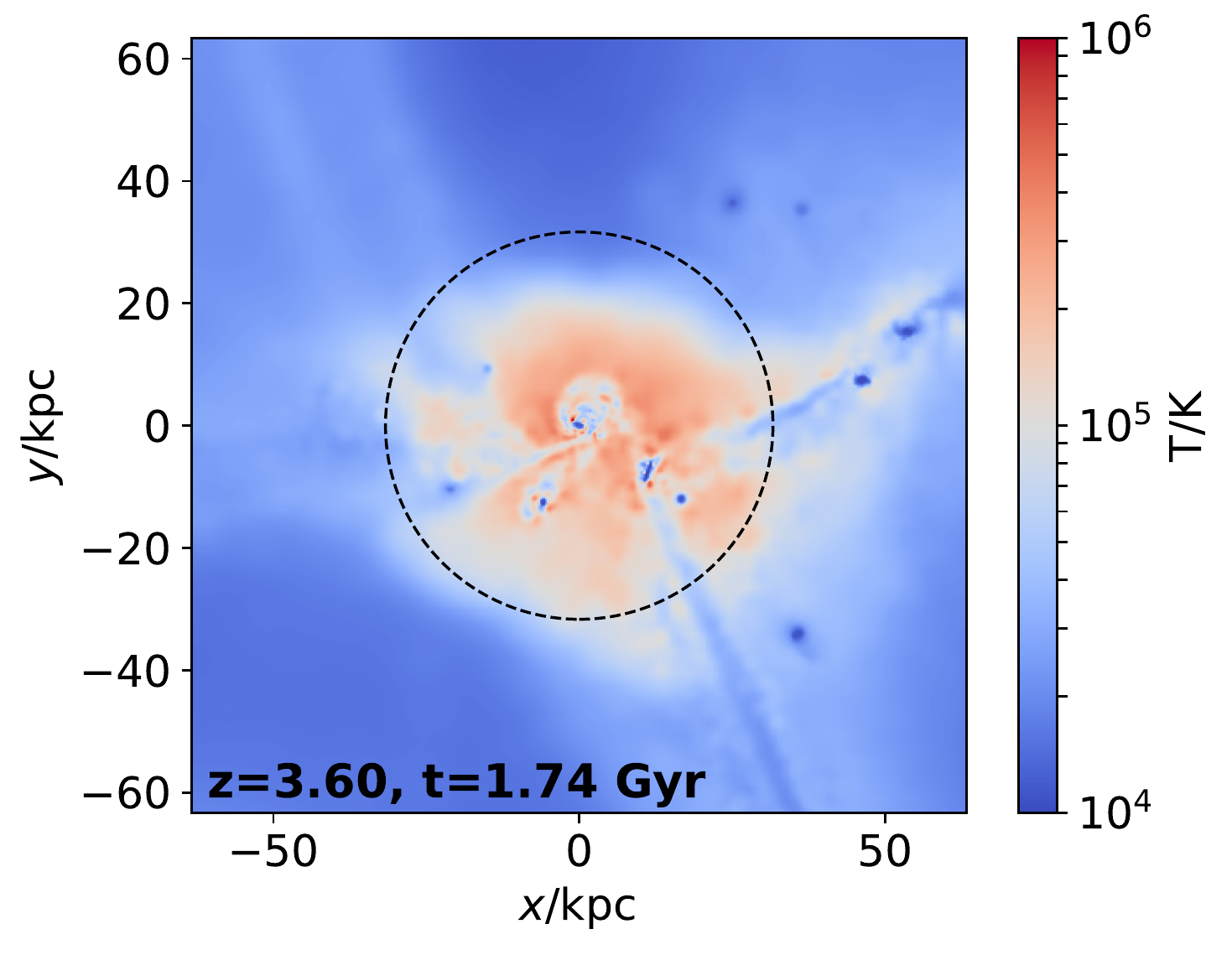}
  \includegraphics[width=0.25\hsize]{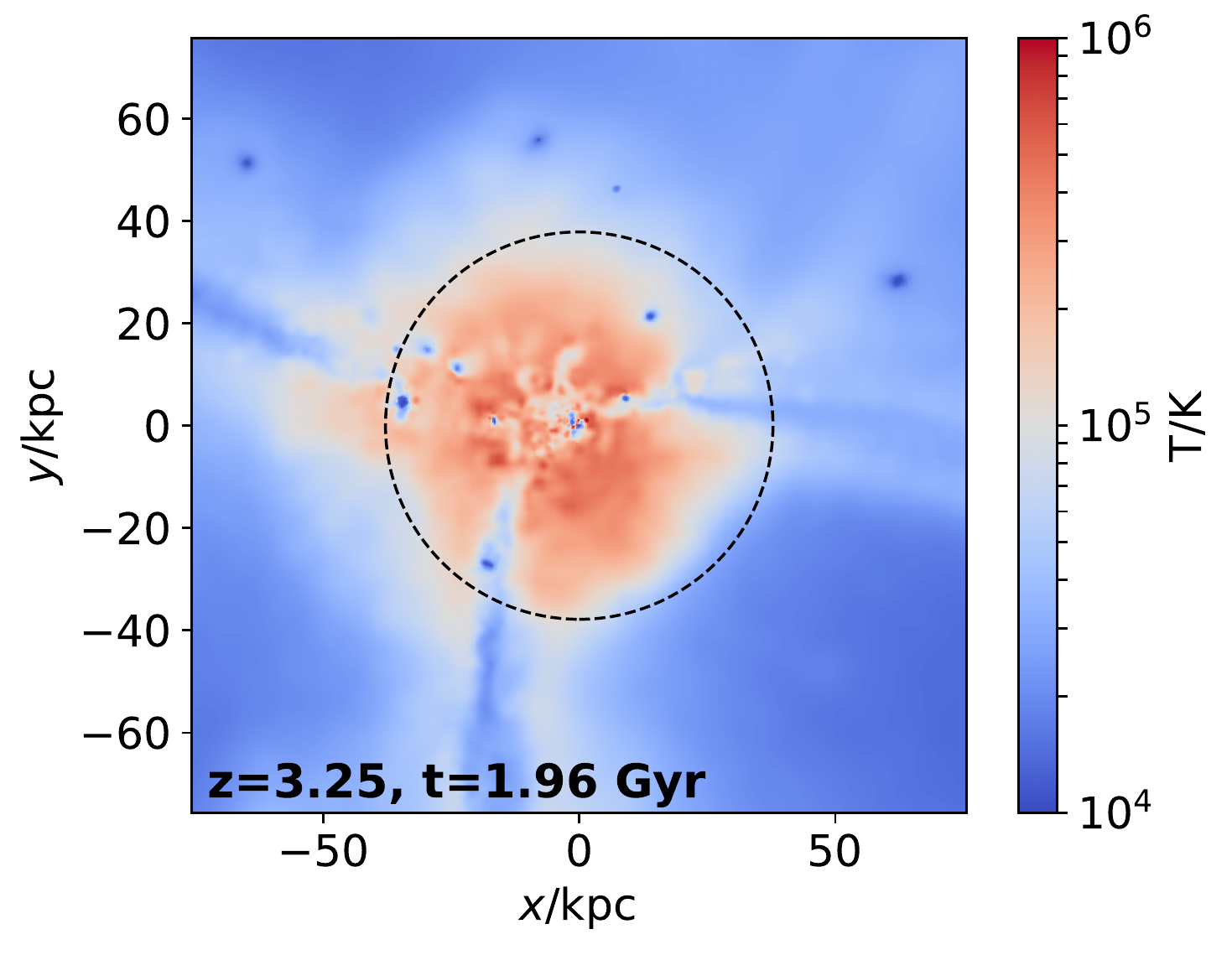}
    \includegraphics[width=0.25\hsize]{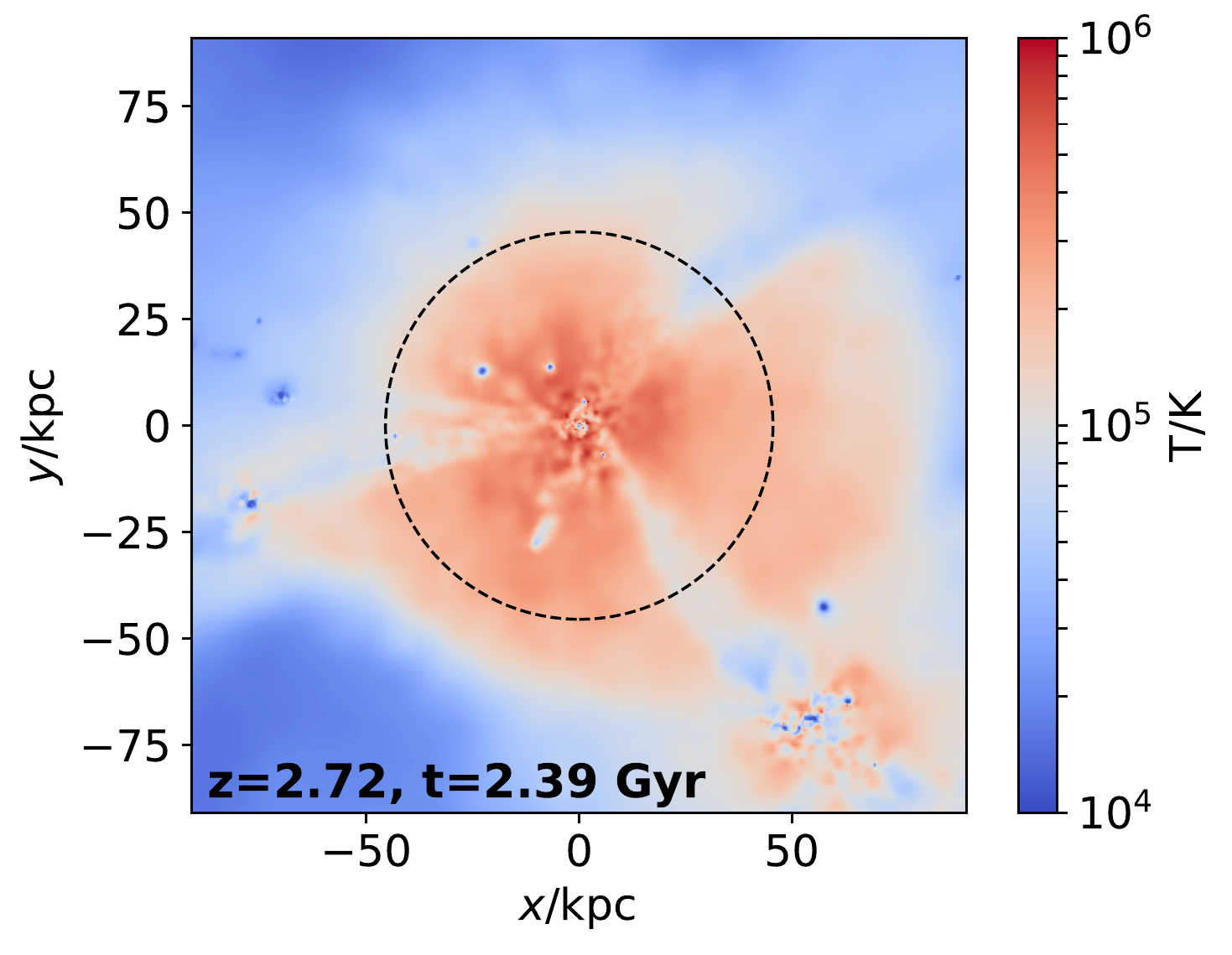}
  \includegraphics[width=0.25\hsize]{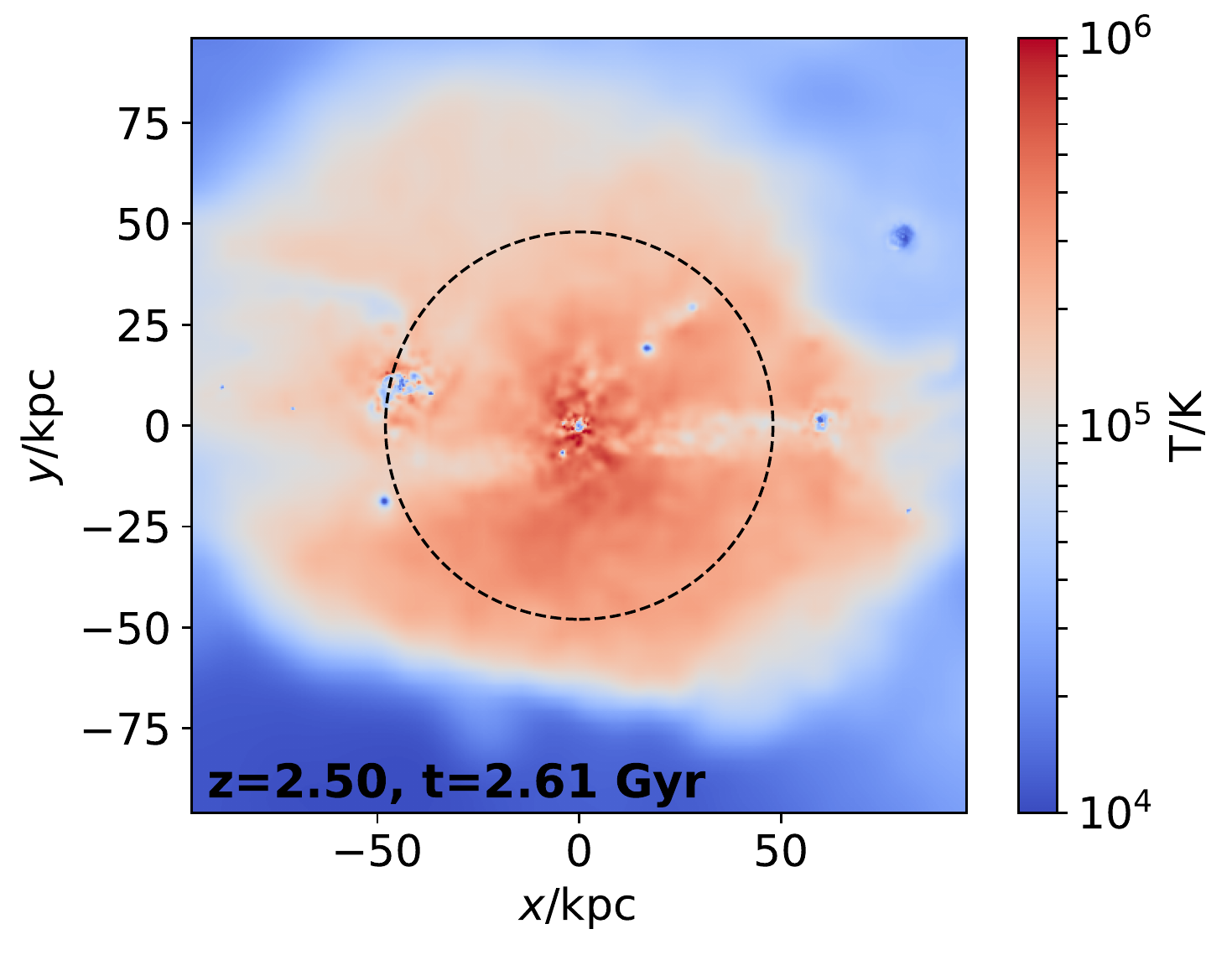}\\
  \includegraphics[width=0.25\hsize]{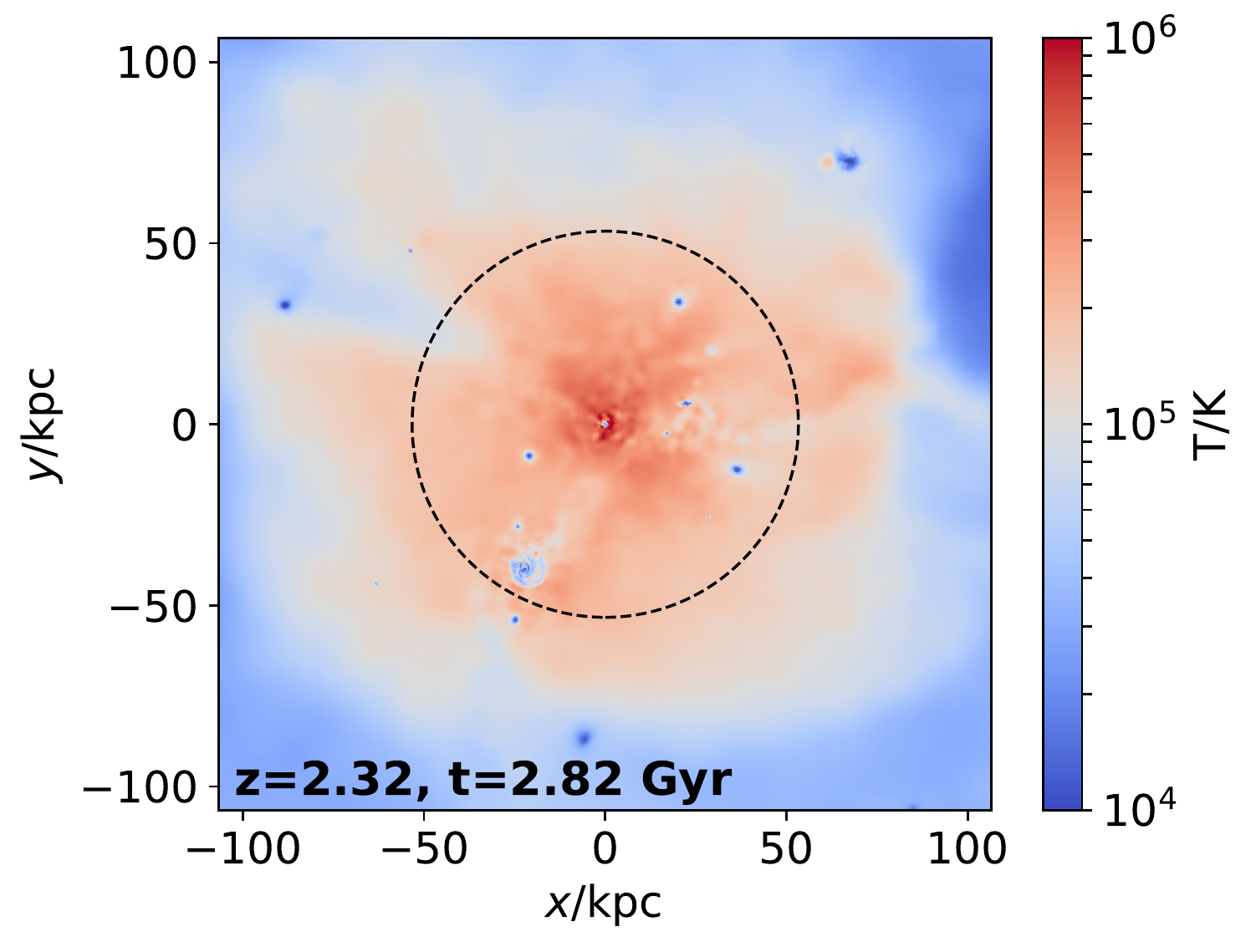}
  \includegraphics[width=0.25\hsize]{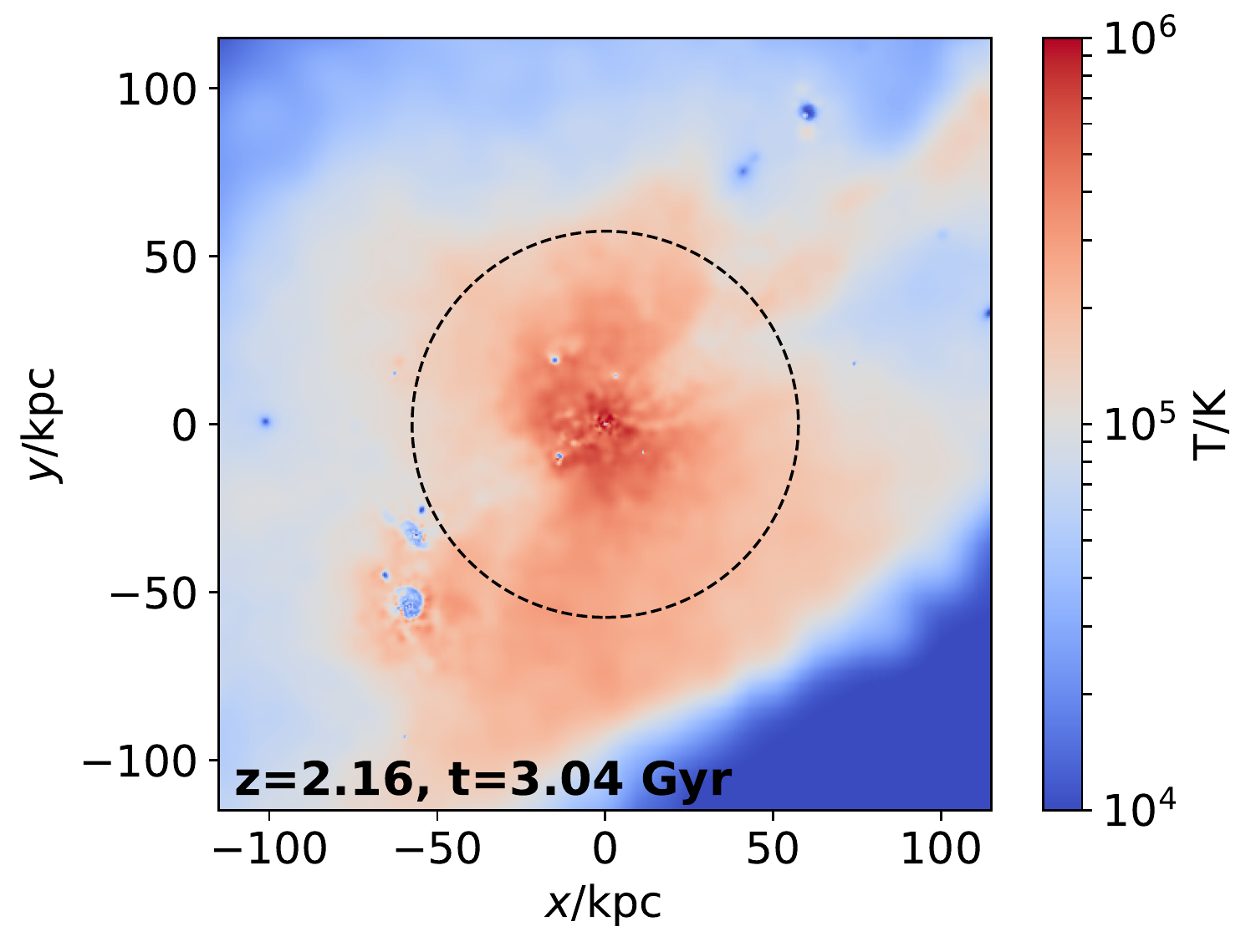}
  \includegraphics[width=0.25\hsize]{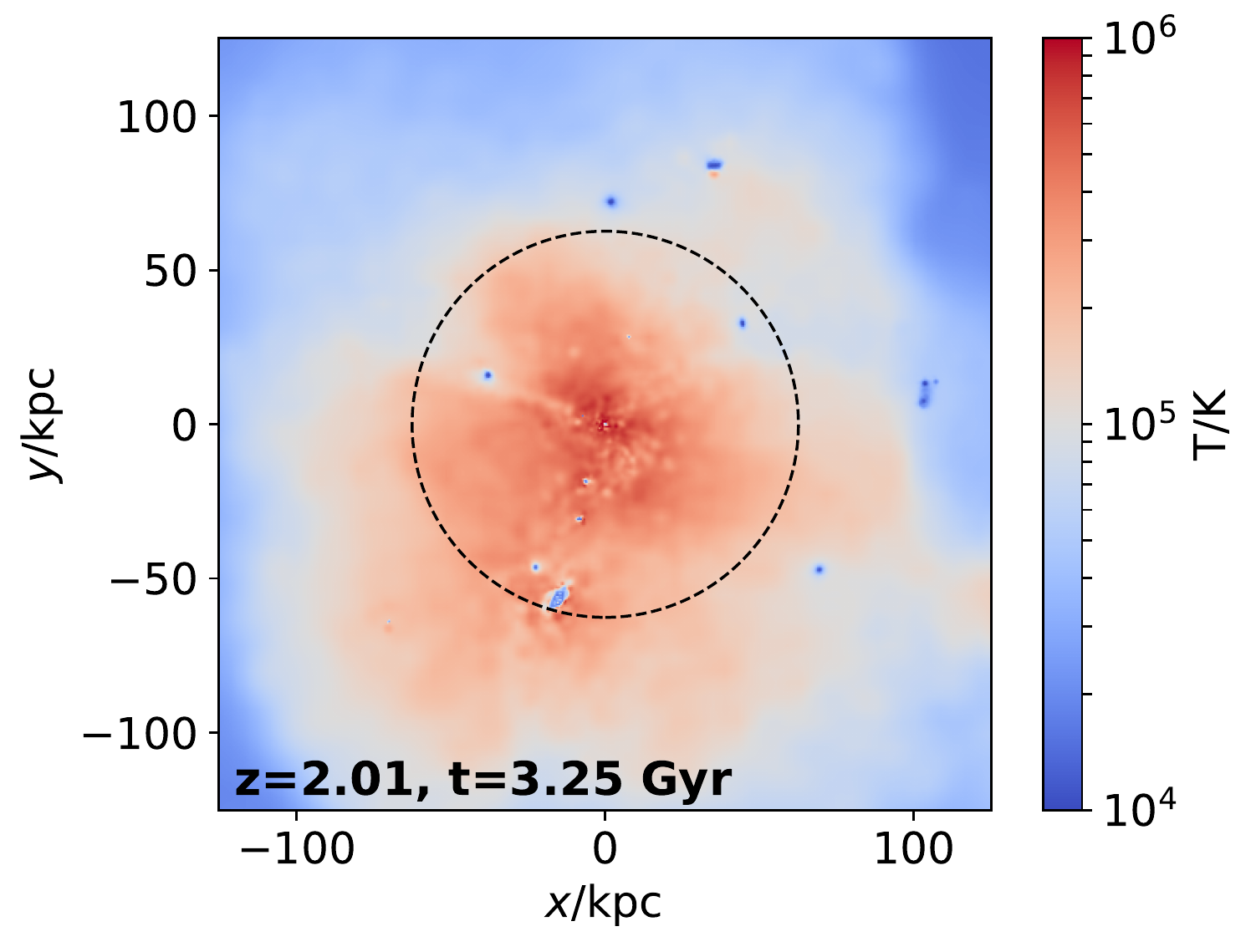}
  \includegraphics[width=0.25\hsize]{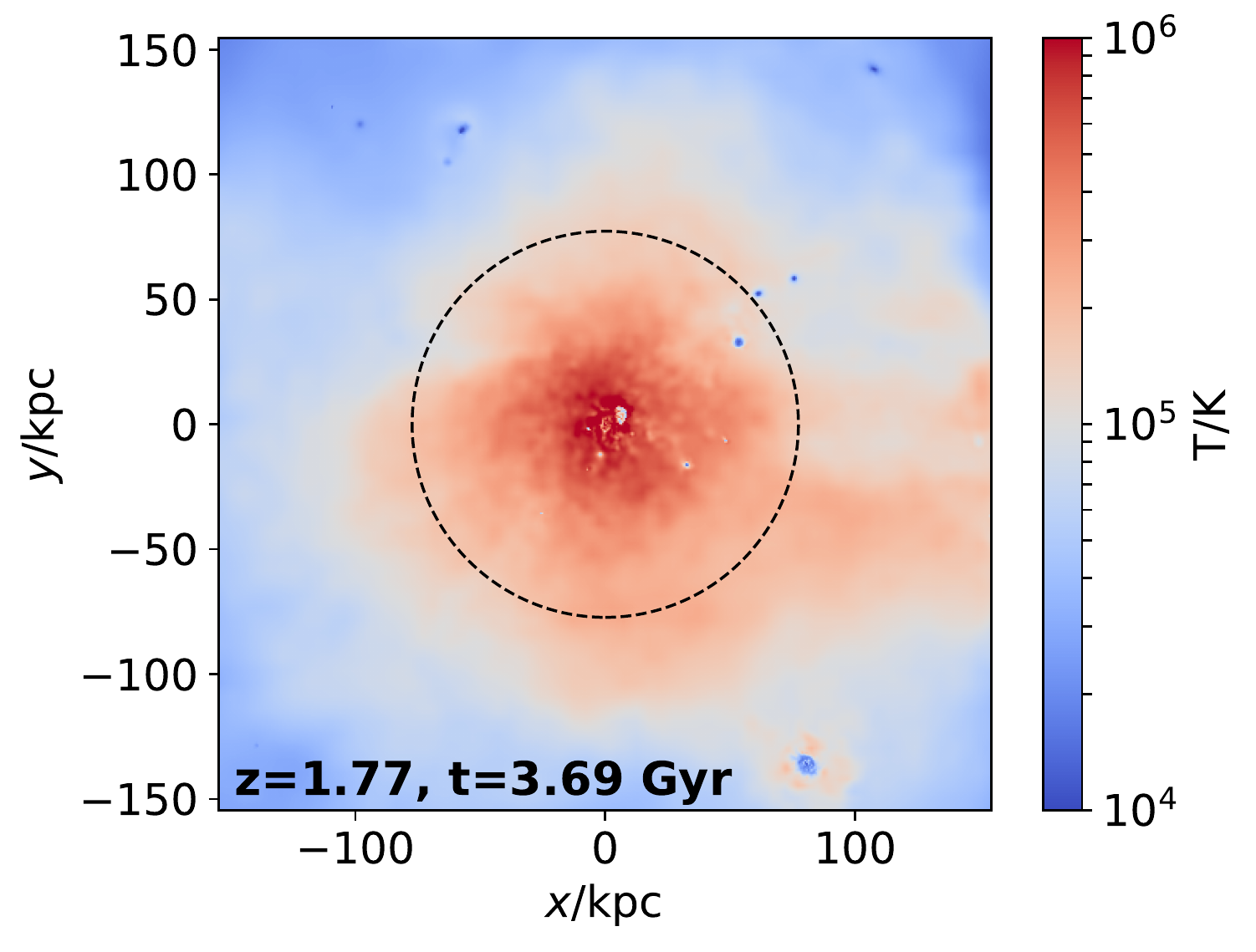}\\
  \includegraphics[width=0.25\hsize]{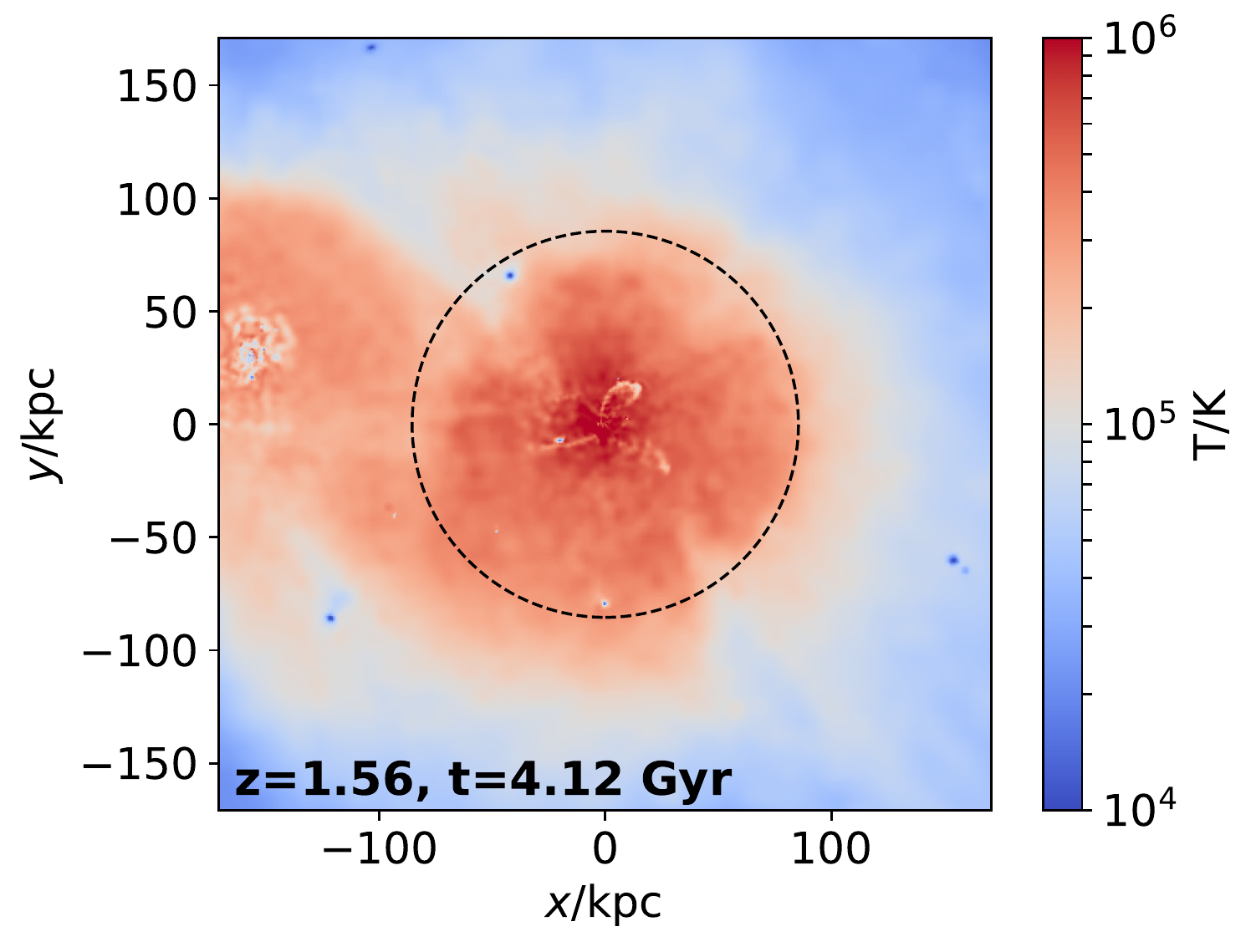}
  \includegraphics[width=0.25\hsize]{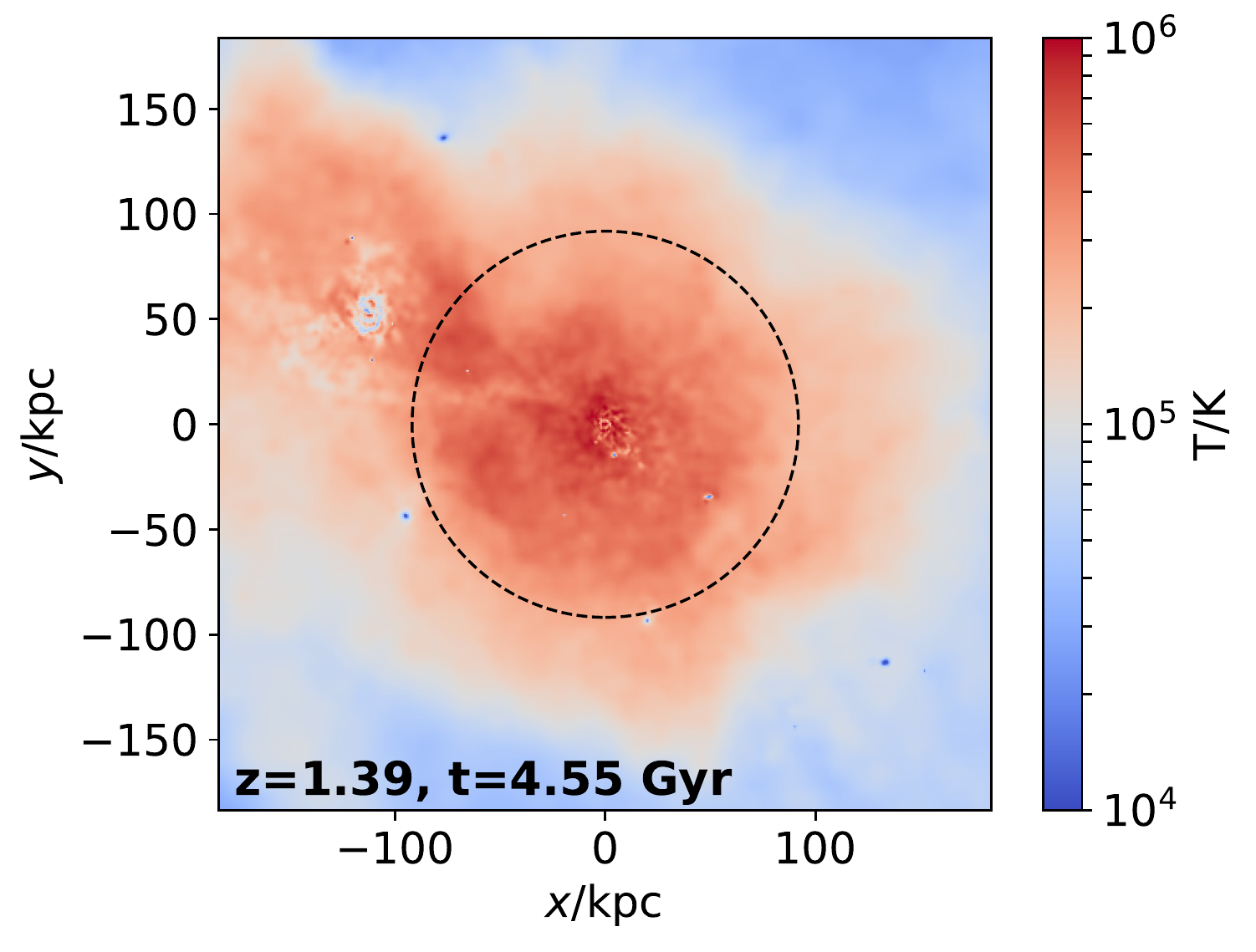}
  \includegraphics[width=0.25\hsize]{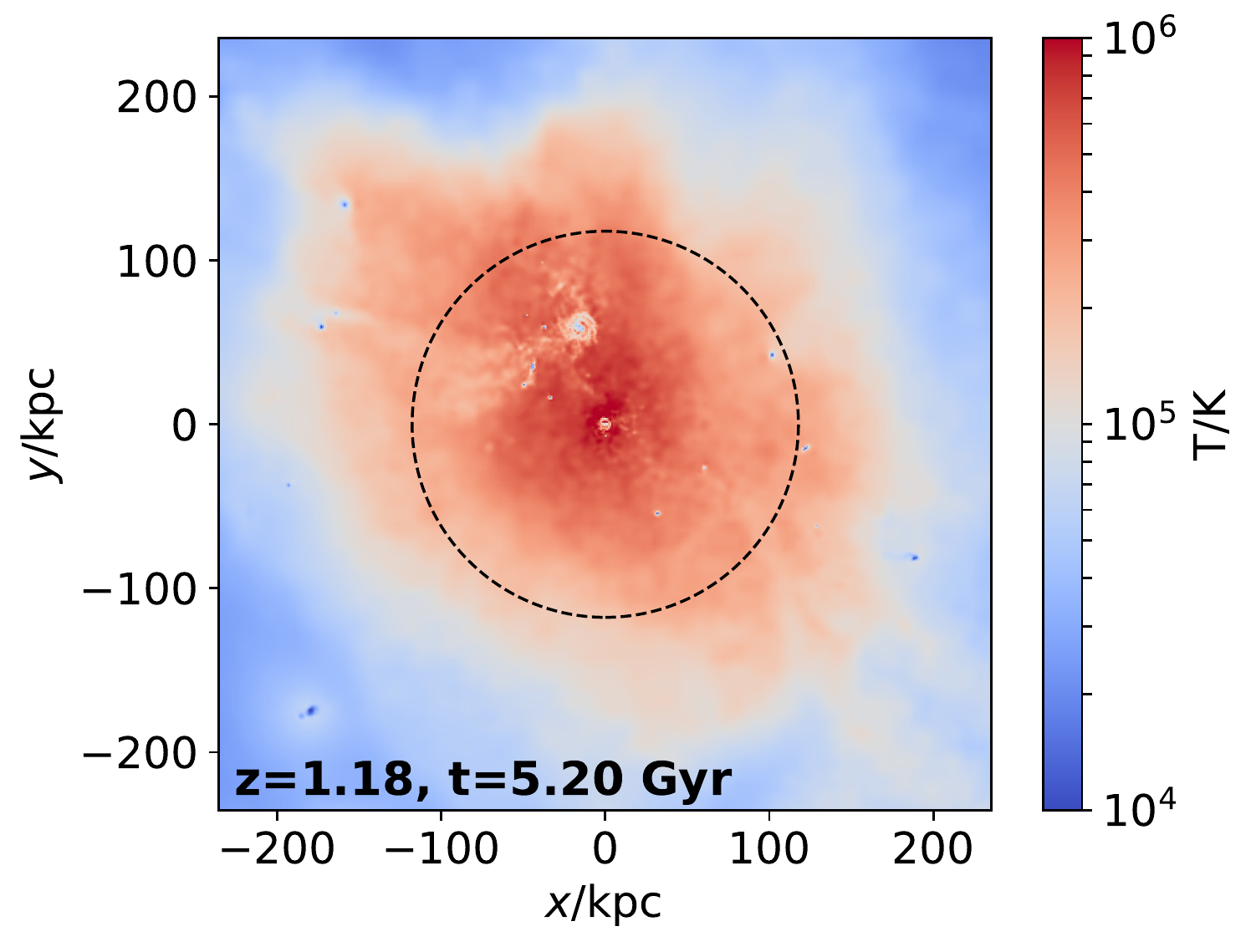}
  \includegraphics[width=0.25\hsize]{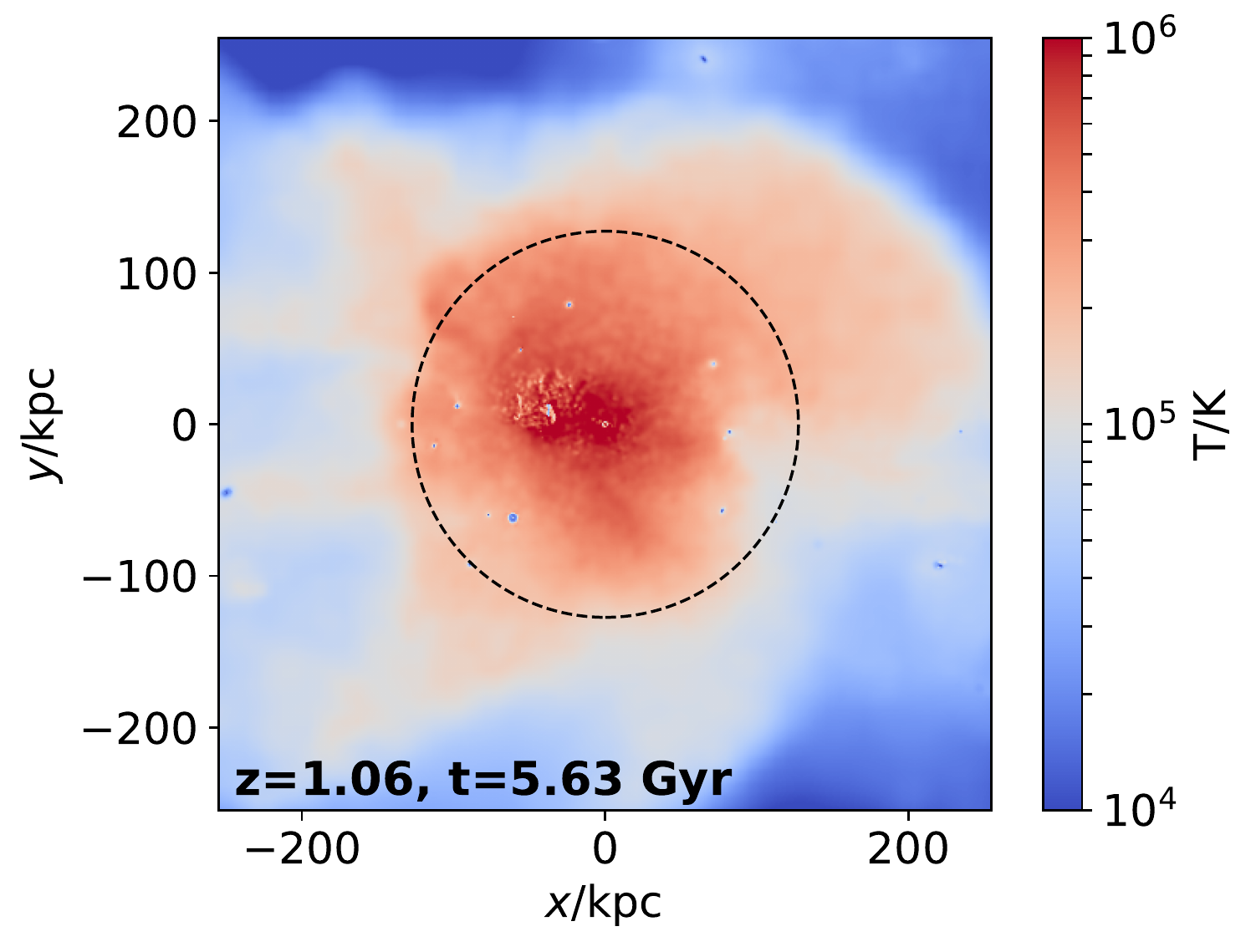}\\
  \includegraphics[width=0.25\hsize]{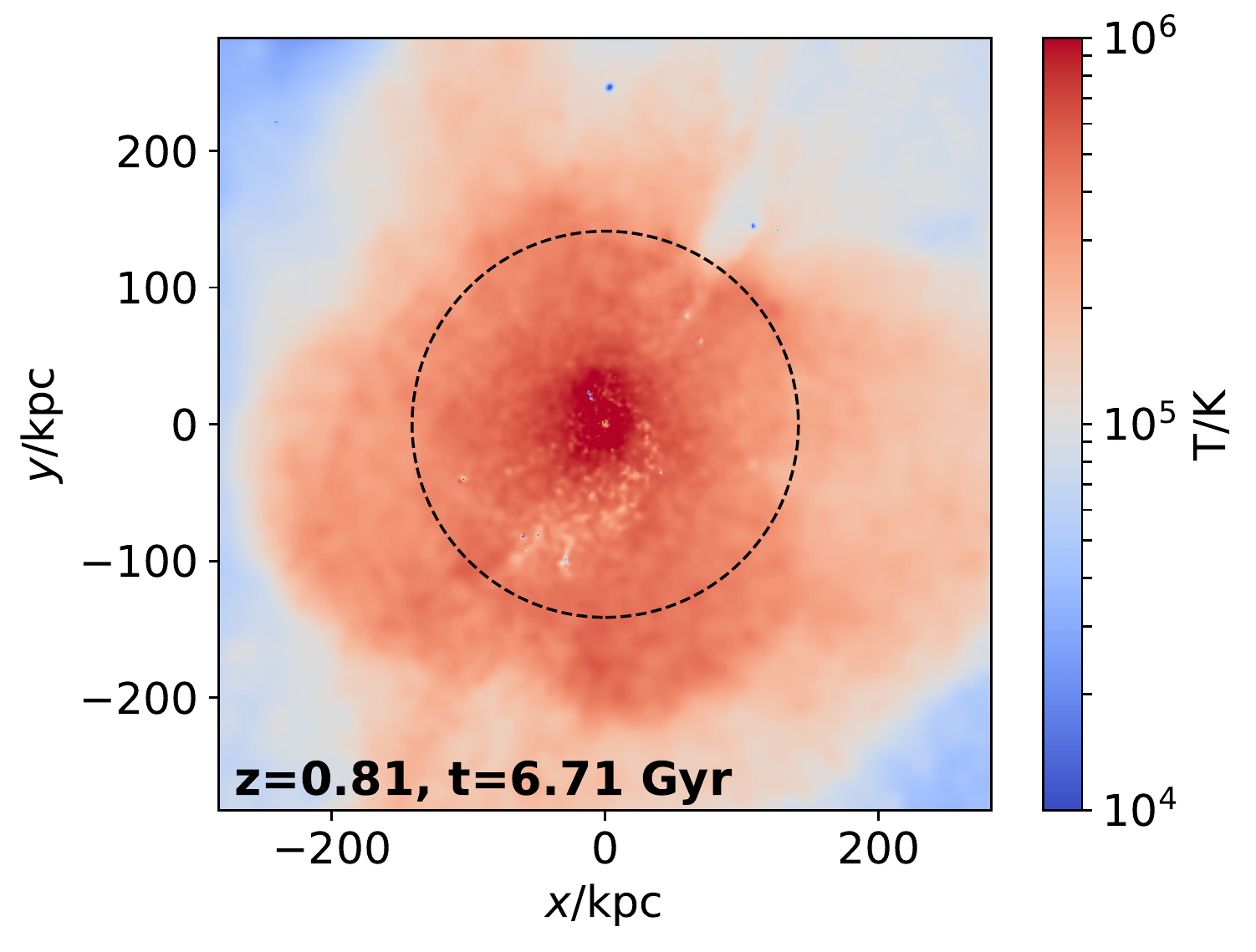}
  \includegraphics[width=0.25\hsize]{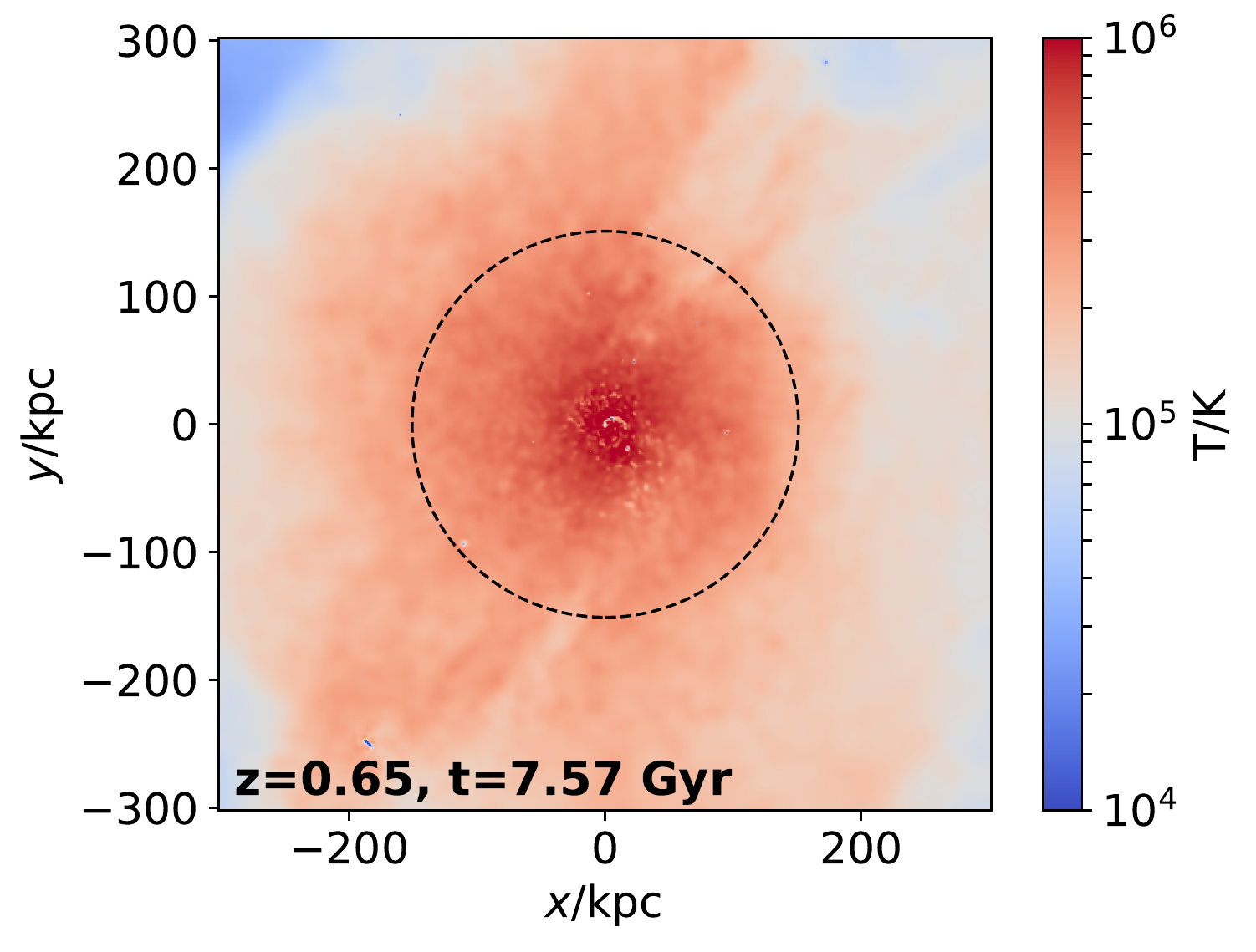}
  \includegraphics[width=0.25\hsize]{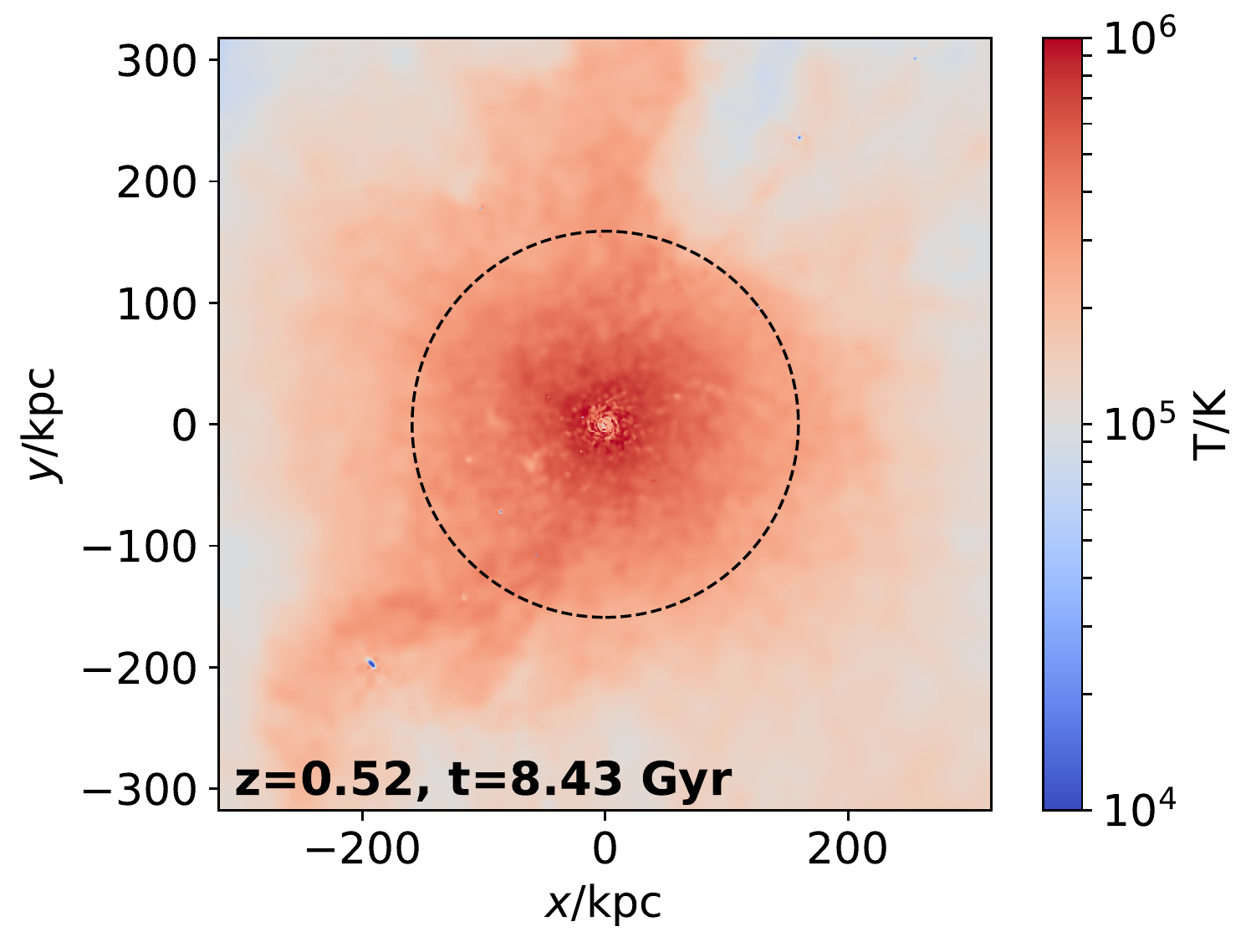}
    \includegraphics[width=0.25\hsize]{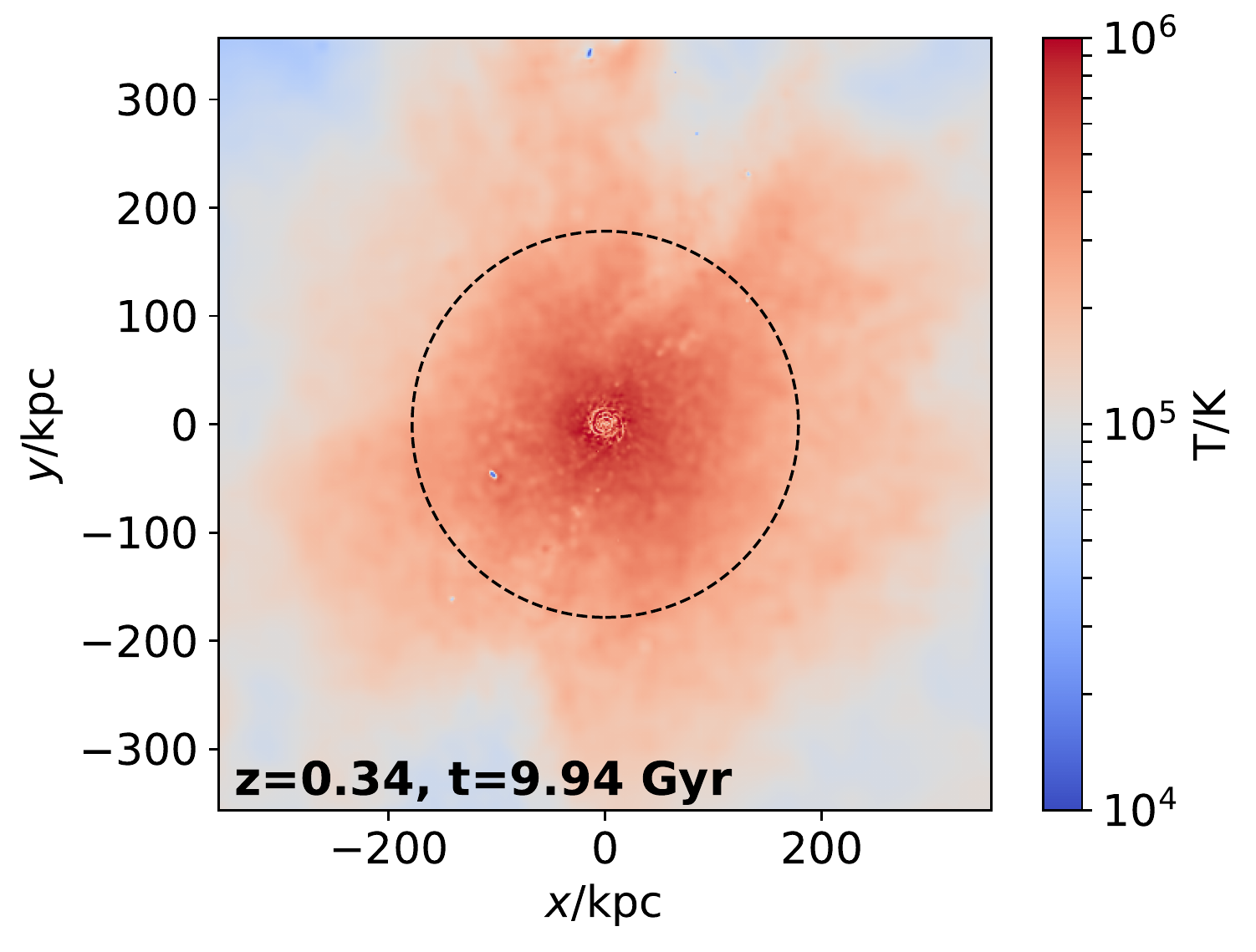}\\
  \includegraphics[width=0.25\hsize]{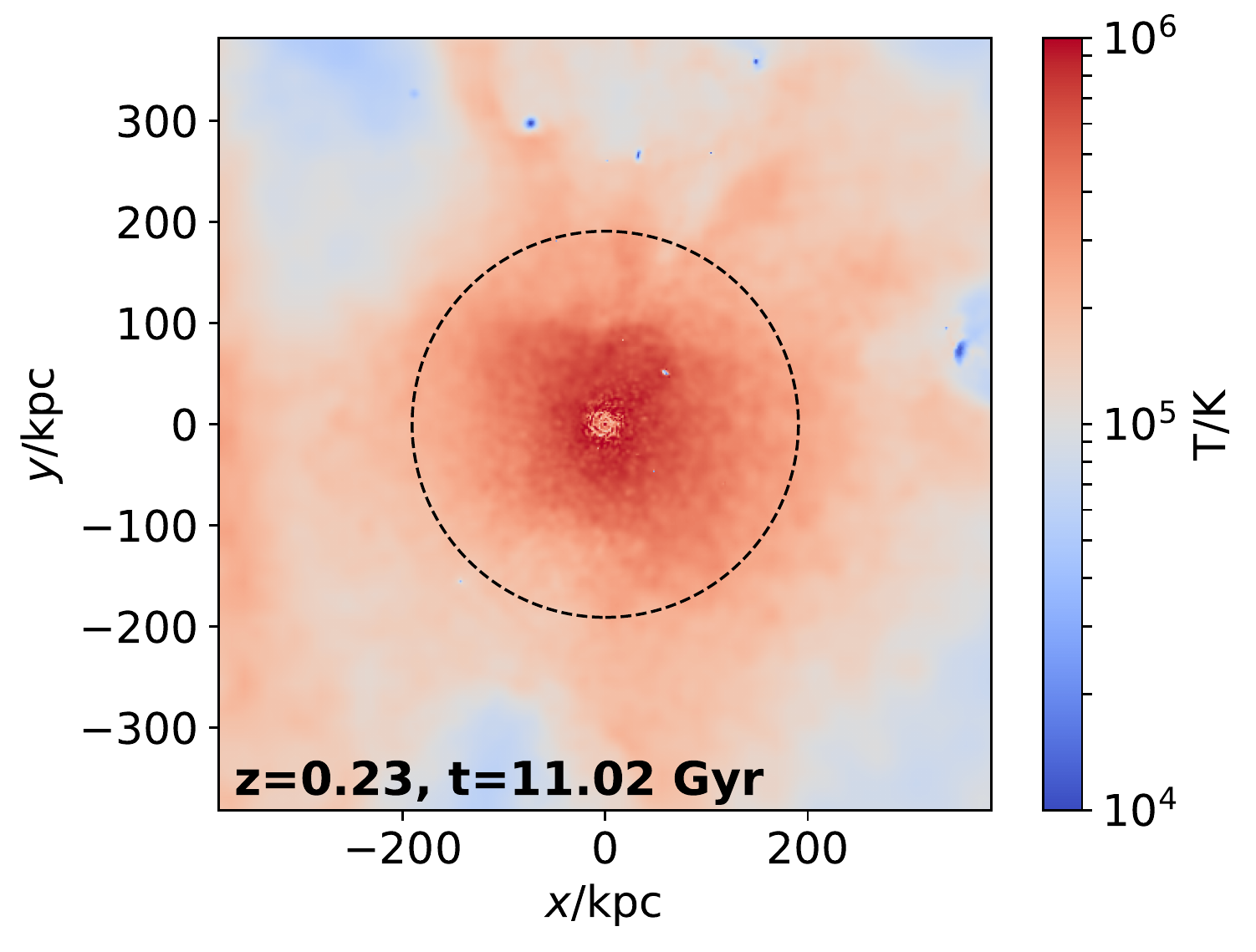}
  \includegraphics[width=0.25\hsize]{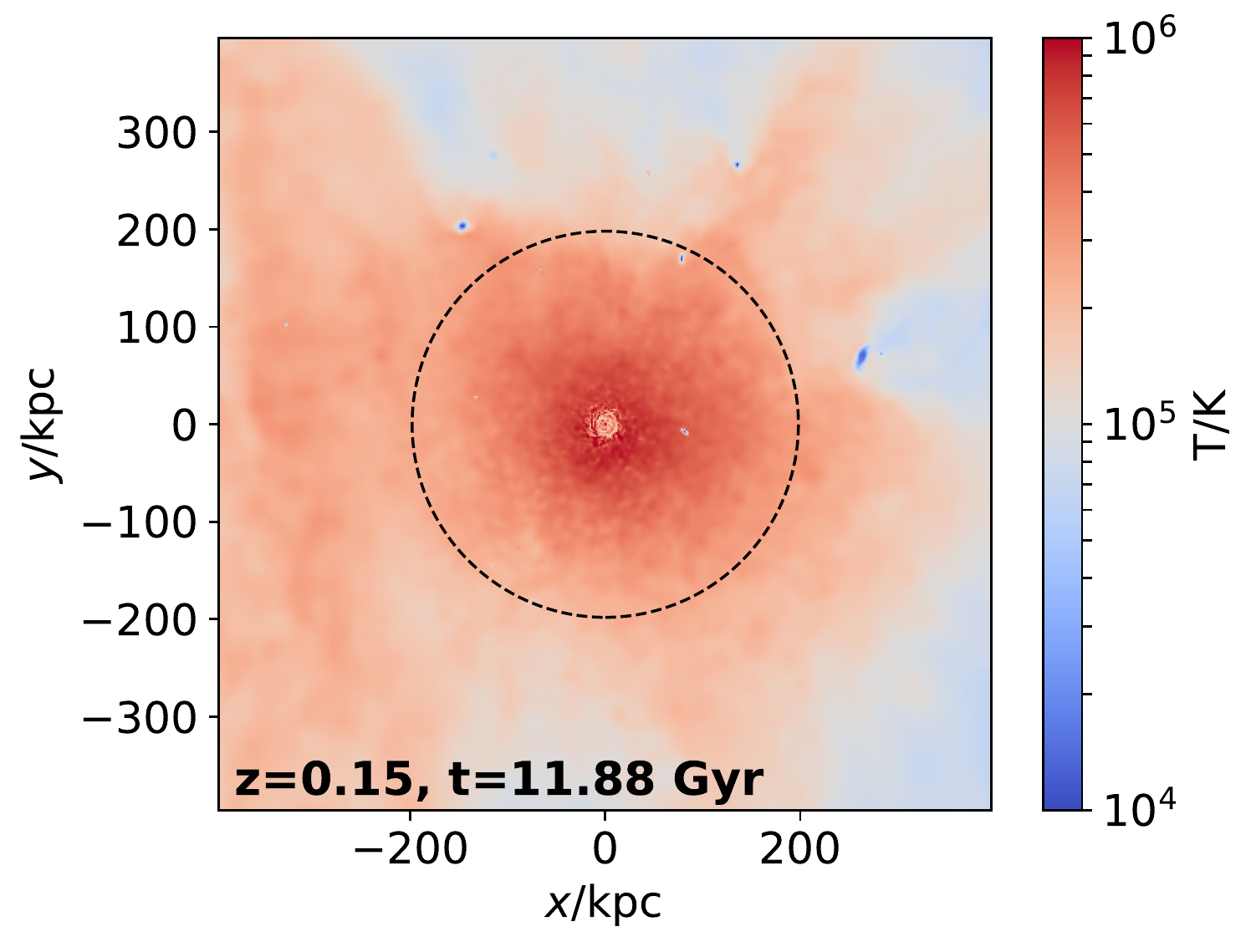}
   \includegraphics[width=0.25\hsize]{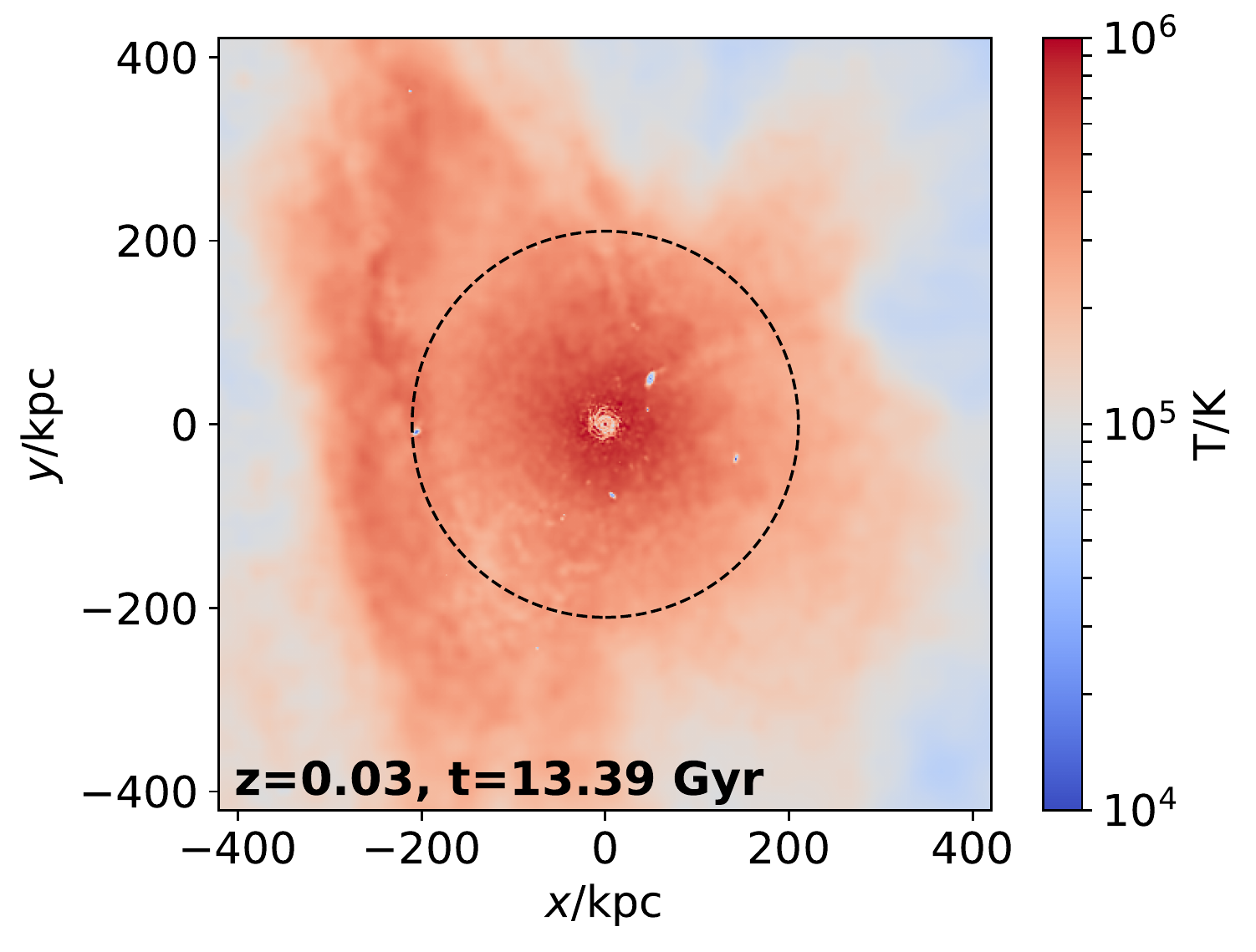}
   \includegraphics[width=0.25\hsize]{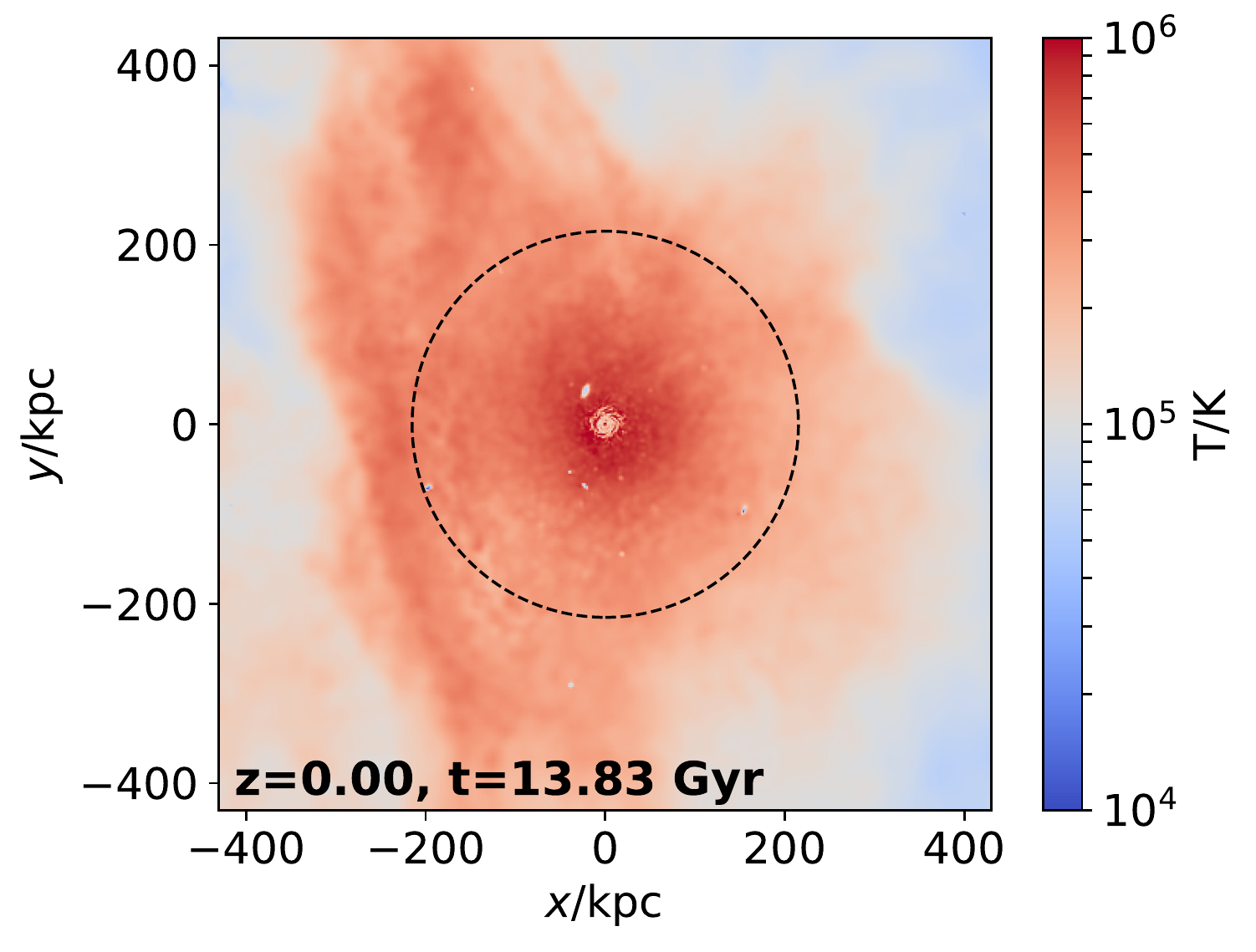}
\end{array}$
\end{center}
\caption{Evolution of the CGM {  of g7.55e11} from $z=6.26$ to $z=0$ {  in the simulation without feedback.} The images show temperature maps. They were taken looking at the central galaxy face-on (from a line of sight perpendicular to the plane of the disc).
The black dashed circles show the virial radius.}
\label{poststamp_images}
\end{figure*} 
\begin{figure*}
\begin{center}$
\begin{array}{cccc}
  \includegraphics[width=0.25\hsize]{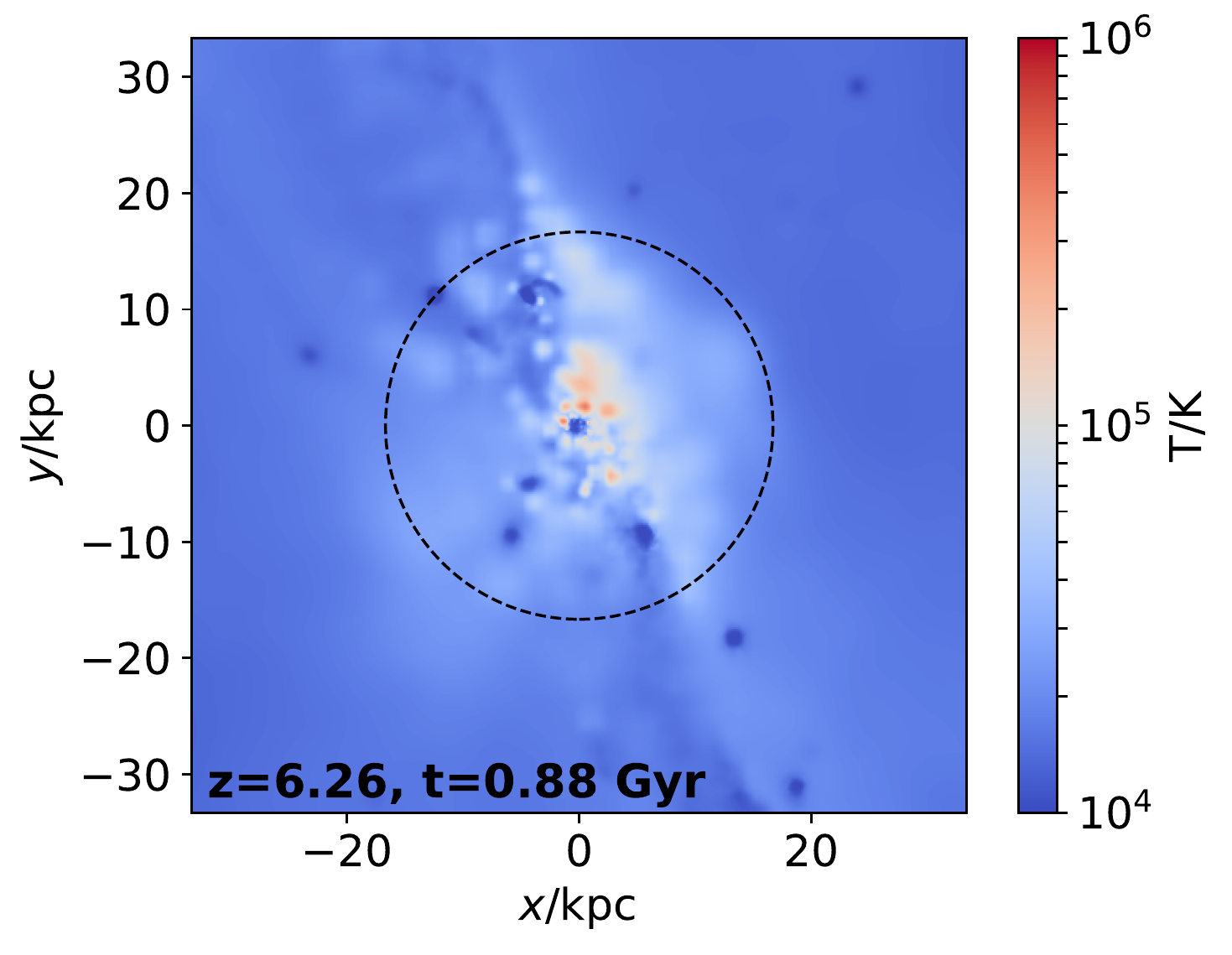} 
  \includegraphics[width=0.25\hsize]{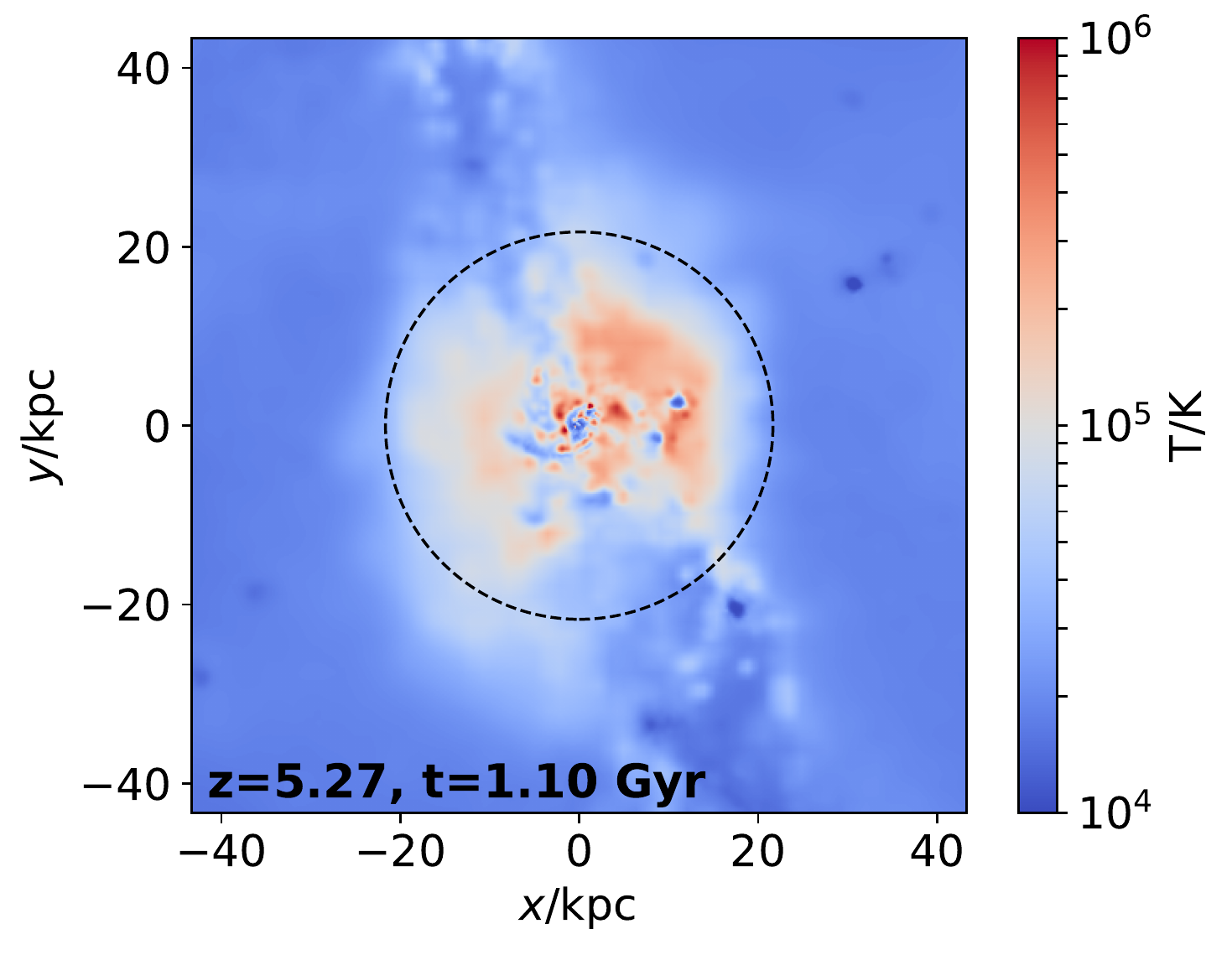}
  \includegraphics[width=0.25\hsize]{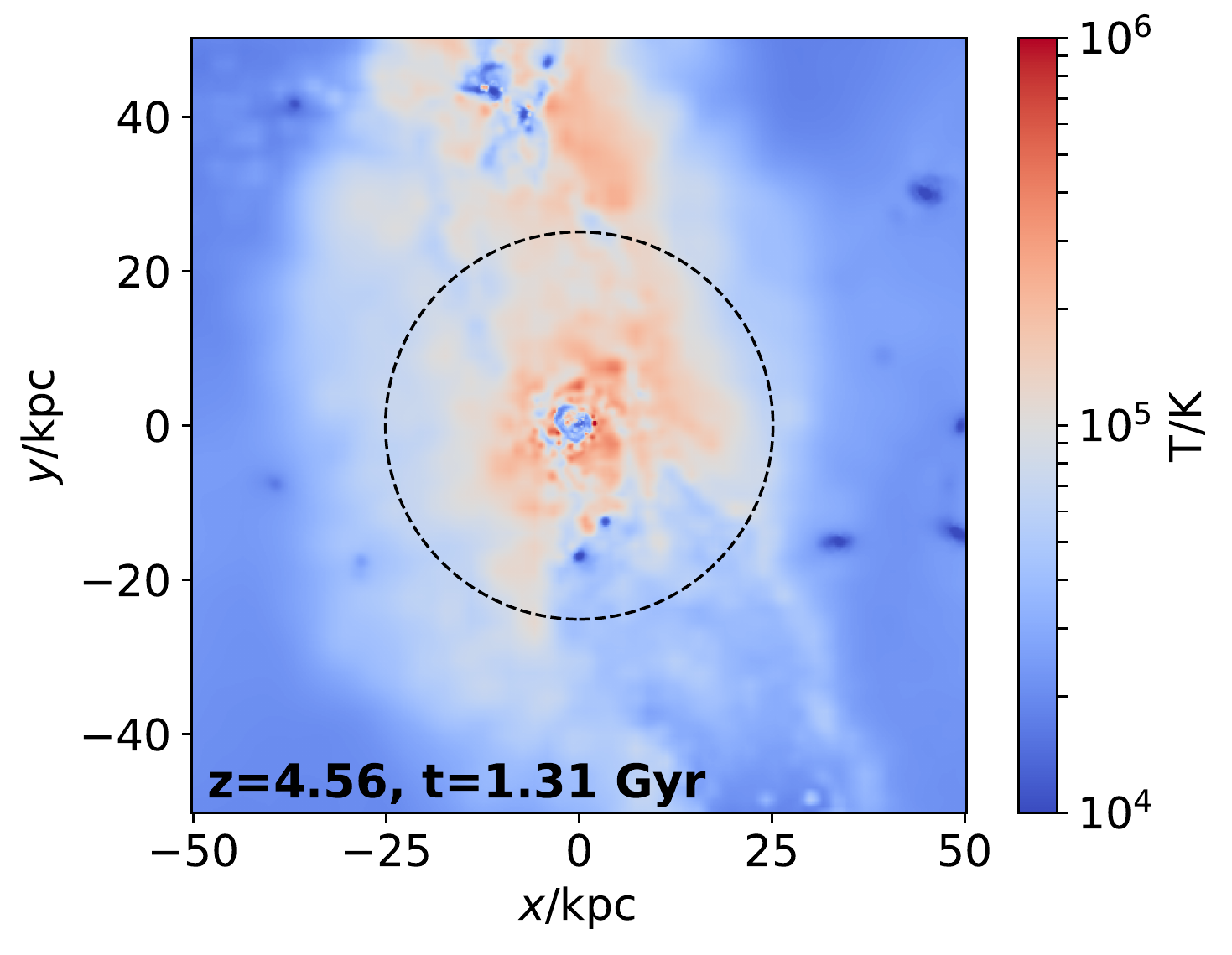}
  \includegraphics[width=0.25\hsize]{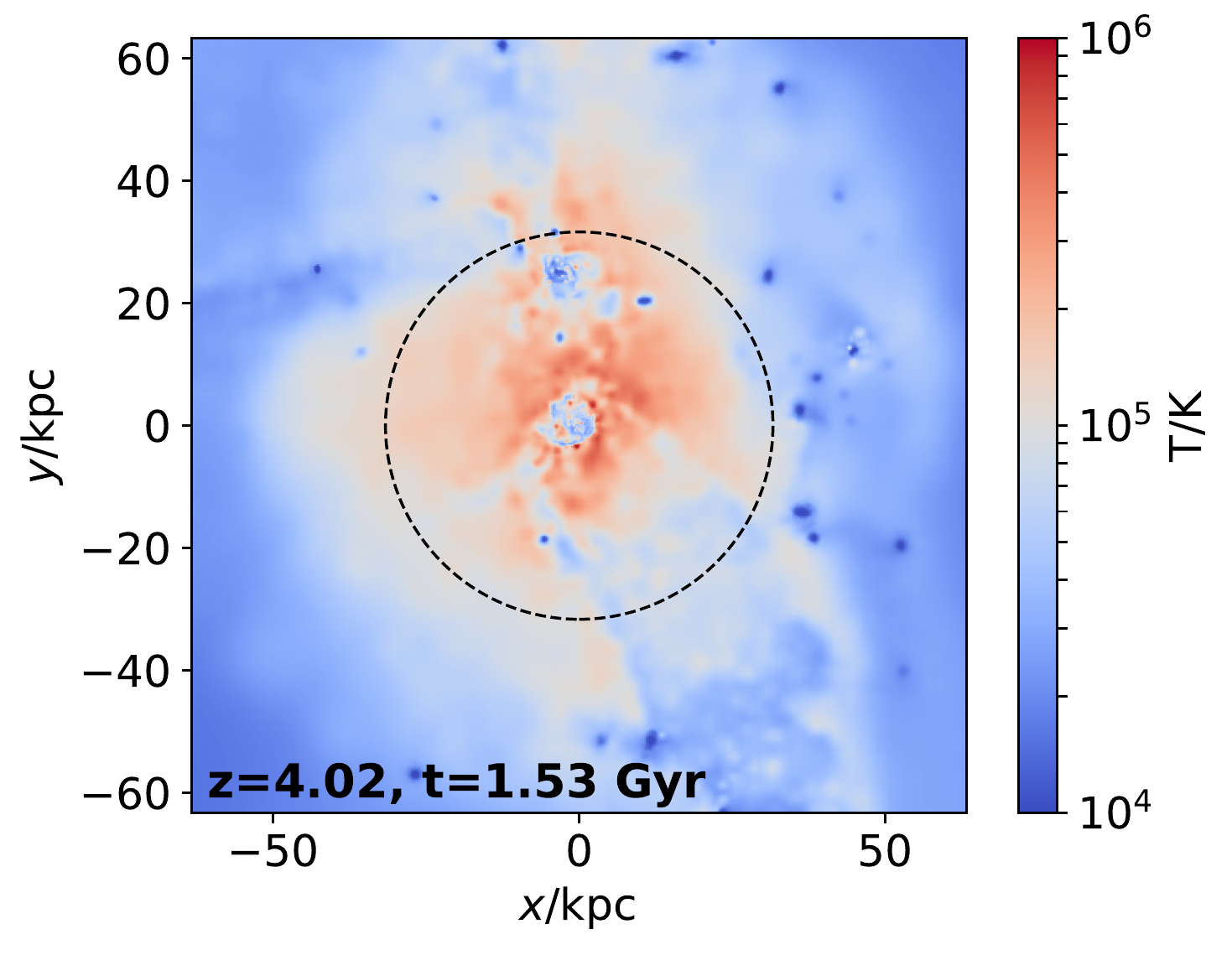}\\
  \includegraphics[width=0.25\hsize]{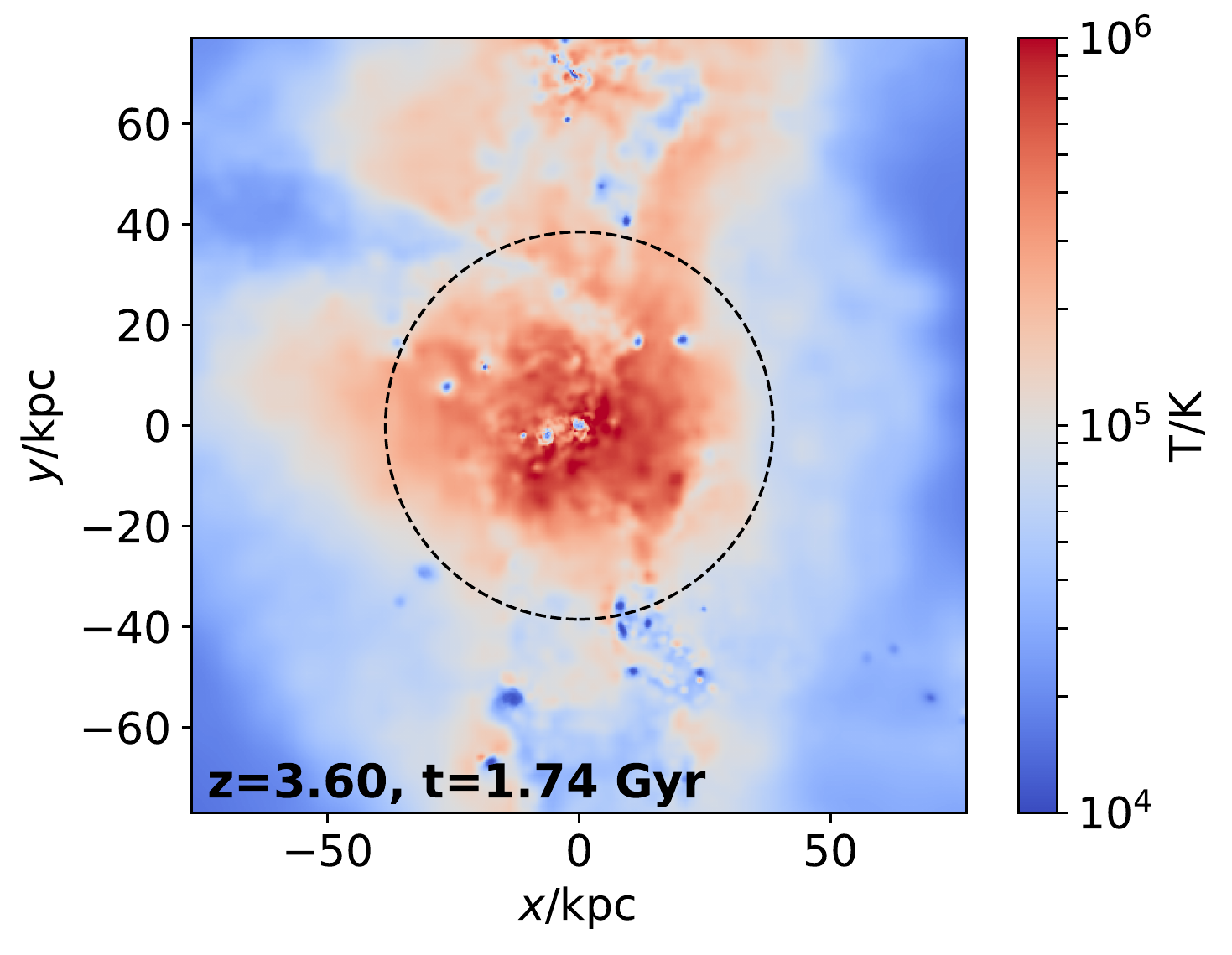}
  \includegraphics[width=0.25\hsize]{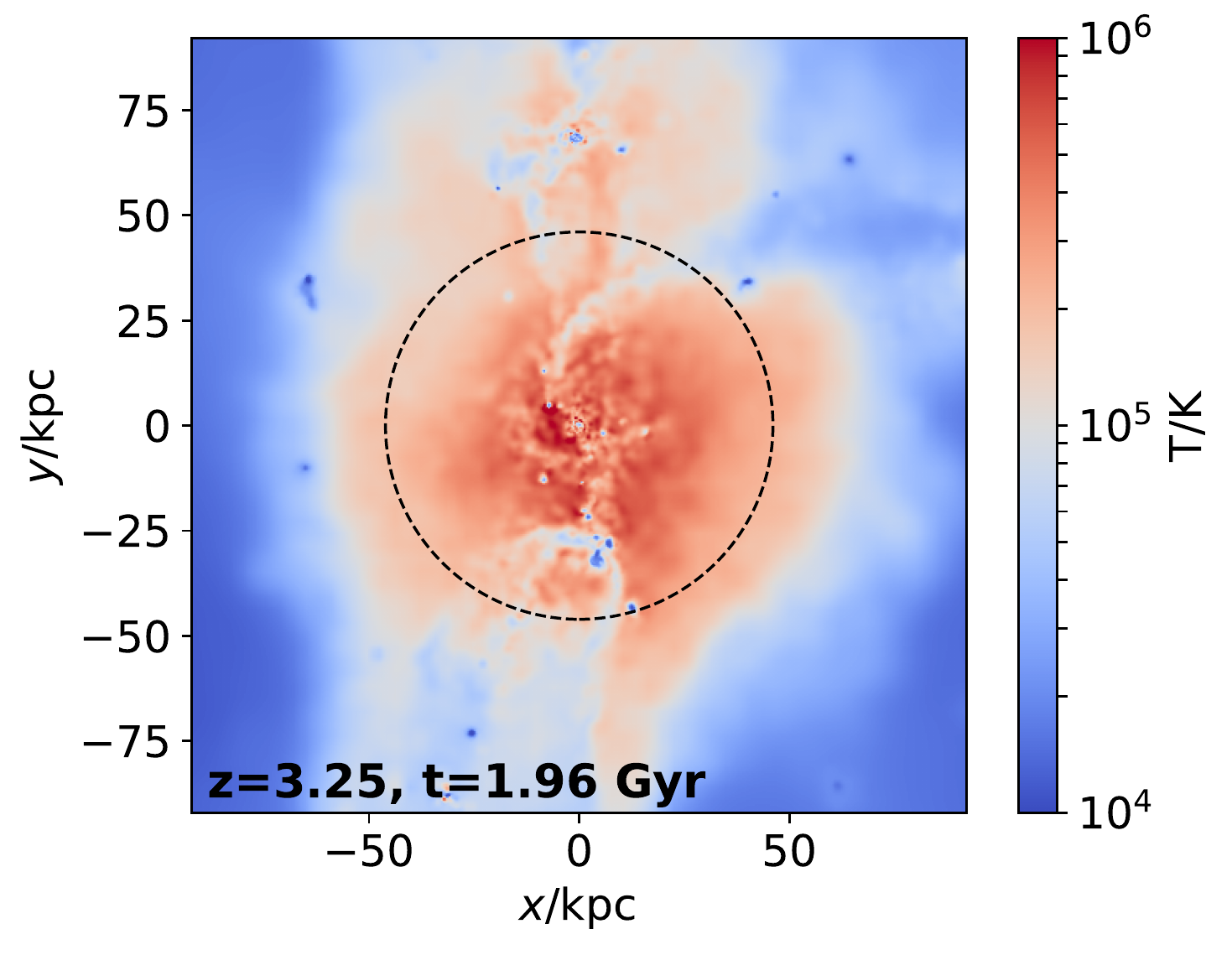}
  \includegraphics[width=0.25\hsize]{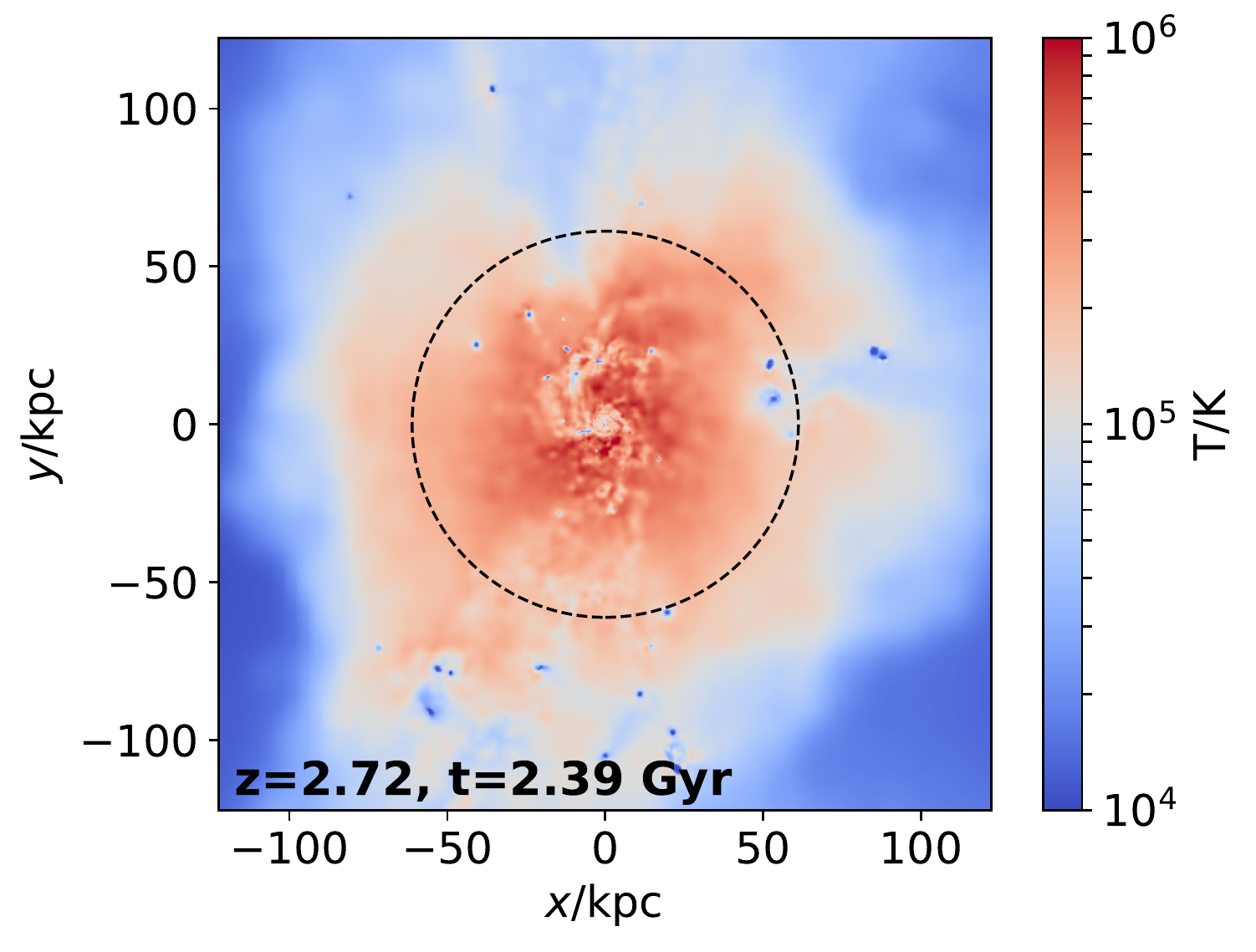}
  \includegraphics[width=0.25\hsize]{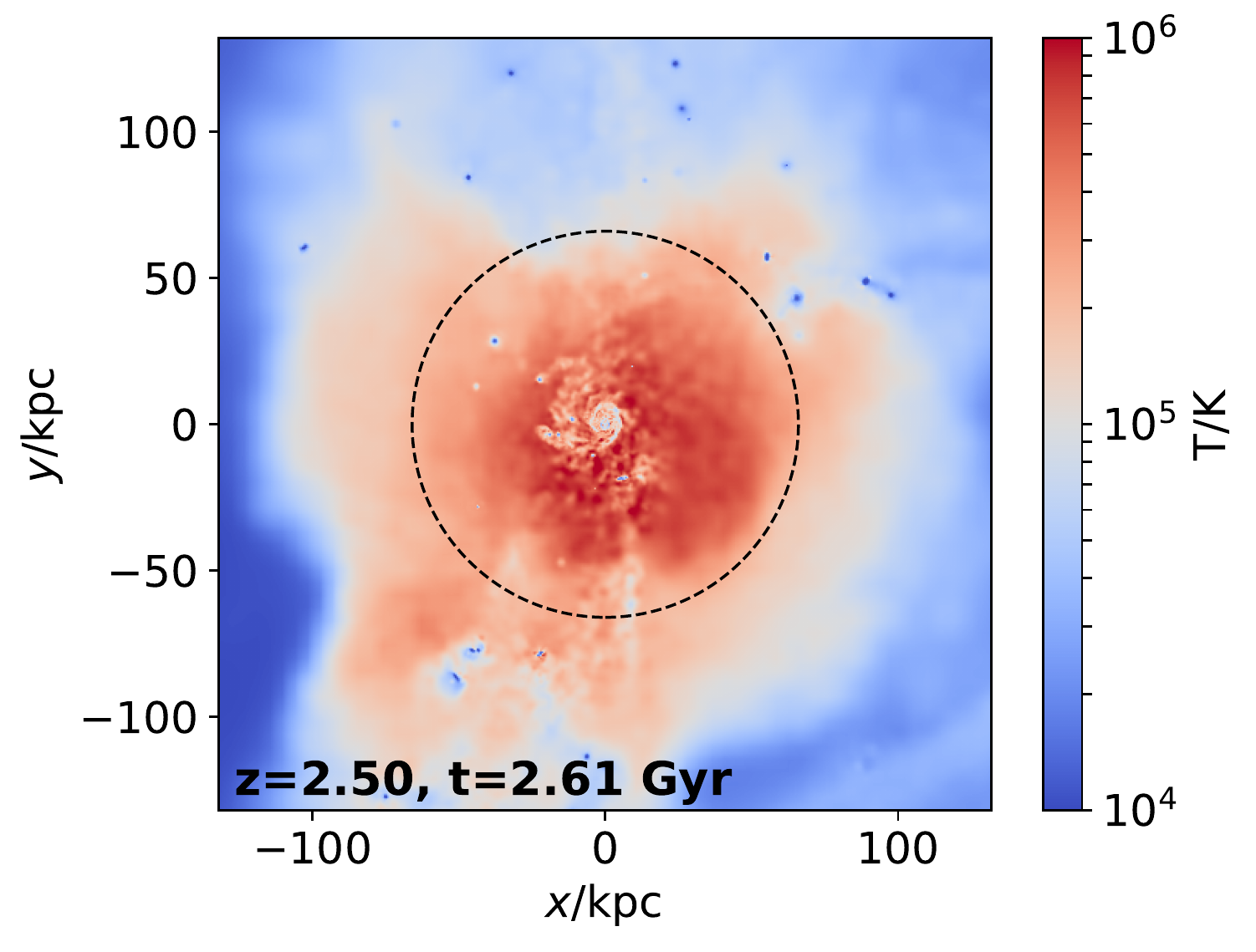}\\
  \includegraphics[width=0.25\hsize]{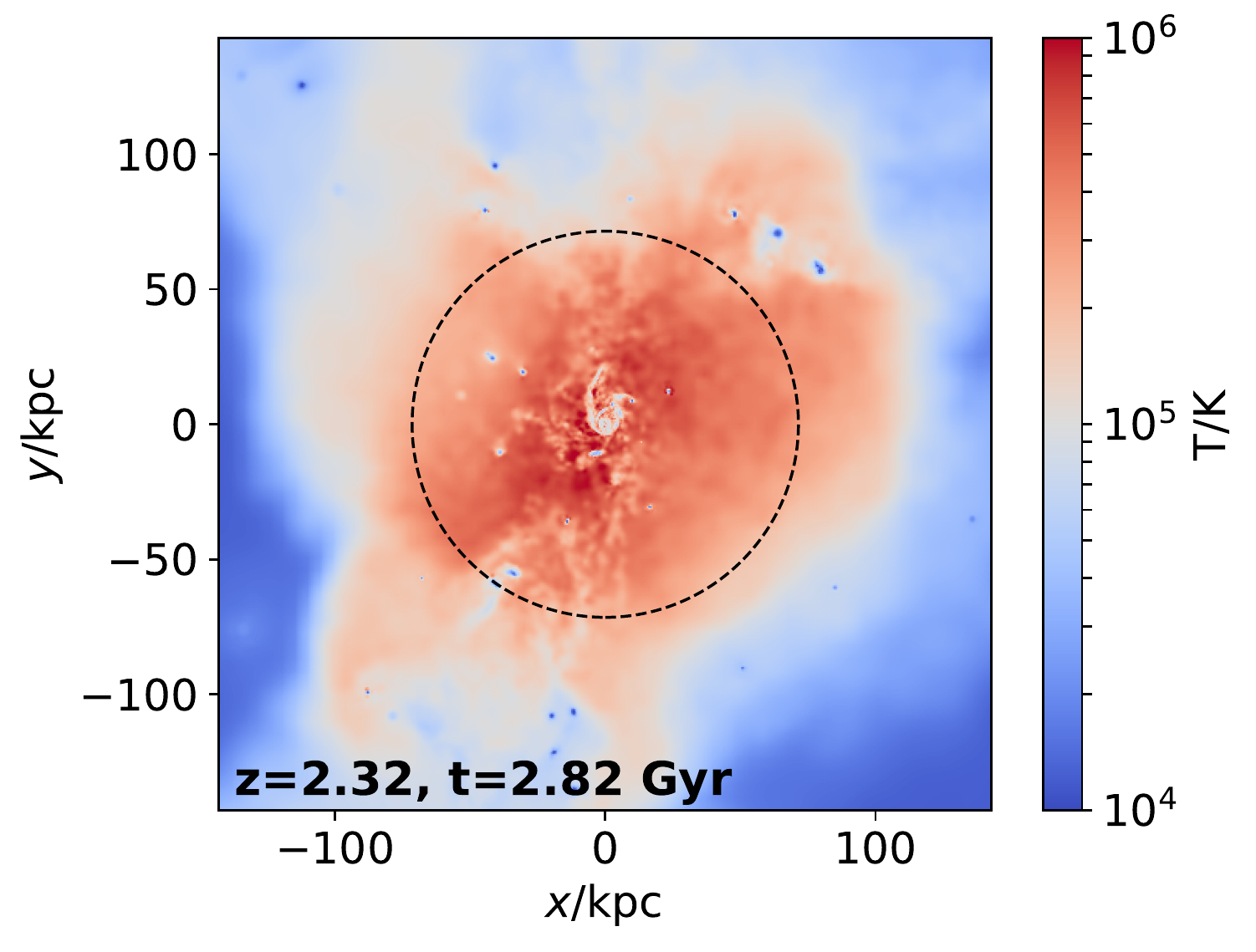}
  \includegraphics[width=0.25\hsize]{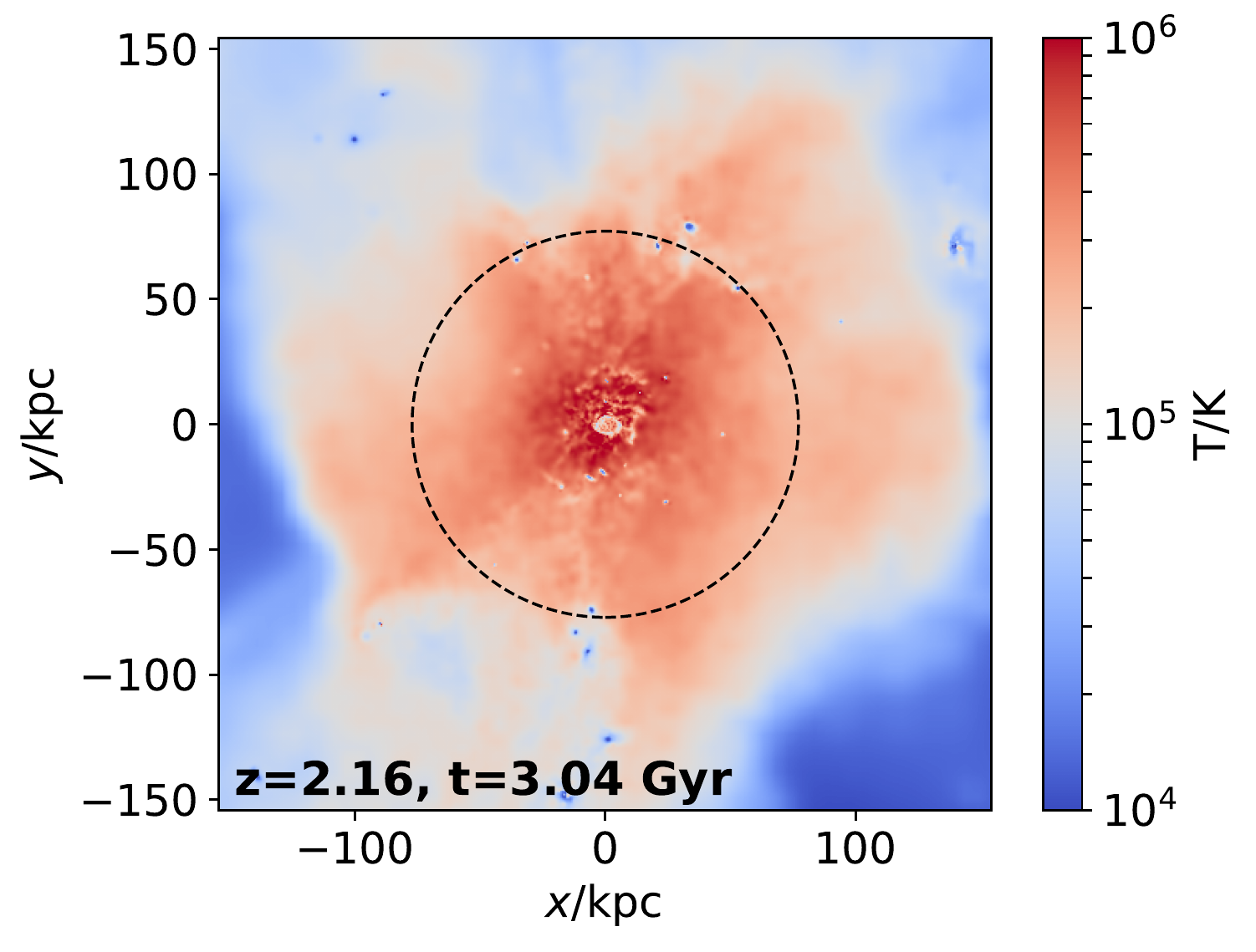}
  \includegraphics[width=0.25\hsize]{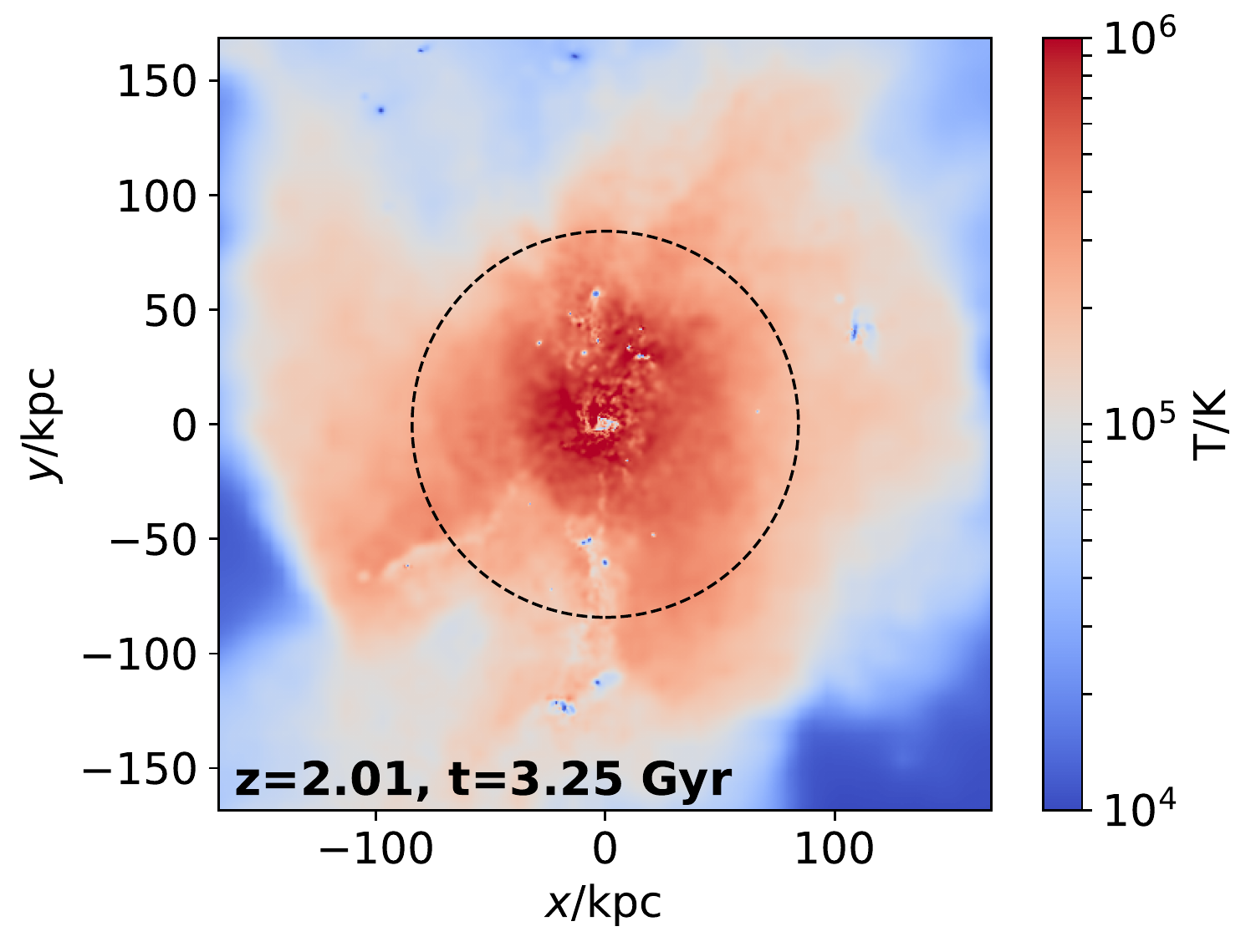}
  \includegraphics[width=0.25\hsize]{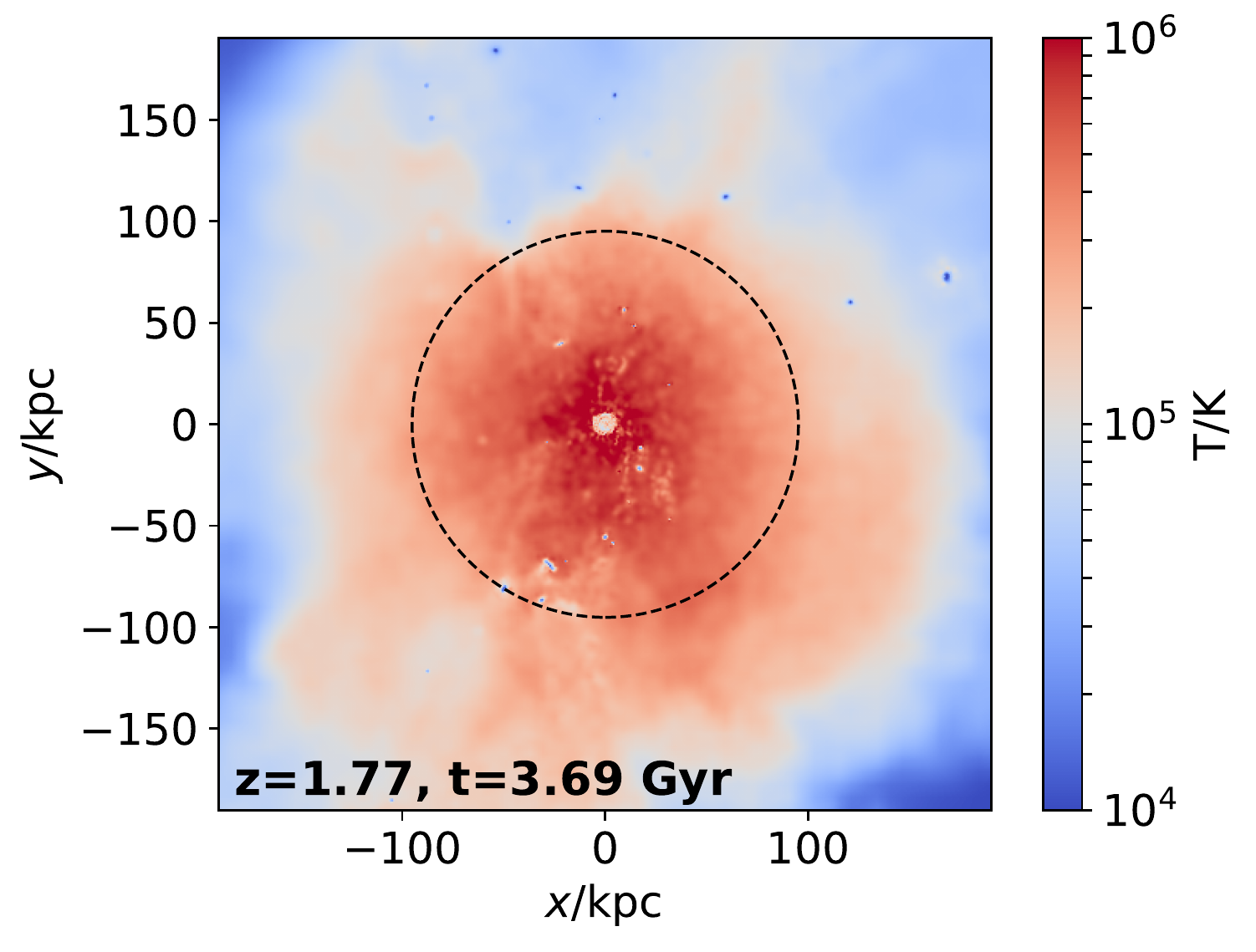}\\
  \includegraphics[width=0.25\hsize]{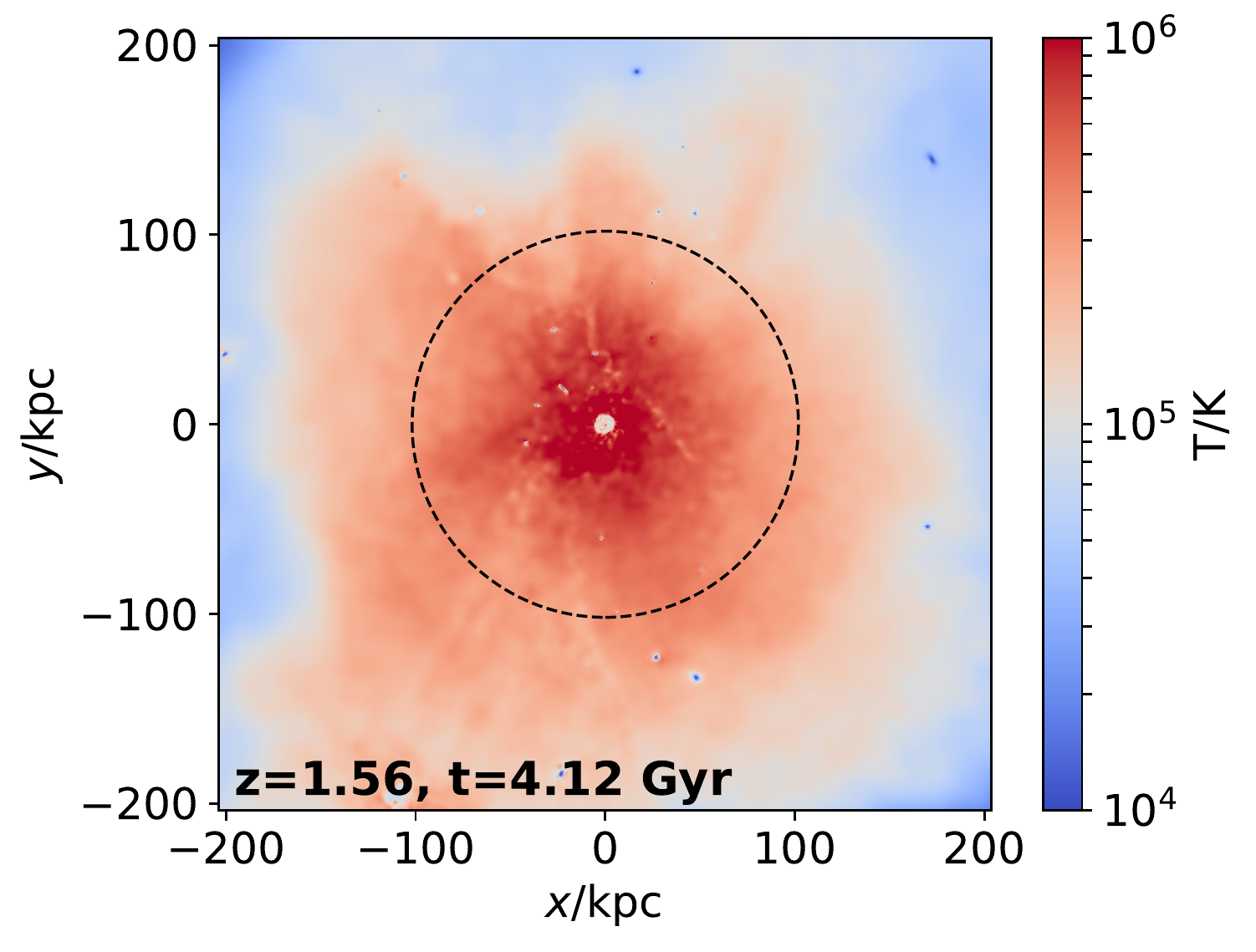}
  \includegraphics[width=0.25\hsize]{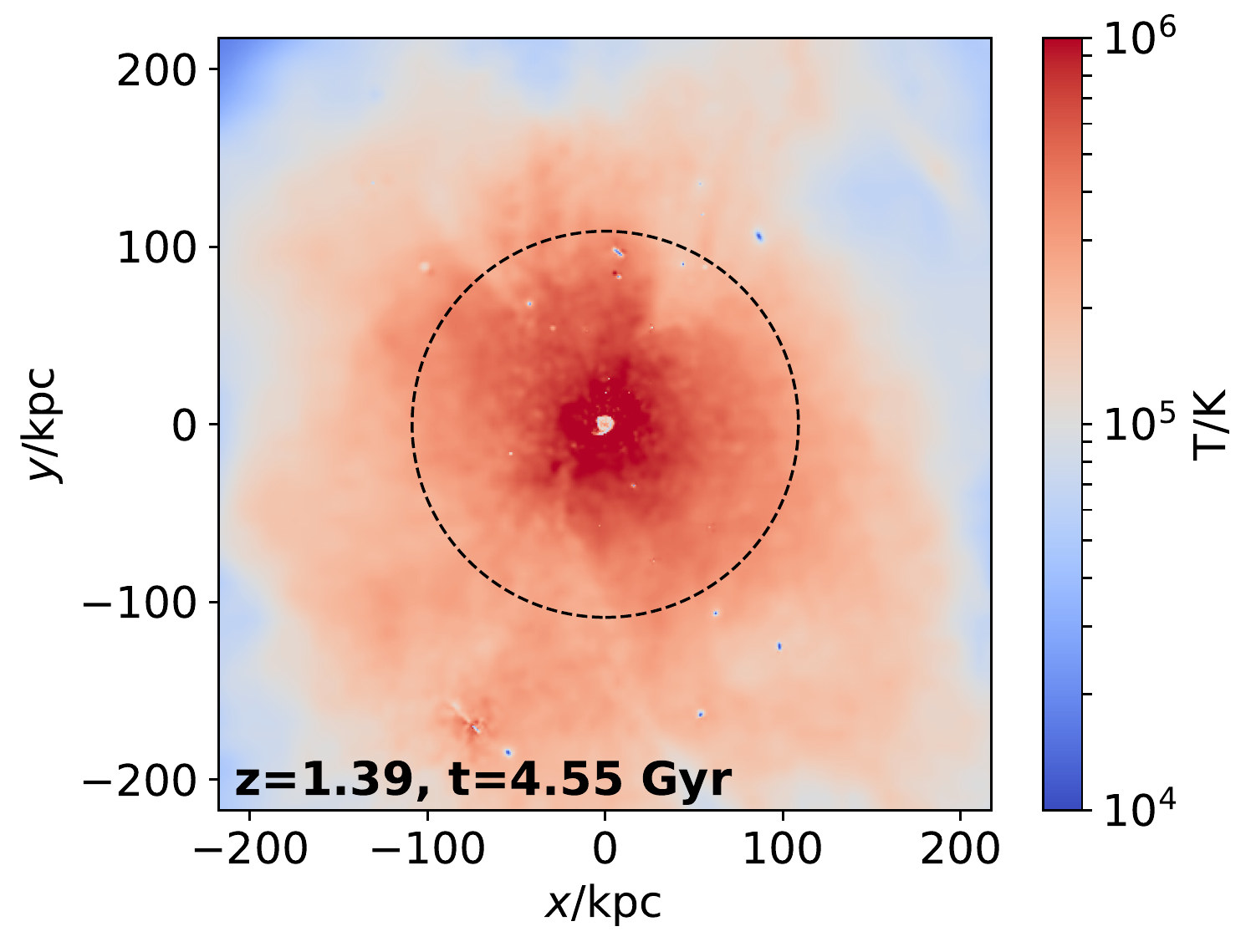}
  \includegraphics[width=0.25\hsize]{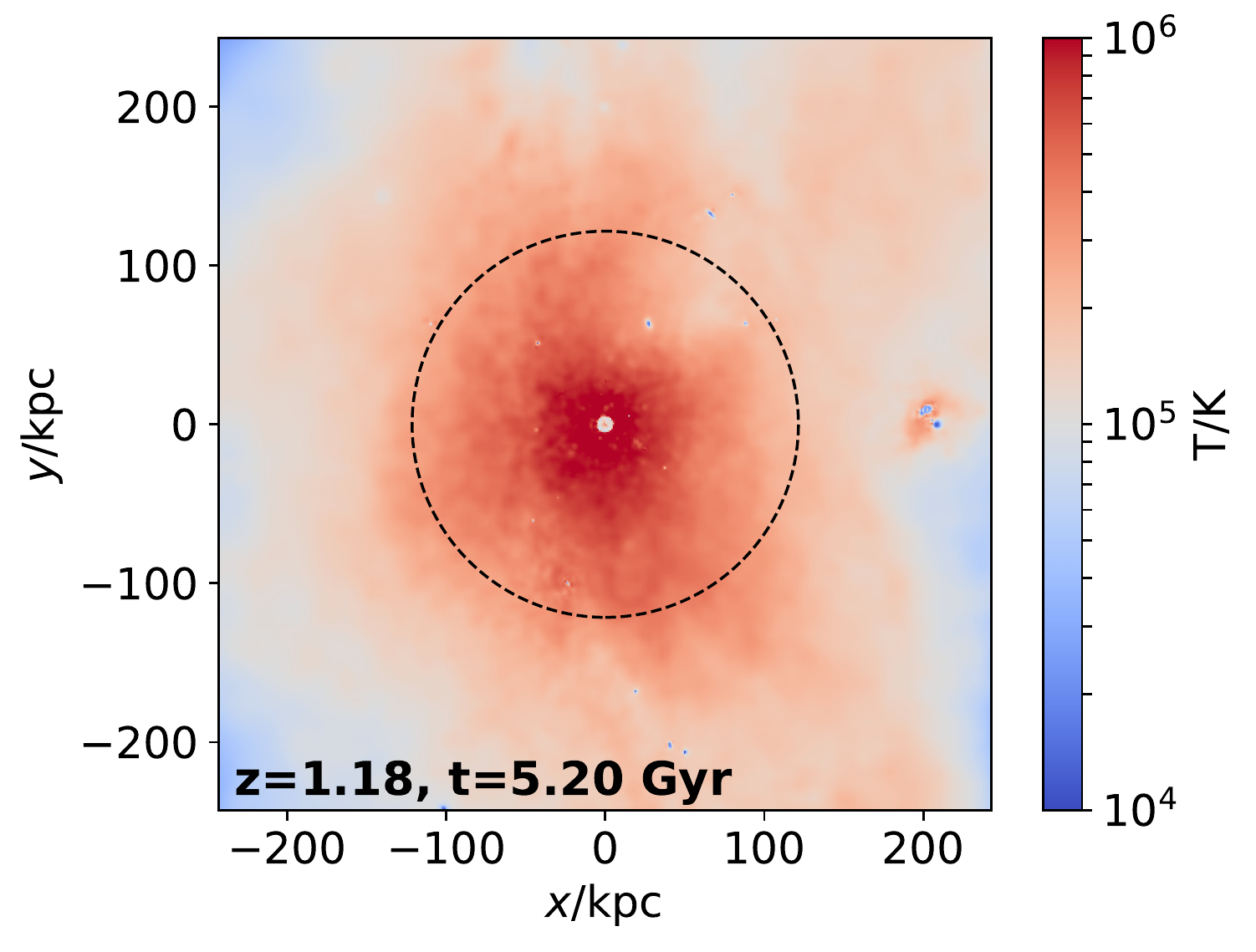}
  \includegraphics[width=0.25\hsize]{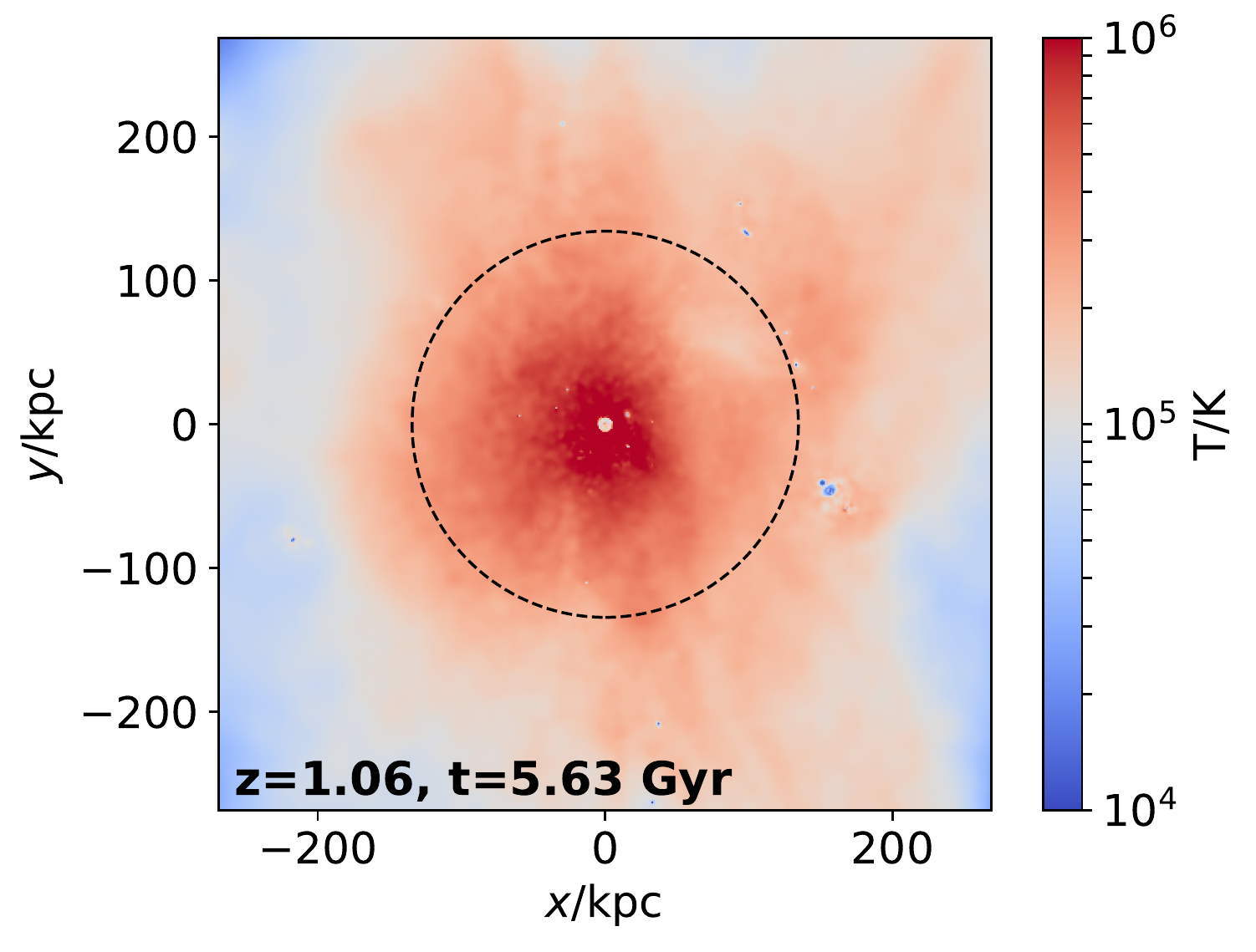}\\
  \includegraphics[width=0.25\hsize]{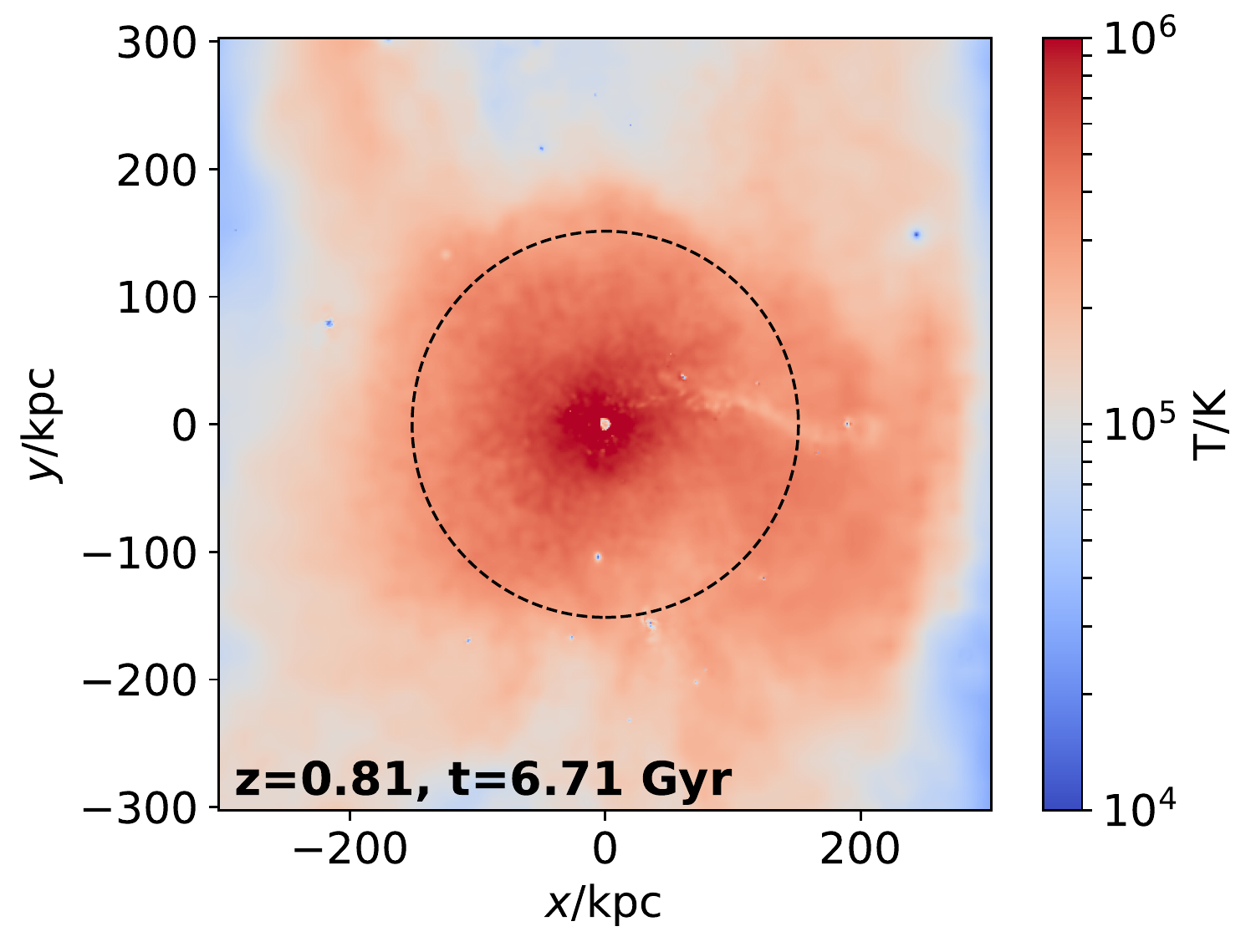}
  \includegraphics[width=0.25\hsize]{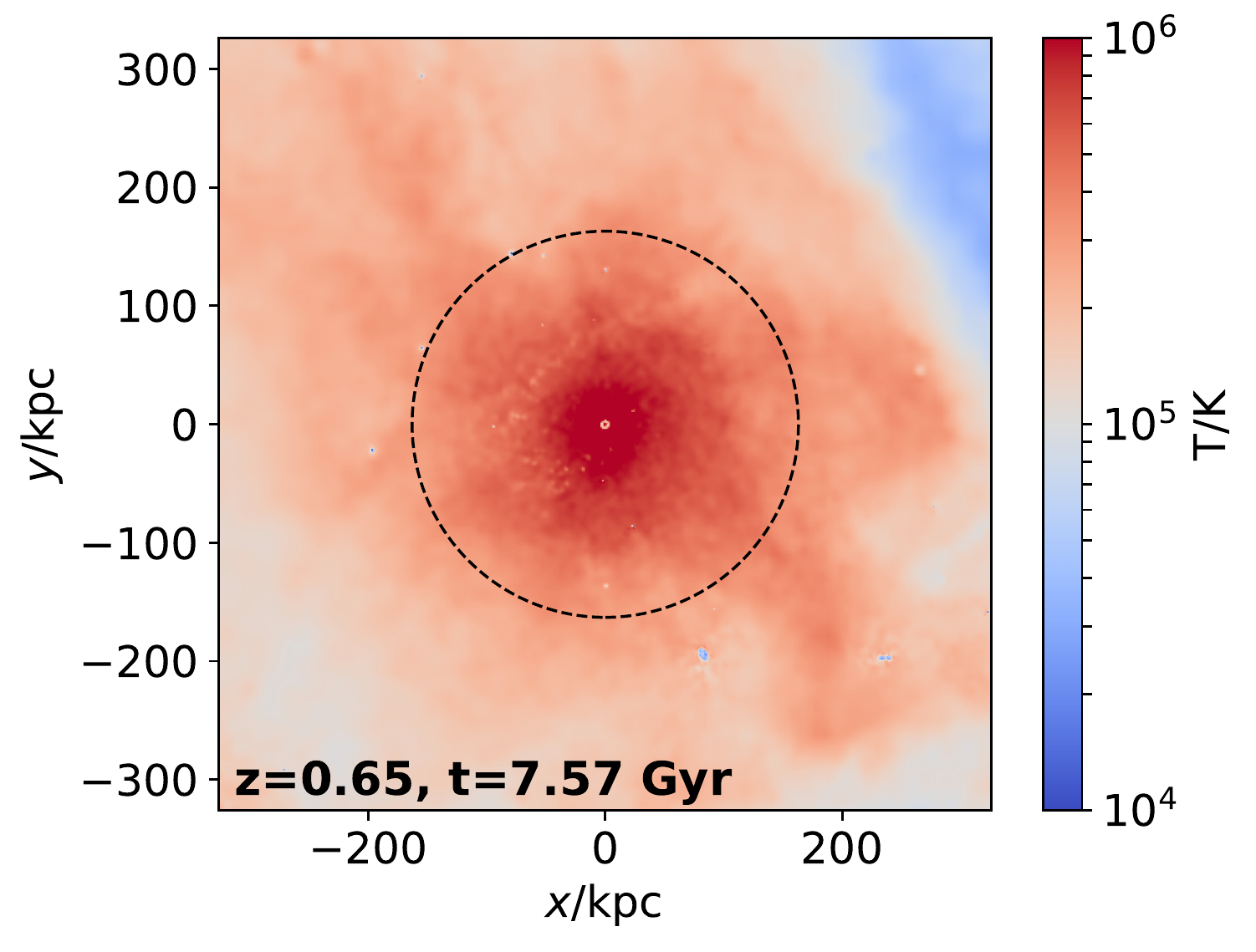}
  \includegraphics[width=0.25\hsize]{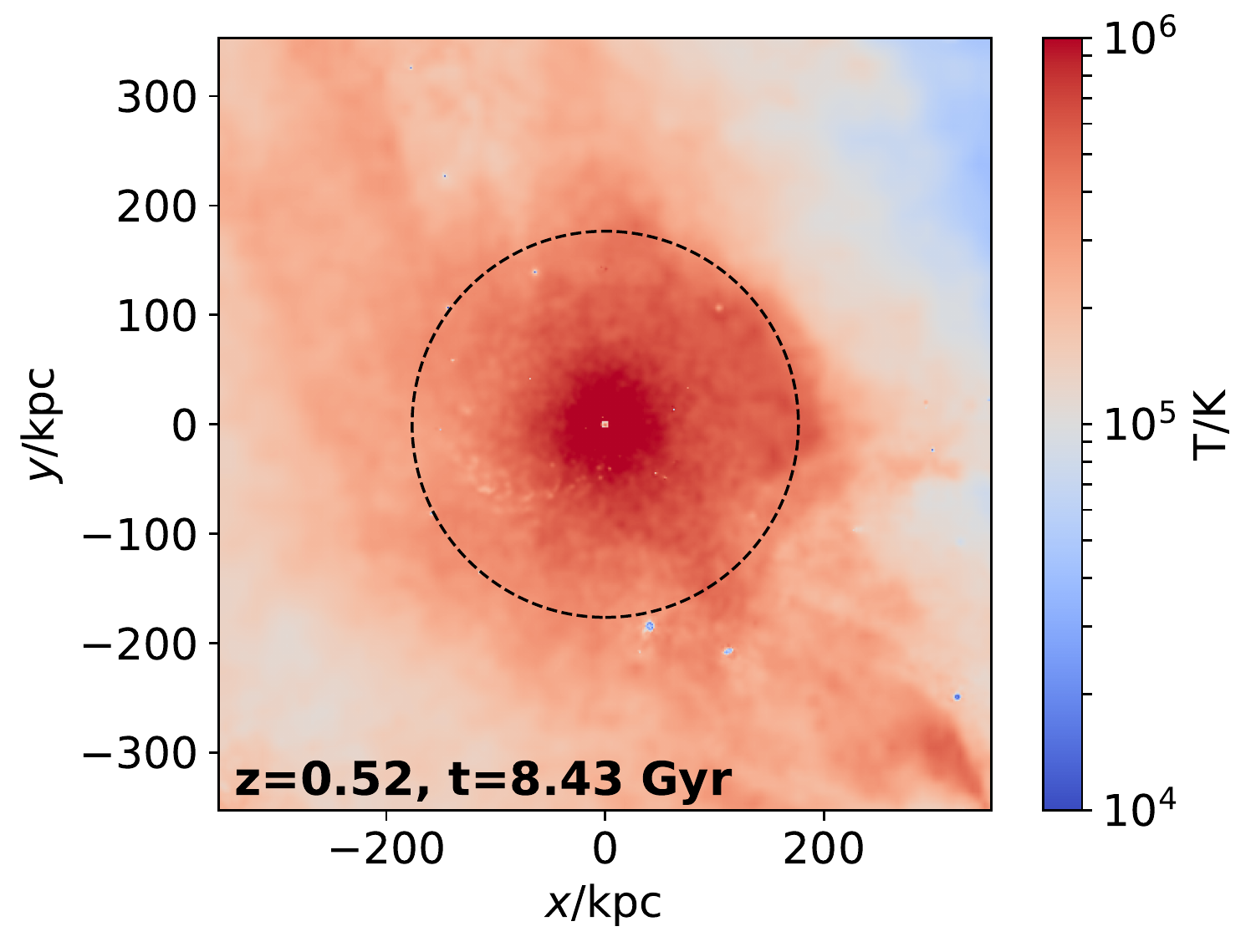}
  \includegraphics[width=0.25\hsize]{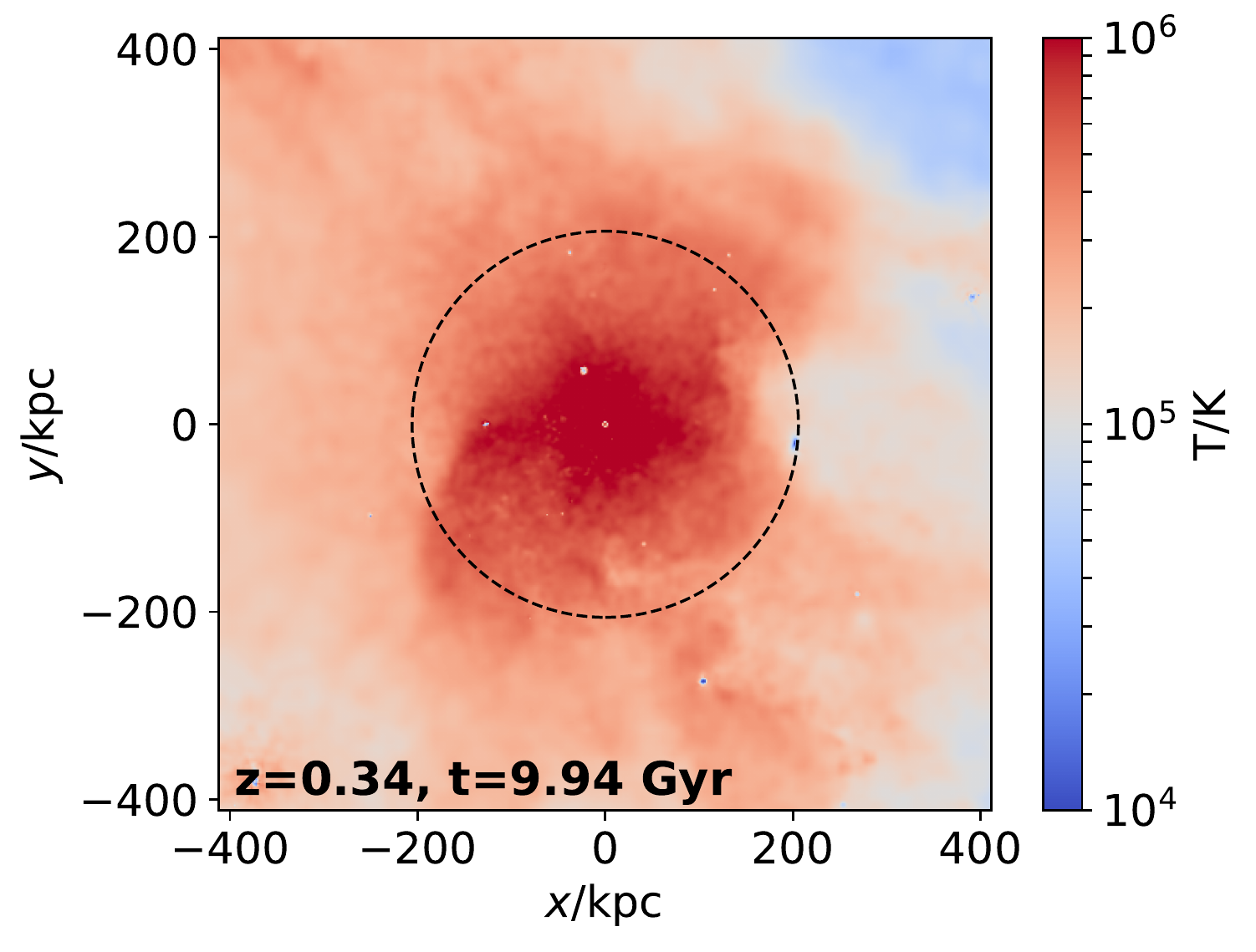}\\
  \includegraphics[width=0.25\hsize]{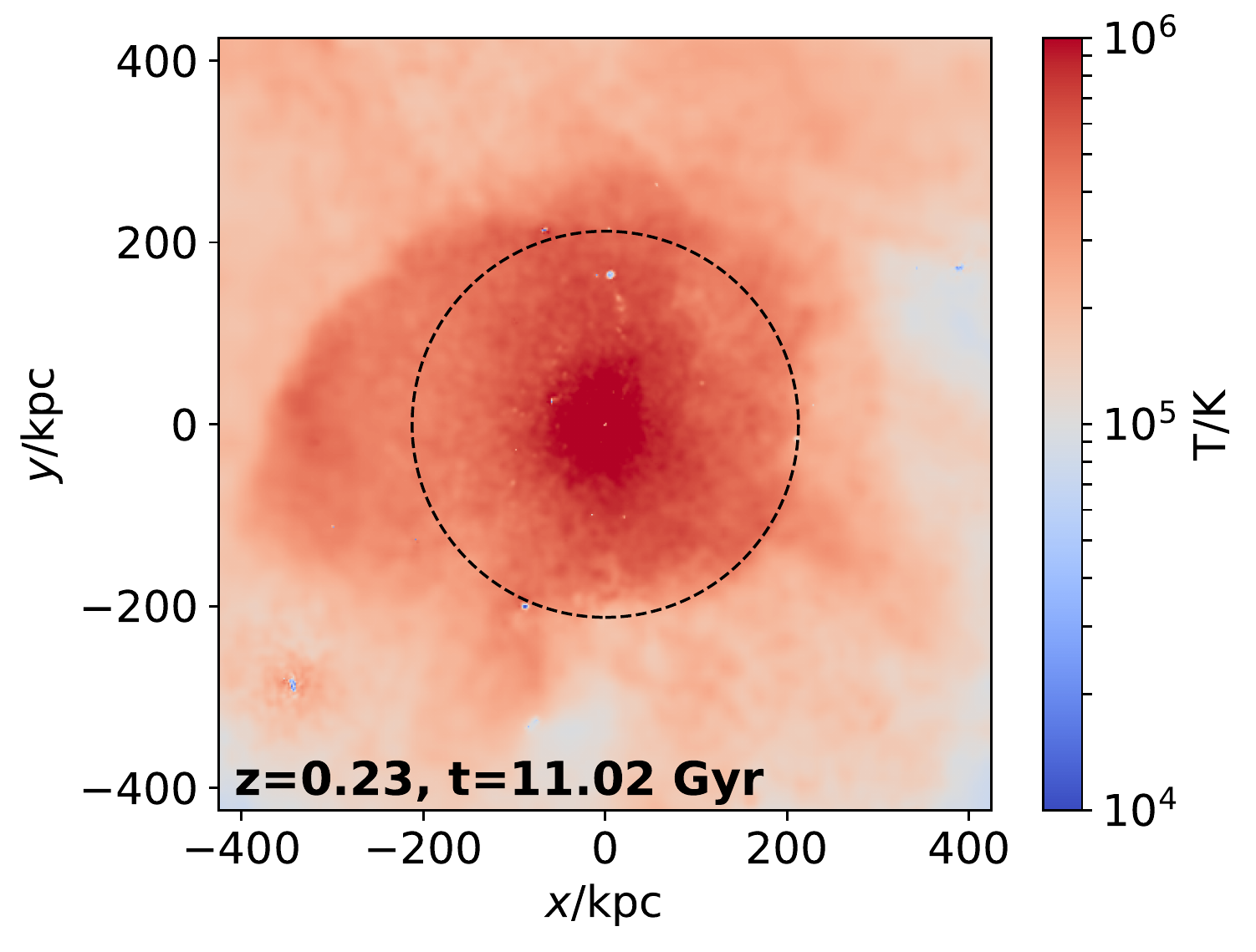}
  \includegraphics[width=0.25\hsize]{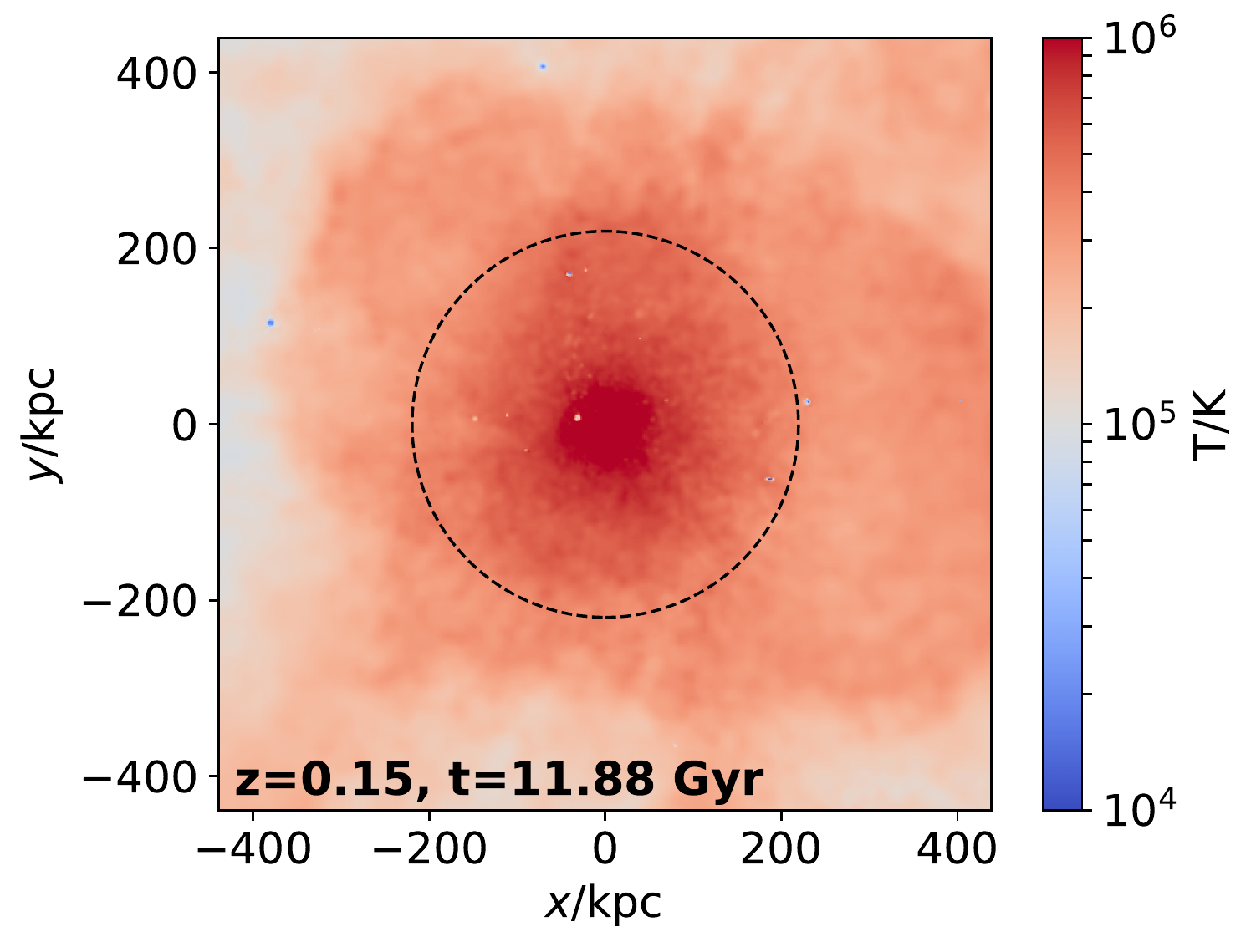}
   \includegraphics[width=0.25\hsize]{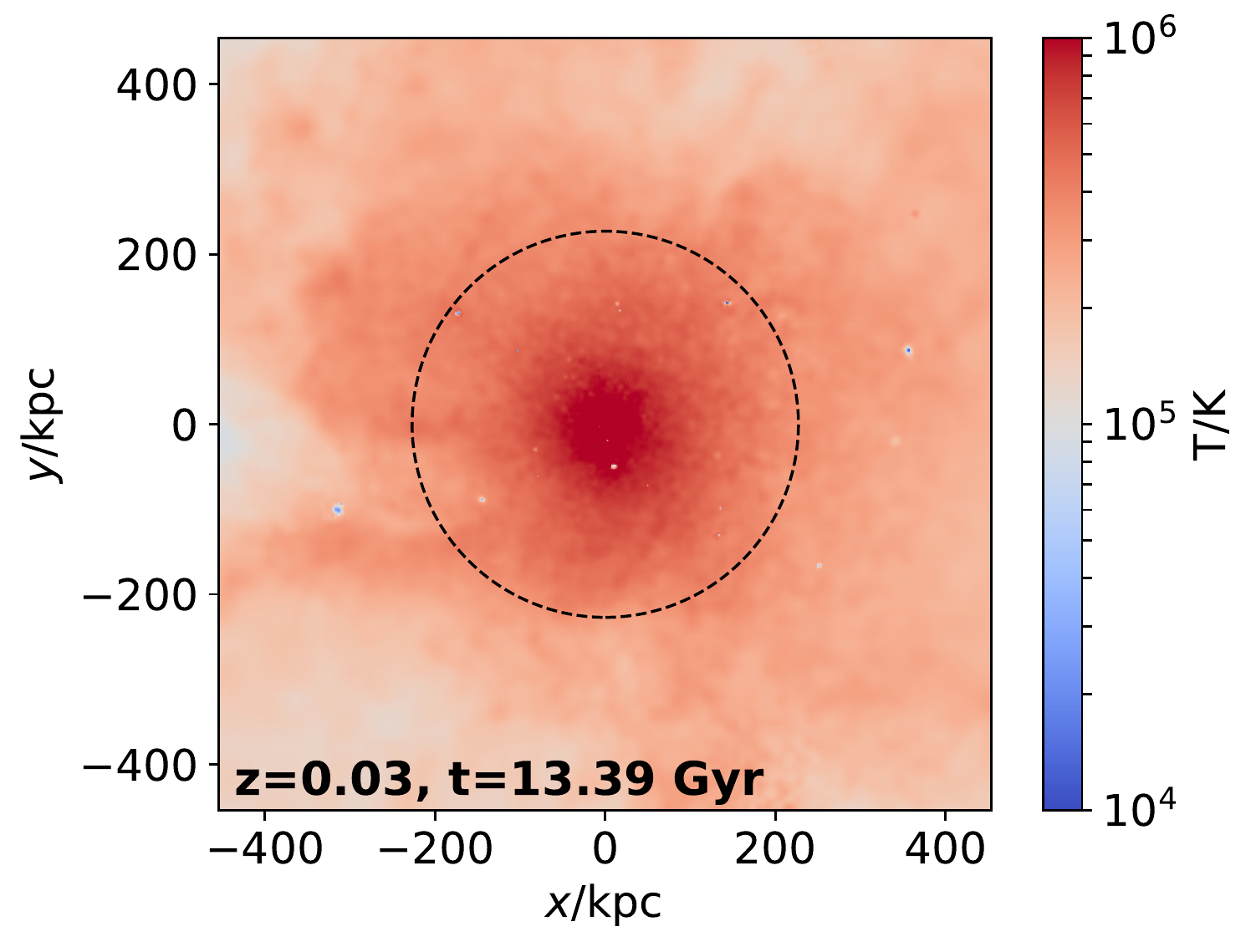}
   \includegraphics[width=0.25\hsize]{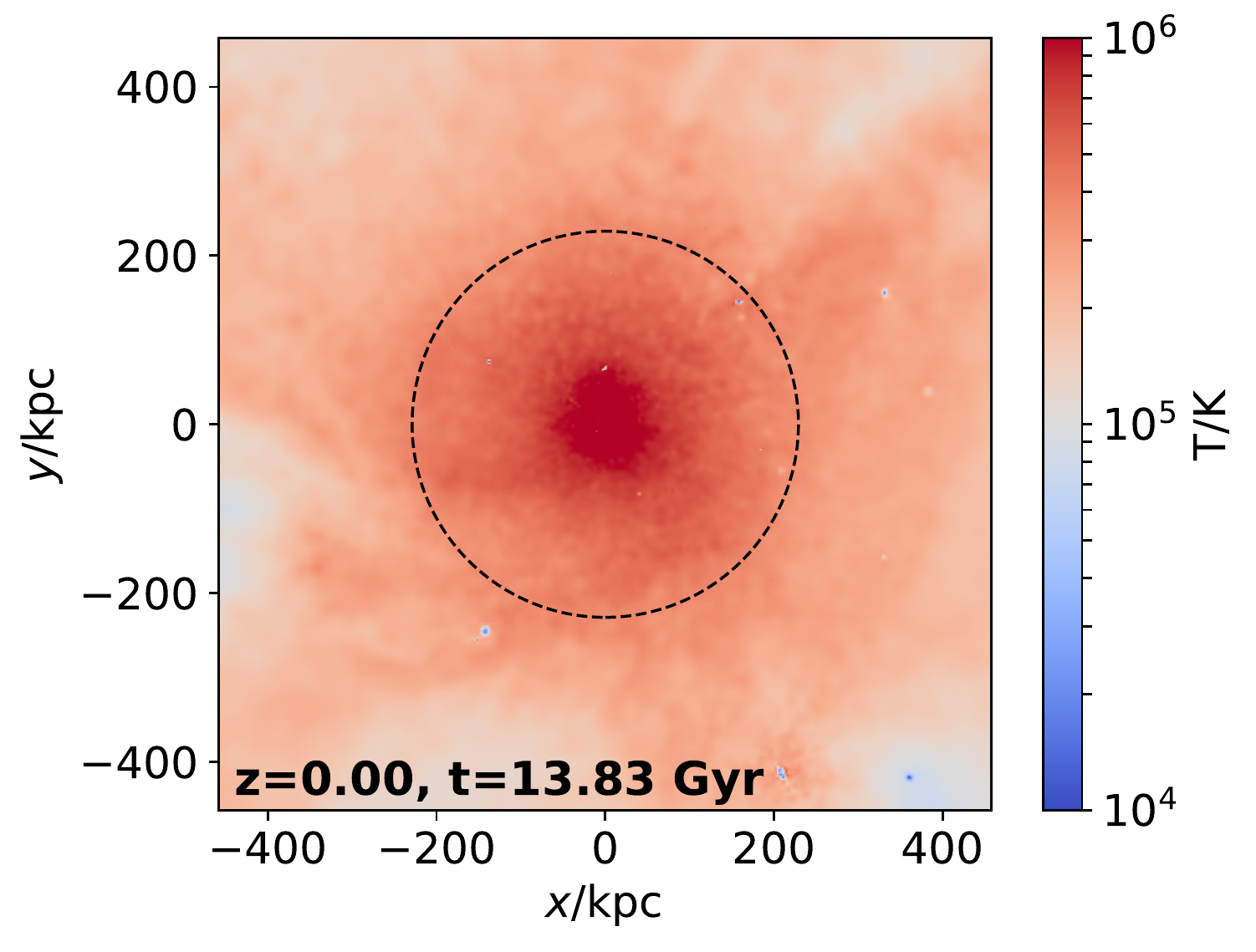}
\end{array}$
\end{center}
\caption{The same as Fig.~\ref{poststamp_images} for g1.12e12. Both figures are for simulations without feedback.}
\label{poststamp_images_S2}
\end{figure*}

{  The gray dashed curves in  Fig.~\ref{massgrowth} show the evolution of $M_{\rm vir}$ with cosmic time $t$ for g7.55e11 and g1.12e22, both with and without feedback.}
$M_{\rm vir}$ has been rescaled by the cosmic baryon fraction $f_{\rm b}=\Omega_{\rm b}/\Omega_{\rm m}\simeq 0.15$, so that it can be compared to the actual baryonic mass 
$M_{\rm b}$ within the virial radius.
Although the NIHAO galaxies were selected discarding all haloes with a
companion of comparable mass within three
virial radii at any time, the impact of mergers  is clearly visible.

 { For g7.55e11 (Fig.~\ref{massgrowth}a),}
the two most significant merging events took place at $3.04{\rm\,Gyr}\lsim t\lsim 3.47{\rm\,Gyr}$ ($2.16\gsim z\gsim 1.88$)
and $4.76{\rm\,Gyr}\lsim t\lsim 5.20{\rm\,Gyr}$ ($1.31\gsim z\gsim 1.18$). 
They were
followed by two
more minor episodes, the first at $9.73{\rm\,Gyr}\lsim t\lsim 9.94{\rm\,Gyr}$  ($0.36\gsim z\gsim 0.34$), the second at
$12.53{\rm\,Gyr}\lsim t\lsim 12.75{\rm\,Gyr}$  ($0.10\gsim z\gsim 0.08$).

{  For g1.12e22 (Fig.~\ref{massgrowth}b), two merging events stand out. The first ($2.17{\rm\,Gyr}\lsim t\lsim 2.39{\rm\,Gyr}$, $2.96\gsim z\gsim 2.72$) results from the simultaneous accretion of two satellite haloes.
$M_{\rm vir}$ increases by about one third. The second  is at $t\simeq 5.20{\rm\,Gyr}$ ($z\simeq 1.18$). The bump in $M_{\rm vir}(t)$ at $9{\rm\,Gyr}\lsim t\lsim 11{\rm\,Gyr}$ is due to two satellites that come in and out
of the virial radius of the main system.

Feedback has very limited impact on the growth history of the dark-matter (DM) halo. The gray dashed curves in panels c and d are almost indistinguishable from those in panels a and b, respectively.}

 {  Comparing $M_{\rm b}$ (gray solid curves) to $f_{\rm b}M_{\rm vir}$ (gray dashed curves) shows that,  without feedback}, $M_{\rm b}$
 is very close to the expectation from the universal baryon fraction ($M_{\rm b}\simeq f_{\rm b}M_{\rm vir}$). 
{  In the simulations with feedback, the gray solid lines are systematically lower than the gray dashed lines by 30 per cent on average, yet $M_{\rm b}$ keeps tracing $f_{\rm b}M_{\rm vir}$
fairly closely.}

{  Without feedback,
cooling and star formation are expectedly efficient.} Most of the baryons within $r_{\rm vir}$ are in stars.
The evolution of the stellar mass $M_\star$ within $r_{\rm vir}$  (green curves) follows
closely that of the DM. 
We note that $M_\star$ is the total stellar mass within $r_{\rm vir}$ and not the stellar mass of the central galaxy.
This is why, during mergers, there is no delay between the growth of $M_{\rm vir}$ and the growth of $M_\star$.
Throughout this article, mergers are halo mergers, not galaxy mergers.

{  One of the most remarkable differences with feedback is that the green curves become much smoother. Feedback greatly reduces the efficiency of star formation in low-mass haloes.
Merger with small satellite haloes contribute to the growth of $M_{\rm vir}$ but have very little impact on $M_\star$.

Figs.~\ref{poststamp_images}--\ref{poststamp_images_S2} 
show temperature  maps of the CGM in the simulations without feedback. The same images with feedback are presented in the supplementary material.
All our temperature maps look at the central galaxy face on, but edge-on views return a very similar picture.

In g7.5e11 (Fig.~\ref{poststamp_images}),} the CGM  at $z=5.27$ is still vastly cold. Circumgalactic shock fronts are clearly visible, but radiative cooling is so effective that the post-shock temperature is always much lower than the virial temperature { ($T_{\rm vir}\sim 2\times 10^5{\rm\,K}$ at $z= 5.27$--$6.26$).}
Some hot gas is already present at 
$t=1.31{\rm\,Gyr}$ ($z=4.56$), but the geometry of the hot phase is very different from that of a quasi-hydrostatic atmosphere.
At $2\lsim z\lsim 4$, cold streams and hot gas coexist.
At lower redshifts, the entire virial volume is filled with hot gas;
the supply of gas through cold filaments is disrupted; the cold blobs
that still come in are satellite galaxies.
 
{  Fig.~\ref{poststamp_images_S2} displays a similar behaviour but the hot gas appears earlier due to g1.12e12's higher mass. 
In g1.12ee12, the hot gas  extends beyond the virial radius already at $t=1.96$ ($z=3.25$)
}

Figs.~\ref{poststamp_images}--\ref{poststamp_images_S2}
confirms that the episodes of rapid growth identified in Fig.~\ref{massgrowth} are indeed linked to mergers.
In the time intervals when $M_{\rm vir}$ grows rapidly, there is always at least one satellite galaxy that enters the virial radius.

\begin{figure*}
\begin{center}$
\begin{array}{cc}
  \includegraphics[width=0.4\hsize]{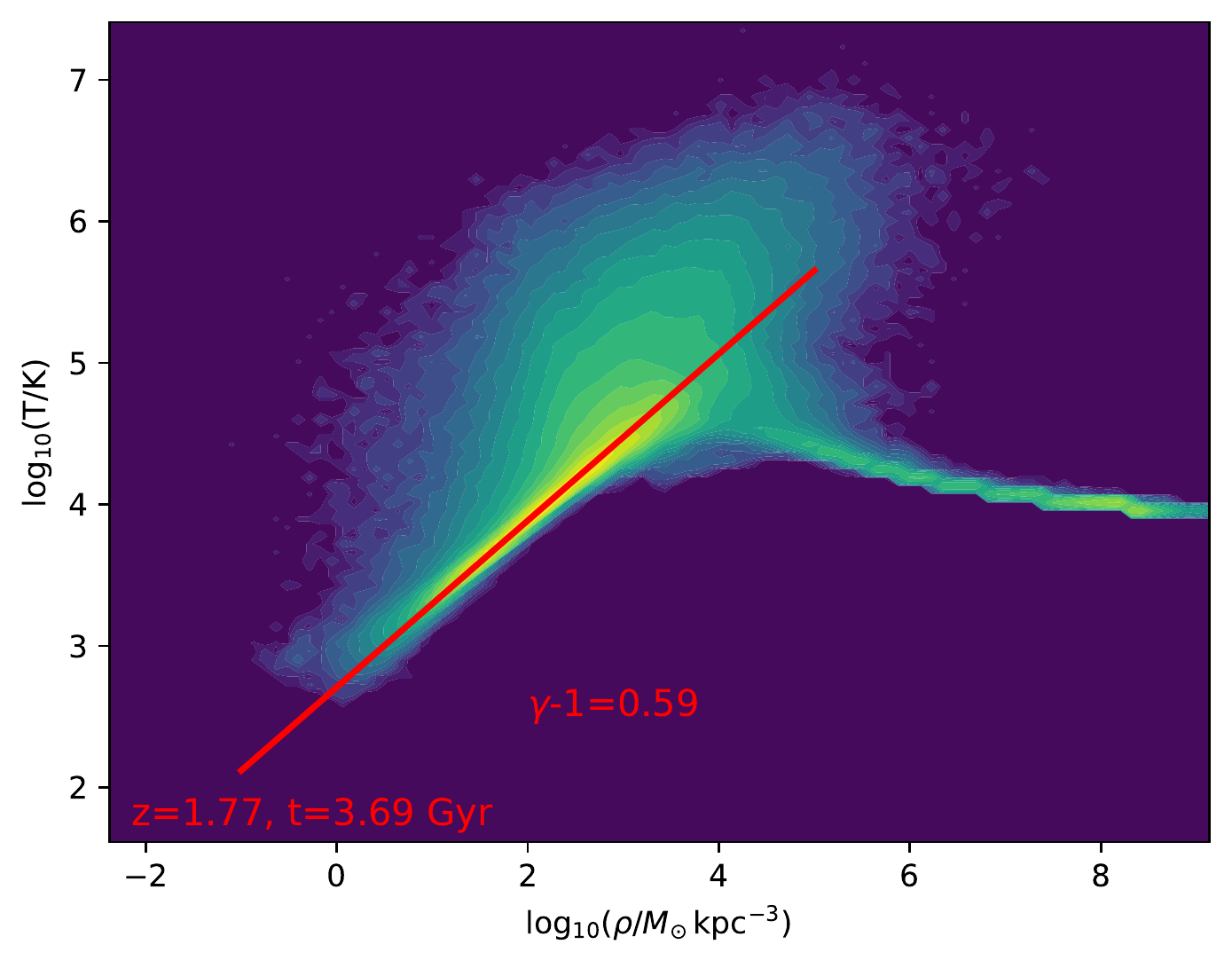}
  \includegraphics[width=0.41\hsize]{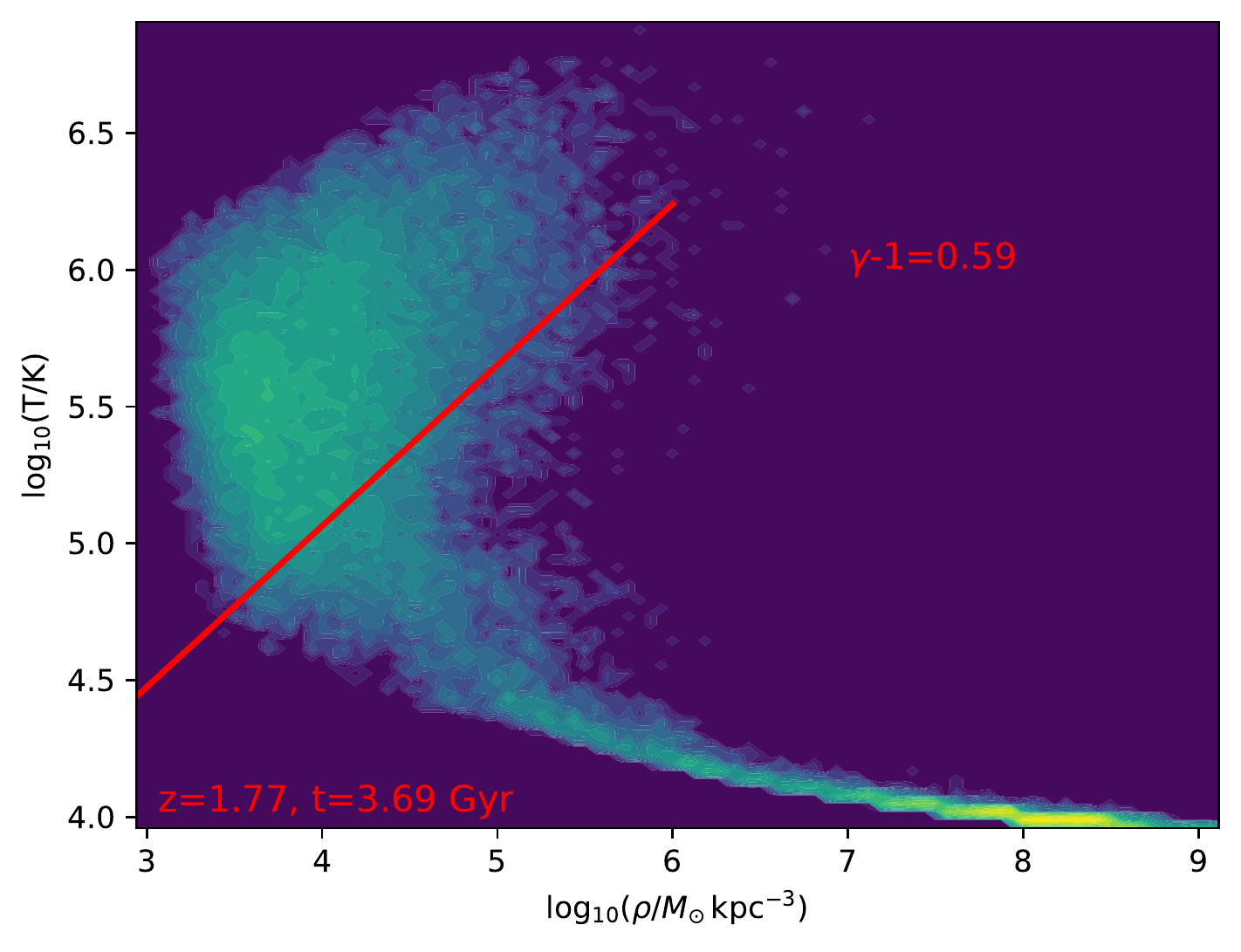}\\
  \includegraphics[width=0.4\hsize]{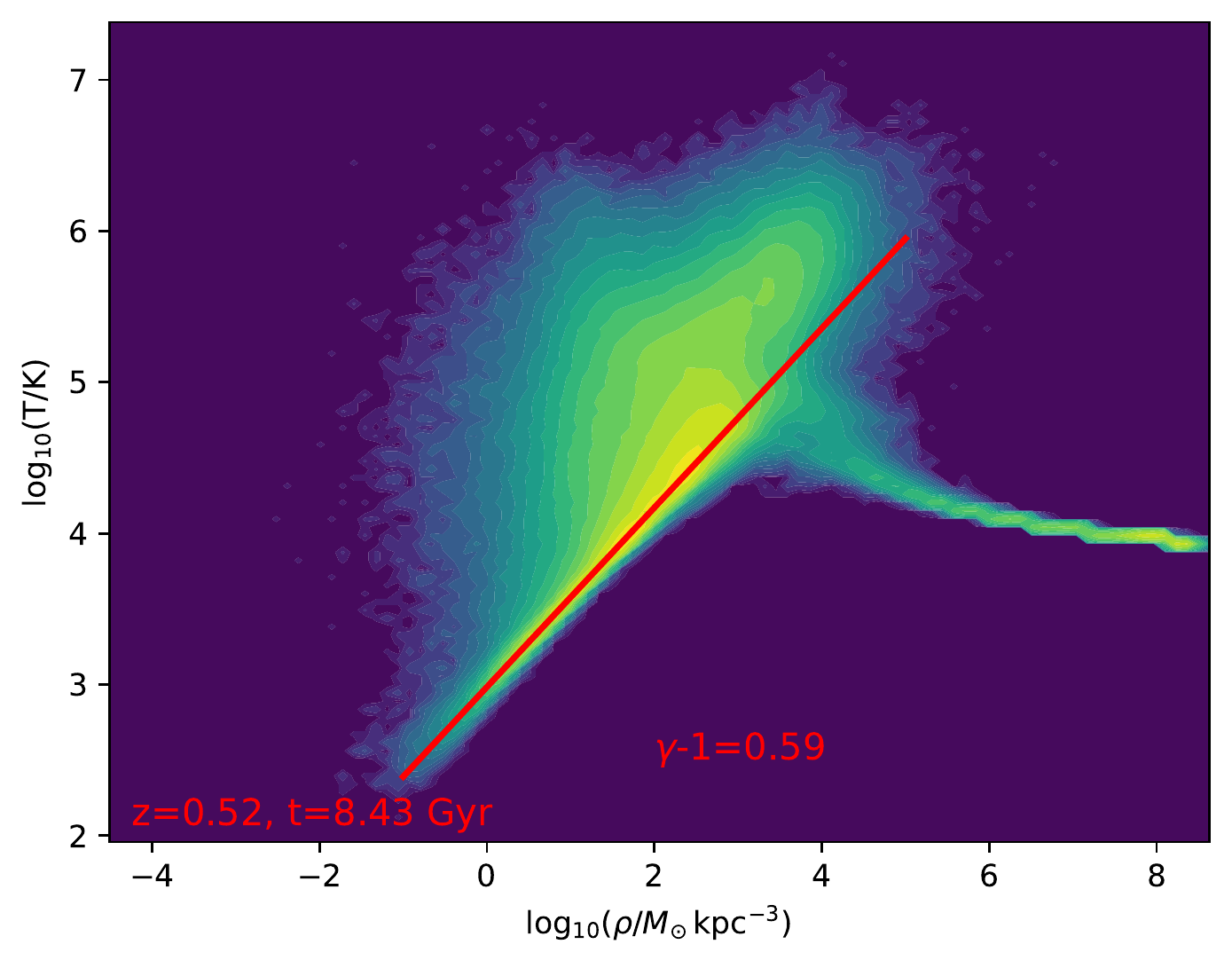}
  \includegraphics[width=0.41\hsize]{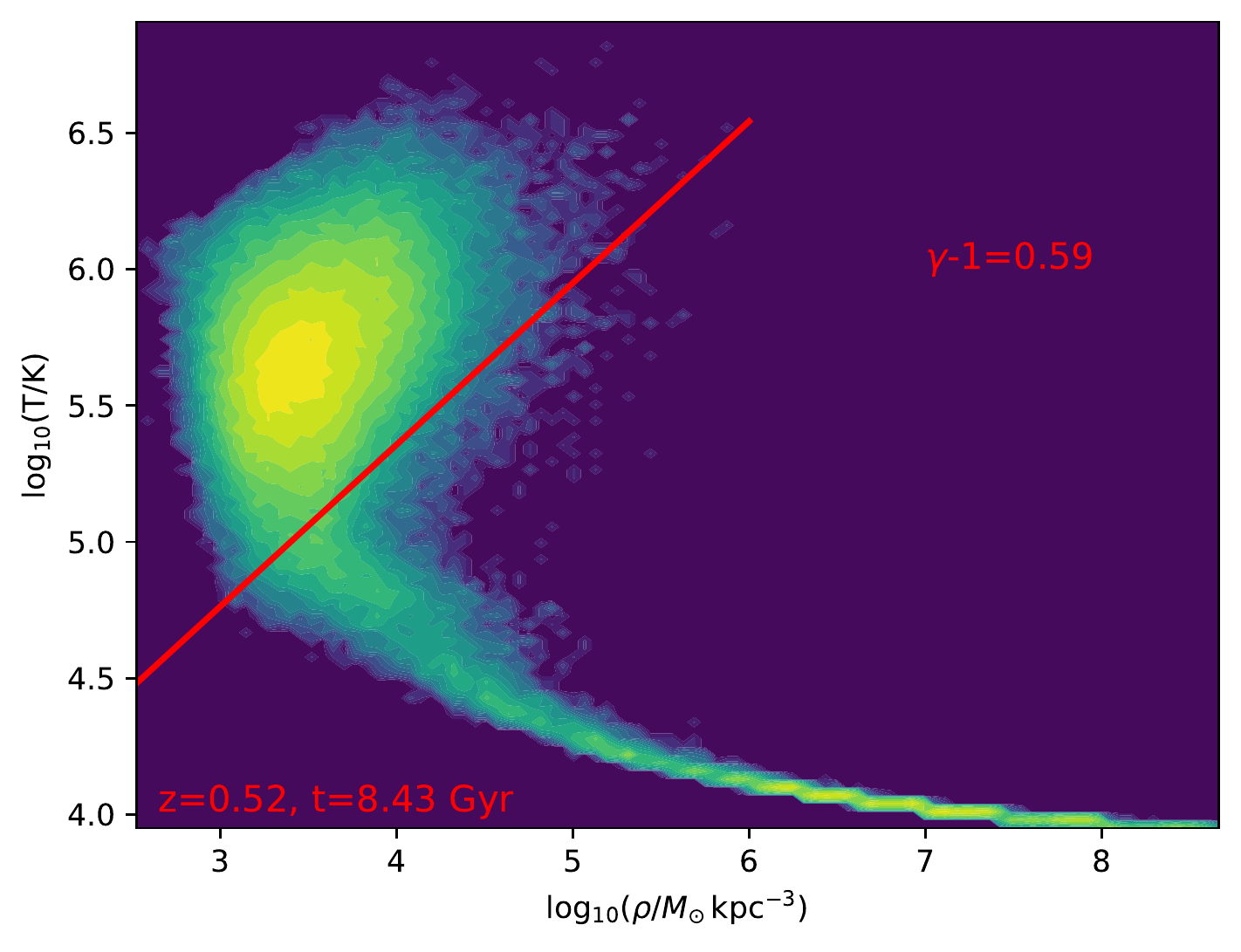}\\
  \includegraphics[width=0.4\hsize]{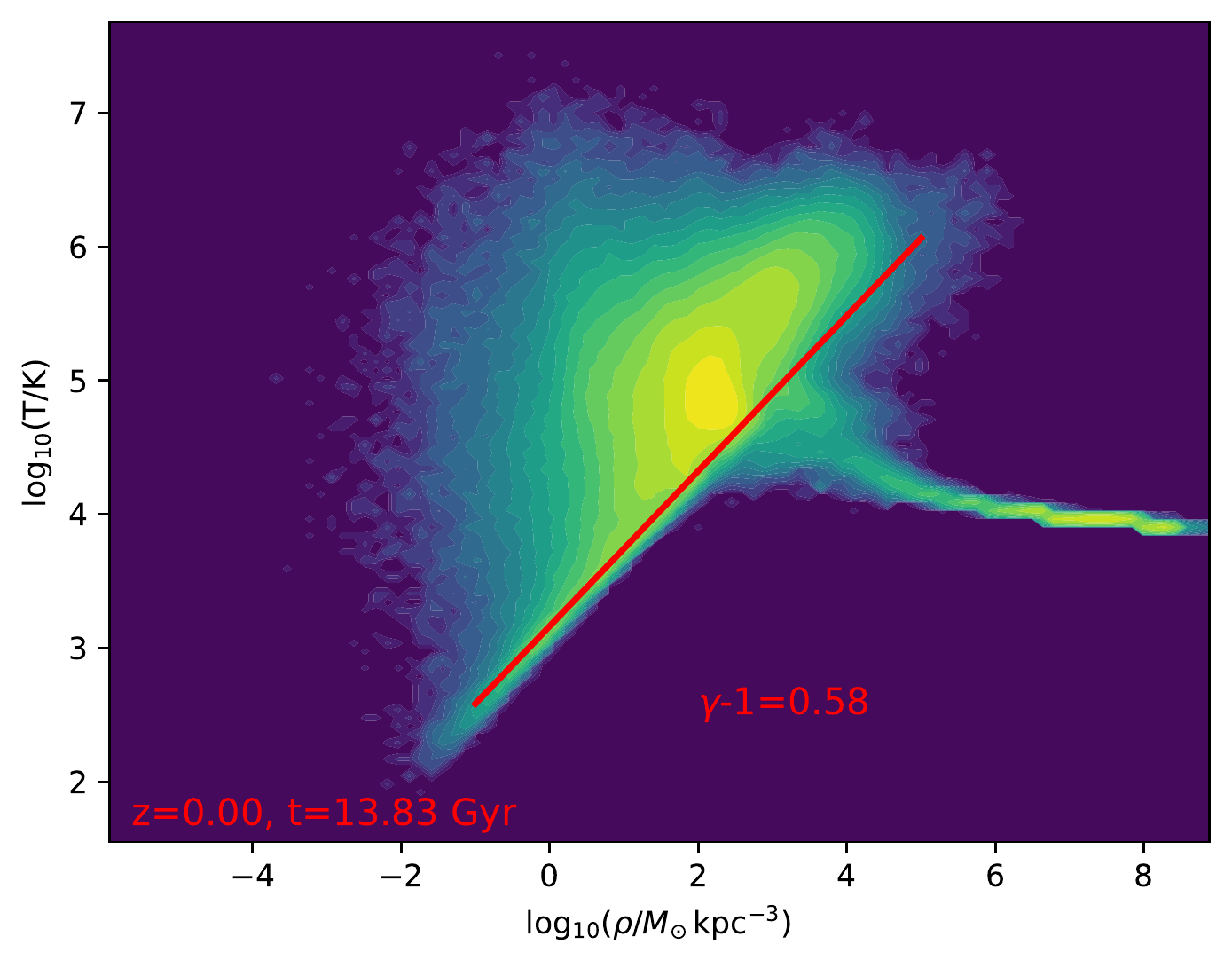}
  \includegraphics[width=0.41\hsize]{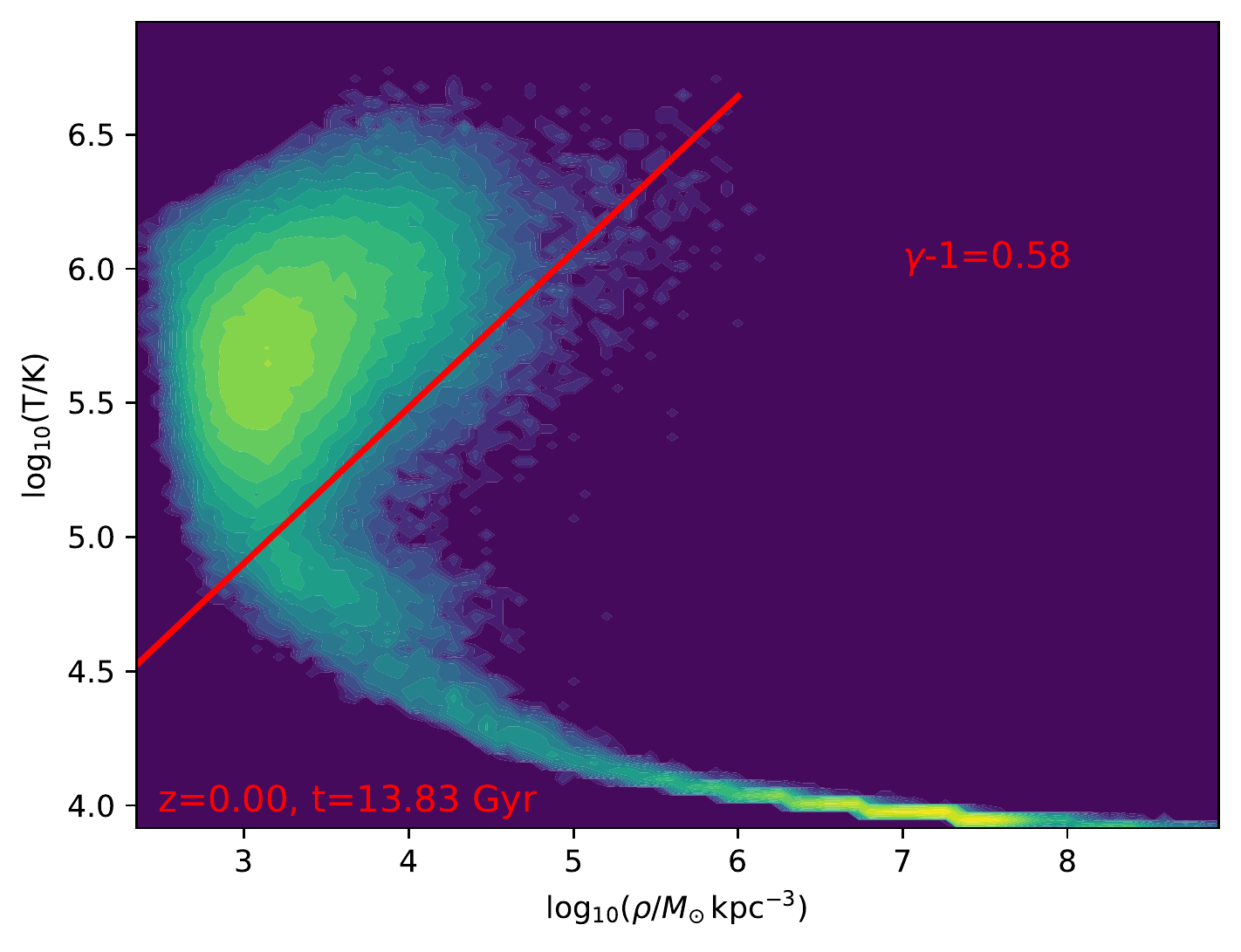}
\end{array}$
\end{center}
\caption{Temperature-density diagram {  for g7.55e11 (simulation without feedback)} at $z=1.77$ (top), $z=0.52$ (middle) and $z=0$ (bottom) for all gas particles (left) and only those at $r<r_{\rm vir}$ (right). 
The IGM is the low-density low-temperature gas that is visible in the left panels and disappears in the right ones. The red lines show the equation of state of the IGM and $\gamma$ is its polytropic index.}
\label{T_rho_diagram}
\end{figure*} 

{  With feedback (supplementary material), shocks are not the only heating mechanism. 
The hot gas appears slightly earlier. Its distribution is less homogeneous, as expected for outflows that are not only anisotropic but also highly turbulent (see, e.g., the X-ray and H$\alpha$ images by \citealp{strickland_etal04}),
and thus shocks are not as prominent on a temperature map. 
At low $z$, the cold gas in galactic discs is more prominent with feedback because its conversion into stars has been less effective.
For the rest, the overall picture is very much the same.}

\section{Phase separation}

In Section~2, we looked at the distribution of the gas in physical space.
Fig.~\ref{T_rho_diagram} shows its distribution on a temperature--density diagram at $z=1.77$, $z=0.52$ and $z=0.00$. 
{  Once again, g7.55e11 without feedback is  the starting point of our analysis, but what follows  also applies to g1.12e12.}

For each $z$, we show two panels: one for all gas particles
within the computational volume (left), the other only for the particles with $r<r_{\rm vir}$ (right).
The gas at low density and temperature is the IGM accreting onto the halo.
The proof is that it disappears when we pass from the panels to the left to those to the right.
 
At  $z\le 3.25$, the effective equation of state of the IGM is a polytrope with $\gamma-1\simeq 0.6$ (red lines) in agreement with forecasts from the asymptotic temperature-density relation \citep{hui_gnedin97}
and observational data. \citet{rudie_etal12} and \citet{bolton_etal14} find
$\gamma-1=0.54\pm 0.11$ at  $z=2.4$, the only redshift at which there are robust observational constraints.
Only at $z\ge 3.60$ does the polytropic index $\gamma$ begin to decrease, as expected when approaching the epoch of cosmic reionisation.
In the NIHAO simulations, cosmic reionisation is modelled based on \citet{haardt_madau12} and occurs at $z\simeq 6.7$.
\citep{onorbe_etal17}.
We find $\gamma-1\simeq 0.3$ at $z=5.27$.

\begin{figure}
\begin{center}
\includegraphics[width=1.00\hsize]{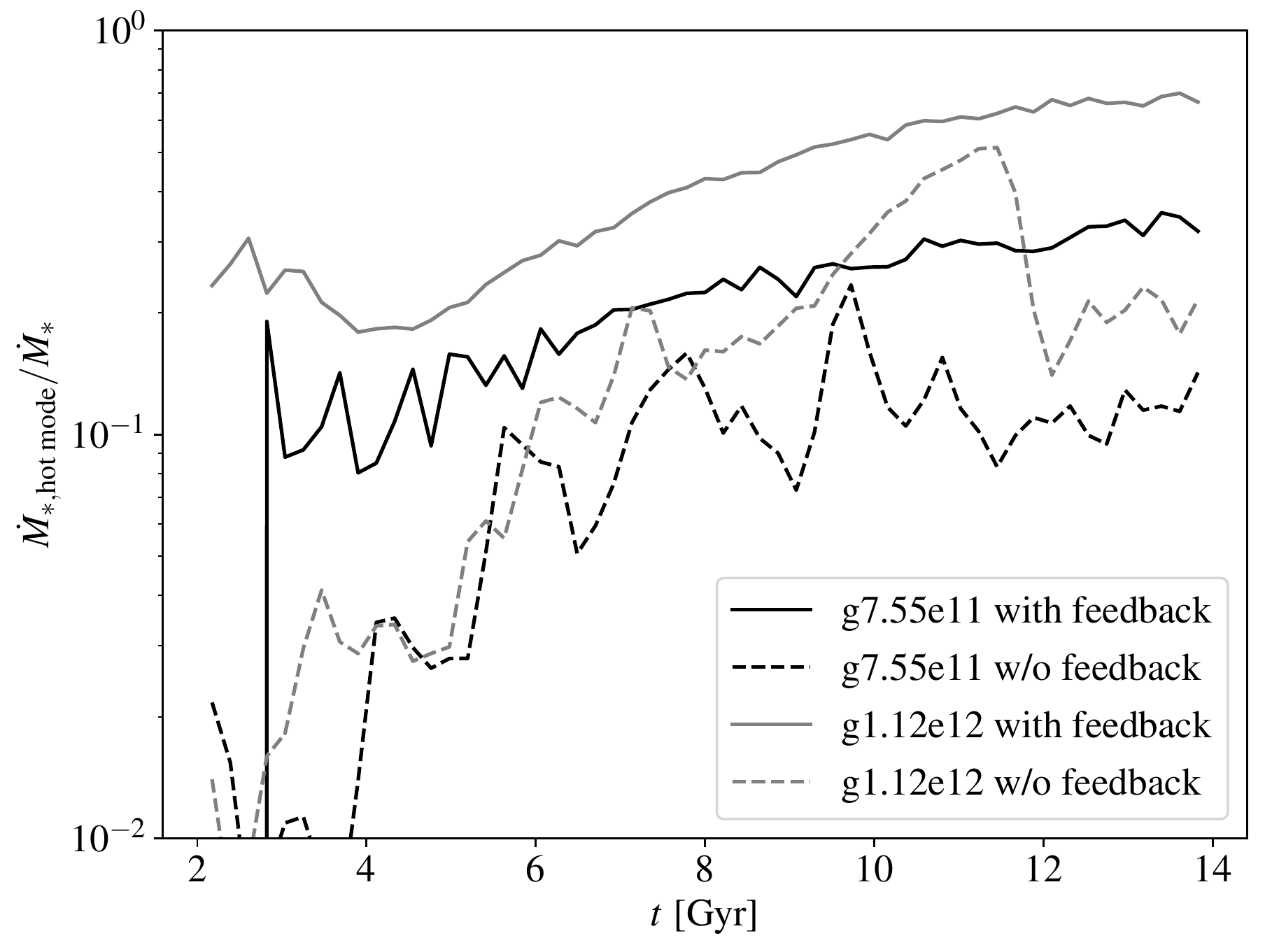} 
\end{center}
\caption{  Fractional contribution of the hot mode (gas that has cooled) to the total  SFR  within $r_{\rm vir}$  with and without feedback (solid and dashed curves, respectively).
The black curves are for g7.55e11. The gray ones are for g1.12e12.}
\label{hot_mode_sf}
\end{figure}

Polytropic indexes lower than the adiabatic one $\gamma=5/3$ are evidence for radiative cooling within the filaments before they enter the virial radius.
Rapid cooling after entering $r_{\rm vir}$ appears as a right turn in
 the trajectory of a particle on the $T$--$\rho$ plane. Shock-heating causes the particle to make a left turn towards higher
 entropies. Fig.~\ref{T_rho_diagram} shows that, within  $r_{\rm vir}$, the higher-entropy gas above the
 polytrope (the red line) and the lower-entropy gas below it form two clearly
 distinct thermodynamic phases. The high-entropy gas is the hot
 CGM. The low-entropy gas comprises the cold CGM and the interstellar
 medium (ISM).
The separation between the ISM and cold CGM is discussed in
 \citet{tollet_etal19} but is irrelevant for this article.

{  In \citet{tollet_etal19}, we have looked at the  $T$--$\rho$ diagram with feedback for the entire NIHAO suite and we have discussed the specific case of g1.12e12 in detail.
The main difference is the appearance of a hot dense phase distinct from and with lower entropy than the hot CGM. This phase corresponds to the hot ISM and hot winds on a galactic scale.
Most of the gas in the hot wind component does not mix with the hot CGM. It cools and falls back onto the galaxy in a galactic fountain.}

The red solid curves in Fig.~\ref{massgrowth} show the mass $M_{\rm hot}$ of the hot CGM.
$M_{\rm hot}(t)$ is the total mass of all gas particles within $r_{\rm vir}$ and above the critical polytrope at $t$.
{  In the case of g7.55e11, two regimes are clearly visible.}
In the first half of the cosmic lifetime (at $t<6\,$Gyr), $M_{\rm hot}$ 
grows rapidly, both in absolute value and as a fraction
 of the total baryonic mass $M_{\rm b}$ within the virial radius. After $t\sim
 6\,$Gyr, $M_{\rm hot}$ stabilises at about a quarter of $M_{\rm b}$.
 {  In the case of g1.12e12, the first regime is less visible because the transition occurs earlier on, at $t=3.47{\rm\,Gyr}$ ($z=1.88$).}
 
 The orange dashed curves in Fig.~\ref{massgrowth} show the total mass of the SPH
 particles that have been in the hot CGM at some time $\le t$. The orange
 dashed curves lie above the red solid curves because there are
 particles that were in the hot CGM at some time $<t$ but are no longer
 there at $t$. Some of these particles cooled and formed stars, but
 most particles that left the hot CGM did not move into the galaxy,
 they moved out of the virial radius.
 
 {  We can prove this by comparing the orange dashed curves with the orange solid curves. The latter show
the mass of the particles  that have been in
 the hot CGM at some time $\le t$ {and} are still in the halo at  $t$. Even without feedback,
 the orange solid curves are at least 30 per cent lower than the orange dashed curves at $z=0$ and over most of the cosmic lifetime.
Many particles that were in the hot CGM at some time $<t$ are no
longer in the halo at $t$ due to post-shock adiabatic expansion (Figs.~\ref{poststamp_images}--\ref{poststamp_images_S2} provide visual confirmation that the shock-heated gas extends beyond $r_{\rm vir}$).
We call this phenomenon {\it spillage}.
The fact that, without feedback, $M_{\rm b}\sim  f_{\rm b}M_{\rm vir}$, even though there is spillage
implies that the integrated accretion rate onto the
halo must be larger than $f_{\rm b}M_{\rm  vir}$. 

The difference between the orange solid curves and the red solid curves is smaller. Few particles left the hot CGM through cooling. Without feedback,
their collective mass is about 5 per cent of the mass of the hot CGM at $z=0$ for both g7.55e11 and g1.12e12.

We can test the contribution of cooling in the hot CGM to the formation of galaxies more directly
by measuring the total star-formation rate  (SFR) $\dot{M}_\star$ within $r_{\rm vir}$ and by comparing
it to the SFR $\dot{M}_{\rm \star, hot\,mode}$ from gas particles that have been in the hot CGM at some previous timestep (Fig.~\ref{hot_mode_sf}).
Let us start from the case without feedback (dashed curves).
In g7.55e11, $\dot{M}_{\rm \star, hot\,mode}/\dot{M}_\star\lsim 0.03$ at $t\sim 5\,$Gyr and  $\dot{M}_{\rm \star, hot\,mode}/\dot{M}_\star\sim 0.1$ at $t\gsim 8\,$Gyr.
In g1.12e12, cooling plays a bigger role and accounts for 20 (instead of 10) per cent of the total SFR at $z=0$.
This finding conforms to the expectation that cooling is more important at higher masses,
even though, with only two galaxies, it is impossible to say how much this is due to a trend with $M_{\rm vir}$ and how much it is random scatter from one object to another.

With feedback, most of the baryons that accrete onto galaxies through cold flows do not immediately form stars.
A lot are ejected. Some mix with the hot CGM and then cool down again.
SN-heated baryons can also cool locally in the ISM. These routes are easily separated because the hot CGM and the hot ISM are different thermodynamic phases.
Both have high temperature, but the hot CGM is  less dense  and has thus higher entropy than the hot IGM. 
As our definition of hot and cold is based on entropy rather than temperature, the hot ISM is not hot in our analysis.
Gas that is heated and cools locally within the galaxy is treated as if it were cold at all time and rightly so, since our focus is on cooling in the hot CGM\footnote{Heating and cooling within the ISM modifies the efficiency of star formation \citep{tollet_etal19}, but that is beyond the scope of the current article.}.

The black solid curves in Fig.~\ref{hot_mode_sf} show the impact of SN feedback on the fraction of star-forming baryons that have been in the hot CGM. In g7.55e11, SNe} 
increase the contribution of cooling to the total SFR increases from  $\sim 3$ to $\sim 10$ per cent at high $z$,
and from  $\sim 10$ to  $\sim 30$ per cent at low $z$.
The integrated stellar mass formed via in-situ cooling grows from less than 7 per cent without feedback to $\sim 20$ per cent with feedback.

{  In g1.12e12, SNe increase the contribution of cooling to the total SFR at $z=0$ from $\sim 20$ per cent without feedback to nearly 70 per cent with feedback.
The integrated stellar mass formed via in-situ cooling grows from 10--20 per cent  of the final stellar mass without feedback to nearly 40 per cent with feedback.

These numbers should be taken with caution. Not only do they fluctuate from one object to another, they also depend on the implementation of SNe in the NIHAO simulations.
However, the picture that emerges from them is robust. 
Without feedback, cooling is negligible at high redshift. Even at low redshift, it contributes to a small of the total SFR.
With feedback, cooling can contribute significantly to the SFR of massive spirals, but that occurs through cooling of gas heated by SNe, and not to through the
cooling of shock-heated gas.

Our definition of the hot CGM  is purely thermodynamic.}
The description of the hot gas in SAMs  is based on the assumption that it can be approximated with a spherical quasi-hydrostatic atmosphere
(see \citealp{somerville_dave15} and \citealp{knebe_etal15} for an overview of the semianalytic approach and the main SAMs, respectively).
As the comparison with SAMs is an important aspect of our work (Section~5), it is important to check that the two pictures are consistent.
{  We present our analysis for g.755e11 without feedback, but we have also looked at the other simulations and the conclusions are very similar.}

We can verify that the hot gas is quasi-hydrostatic by measuring  the mean bulk speed $|{\bf  v}|$ and the mean sound speed $c_{\rm s}$ of the hot CGM's SPH particles (both mass-weighted within the virial sphere), and by comparing them
to the virial velocity $v_{\rm vir}$ of the DM halo (Fig.~\ref{characteristic_speeds}).
For a hydrostatic singular isothermal sphere, ${\bf  v}=0$ everywhere and $c_{\rm s}^2=(5/6)v_{\rm vir}^2$.
For the \citet{komatsu_seljak01} solution (the exact solution for a polytropic gas in hydrostatic equilibrium with the gravitational potential of an NFW halo), the value of $c_{\rm s}^2$ is about twice $v_{\rm vir}^2$ at the centre
(the precise factor depends on the concentration of the DM halo) and decreases at larger radii.
Therefore, hydrostatic equilibrium requires $c_{\rm s}\sim v_{\rm vir}$ and $\langle{|{\bf  v}|}\rangle\ll v_{\rm vir}$.
Fig.~\ref{characteristic_speeds} confirms that $c_{\rm s}$ (red solid curve) and $v_{\rm vir}$ (black dashed curve) are indeed comparable and shows that the evolution of $c_{\rm s}$ follows that of $v_{\rm vir}$.
{  We also remark that $c_{\rm s}$ is much more stable than $\langle{|{\bf  v}|}\rangle$. That is even truer for g1.12e12, where the red curve is almost flat.}

Comparable does not mean that $c_{\rm s}$ attains the values required for a perfectly hydrostatic configuration.
The blue solid curve ($\langle{|{\bf  v}|}\rangle$) shows that bulk motions are not negligible, especially during mergers, where $\langle{|{\bf  v}|}\rangle\sim c_{\rm s}$.
Indeed, at the second peak of the blue curve ($t=5.63\,$Gyr, $z=1.06$), the hot CGM does not look like a static spherical atmosphere (Fig.~\ref{poststamp_images}).
The ratio of the bulk speed to the sound speed decreases at $t>6\,$Gyr. However, even at $z=0$, where the hot CGM is approximately hydrostatic,
 turbulence accounts for almost one third of the total pressure. We have verified that the simulations with feedback return a very similar picture.
 In conclusion, hydrostatic equilibrium is a spherical-cow assumption: reasonable, as long as one understands its limitation.
 
For comparison, \citet{faerman_etal17,faerman_etal21} probed the hot CGM of the Milky Way with oxygen lines and found that the thermal pressure at the Solar radius accounts for only a third of the total pressure required for hydrostatic equilibrium.
Assuming hydrostatic equilibrium, they proposed a model in which the pressure support is one third thermal, one third turbulent, and the last third from magnetic fields and cosmic rays (on which we cannot comment, since they are not
included in our work).

\begin{figure}
\begin{center}$
\begin{array}{c}
\includegraphics[width=0.95\hsize]{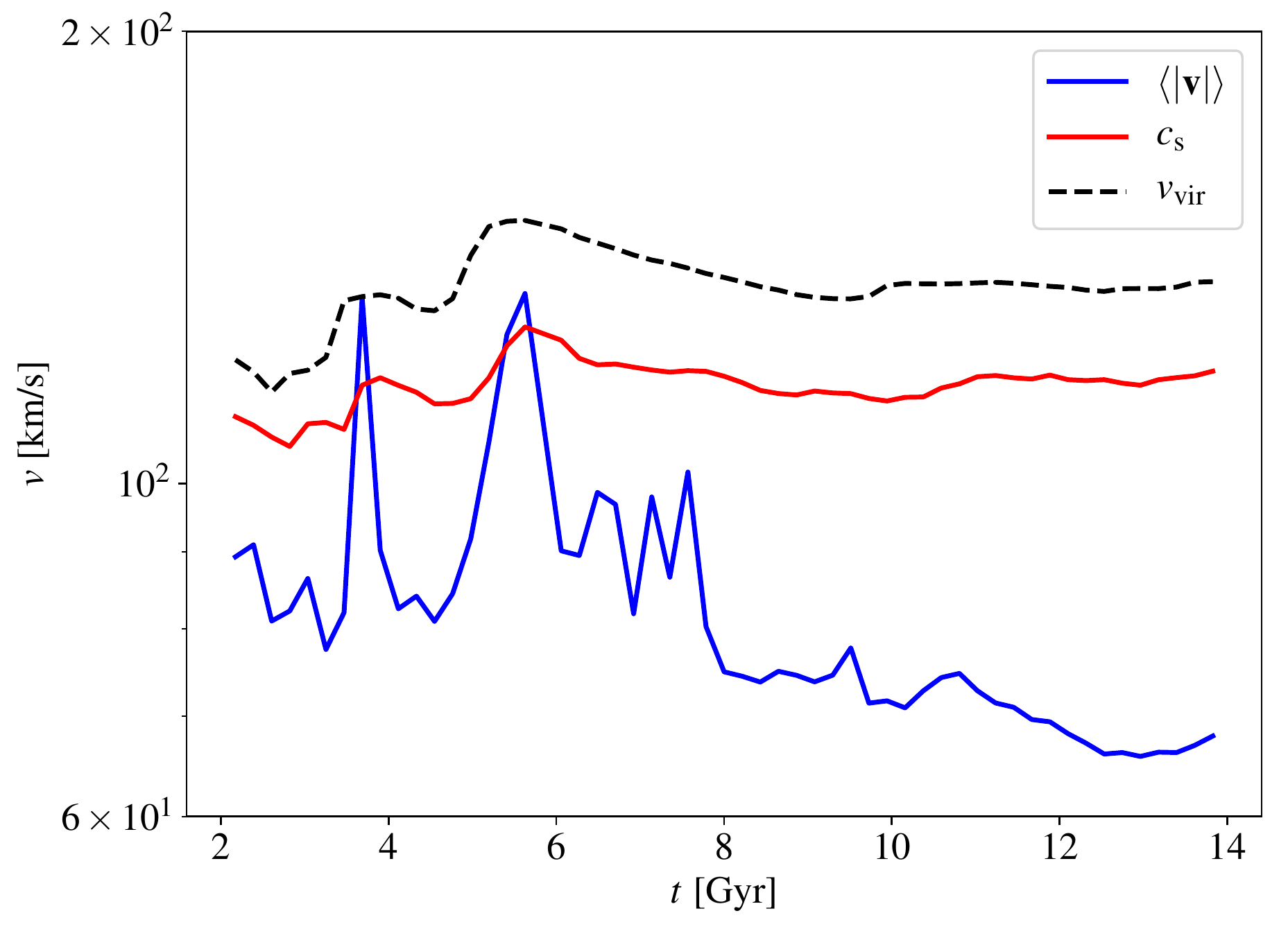} 
\end{array}$
\end{center}
\caption{Evolution with cosmic time $t$ of the virial velocity $v_{\rm vir}$ (black), the speed of sound $c_{\rm s}$ (red) and the average gas speed $\langle{|{\bf  v}|}\rangle$ (blue);
$c_{\rm s}$ and $\langle{|{\bf  v}|}\rangle$ are mass-weighted averages over all the particles in the hot CGM. {  This figure is for g7.55e11 without feedback,
but  the other simulations return a similar picture.}}
\label{characteristic_speeds}
\end{figure}

\begin{figure*}
\begin{center}$
\begin{array}{ll}
\includegraphics[width=0.45\hsize]{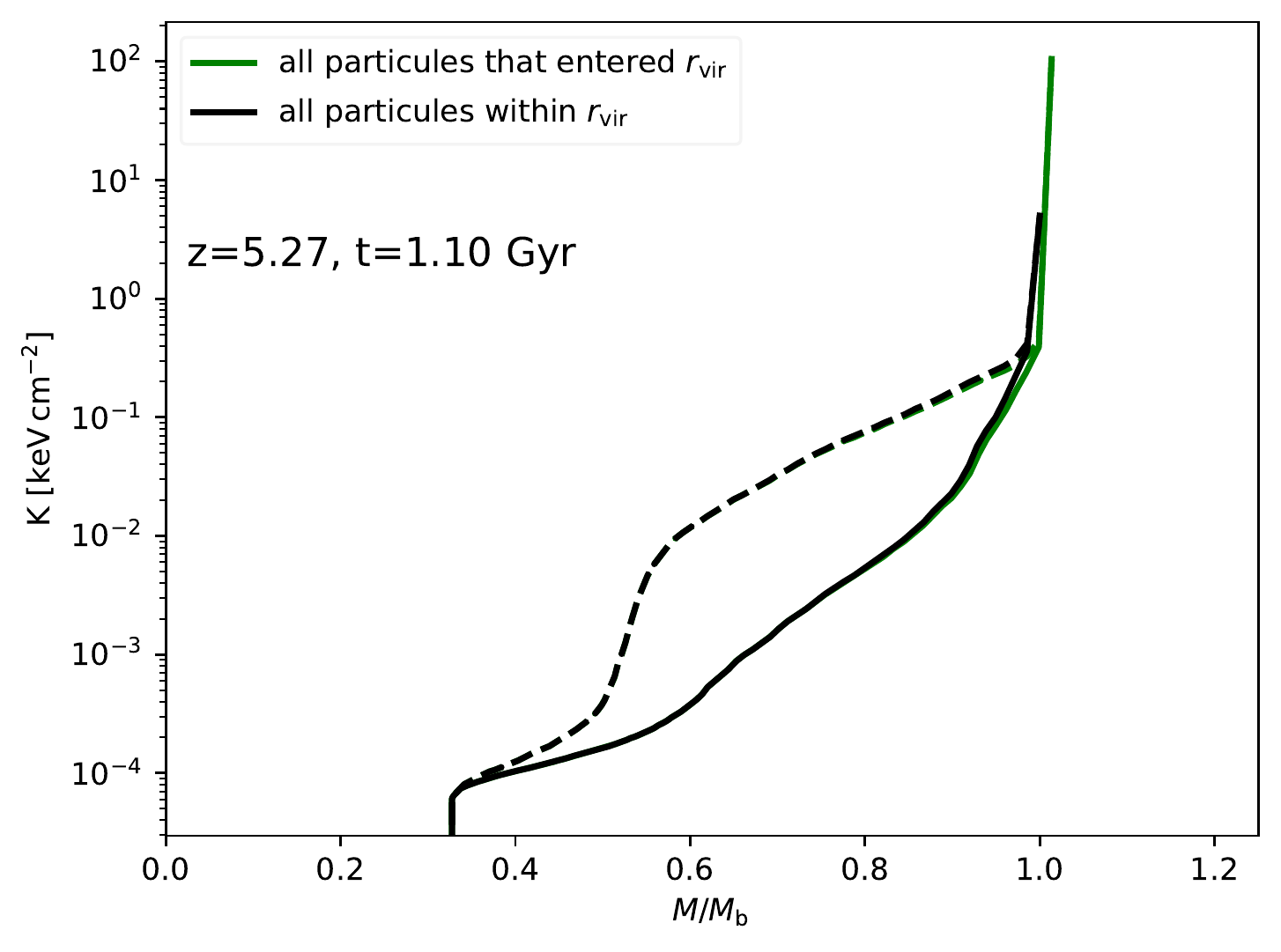}
\includegraphics[width=0.45\hsize]{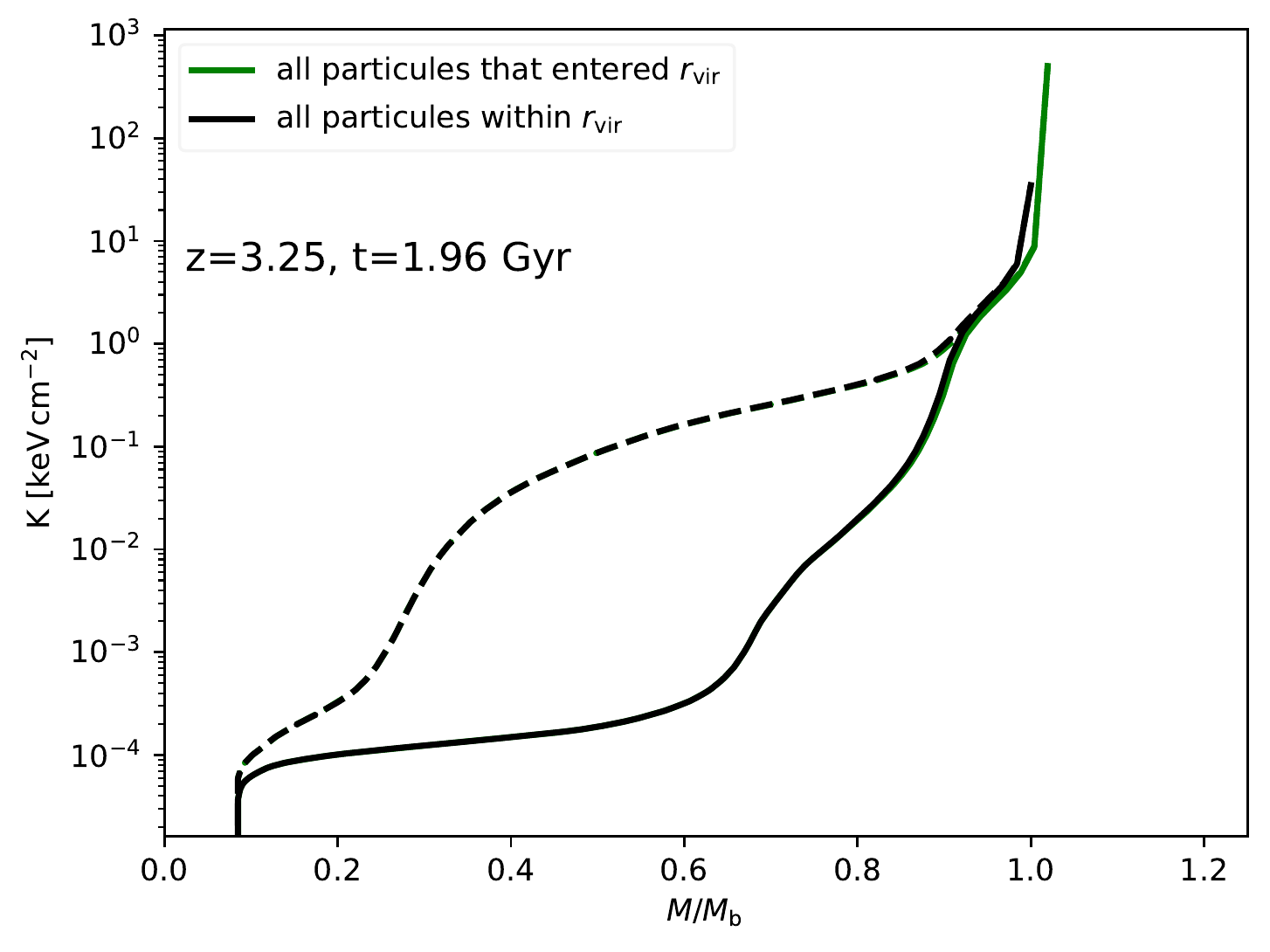}\\
\includegraphics[width=0.45\hsize]{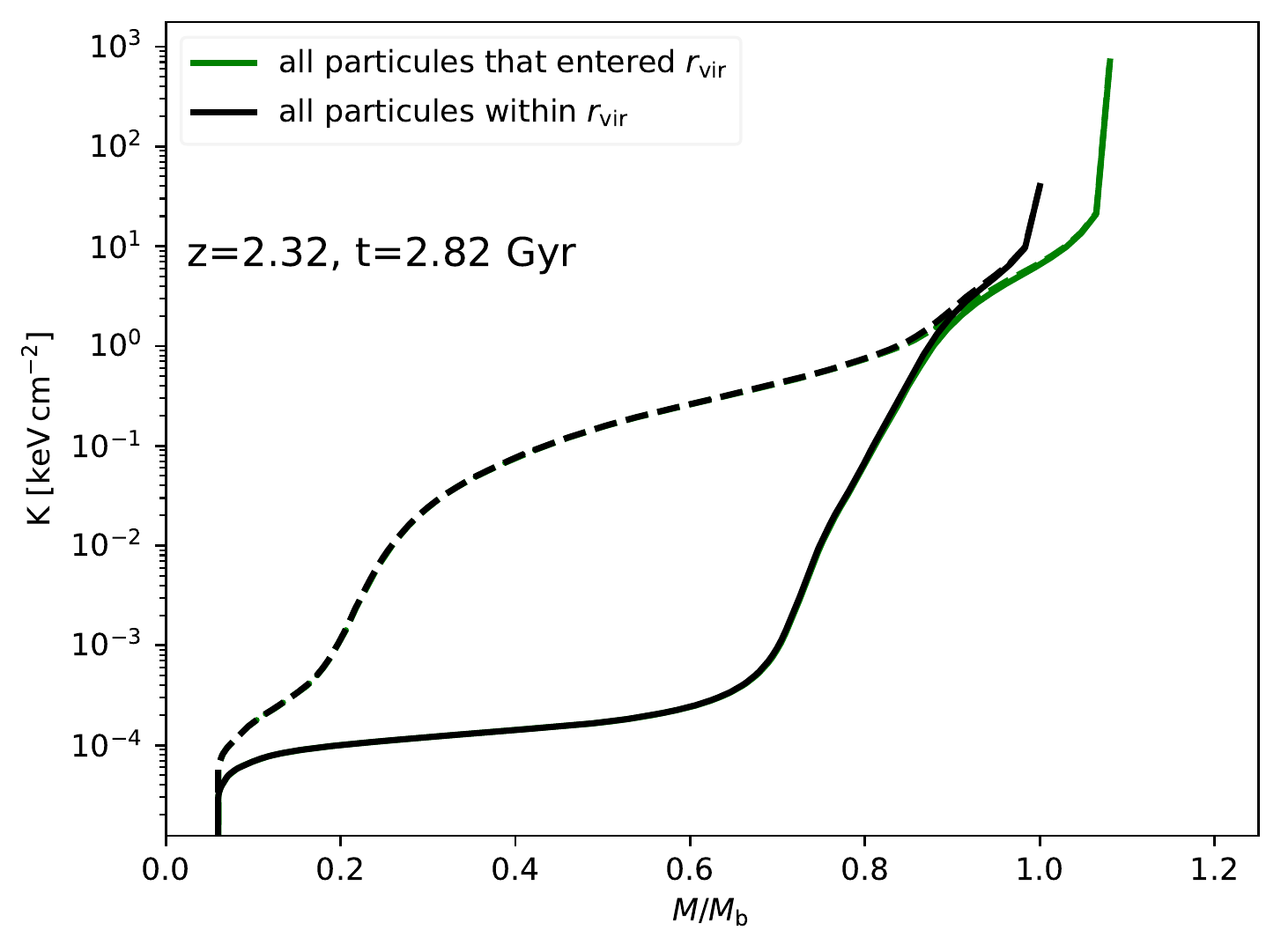}
\includegraphics[width=0.45\hsize]{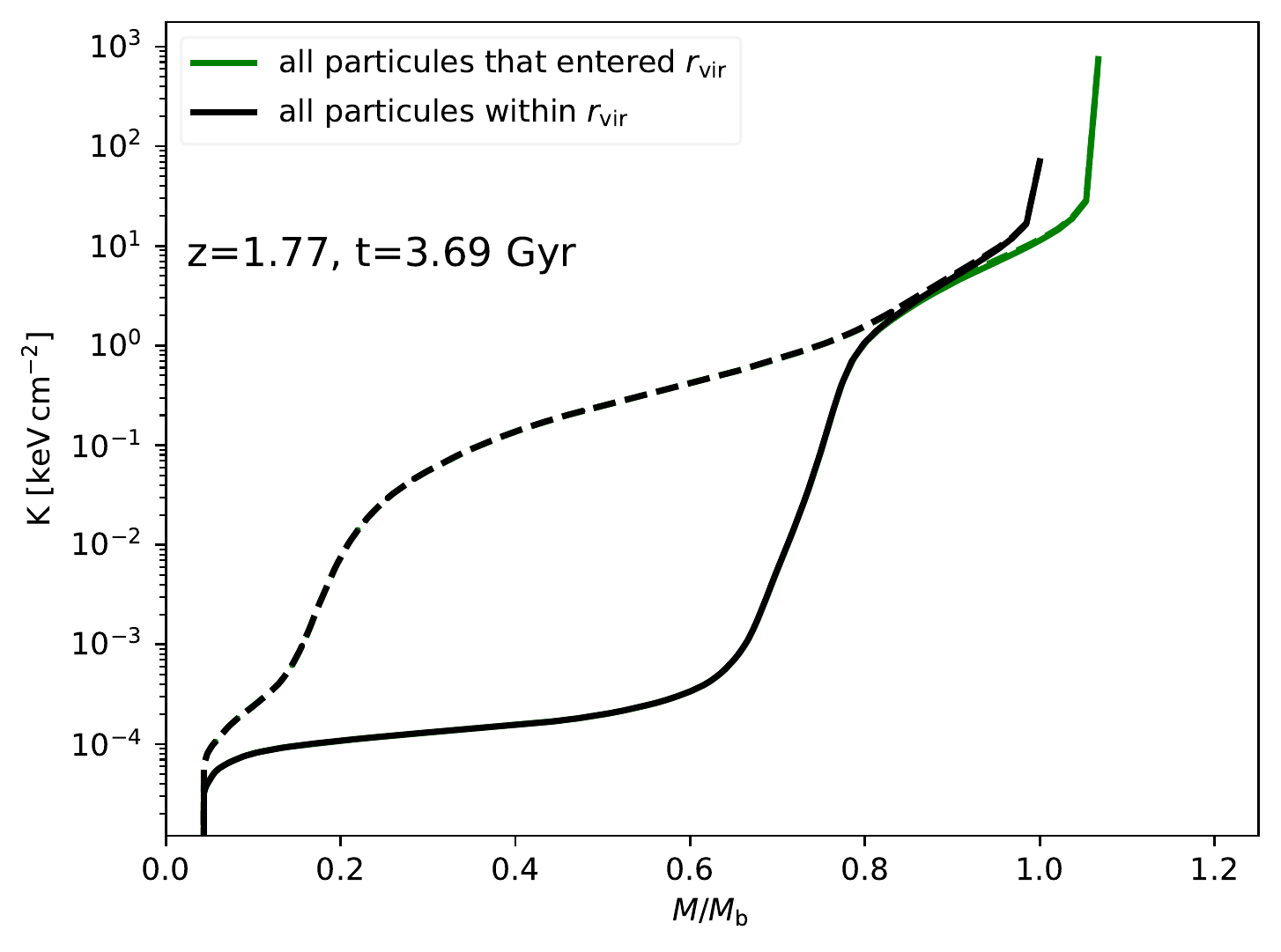}\\
 \includegraphics[width=0.45\hsize]{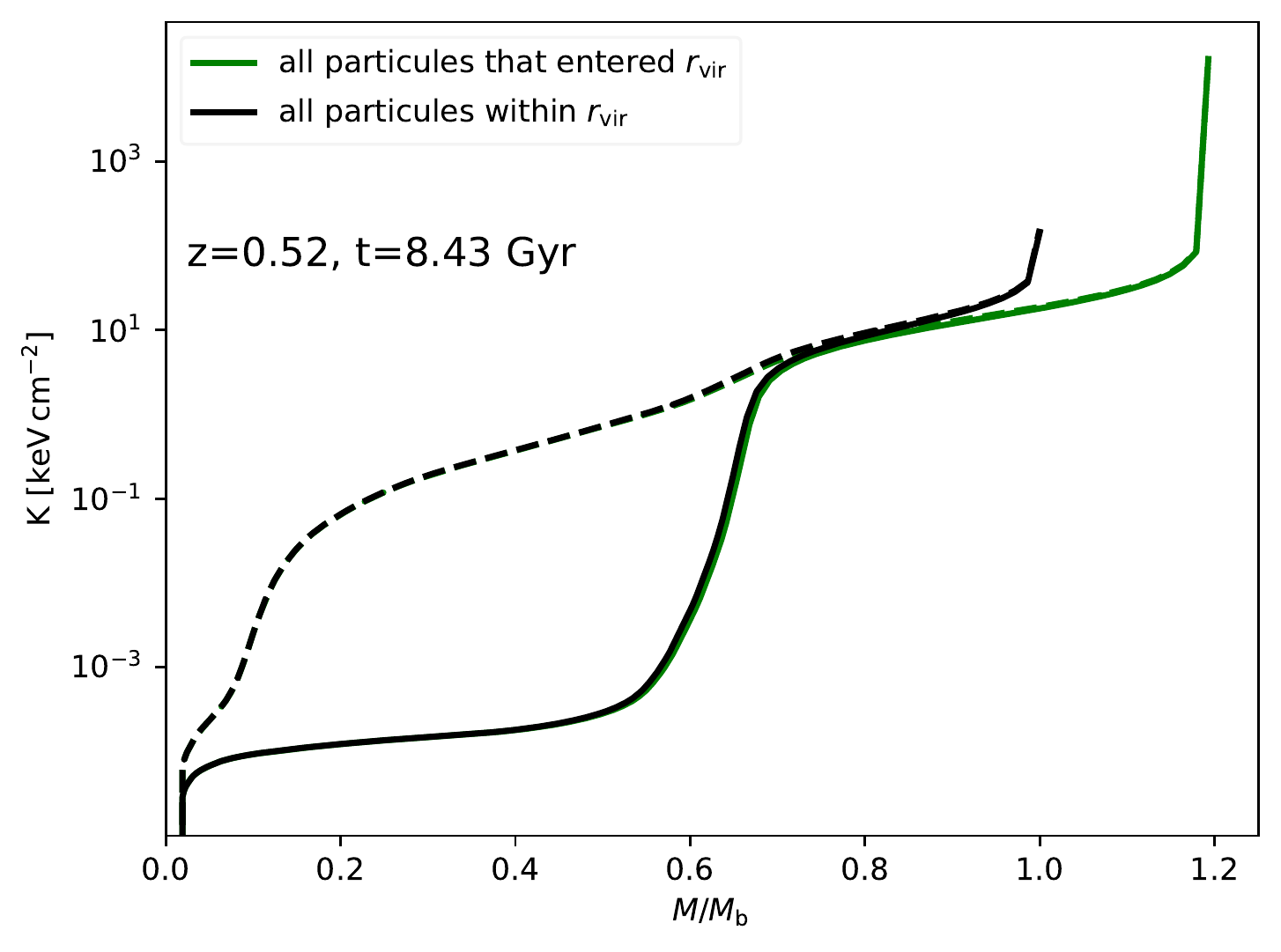}
 \includegraphics[width=0.45\hsize]{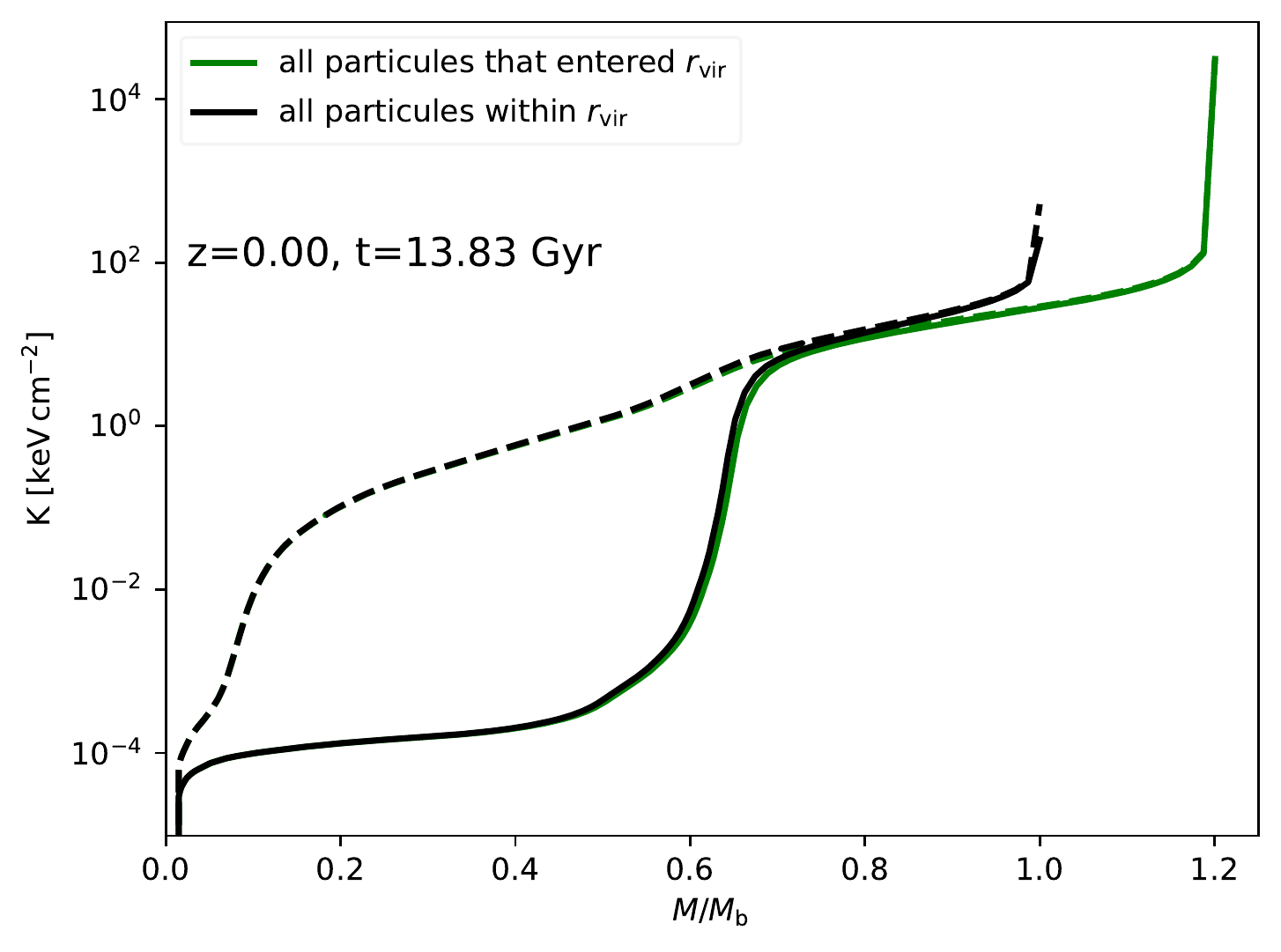}
 \end{array}$
\end{center}
\caption{{  g7.55e11 without feedback:} entropy distribution $K(M)$ (solid curves) and maximum entropy distribution $K_{\rm max}(M)$ (dashed curves) for all particles within $r_{\rm vir}$ at redshift $z$ (black curves) and
all particles that have been within $r_{\rm vir}$ at some redshift $\ge z$ (green curves). Baryonic masses are shown relative to the total mass $M_{\rm b}$ of all baryonic particles within
$r_{\rm vir}$ at redshift $z$.}
\label{ent_dist1}
\end{figure*} 
 
\begin{figure*}
\begin{center}$
\begin{array}{ll}
 \includegraphics[width=0.45\hsize]{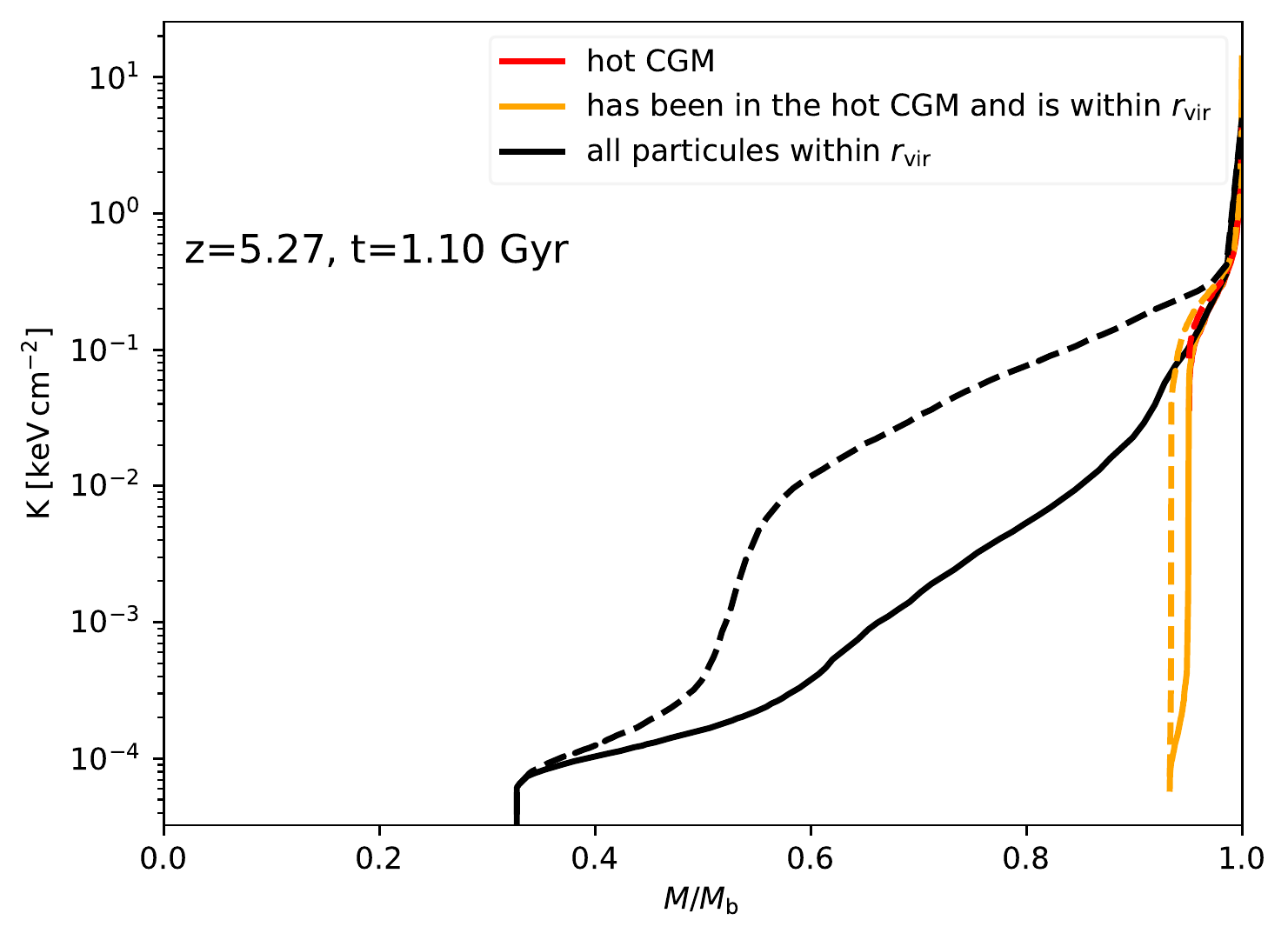}
  \includegraphics[width=0.45\hsize]{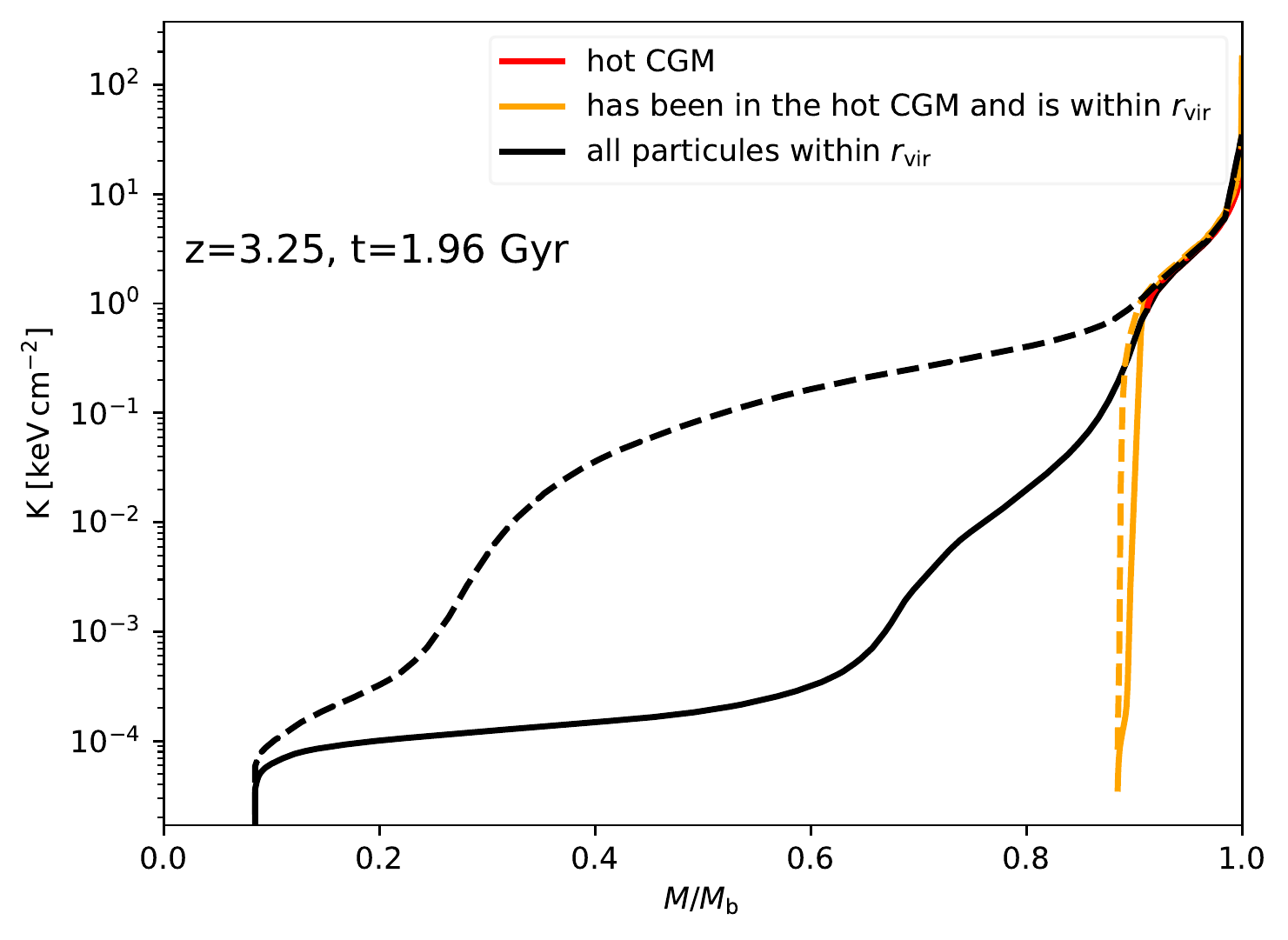}\\
\includegraphics[width=0.45\hsize]{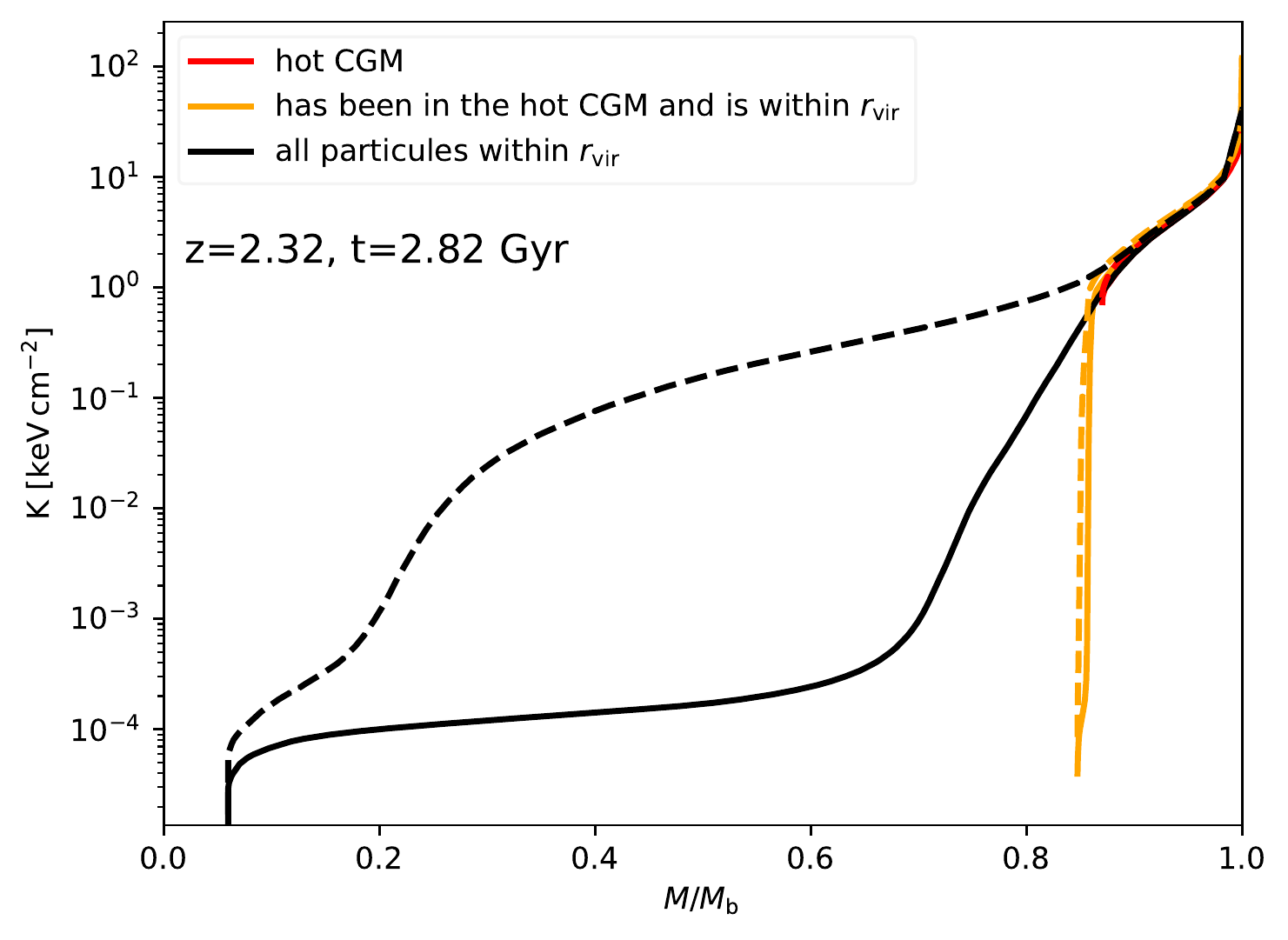}
  \includegraphics[width=0.45\hsize]{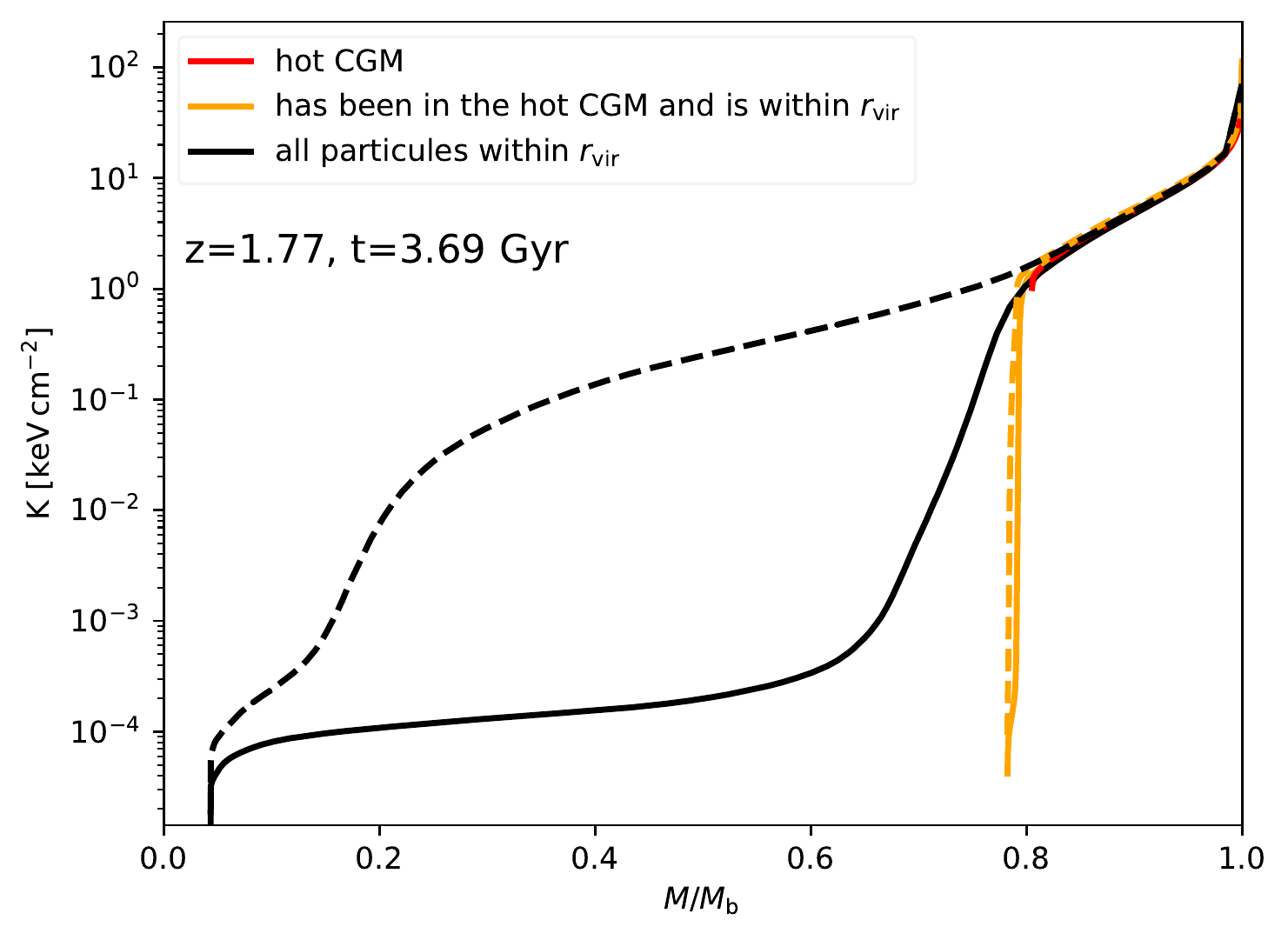}\\
  \includegraphics[width=0.45\hsize]{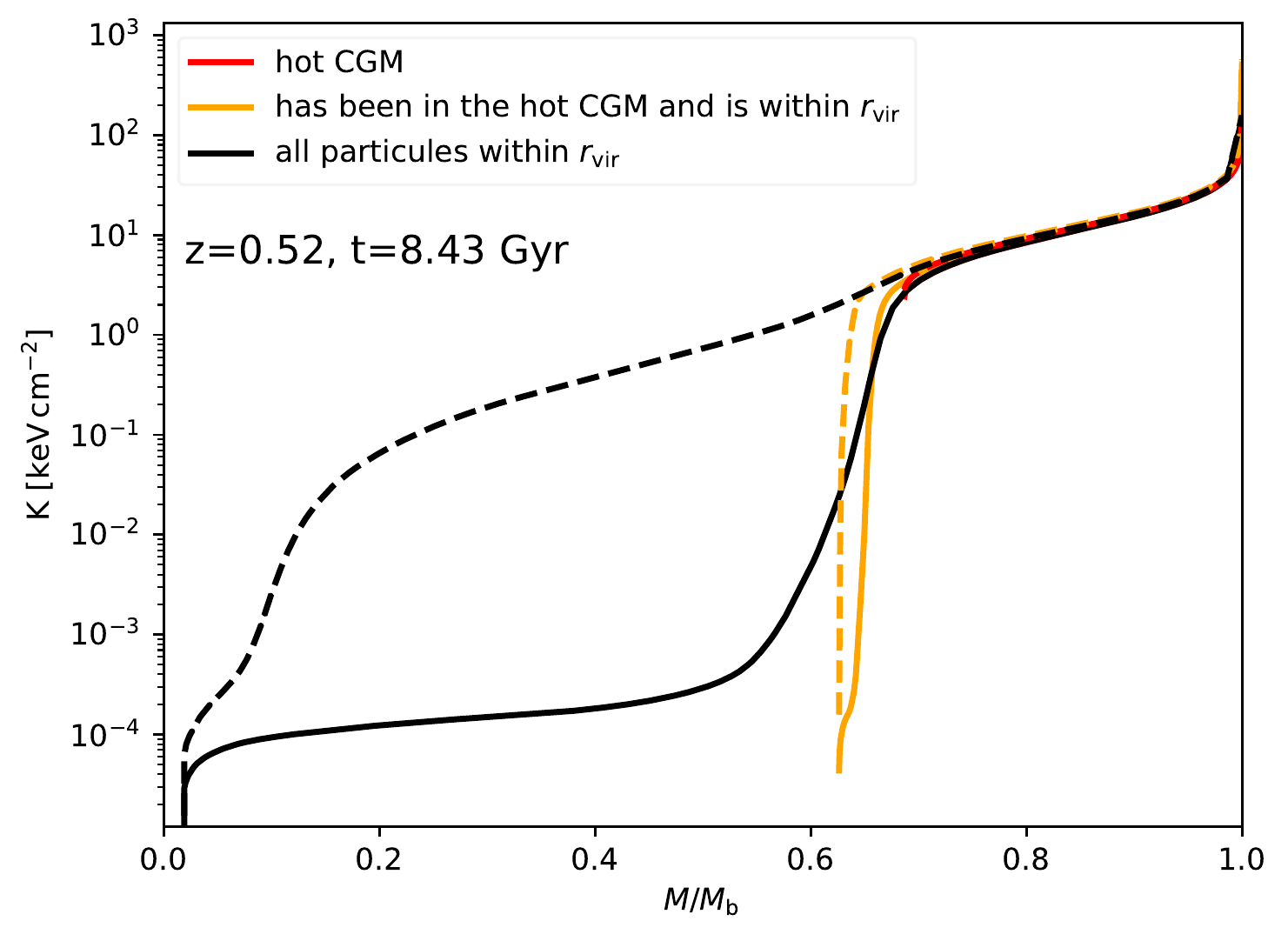}
  \includegraphics[width=0.45\hsize]{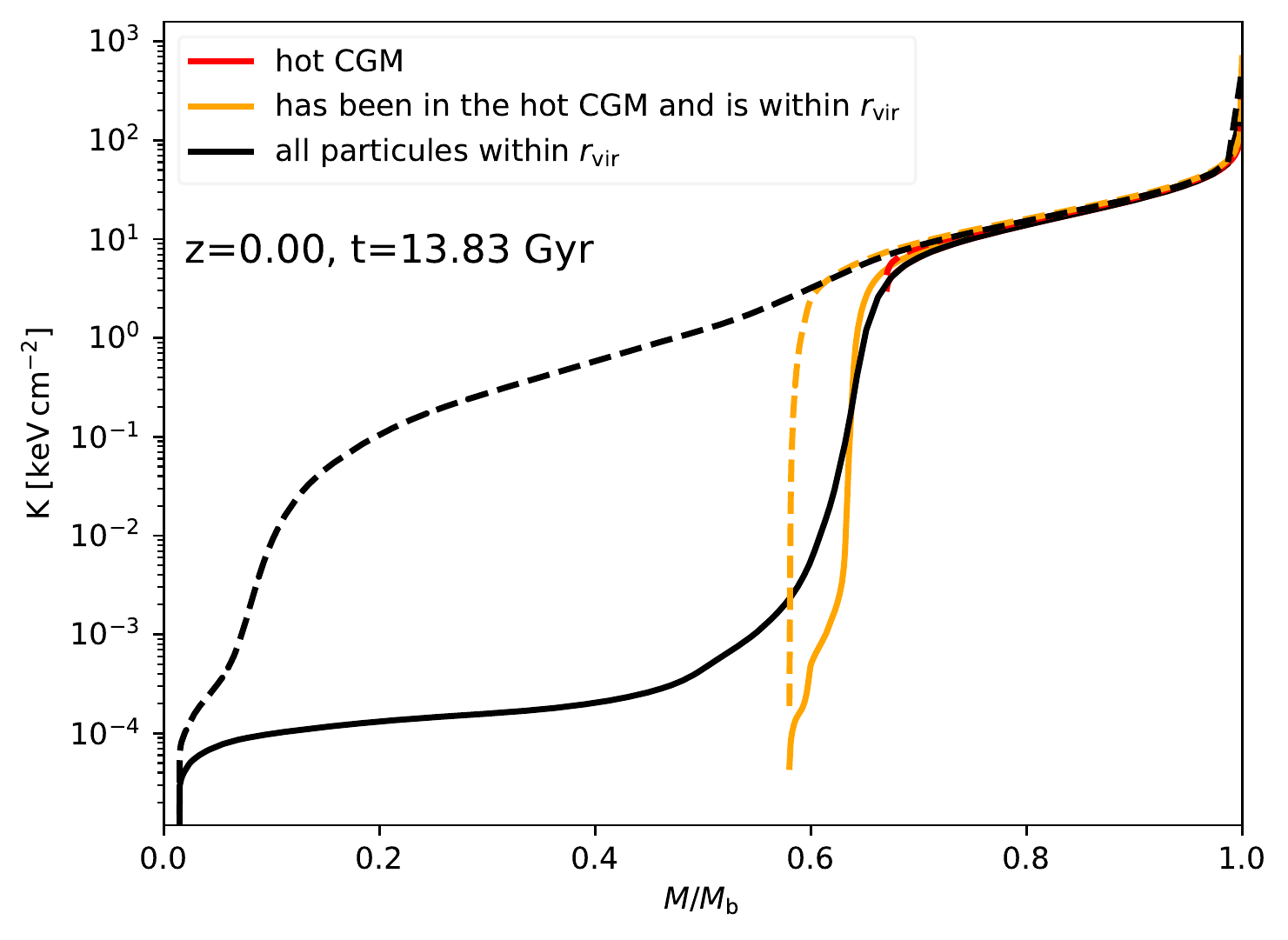}
\end{array}$
\end{center}
\caption{{  g7.55e11 without feedback:} entropy distribution $K(M)$ (solid curves) and maximum entropy distribution $K_{\rm max}(M)$ (dashed curves) for all particles within $r_{\rm vir}$ at redshift $z$ (black curves),
all particles within $r_{\rm vir}$ that have been in the hot CGM at some redshift $\ge z$ (orange curves) and all particles within $r_{\rm vir}$ that are in the hot CGM at  redshift $z$ (red curves).}
\label{ent_dist2}
\end{figure*} 

\begin{figure*}
\begin{center}$
\begin{array}{ll}
 \includegraphics[width=0.45\hsize]{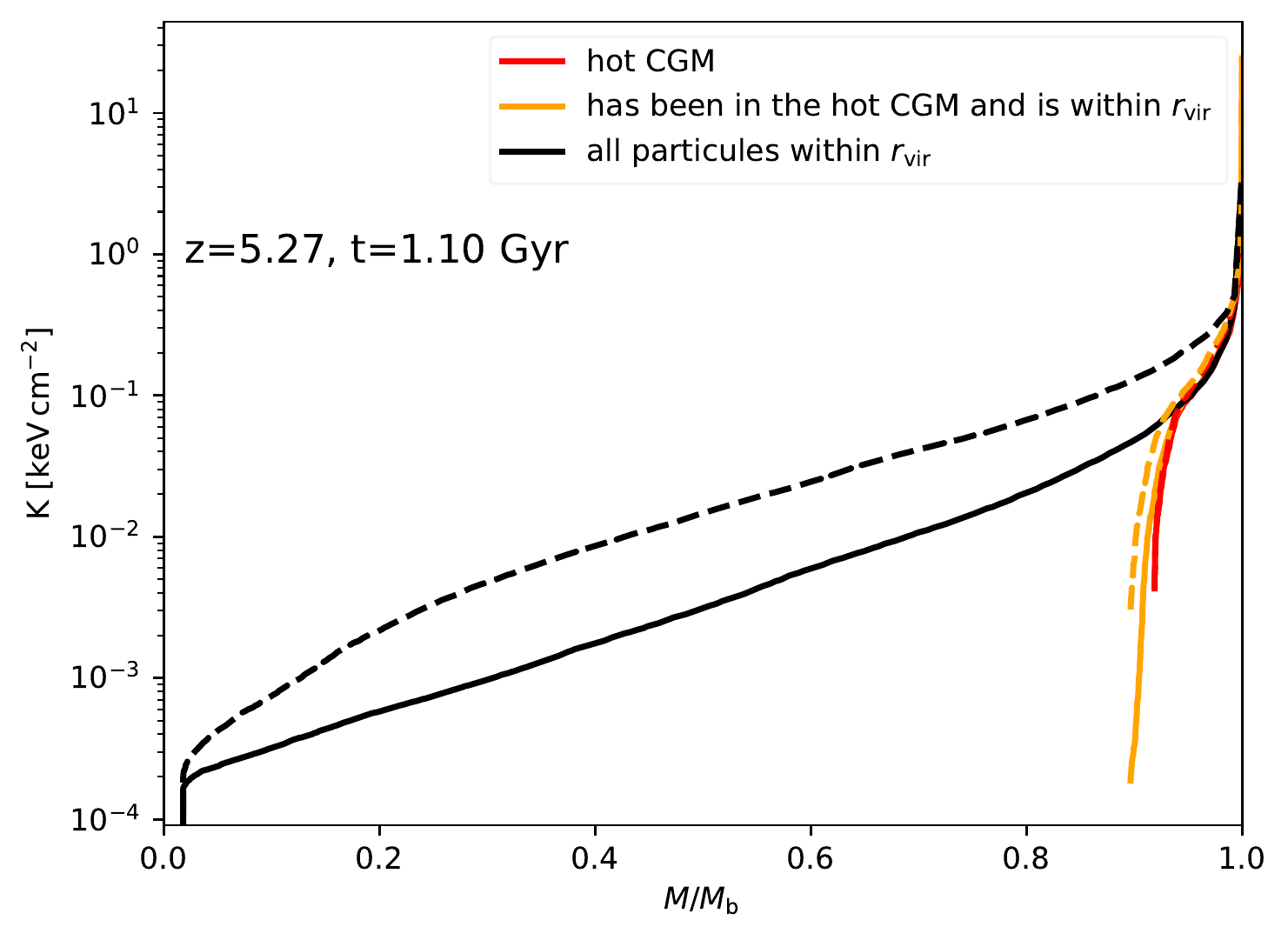}
  \includegraphics[width=0.45\hsize]{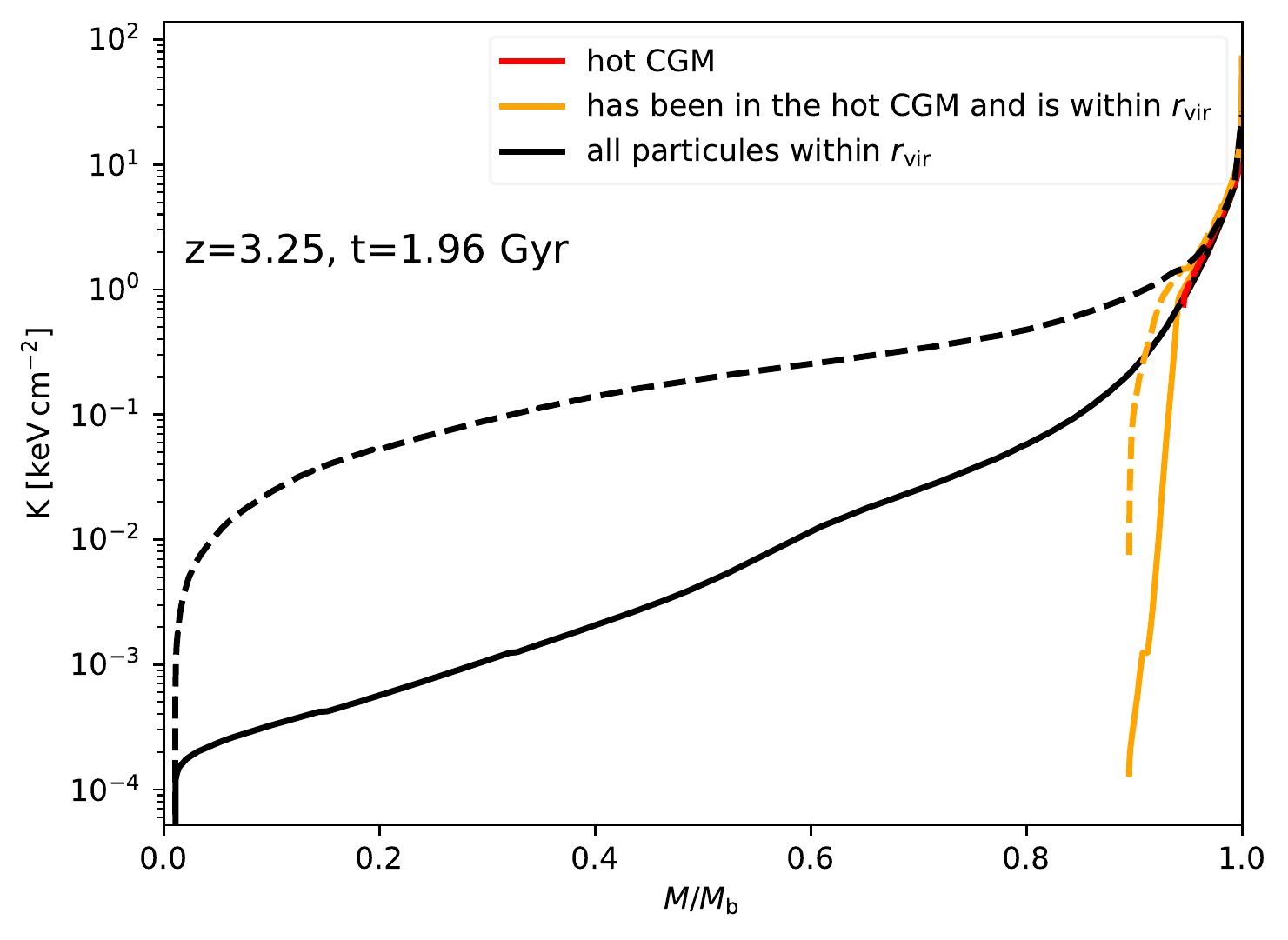}\\
\includegraphics[width=0.45\hsize]{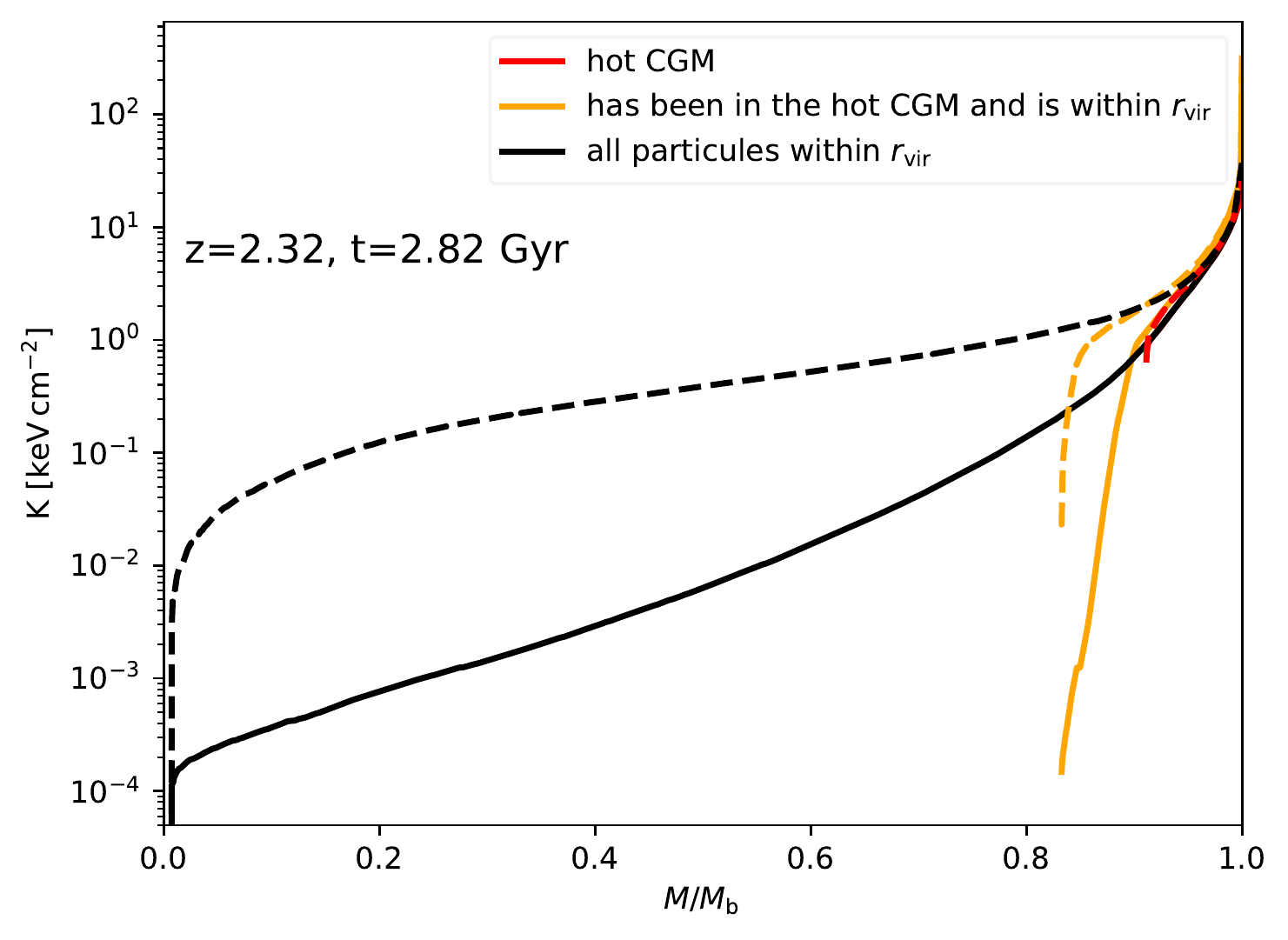}
  \includegraphics[width=0.45\hsize]{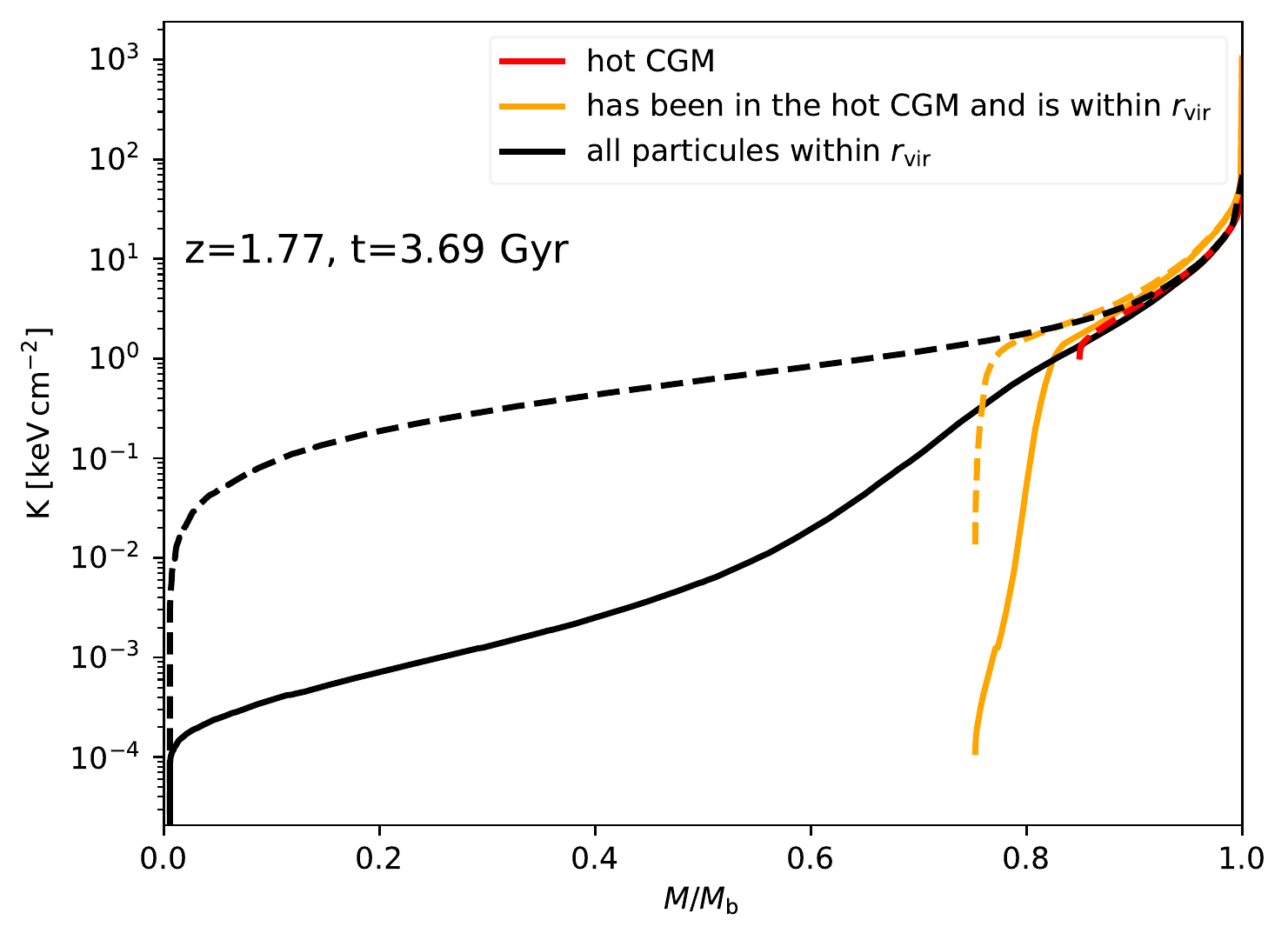}\\
  \includegraphics[width=0.45\hsize]{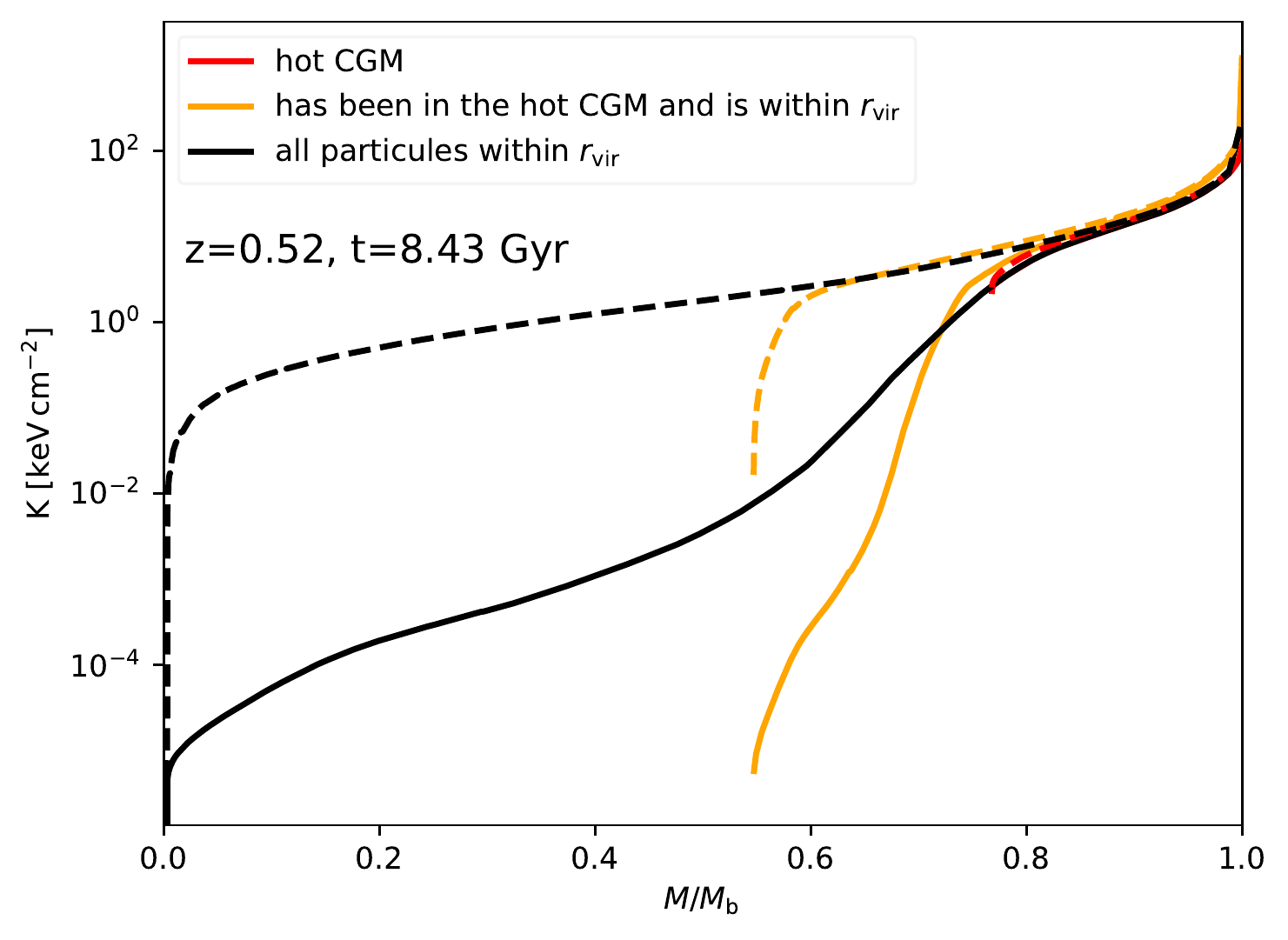}
  \includegraphics[width=0.45\hsize]{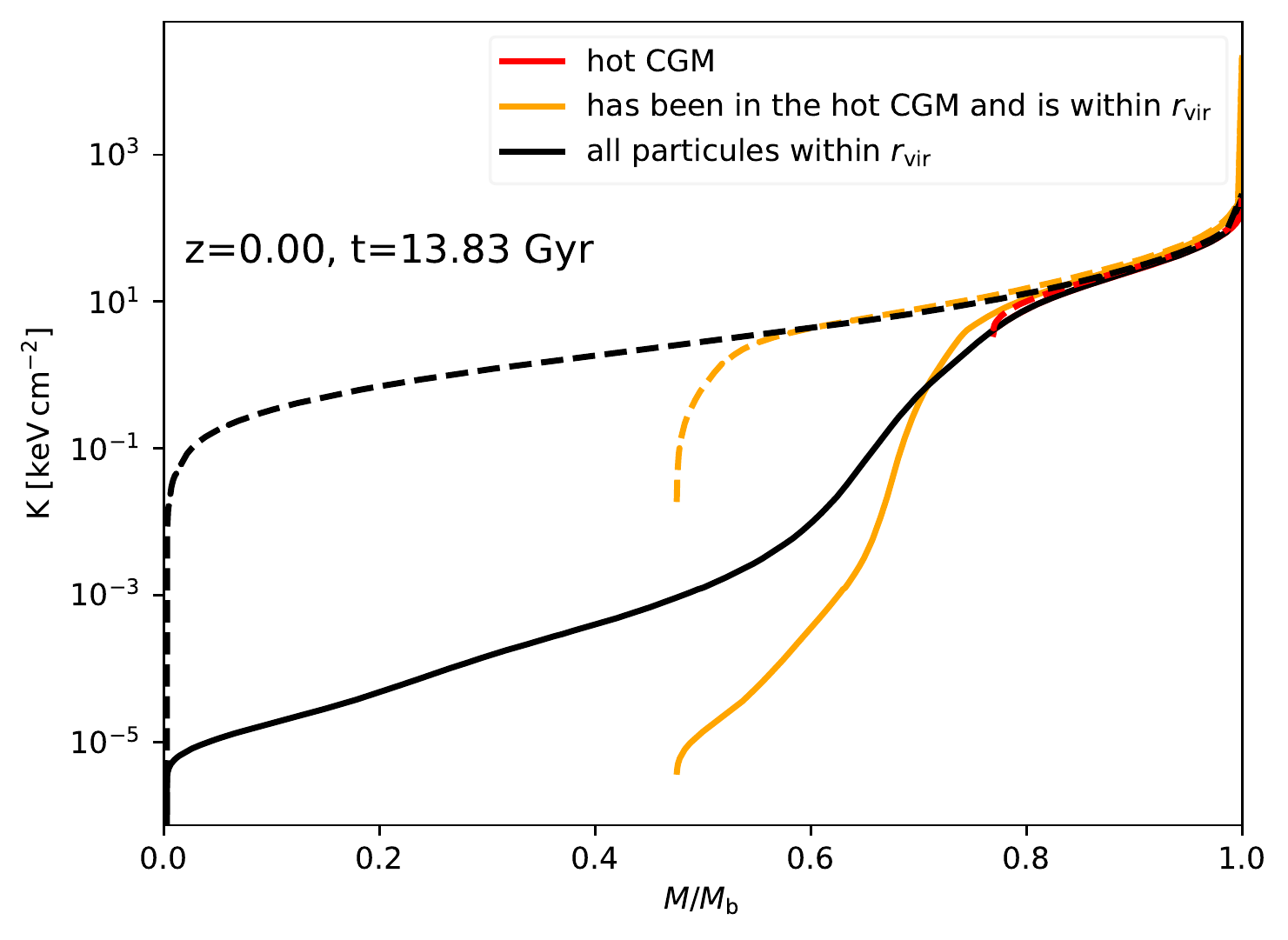}
\end{array}$
\end{center}
\caption{The same as Fig.~\ref{ent_dist2} but this time with SN feedback.}
\label{ent_dist3}
\end{figure*}

\section{Entropy distribution}

The thermodynamic entropy per particle of a fully ionised monatomic gas is
\begin{equation}
s= {3\over 2}k\ln K,
\label{thermo_s}
\end{equation}
where
\begin{equation}
  K=kTn_{\rm e}^{-{2\over 3}},
\label{K_const}  
\end{equation}
$k$ is the Boltzmann constant, $T$ is the temperature and
$n_{\rm e}$ is the electron density, which is proportional to the
total number density of particles ($s$ is defined up to an additive
constant). For a
plasma that is three quarters
  hydrogen and one quarter helium in mass:
\begin{equation}
  n_{\rm e}={14\over 16}{\rho\over m_{\rm p}},
  \label{ne}
\end{equation}
where $\rho$ is the mass density and $m_{\rm p}$ is the proton mass
(12 hydrogen atoms and 1 helium atom make 14 electrons and 16 baryons).
The quantity $K$ in  Eq.~(\ref{K_const}) is often referred to as the entropy in
astrophysics.
We  follow this definition in this article, even though it is not
the standard one in thermodynamics, and use Eq.~(\ref{K_const}) to assign an entropy to SPH gas particles
(Eq.~\ref{thermo_s} give the thermodynamic entropy of a real physical
particle, that is, an electron or an atomic nucleus, assuming equipartion).
 
 Stellar SPH particles have no temperature. Hence, we cannot use Eq.~(\ref{K_const}) to compute their entropies.
 We could use the entropy of the gas out of which they formed, but the outputs of the simulations do not contain the information required to track which gas particle a stellar particle came from.
 A statistical approach is nonetheless possible and statistics are all we care about.
 At each output, we know how many gas particles have disappeared and how many stellar particles have formed.
 We know the entropy distribution of the gas that has been converted into stars because, for each gas particle $i$, we have saved its current entropy $K_i$ as well as its maximum entropy $K_i^{\rm max}$ over its entire past history.
$K_i^{\rm max}(t)$ is the maximum of $K_i$ over all outputs at any cosmic time $\le t$.
 We can thus random sample the distributions for $K$ and $K_{\rm max}$ and assign values of  $K$ and $K_{\rm max}$ to each newly formed stellar particle 
so that the newly formed stellar particles have the same entropy distribution as the gas particles that have been converted into stars.
The exception are the stars that had already formed at $z=6.26$. Since we begin our analysis at  $z=6.26$, these stars have no history.
We assign them $K_{\rm max}=K=0$.

  Let us sort all {  baryonic} SPH particles within $r_{\rm vir}$ at redshift $z$ by growing $K_i$, let us define $M_i=\sum_{j=1}^im_j$, where $m_j$ 
  is the mass of the particle $j$, and let us plot $K_i$ versus $M_i$.
  We have so many particles that the result
  appears as a continuous curve.
  {  Hence, we can drop the subscript $i$ and write $K=K(M)$, where $K(M)$ is the entropy distribution of the baryons within the halo.
  The black curves in Fig.~\ref{ent_dist1} show $K(M)$ for g7.55e11 without feedback. The effects of feedback will be discussed later on. 
  We have also looked at the entropy distribution for g1.12e12. We do not show it because the discussion of  g1.12e12 would not add anything to what we learn from  g7.55e11.}
  
  If we sort the particles by growing $K_i^{\rm max}$ rather than $K_i$, we get the black dashed curves.
  The difference between $K(M)$ and $K_{\rm max}(M)$ shows the extent to which radiative cooling has modified the entropy of the baryons within the halo.

Let $M_{\rm \star in}$ be the stellar mass within the virial radius at $z=6.26$. $M_{\rm \star in}$ is the mass at which the black curves start because 
$K(M)=0$ for $M\le M_{\rm \star in}$ and
$K(M)>0$ for $M>M_{\rm \star in}$. $M_{\rm \star in}/M_{\rm b}$ decreases with cosmic time because $M_{\rm b}$ grows.
Only a small fraction of the stars within $r_{\rm vir}$ at $z=0$ formed at $z\ge 6.26$.
The starting point of the black curves depends on our arbitrary choice of the redshift at which we decided to begin our analysis. It has no physical meaning per se.

Moving towards {   larger baryonic masses}, the second main feature of the black solid curves corresponds to the mass $M\sim 0.6M_{\rm b}\sim M_\star$, above which $K(M)$ becomes much steeper
{  (stars are the baryons with the lowest entropies).}
Gas with $K$ lower than a few $10^4{\rm\,keV\,cm}^{2}$ has either formed stars or is in the dense star-forming ISM. The wide gap between the black solid curves and the black dashed curves at $M< 0.6M_{\rm b}$
shows that the gas must dissipate most of its initial entropy in order to form stars.

Above the steep slope, there is a shallower high-entropy region, where the solid black curves join the dashed black curves. This region corresponds to the hot CGM. It is absent in the panel at $z=5.27$, where 
there is some shock-heated gas, but a hot atmosphere has not developed yet (Fig.~\ref{poststamp_images}).

The solid black curve and the dashed black curve in Fig.~\ref{ent_dist1} become  the  solid green curve and the dashed green curve, respectively, if we consider all the particles that have 
been inside $r_{\rm vir}$ at some redshift $\ge z$ even if they are no longer within $r_{\rm vir}$ at redshift $z$.
The black curves stop at $M=M_{\rm b}$ by construction, since $M_{\rm b}$ is the total mass of all baryonic particles within $r_{\rm vir}$ at redshift $z$.
The green curves are almost identical to the black curves  and disappear under them 
at $M\lsim 0.8M_{\rm b}$, but they
extend to $M>M_{\rm b}$ because {  of the spillage phenomenon discussed in Section~3.}

The black curves in Fig.~\ref{ent_dist2} are the same as the black curves in Fig.~\ref{ent_dist1} at the same $z$.  The new elements in Fig.~\ref{ent_dist2} are the red and the orange curves.
The black curves are for all baryonic particles within $r_{\rm vir}$.
The red curves are for particles in the hot CGM at redshift $z$. 
 The orange curves are for particles that are  within $r_{\rm vir}$ at redshift $z$ and have been in the hot CGM at some redshifts $\ge z$.
  The particles used to compute the red curves are as a subsample of the particles used to compute the orange curves, which are a subsample of the particles used to compute the black curves.
  For all three colours (red, orange, black),   the solid curves and the dashed curves correspond to $K$ and $K_{\rm max}$, respectively.
    
Following our procedure for computing $K(M)$ and $K_{\rm max}(M)$, the red curves should stop at $M_{\rm hot}<M_{\rm b}$. The orange curves should stop at a mass equal to the total mass of all particles that have been in the hot CGM and are inside the virial radius; this mass is necessarily  greater or equal to $M_{\rm hot}$ and smaller or equal to $M_{\rm b}$.
  
 We have translated the red curves and the orange curves to high masses, so that all the curves in Fig.~\ref{ent_dist2} end at  $M=M_{\rm b}$.
 The logic for that is that the hot CGM should correspond to the highest-entropy gas within the halo.
 Even though all curves  in Fig.~\ref{ent_dist2} end at $M=M_{\rm b}$ by construction, that does not guarantee that they all overlap at high masses.
  The fact that they do so validates the correctness of our assumption that the hot CGM corresponds to the highest-entropy baryons.
 
As the critical polytrope introduced in Section~3  to separate the cold phase from the hot phase has a polytropic index only a little lower than the adiabatic value $\gamma=5/3$,
 the criterion introduced in Section~3 is approximately an entropy criterion with a critical entropy  of a few ${\rm keV\,cm}^{2}$. 
 The red and orange curves  in Fig.~\ref{ent_dist2} depend on this critical entropy, but the black curves do not. 
 The presence of a feature in the black dashed curves at a $K$ of a few ${\rm keV\,cm}^{2}$ and at the mass
 where the orange curves begin 
  confirms  the robustness of our analysis and the importance of this  entropy scale. 
   
  The point is clearer at the two lowest redshifts ($z=0.52$ and $z=0.00$). Between $M=0.75M_{\rm b}$ and $M=0.95M_{\rm b}$, the black curves are almost straight lines on a $\log K$--$M$ diagram.
 If we extrapolate these straight lines at  low masses, 
  we see that, at $M\lsim  0.6M_{\rm b}$, the extrapolated curves
   fall above the real black dashed curves at $M\lsim 0.6M_{\rm b}$.
There   is a clear physical distinction between
 the 60 per cent of the baryons that never reached entropies above a few ${\rm keV\,cm}^{2}$ and the 40 per cent that did.

 Radiative cooling moves the hot CGM'S lowest-entropy gas to the cold phase (the transition from the orange dashed curves to the orange solid curves in Fig.~\ref{ent_dist2}). Negligible at $z>1$, the phenomenon is clearly visible at lower redshifts, even
 though its quantitative importance remains small. At $z=0$, the orange curves start at $M\simeq 0.60M_{\rm b}$ (about 40 per cent of the baryons have been in the hot CGM).
 The solid one drops dramatically at $M\simeq 0.64M_{\rm b}$: about 4 per cent of the baryons have been in the hot CGM but have cooled later on.
 This 4 per cent pales compared to the 60 per cent accreted in the cold mode. 
 Through this analysis, we rediscover what we had already found from Figs.~\ref{massgrowth} and \ref{hot_mode_sf}.
 The hot mode makes a small contribution to the growth of the central galaxy.
   
 The orange solid curve and the orange dashed curve part below $M\simeq 0.67M_{\rm b}$, the mass at which the red curves begin. As $67-64=3$, it follows that
 $\sim 3$ per cent of baryons have been in the hot CGM, have not cooled yet, but are in the process of doing so (we could say that they are in a cooling flow).

 The remaining $33$ per cent of the baryons were shock-heated and are still in the hot CGM with $K\simeq K_{\rm max}$ (in Fig.~\ref{ent_dist2}, 
 the red solid curves and the red dashed curves are almost indistinguishable).
 Our interpretation of this finding is that the baryons that are still in the hot CGM have not been affected by radiative cooling 
 because they are those with the highest entropies and thus the longest cooling times.
 
 Fig.~\ref{ent_dist3} is the same as Fig.~\ref{ent_dist2} but this time for the simulation with SN feedback. 
Three differences stand out.
 
 First, feedback delays star formation. In the simulation with feedback, fewer stars formed at very high redshift ($z\ge 6.26$).
 
 Second, feedback heats the gas, albeit transiently. There are much fewer baryons with low $K_{\rm max}$ with feedback than without it.
We note, however, that most of the gas heated by SNe is heated to fairly low entropies corresponding to a warm-hot phase.
 The mass that crosses the critical polytrope and becomes part of the hot CGM is fairly low.
  This is why the $M$ at which the orange dashed curves start is only slightly lower in Fig.~\ref{ent_dist3} than in Fig.~\ref{ent_dist2}.
   
 Third, feedback promotes cooling by replenishing the hot CGM with lower-entropy gas that has short cooling times. 
  In Fig.~\ref{ent_dist2}, the lowest entropy associated with the hot CGM at $z=0$ is  $K\sim 2{\rm\,keV\,cm}^{2}$.
 The gas with $ 2{\rm\,keV\,cm}^{2}\lsim K\lsim 4{\rm\,keV\,cm}^{2}$ has cooled leaving behind the gas with $K>4{\rm\,keV\,cm}^{2}$, which has long cooling times.
 In Fig.~\ref{ent_dist3}, feedback has replenished the hot CGM with gas that has entropies as low as  $K\sim 0.8{\rm\,keV\,cm}^{2}$.
 This gas cools rapidly leaving behind, once again, the baryons with $K>4{\rm\,keV\,cm}^{2}$.

 {  In summary and converting all percentages to the total accreted mass (rather than the mass $M_{\rm b}$ within the virial radius)\footnote{Restricting our attention the baryons within $r_{\rm vir}$ at $z=0$ gives the false impression that feedback reduces the mass of the hot CGM. In reality, the mass of the hot CGM increases slightly, but the increase is masked by the expansion of the hot gas
 at $r\gsim r_{\rm vir}$.}, in g7.5e11 without feedback,
 $\sim 50$ per cent of the baryons  have been accreted in the cold mode, $\sim 6$ per cent have cooled or are in a cooling flow, the rest
 are in the hot CGM and cooling has not affected their entropies.
 With feedback, the fraction of the baryons that has never been in the hot CGM decreases to $\sim 30$ per cent. Its decrease is compensated by
 a much larger fraction ($\sim 20$ per cent) of baryons that have cooled or are about to do so, while the mass of of high-entropy baryons that have not been affected by cooling is more or less the same,
although a higher fraction of the high-entropy extends beyond $r_{\rm vir}$ with feedback than without it.
  
 In g1.12e12, the percentages without feedback are $\sim 40$ per cent for the baryons accreted in the cold mode and $\sim 4$ per cent for the baryons that  have cooled or are in a cooling flow.  The percentages with feedback are $\sim 26$ and $\sim 15$ cent, respectively. The main difference is that g1.12e12 has a higher fraction of high-entropy baryons
 (56--59 per cent) and also more spillage outside $r_{\rm vir}$.

In both cases, a small fraction ($\sim 4$--6 per cent) of the shock-heated baryons are able to cool. A larger sample may change this result quantitatively but is highly unlikely to modify its qualitative aspect.
 SNe can increase the importance of cooling by replenishing the hot CGM with lower entropy gas that has a short cooling time, but this has little effect on the mass of the high-entropy shock-heated gas. }
  
{Before concluding this section, it is interesting to compare the results of NIHAO with the probability-density function (PDF) for the  entropy of the gas particles
 in the EAGLE simulations \citep{correa_etal18},
 which is unimodal at low masses but bimodal for  $M_{\rm vir}\gsim 10^{12}{\rm\,M}_\odot$,
 with a low-entropy peak and a high-entropy peak separated by a minimum  around $10^{7.2}{\rm K\,cm}^2$ ($\sim 1{\rm\,keV\,cm}^2$ in our units), although the precise value depends on halo mass. This is the critical entropy that  \citet{correa_etal18} used to select the hot gas when applying an entropy criterion.
  For $M_{\rm vir}\sim 10^{12}{\rm\,M}_\odot$,  the PDF of the hot gas in EAGLE peaks at $10^{8.6}{\rm K\,cm}^2$, which corresponds to  $K\sim 30{\rm\,keV\,cm}^2$ in our units.

 Our entropy distribution are in an integral form. Hence, the maxima and minima of the  PDF appear  in Figs.~\ref{ent_dist1}--\ref{ent_dist3}
 as changes of concavity in the black curves.
 The general behaviour is nevertheless the same. The minimum that separates the low-entropy gas from the high-entropy gas (the switch from $K''>0$  to $K''<0$) is at
 $K\sim 1{\rm\,keV\,cm}^2$.
  The high entropy peak  (the switch from $K''>0$  to $K''<0$) is 
 at $K\sim 20{\rm\,keV\,cm}^{2}$. 
 
 These figures are consistent with those found in EAGLE, but they are for the simulations without feedback (Fig.~\ref{ent_dist2}; the results for g1.12e12 are substantially similar).
 With feedback, in g7.55e11 the minimum that separates the gas with low and high entropies descends to $K\sim 0.1{\rm\,keV\,cm}^2$ (Fig.~\ref{ent_dist3})
 because SNe have replenished the hot CGM with a lot of gas at $0.1{\rm\,keV\,cm}^2\lsim K \lsim 1{\rm\,keV\,cm}^2$. In g1.12e12, this lower-intermediate entropy gas stands out as a third local maximum of the entropy distribution. 
 In both galaxies, feedback makes the differences with EAGLE more significant.

We cannot be sure of the origin of this discrepancy, but we strongly suspect that it derives from how feedback is implemented in the SPH codes used to run the EAGLE and NIHAO simulations.
Both include thermal SN feedback and chemical enrichment, but the ways they do it differ in both the numerical implementation
and the inclusion of specific physical processes (e.g., there is no metal diffusion in EAGLE). 
EAGLE also includes AGN feedback, which has a strong heating effect in massive galaxies and may explain why the PDF of the hot gas peaks at
$K\sim 30{\rm\,keV\,cm}^2$ in EAGLE and $K\sim 20{\rm\,keV\,cm}^2$ in the NIHAO simulations without feedback. 
 
 The sensitivity of the entropy distribution to the feedback model conforts our choice to select the shock-heated gas based on the asymptotic equation of state of the IGM, which does not depend on it.
 In practice, however, our criterion corresponds to an entropy threshold of $\sim 2{\rm\,keV\,cm}^2$, which is very similar the value of $1.2{\rm\,keV\,cm}^2$  used by \citet{correa_etal18}. 

 }

\section{Predicting the accretion mode}

In this section, we address the question of how robustly  a simple analytic model can predict whether an individual halo is in the cold mode or the hot mode at a specific time.
However, before we describe our model, let us
determine how the shock-heated fraction $f_{\rm hot}$ evolves with cosmic time  in the NIHAO simulations.

\begin{figure}
\begin{center}$
\begin{array}{c}
\includegraphics[width=0.95\hsize]{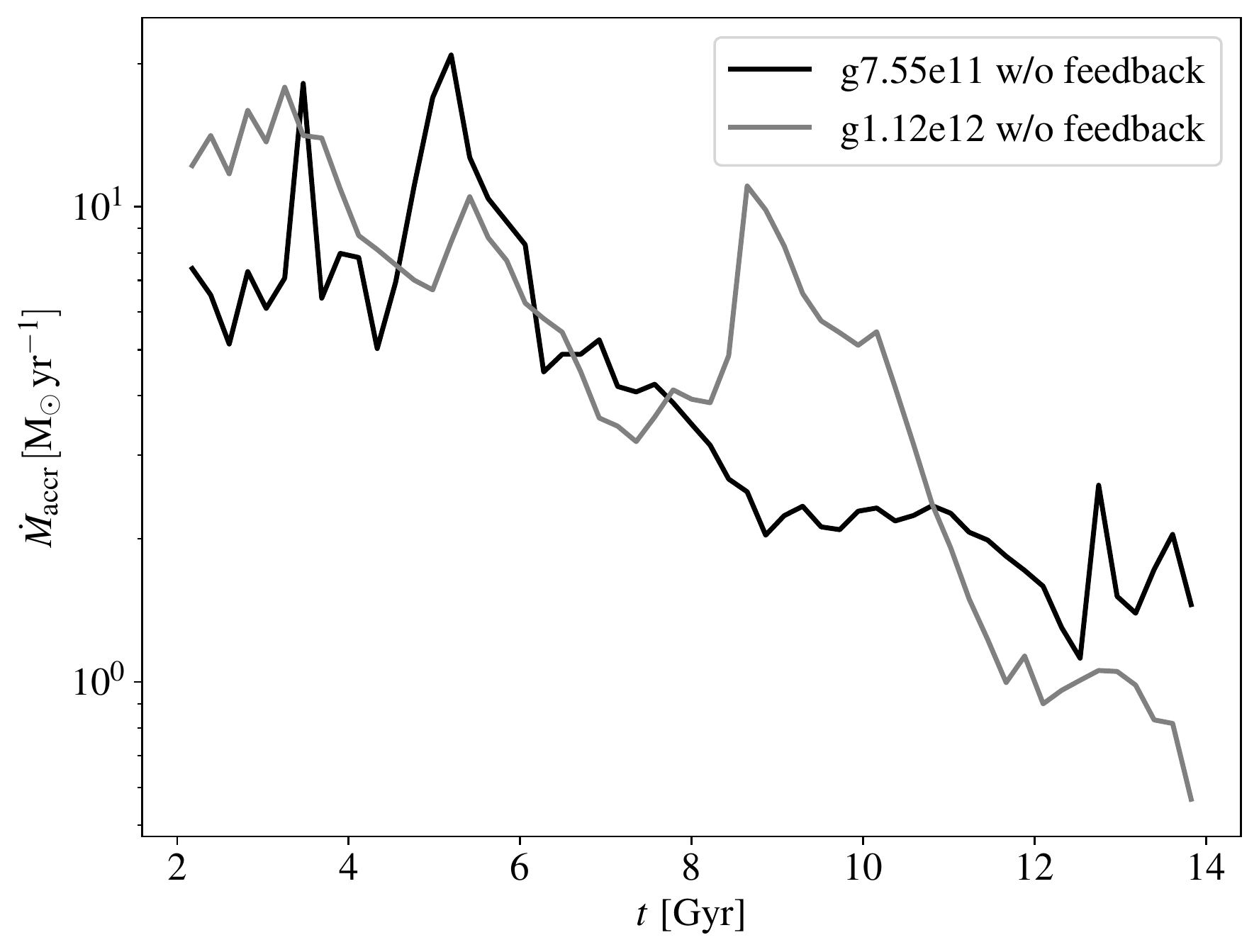} 
\end{array}$
\end{center}
\caption{Baryonic accretion rate onto the halo as a function of cosmic time. 
The two peaks at $t<6\,$Gyr for g7.55e11 (black curve) correspond to the times of rapid halo growth in Fig.~\ref{massgrowth}a.
The peak at t $t\sim 9\,$Gyr for g1.12e12 (gray curve) corresponds to the beginning of the long fly-bye visible in Fig.~\ref{massgrowth}b.}
\label{Maccr}
\end{figure}

We define the baryonic accretion rate $\dot{M}_{\rm accr}$ onto the halo 
as the mass of the baryonic particles that enter $r_{\rm vir}$ for the first time between two consecutive output timesteps divided by the time interval between them. 
{  Fig.~\ref{Maccr} shows  $\dot{M}_{\rm accr}(t)$ for g7.55e11 and g1.12e12
in the simulations without feedback, but feedback has little impact on 
 $\dot{M}_{\rm accr}$ in massive spirals (but not in dwarf galaxies; \citealp{tollet_etal19}).}
 
The shock-heated fraction $f_{\rm hot}$ is the fraction of the accreted mass that ends up in the hot CGM, independently of whether it remains there or it cools at same later time.
We do not require that the gas be hot at the time of accretion because shocks may occur deep into the halo and the gas that enters the halo may take a freefall time $t_{\rm ff}\sim r_{\rm vir}/v_{\rm vir}$
before it reaches the shock radius $r_{\rm s}$. We note that $t_{\rm ff}$ is short compared to the age of the Universe but  longer than the time intervals between output timesteps.

Fig.~\ref{fhot} shows the evolution of $f_{\rm hot}$ with cosmic time in NIHAO (red solid curves). Let us start from the case without feedback.
{  In g7.55e11 (Fig.~\ref{fhot}a),} at $t<4\,$Gyr, a small fraction of the accreted gas is shock-heated (30 per cent on average).
The halo is in the cold mode. $4{\rm\,Gyr}<t<6{\rm\,Gyr}$ is a transition epoch. After $t=6\,$Gyr, the hot mode is clearly dominant with $f_{\rm hot}\sim 0.9$ on average if we exclude an ephemeral comeback 
of some cold accretion around $t=12.5\,$Gyr.
{  In g.12e12 (Fig.~\ref{fhot}b), the shock-heated fraction grows from $f_{\rm hot}\sim 0.3$ at $t\sim 2\,$Gyr to $f_{\rm hot}\sim 0.8$ at $t\sim 4\,$Gyr.
Hence, the evolution is similar, even though the growth of $f_{\rm hot}$ is more gradual.
}

\begin{figure*}
\begin{center}$
\begin{array}{cc}
\includegraphics[width=0.455\hsize]{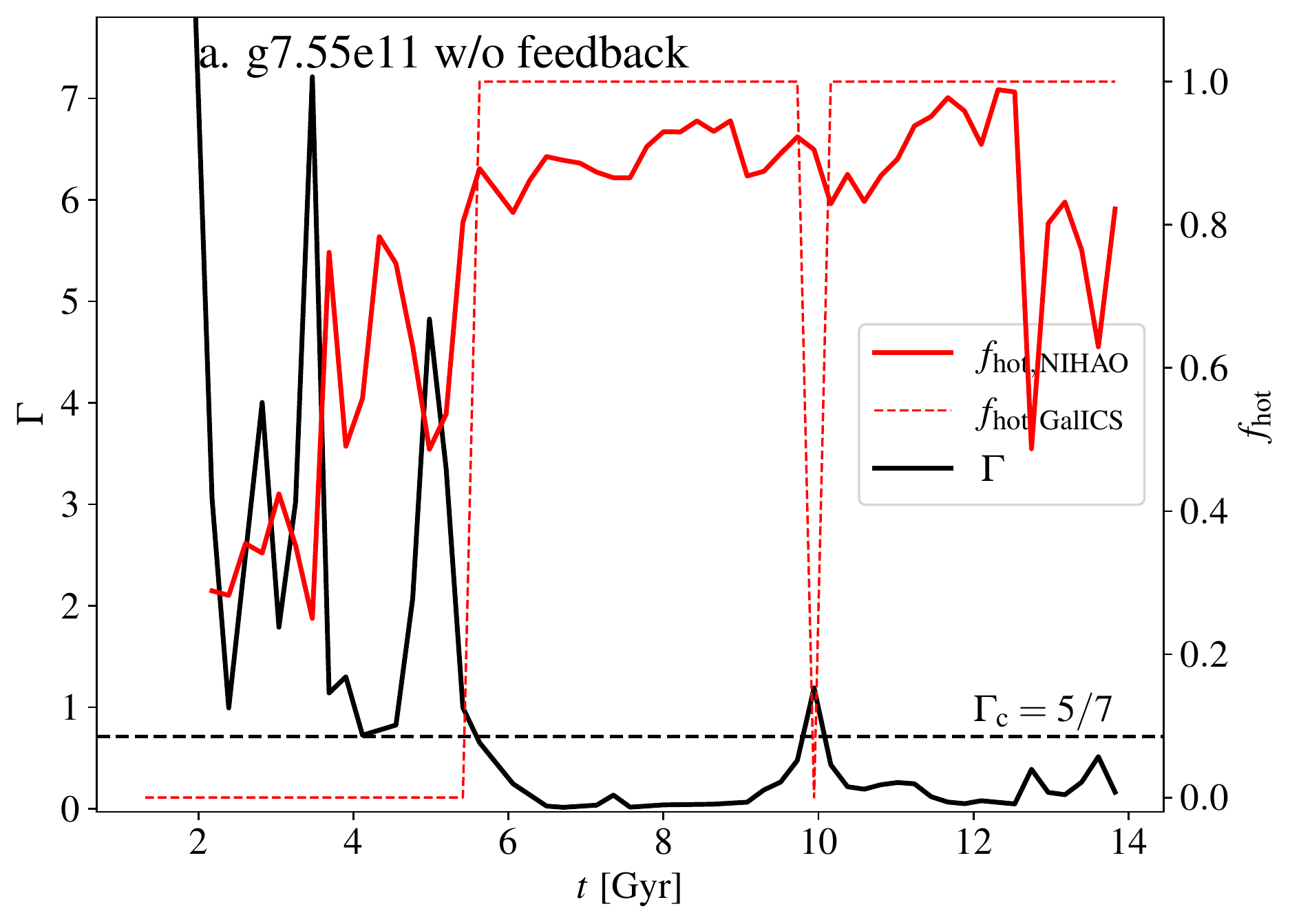} 
\includegraphics[width=0.455\hsize]{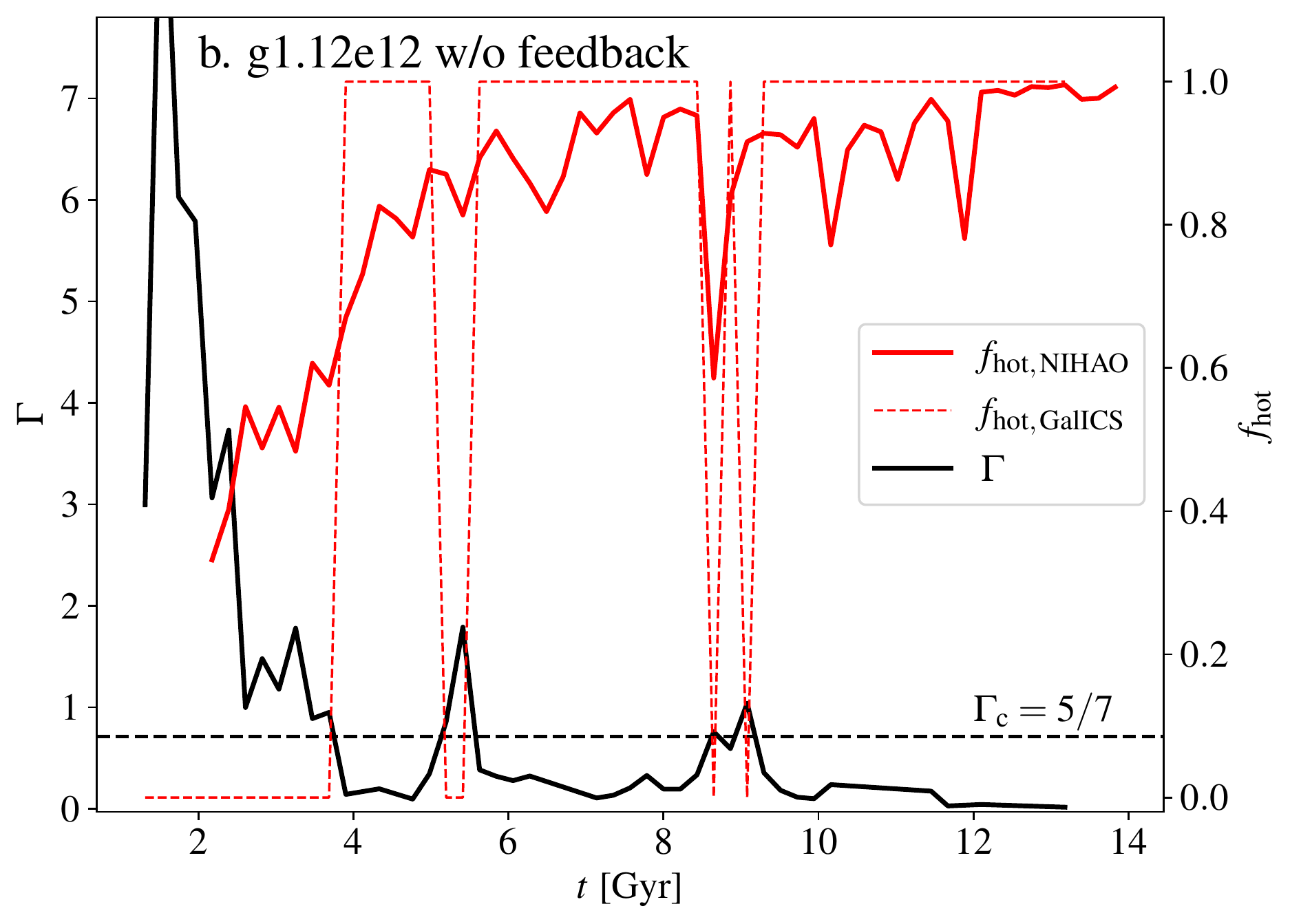} \\
\includegraphics[width=0.455\hsize]{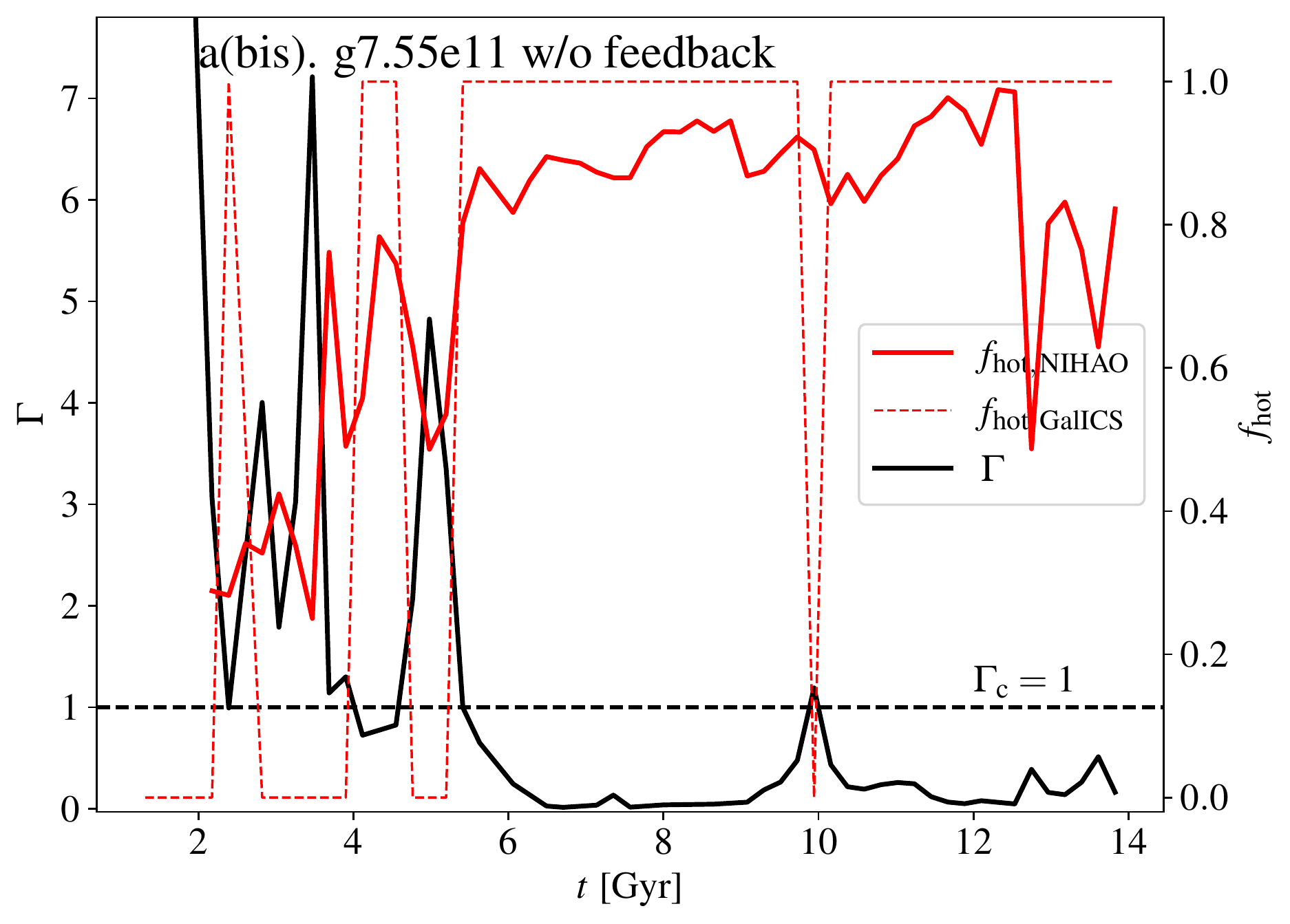} 
\includegraphics[width=0.455\hsize]{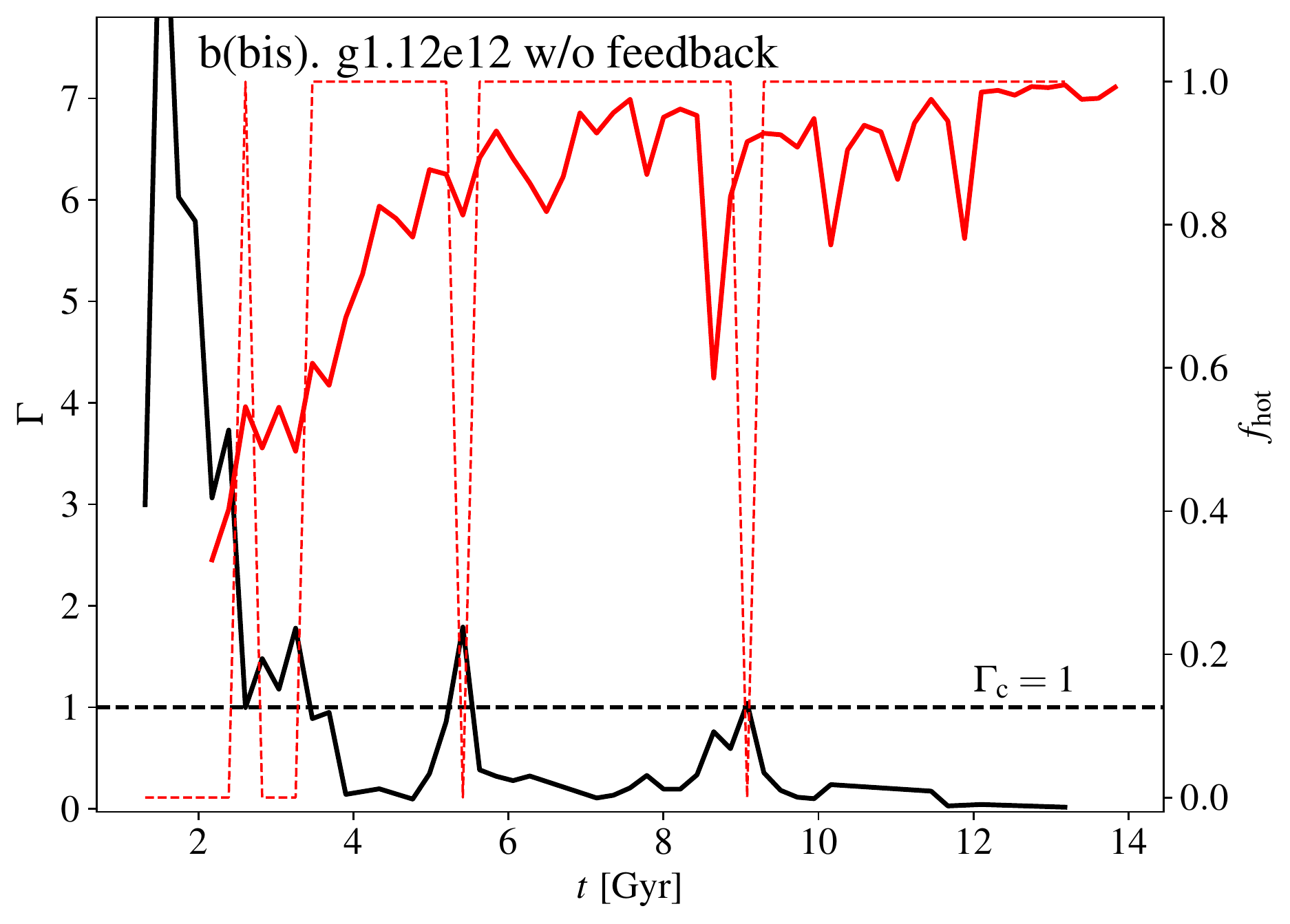} \\
\includegraphics[width=0.455\hsize]{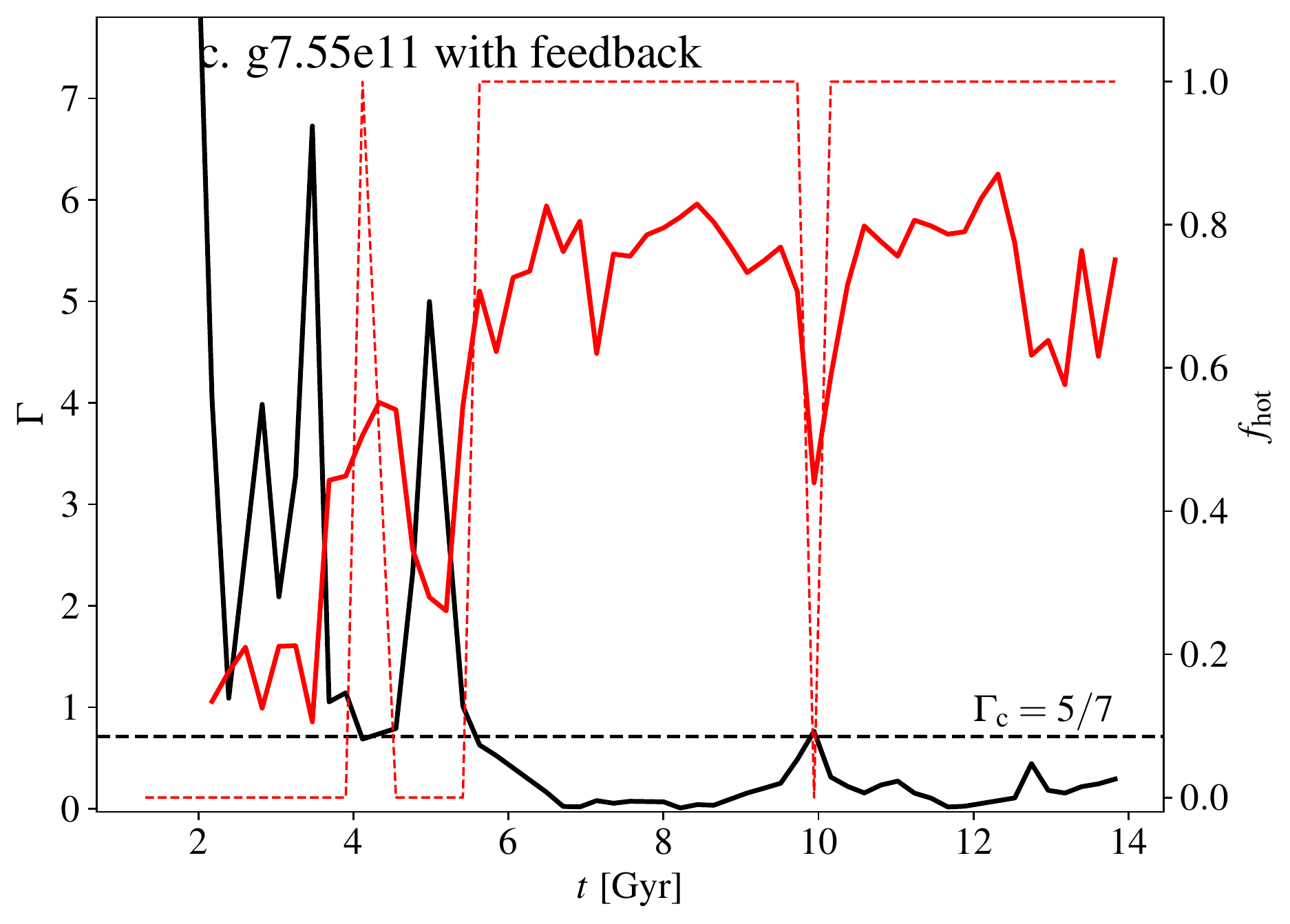}
\includegraphics[width=0.455\hsize]{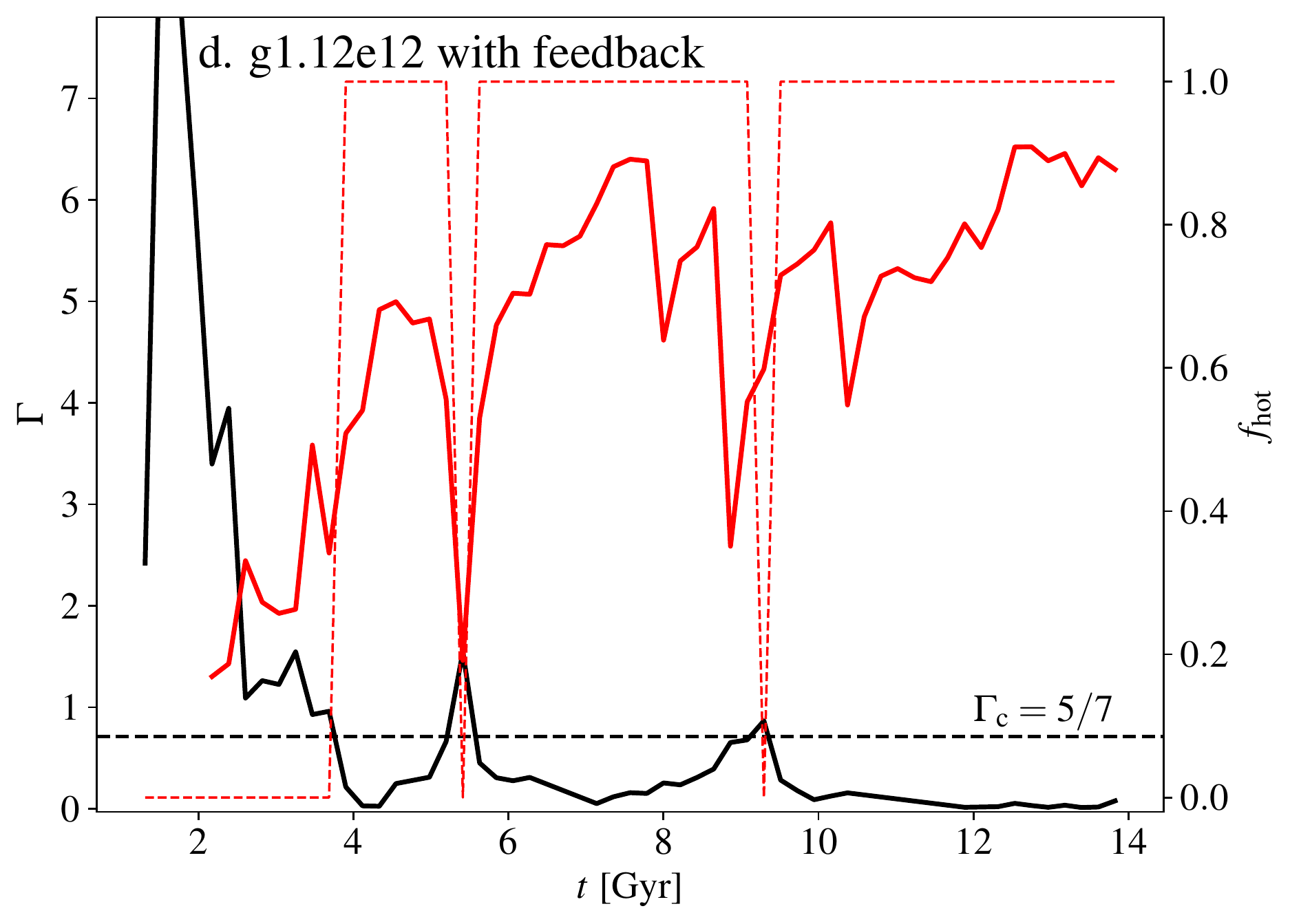}\\ 
\end{array}$
\end{center}
\caption{Evolution with cosmic time $t$ of
the shock-heated fraction $f_{\rm hot}$ in NIHAO  (red solid curves) and of the $\Gamma$ variable in the {\sc GalICS~2.1} SAM (black solid curves) {  for g7.55e11 (left) and g1.12e12 (right).}
Both $f_{\rm hot}$  and  $\Gamma$ are dimensionless (notice their different ranges of ordinates).
The black dashed horizontal lines corresponds to the critical $\Gamma$, below which {\sc GalICS~2.1} predicts that the accreted gas should be shock-heated. 
{  For the simulations without feedback, we have shown two cases:  $\Gamma_{\rm c}=5/7$ (top row)  and $\Gamma_{\rm c}=1$ (middle row).
For the simulations with feedback, we have shown only the case $\Gamma_{\rm c}=5/7$ (bottom row).}
The red dashed curves show the predictions of {\sc GalICS~2.1}: $f_{\rm hot}=0$ for $\Gamma>\Gamma_{\rm c}$, $f_{\rm hot}=1$ for $\Gamma<\Gamma_{\rm c}$.
}
\label{fhot}
\end{figure*}

\citet{birnboim_dekel03} and \citet{dekel_birnboim06} performed a shock-stability analysis of the transition between the two accretion modes.
The outcome of a shock depends on two timescales: the post-shock radiative cooling time $t_{\rm cool}$ and the compression time $t_{\rm comp}=r_{\rm s}/u_2$,
where $r_{\rm s}$ is the shock radius and $u_2$ is the post-shock infall speed.
{  The condition for a stable shock is: 
\begin{equation}
{t_{\rm comp}\over t_{\rm cool}}\le\Gamma_{\rm c}
\label{stab_cond}
\end{equation}
with $\Gamma_{\rm c}=5/7$. The left-hand side of Eq.~(\ref{stab_cond}) depends on radius. The shock radius $r_{\rm s}$ is the largest radius for which Eq.~(\ref{stab_cond})  is satisfied.
\citet{cattaneo_etal20} implemented this condition in the {\sc GalICS}~2.1 SAM by assuming that there is shock heating when:
\begin{equation}
\Gamma=\min\left({t_{\rm comp}\over t_{\rm cool}}\right)_{r\le r_{\rm vir}}\le\Gamma_{\rm c}
\label{stab_cond2}
\end{equation}
and used this  criterion to separate cold and hot accretion.}

{\sc GalICS~2.1} cannot compute  $f_{\rm hot}$ for an individual halo. It assumes $f_{\rm hot}=0$ for haloes with $\Gamma>\Gamma_{\rm c}$
and $f_{\rm hot}=1$ for haloes with $\Gamma\le\Gamma_{\rm c}$.
Only on a population basis can $f_{\rm hot}$ take all intermediate values between zero and unity.
$\Gamma$ is determined by the post-shock density, temperature and metallicity used to compute $t_{\rm cool}$, and by the
the post-shock infall speed $u_2$ used to  compute $t_{\rm comp}$. 

Without feedback, there is no mechanism to enrich the IGM. 
Even with feedback, however, the metallicity of the filaments is usually quite small \citep{ocvirk_etal08,rafelski_etal12,rafelski_etal14,berg_etal16}.
{  In NIHAO, \citet{roca_etal19} investigated the metallicity of the CGM out to $r=1.5r_{\rm vir}$. The mean metallicity of the cold gas ($T<10^{4.5}\,$K) at $r=1.5r_{\rm vir}$ increases from $Z\simeq 10^{-2.5}{\rm\,Z}_\odot$ at $z=2.3$ to $Z\simeq 10^{-1.8}{\rm\,Z}_\odot$ at $z=0$, but there is a lot of scatter. The metallicity is significantly lower for inflows than it is for outflows. Hence, those are upper limits for the metallicity of the IGM in NIHAO. 
At $z=0$, the cold inflowing gas spans the metallicity range $10^{-2.4}{\rm\,Z}_\odot\lsim Z \lsim 10^{-1.8}{\rm\,Z}_\odot$.
Assuming primordial cooling is a good approximation for $Z\lsim 10^{-3}{\rm\,Z}_\odot$ and continues to be acceptable as long as $Z< 10^{-2}{\rm\,Z}_\odot$ 
Hence, the IGM is effectively primordial over most of the cosmic lifetime and that will be our assumption for the semi-analytic calculations in this article.
A cautionary note is that the actual metallicity of the post-shock gas may be higher than the metallicity of the IGM if the filaments are
enriched by mixing with the CGM within the halo.}

For a strong shock, the post-shock density $\rho_2$ is four times the pre-shock density $\rho_1$, the post-shock speed $u_2$ is one fourth of the pre-shock speed $u_1$, and the post-shock temperature is easily calculated
from the value of $u_1$ (see \citealp{cattaneo_etal20} for details).
Hence, the calculation  of $\Gamma$ reduces to that of two variables: $\rho_1$ and $u_1$.

The scaling of $t_{\rm comp}/t_{\rm cool}$ with  $\rho_1$ and $u_1$ follows from $t_{\rm comp}\propto u_1^{-1}$ and $t_{\rm cool}\propto\rho_1^{-1}T/\Lambda(T)$, where $T$ is the post-shock temperature and
$\Lambda(T)$ is the cooling function.
In the relevant temperature range (several $10^5\,$K; Figs.~\ref{poststamp_images}--\ref{poststamp_images_S2}), 
$\Lambda\propto T^{-1/2}$. Since $T\propto u_1^2$, by combining these relations, we find:
\begin{equation}
{t_{\rm comp}\over t_{\rm cool}}\propto{\rho_1\over u_1^4}.
\label{tcomp_over_tcool}
\end{equation}
Dense slow inflows favour cold accretion. Low densities and high speeds promote shock heating.

The rest of this section can be broken down into three parts:
the calculation of $u_1$, the calculation of $\rho_1$, and the use of $\Gamma(t)$ to predict the halo's accretion mode.

\subsection{The pre-shock infall speed}

\begin{figure*}
\begin{center}$
\begin{array}{ccc}
\includegraphics[width=0.33\hsize]{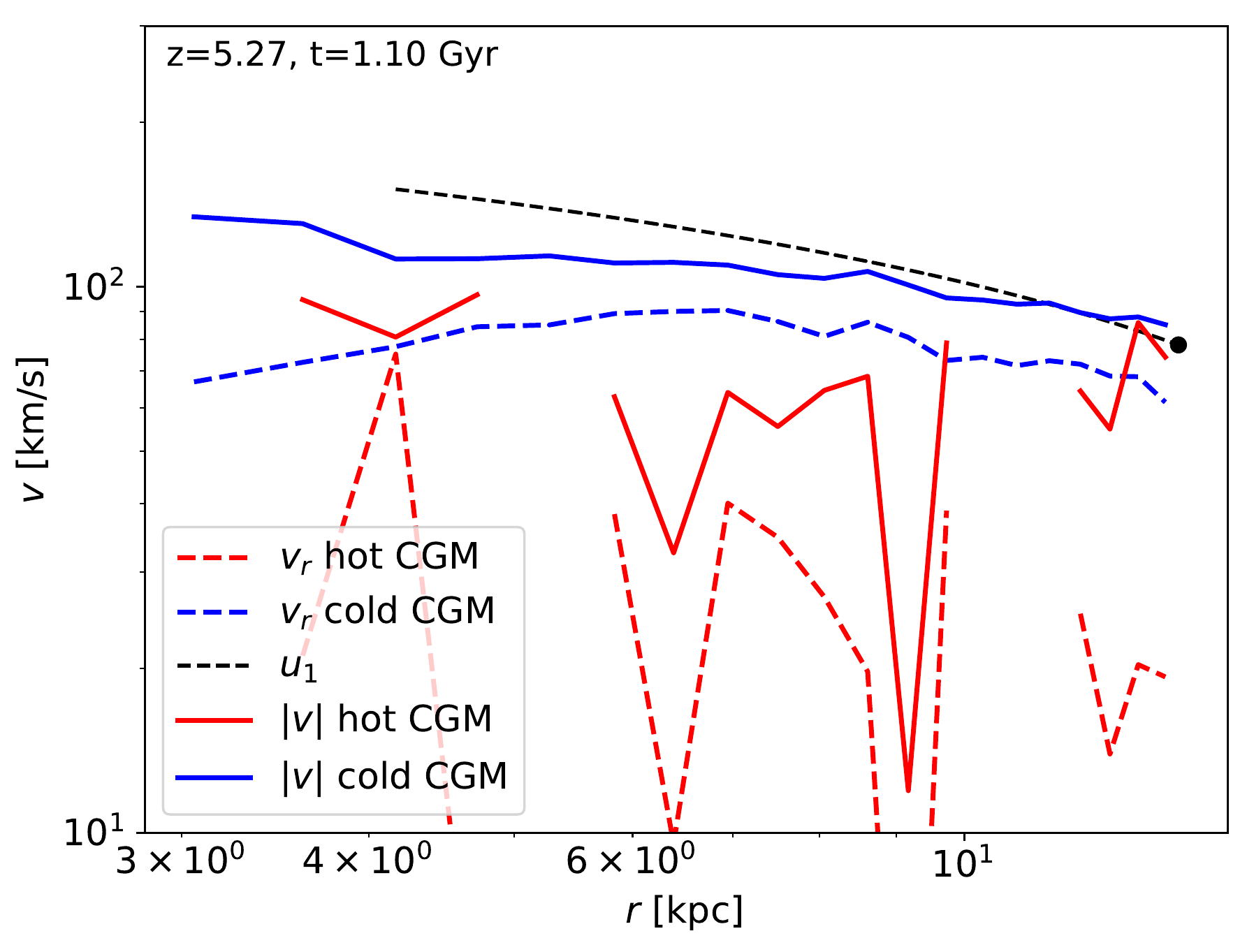}
\includegraphics[width=0.33\hsize]{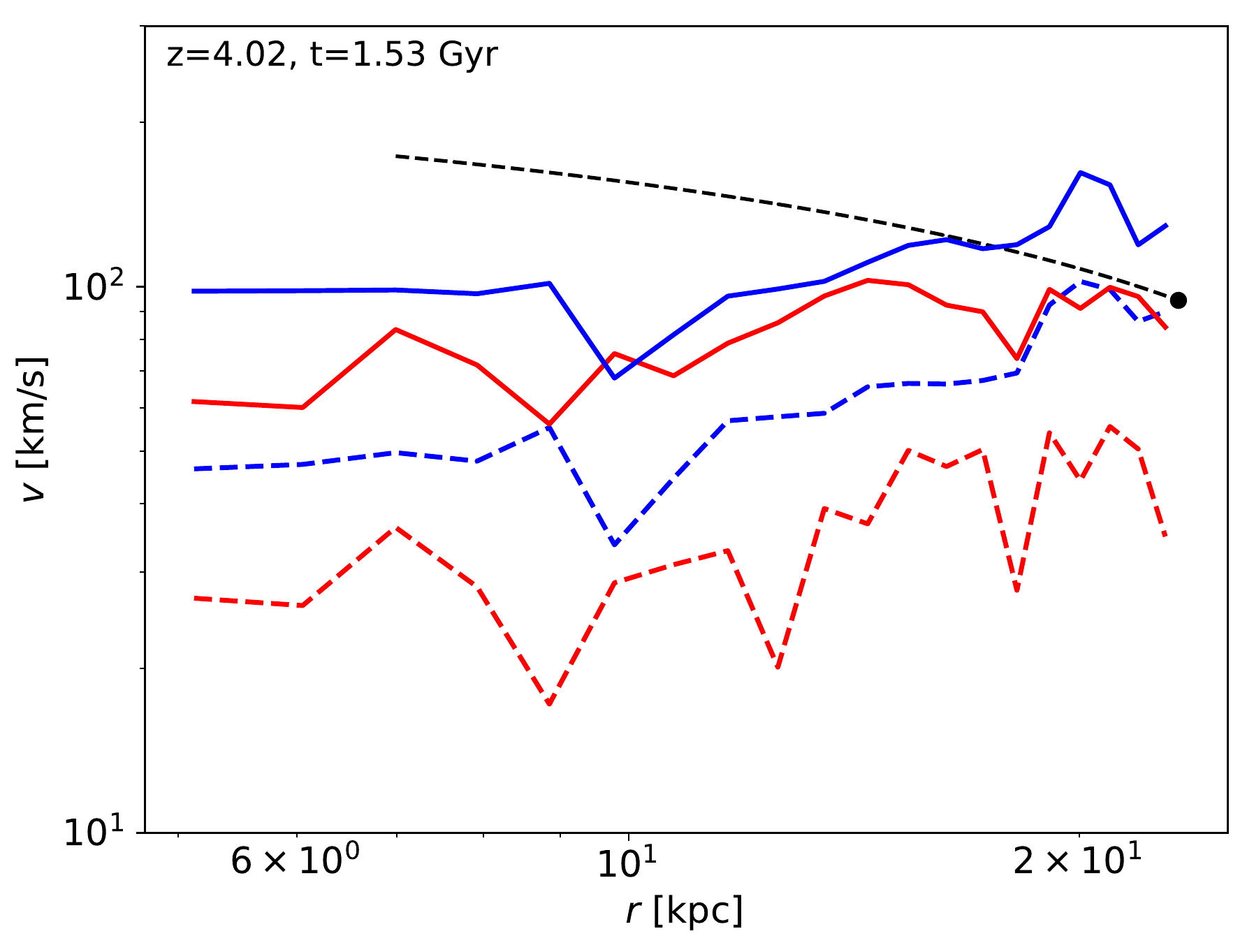}
\includegraphics[width=0.33\hsize]{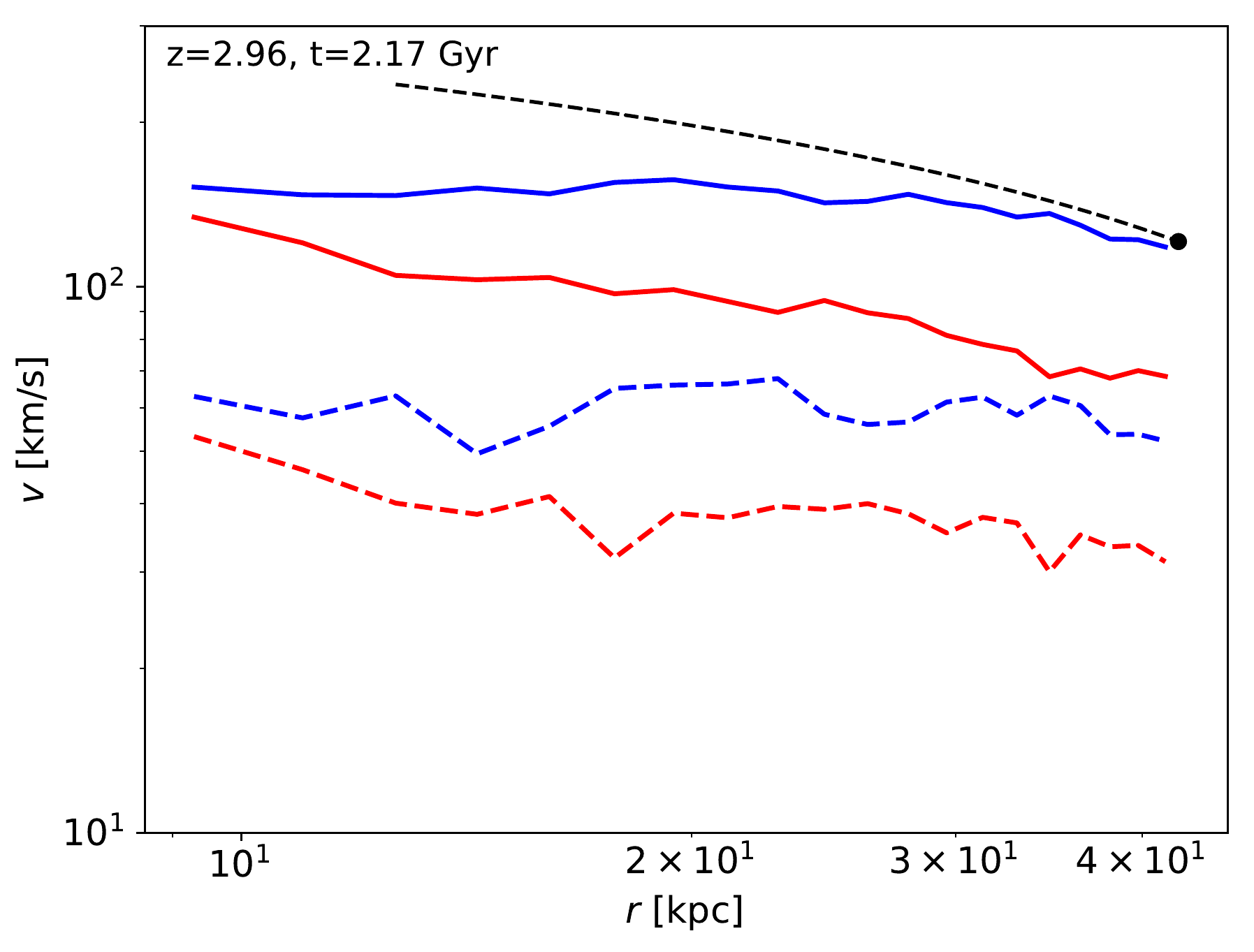}\\
  \includegraphics[width=0.33\hsize]{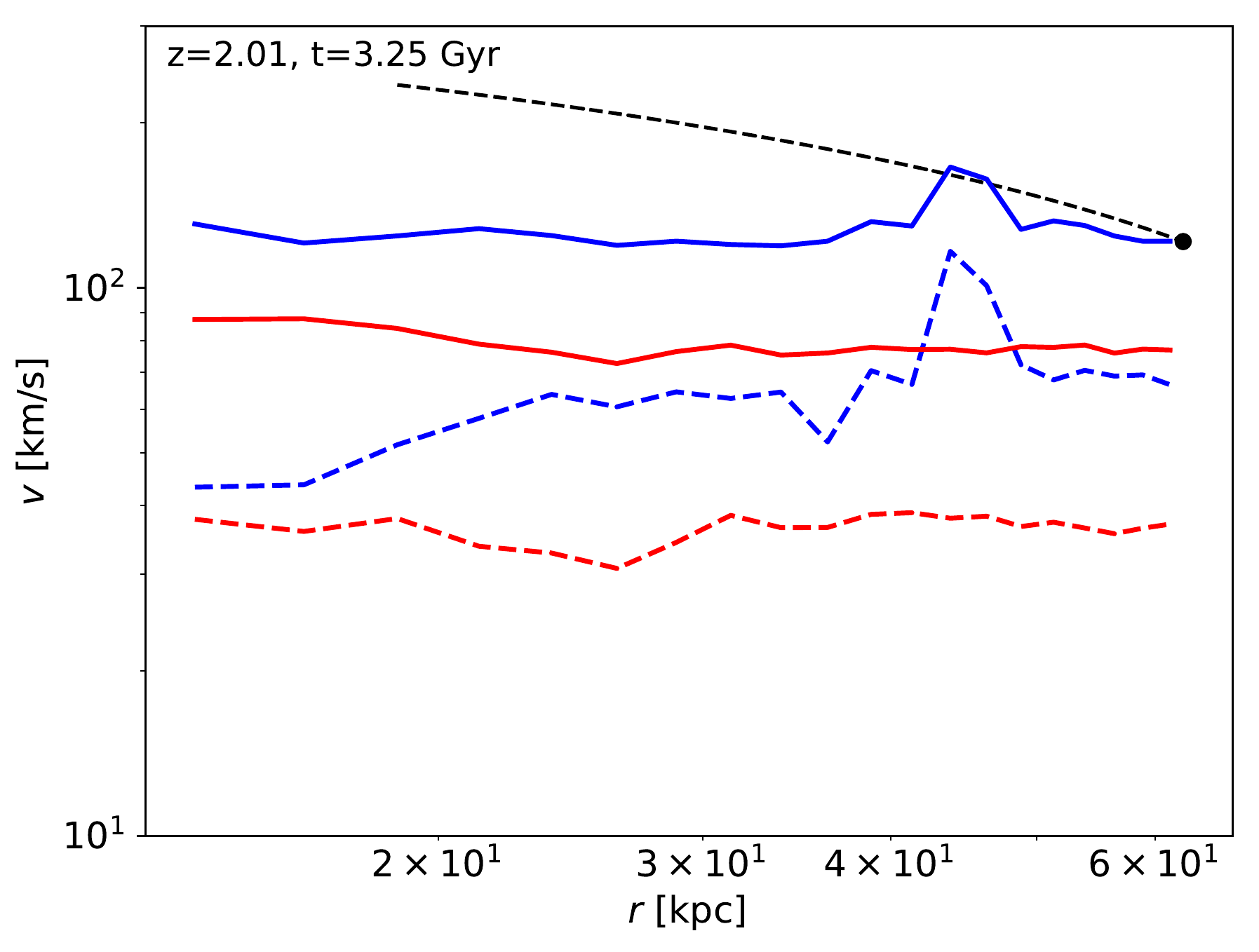}
  \includegraphics[width=0.33\hsize]{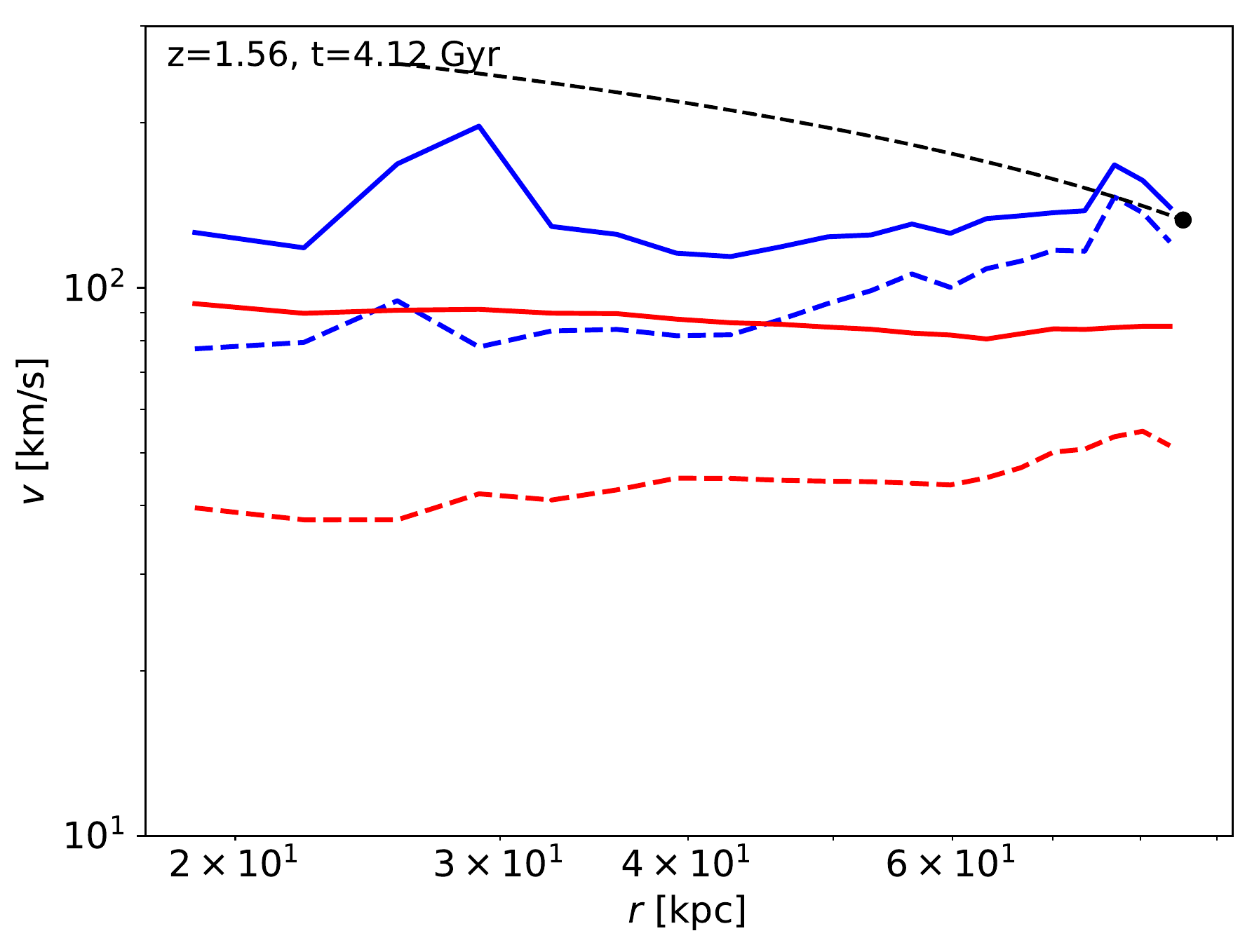}
 \includegraphics[width=0.33\hsize]{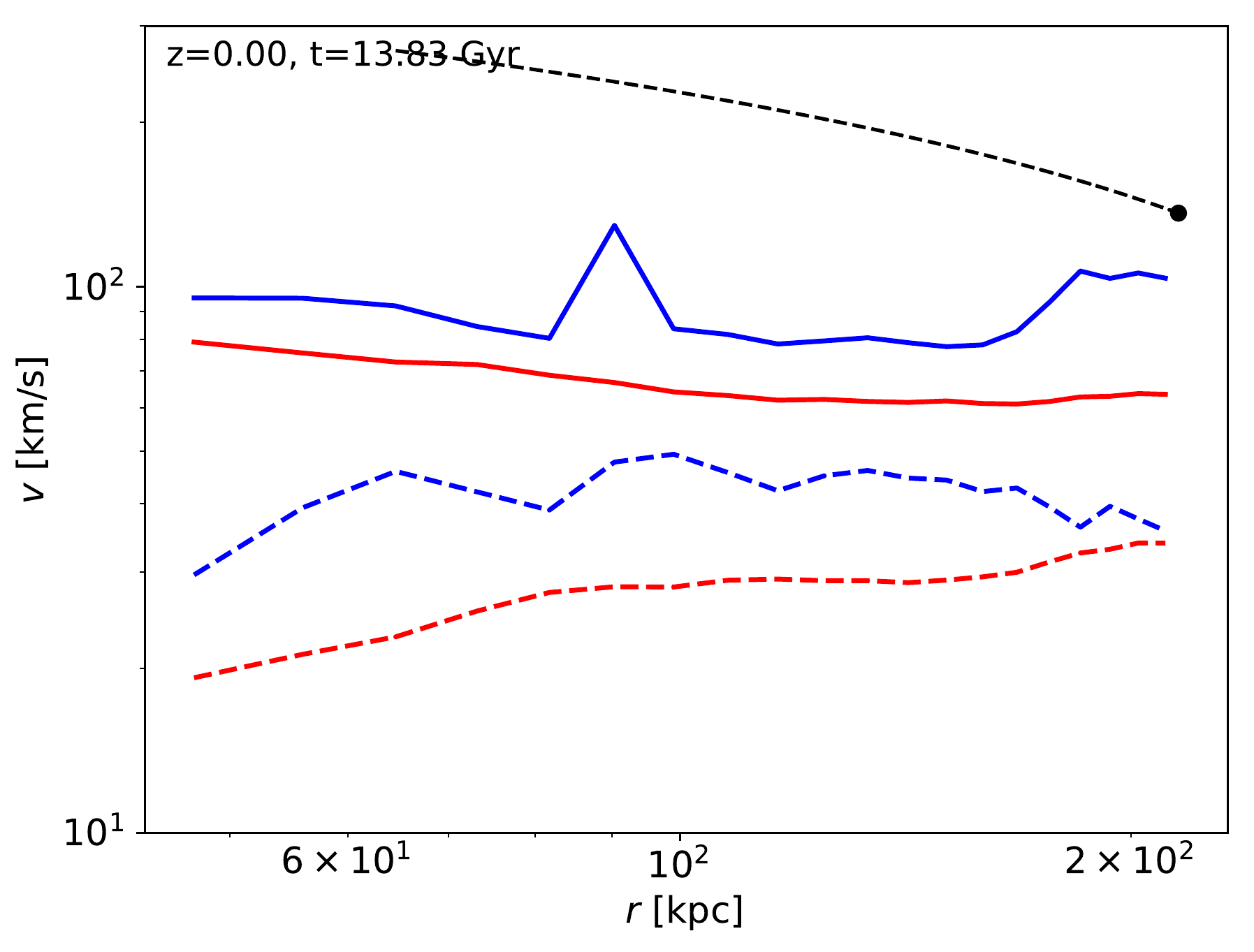}
 \end{array}$
\end{center}
\caption{{  Galaxy g7.55e11 without feedback: $\langle |{\bf  v}(r)|\rangle$ (blue and red solid curves)
and $\langle v_r(r)\rangle$  (blue and red dashed curves) at six different  $z$ (the corresponding redshift is indicated on each panel).} The blue/red curves are for the cold/hot CGM, respectively.
The averages are mass-weighted in spherical shells. 
The black circles have coordinates $(r_{\rm vir},v_{\rm vir})$.
The black dashed curves  show the freefall speed $u_1(r)$ computed with Eq.~(\ref{freefall}).}
\label{u1}
\end{figure*} 

{\sc GalICS}~2.1 computes $u_1$ by assuming that the pre-shock gas (the cold gas in the filaments) is in free fall.
The infall can be broken down into two parts.

From the turn-around radius, where the gas separates from the Hubble flow, to the virial radius, the gas is in free fall with the DM, so that it enters $r_{\rm vir}$ at the infall speed:
\begin{equation}
u_1(r_{\rm vir})= v_{\rm vir}.
\label{freefall_rvir}
\end{equation}
Eq.~(\ref{freefall_rvir}) is exact for pressureless spherical collapse in a universe without cosmological constant.
The cosmological constant generates a repulsive force that reduces the infall speed but its effect on $u_1$ is less than 2 per cent and is negligible compared to those of hydrodynamic phenomena, such as shocks along the filaments.

From $r_{\rm vir}$ to $r_{\rm s}$, the DM is static (in virial equilibrium); the gas sinks into the gravitational potential $\phi_{\rm NFW}(r)$ of the DM halo, 
which we assume to follow \citet*{navarro_etal97}'s  density distribution. Hence:
\begin{equation}
u_1(r_{\rm s})=\sqrt{ v_{\rm vir}^2+2[\phi_{\rm NFW}(r_{\rm vir})-\phi_{\rm NFW}(r_{\rm s})]}.
\label{freefall}
\end{equation}
Eq.~(\ref{freefall}) is relevant only for $r_{\rm s}<r_{\rm vir}$. For $r_{\rm s}\ge r_{\rm vir}$,
the gas that enters $r_{\rm vir}$  is slowed down by the resistance of the hot CGM. We use Eq.~(\ref{freefall_rvir}) in that regime.

To check these assumptions, we have plotted 
$|{\bf  v}(r)|$ and $v_r(r)$ for the cold CGM, where  $|{\bf  v}|$ and $v_r$  are mass-weighted over spherical shells.
Here $v_r$ is the mean radial infall speed taken with a positive sign for infalling particles and discarding any outflowing particles.
The reason why we need both $|{\bf  v}|$ and $v_r$ is that $v_r$ underestimates the real infall speed if the filaments enter the virial sphere on a trajectory that is not purely radial,
while $\langle|{\bf  v}(r)|\rangle$ often contain peaks that are associated with satellite galaxies and do not measure the infall speed of the cold CGM.

{  The plots shown in Fig.~\ref{u1} are for g7.55e11 only and without feedback, but the differences from one output timestep to another are as large as the differences from one simulation to another.
Globally, g7.55e11 and g1.12e12 exhibit the same qualitative behaviour, which we can summarise in three points. First, $v_r<|{\bf  v}|$. Hence, the infall is not purely radial.
Second, $|{\bf v}(r_{\rm vir})|\simeq v_{\rm vir}$ is usually a good assumption at all but the lowest $z$.
Third, $|{\bf  v}|\simeq u_1$ is  a reasonable approximation only at high $z$, when $r_{\rm s}\ll r_{\rm vir}$. At later epochs  $|{\bf  v}(r)|$ is flat or decreases towards low $r$, as expected if hot CGM resists penetration by the cold filaments.

Our assumptions are nevertheless reasonably correct at high $z$ before a stable shock appears and a lower $z$ after $r_{\rm s}$ reaches $r_{\rm vir}$.
Only in the intermediate regime is there a possibility that Eq.~(\ref{freefall}) may substantially overestimate the infall speed.}

The red curves in Fig.~\ref{u1} show $\langle|{\bf  v}(r)|\rangle$ and $\langle v_r(r)\rangle$ for the hot CGM. They confirm our expectation that the hot phase has lower bulk speeds that the cold one, 
but they also show a certain degree of correlation between the kinematics of the hot and the cold CGM.

\begin{figure}
\begin{center}$
\begin{array}{c}
\includegraphics[width=0.95\hsize]{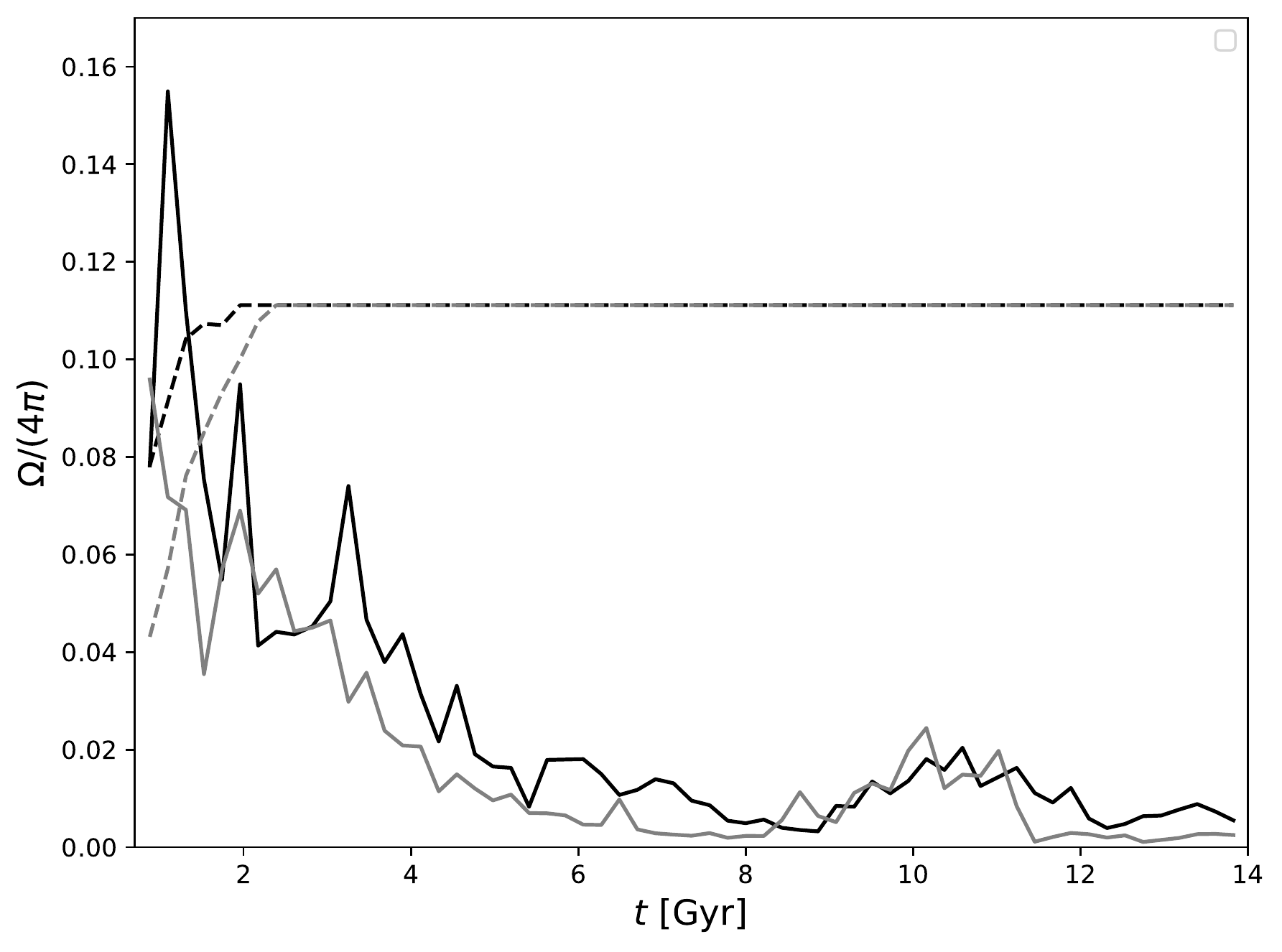} 
\end{array}$
\end{center}
\caption{Fraction of the total solid angle covered by the filaments as a function of cosmic time {  in NIHAO for g7.55e11 (black solid curve) and g1.12e12 (gray solid curve).
The black and gray dashes show the predictions of {\sc GalICS}~2.1 for g7.55e11 and g1.12e12, respectively.}}
\label{Omega}
\end{figure}

\subsection{The pre-shock density}

The pre-shock density  $\rho_1$ is determined from the continuity equation for a stationary flow:
\begin{equation}
\dot{M}_{\rm accr}=\Omega r_{\rm s}^2\rho_1u_1,
\label{continuity}
\end{equation}
where $\Omega$ is the solid angle from which the gas is accreted ($\dot{M}_{\rm accr}$ varies on cosmological timescales,
but variations on timescales of the order of $t_{\rm ff}$ are small; Fig.~\ref{Maccr}). For the same values of $\dot{M}_{\rm accr}$, $r_{\rm s}$ and $u_1$, the filaments are 
denser if the gas is accreted from a narrower solid angle.

$\Omega$  is one of our greatest uncertainties. {  \citet{cattaneo_etal20} assumed that the radius $r_{\rm fil}$ of the DM filaments is the virial radius of a halo with mass equal to the non-linear mass at the redshift of interest.
The filaments that accrete onto a halo are cylindrical with radius $r_{\rm fil}$ at $r>r_{\rm vir}$ and conical at $r<r_{\rm vir}$.
The solid angle $\Omega_{\rm fil}$ covered by a DM filament is determined by $r_{\rm fil}/r_{\rm vir}$.
We follow \citet{codis_etal18} to compute the number of filaments $n_{\rm fil}$ as a function of $M_{\rm vir}$ and $z$ (see \citealp{cattaneo_etal20}, Appendix~A, for details).
The total solid angle covered by all DM filaments is the maximum between $n_{\rm fil}\Omega_{\rm fil}$ and $4\pi$.
The gas filaments are, however, more concentrated than the DM ones \citep{ramsoy_etal21}.
 \citet{cattaneo_etal20} treated the uncertain concentration factor as a free parameter of the SAM. The gaseous filaments cover a fraction $1/C$ of the total solid angle covered by the DM filaments.}

With the parameterisation above, Eq.~(\ref{continuity}) gives $\rho_1\propto u_1^{-1}$, and Eq.~(\ref{tcomp_over_tcool}) becomes:
\begin{equation}
{t_{\rm comp}\over t_{\rm cool}}\propto{C\over u_1^5}.
\label{tcomp_over_tcool2}
\end{equation}
Higher $C$ (denser filaments) give more efficient cold-mode accretion,  and thus higher stellar masses, but in {\sc GalICS~2.1} we can increase $C$ and still reproduce the same observational data
if the efficiency of SN feedback, the true value of which is considerably uncertain, is increased simultaneously.
{   \citet{cattaneo_etal20} broke this degeneracy by calibrating $C$ on cosmological hydrodynamic simulations without feedback.
They   found the best agreement with the galaxy stellar mass function of  \citet{keres_etal09} for a $C=9$. }

In NIHAO, we compute $\Omega$ by selecting all infalling cold (below the critical polytrope) gas particles with $r_{\rm vir}<r<1.05r_{\rm vir}$  that have never been at $r\le r_{\rm vir}$.
To each particle, we can associate a volume equal to its mass divided by its SPH density.
 The filaments' covering fraction $\Omega/(4\pi)$ is then the ratio between the volume occupied by these particles and the total volume of the spherical shell.
 {  This procedure measures the isotropy of cold accretion on the scale of the virial radius. $\Omega$ would be the same at all radii if the filaments were conical. That is obviously an approximation, but
 Fig.~\ref{poststamp_images} shows that it is not too far from reality, at least at high $z$, where filaments are clearly visible.}
 
Fig.~\ref{Omega} shows  the covering factors  computed with this method  {  for g7.55e11 and g1.12e12 (solid curves).
In NIHAO, $C$ is not constant.} As we can also see in Figs.~\ref{poststamp_images}--\ref{poststamp_images_S2}, the accretion is  more isotropic at high $z$ and less so at low $z$.

{  This behaviour is very different from the one expected from our SAM. As the non-linear mass grows more rapidly than $M_{\rm vir}$, our SAM predicts that $\Omega$ should increase with time until it converges 
$4\pi/C$. (Fig.~\ref{Omega}, dashed curves). The discrepancy arises because our calculation of $\Omega$ is purely cosmological.
We have also assumed that $C$ is constant while it may depend on the gaseous environment.
In reality, once a hot CGM has developed, the filaments are confined by it.
The values of $\Omega$ measured at low $z$ should be taken with the greatest caution because $\Omega$  is not very meaningful when there are no cold flows and most of the cold gas comes in through satellite galaxies (Fig.~\ref{poststamp_images}, $0.03<z<0.15$). 

The  relevant question for our SAM is whether the $\Omega$ predicted by {\sc GalICS}~2.1 at high $z$ before shock heating agree with NIHAO.
The answers is that they do within a factor $<2$. In reality, it is likely that there is a compensation. Fig.~\ref{u1} shows that, by assuming  free fall, we tend to overestimate the infall  speed. Eq.~(\ref{tcomp_over_tcool2}) shows that $\Gamma\propto C/u_1^5$.
Hence, it is logical that our SAM tends to exaggerate $C$ to compensate the error on $u_1$ and get the stellar mass function of galaxies right. }

\subsection{From $\Gamma$ to $f_{\rm hot}$}

{\sc GalICS}~2.1 computes $\Gamma$ in the context of a SAM  that 
follows the evolution of baryons within DM merger trees  from a cosmological N-body simulation.
$M_{\rm vir}$, $r_{\rm vir}$ and $c$ are used to compute $u_1$. $\dot{M}_{\rm vir}$ is used to compute $\dot{M}_{\rm accr}$.

Fig.~\ref{fhot} (black solid curves) shows the $\Gamma(t)$ predicted by {\sc GalICS}~2.1 when we extract $M_{\rm vir}$, $\dot{M}_{\rm vir}$, $r_{\rm vir}$ and $c$ from NIHAO and 
we feed them to {\sc GalICS}~2.1 with $C=9$. 
The halo is predicted to be in the cold mode when $\Gamma>\Gamma_{\rm c}$
($\Gamma_{\rm c}=5/7$ in our default model).

A discrete function that takes only two values, zero and unity (the red dashed curve) is inevitably a crude approximation to one that varies continuously between these two extreme (the red solid curve),
but the fundamental behaviour is reproduced reasonably well. 
{  Let us start from the case without feedback.

In g7.55e11, {\sc GalICS~2.1} predicts a transition to the hot mode at $t=5.63\,$Gyr (Fig.~\ref{fhot}$a$, red dashed curve);
$t=5.4$--$5.6\,$Gyr  is indeed the time  after which the shock-heated fraction in NIHAO is steadily above $f_{\rm hot}=0.8$, although a first shock-heating episode has already occurred at $t=3.69\,$Gyr.
In g1.12e12,  shocks on a scale of $\sim 0.5r_{\rm vir}$ are already present at $t=1.10\,$Gyr (Fig.~\ref{poststamp_images_S2}).
The hot CGM fills the virial sphere by $t=1.96\,$Gyr, even though at $t\lsim 2\,$Gyr the hot mode accounts for less than a third of the baryonic accretion rate  (Fig.~\ref{fhot}$b$, red solid curve).
The shock-heating time according to {\sc GalICS~2.1} corresponds to the time
$t\sim 4\,$Gyr, after which the shock-heated fraction in NIHAO increases rapidly from $f_{\rm hot}\sim 0.4$ to $f_{\rm hot}\sim 0.8$.}

The predictions of {\sc GalICS~2.1} are based on
$\Gamma$.
{  If we compare $f_{\rm hot}$  (the red solid curve) and $\Gamma$ (the black solid curve) in Fig.~\ref{fhot}$a$, we do see a strong anti-correlation.
The maxima of $\Gamma$ at $t=\,$2.8, 3.5, 3.9, 5.0$\,$Gyr and its minima at $t=\,$3.0, 3.7, 5.0$\,$Gyr correspond to minima and maxima of $f_{\rm hot}$, respectively.
Only the minimum  at $t=4.1\,$Gyr does not correspond to a maximum of $f_{\rm hot}$, but  $f_{\rm hot}$ has started to grow rapidly and has a maximum at the next output timestep.
At $t=5.6\,$Gyr, where $f_{\rm hot}$ has a maximum, $\Gamma$ has a not reached a minimum yet but its value has dropped dramatically.
The same consideration applies  to the other panels of Fig.~\ref{fhot} and demonstrates that $\Gamma$ is a good predictor of the behaviour of $f_{\rm hot}$.}

However, the predictions of {\sc GalICS~2.1} depend not only on $\Gamma$ but also on $\Gamma_{\rm c}$.
{{\sc GalICS~2.1} would have reproduced {  the two shock-heating episodes (i.e. the two peaks of the  red solid curve) at $t=3.69\,$Gyr and $t=4.33\,$Gyr } in Fig.~\ref{fhot}a if we had used $1.09<\Gamma_{\rm c}<1.18$ instead of $\Gamma_{\rm c}=5/7$.

This point is important because $\Gamma_{\rm c}$ depends on assumptions about the equation of state of the post-shock gas and the geometry of the shocks \citep{dekel_birnboim06}.
$\Gamma_{\rm c}=5/7$ follows from \citet{birnboim_dekel03}'s assumption that the temperature of the post-shock gas is constant across a shell.
Assuming a polytropic equation of state with a constant polytropic index would give  $\Gamma_{\rm c}=1$. 
{  Figs.~\ref{fhot}$a$(bis) and $b$(bis) show that using $\Gamma_{\rm c}=1$ instead of $\Gamma_{\rm c}=5/7$
can substantially move forward the time of the first shock-heating episode.

With only two galaxies, we cannot use our results to draw any conclusion on the  value of $\Gamma_{\rm c}$.
A larger sample would not automatically solve the problem, however, because only the ratio $\Gamma/\Gamma_{\rm c}$ matters for our SAM
and there may be systematic errors in the estimate of $\Gamma$, notably because of the uncertainties on $u_1$ (Section~5.1) and $\Omega$  (Section~5.2).
The best-fit $\Gamma_{\rm c}$ may thus compensate these errors rather than return the real physical value of the critical $t_{\rm comp}/t_{\rm cool}$ ratio above which
the gas is shock-heated.}

Even with these uncertainties, {\sc GalICS}~2.1 robustly predicts that cold accretion should be important when $\Gamma\gg 1$ and that our system should be in the hot mode at $t\ge 5.63\,$Gyr.
Both predictions are verified. It is the detailed behaviour for $\Gamma\sim 1$ that is uncertain. 

{  We conclude our detailed discussion of Fig.~\ref{fhot}$a$ by analysing two episodes where {\sc GalICS}~2.1 seems to fail dramatically for both $\Gamma_{\rm c}=5/7$ and $\Gamma_{\rm c}=1$.
At $t=9.94\,$Gyr,  $f_{\rm hot}$ drops dramatically in {\sc GalICS}~2.1 but not  in NIHAO. At $t=12.75\,$Gyr, the problem is the opposite: $f_{\rm hot}$ drops dramatically  in NIHAO but not in {\sc GalICS}~2.1.
Let us start from the first episode. }

Fig.~\ref{poststamp_images} shows no evidence for cold streams at $t=9.94\,$Gyr on a face-on view, but filaments are visible if we look at the central galaxy edge-on. Most of this gas will not make it into the central galaxy, however, and this is the key point to reconcile {\sc GalICS}~2.1  with NIHAO.

{  {\sc GalICS}~2.1 applies Eq.~(\ref{stab_cond2}) to decide whether accretion {\it onto the halo} is in the cold or the hot mode.
This decision is made independently of the presence of a massive hot CGM, while,
in reality, its presence does affect the ability of cold streams to reach the central galaxy.
In NIHAO, all particles that end up in the hot CGM contribute to the hot mode, independently of whether they do it through shock-heating or  through mixing
at the boundary surface between the cold and the hot phase.

Despite the apparent discrepancy, the final predictions of {\sc GalICS}~2.1 are not inconsistent with the behaviour  measured in NIHAO.
Even if the surge of $\Gamma$ at $t=9.94\,$Gyr results in some cold accretion onto the halo,  the Kelvin-Helmholtz instability destroys the cold filaments on a timescale proportional to $(M_{\rm fil}/M_{\rm hot})^{1/2}$  \citep[Appendix~B]{cattaneo_etal20} that is quite short when $M_{\rm fil}\ll M_{\rm hot}$.
The same argument solves the problem of similar episodes in  Fig.~\ref{fhot}$b$.
A careful inspection of Fig.~\ref{poststamp_images_S2} shows evidence  for cold gas flowing into the virial radius but this cold gas disappears from one snapshot to another, it does not reach 
 the central galaxy.
 
 The sudden drop of $f_{\rm hot}$ measured in NIHAO {  for g7.55e11} at $t=12.75\,$Gyr (Fig.~\ref{fhot}$a$, red solid curve) comes from
 a gas-rich satellite galaxy that enters $r_{\rm vir}$ between  $t=11.88\,$Gyr and  $t=13.39\,$Gyr (Fig.~\ref{poststamp_images}).
 {\sc GalICS}~2.1 has correctly assessed that there is no accretion via cold filaments in that time interval.}

Our discussion has been based on the simulations without feedback, which are more easily interpreted.
Figs.~\ref{fhot}$c$ and $d$ show that SNe do not modify any of our conclusions.
Interestingly, although possibly fortuitously,
when we include SN feedback, the agreement between {\sc GalICS~2.1} and NIHAO becomes better, not worse.

\section{Density of the hot CGM}

One of our main results is how inefficiently the shock-heated gas cools. To exclude the possibility that this is due to unrealistically low densities  in NIHAO,
 we have measured the electron density distribution $n_{\rm e}(r)$ of the hot CGM  in the simulations
and compared it to the observational data for the the Milky Way compiled by \citet[Fig.~\ref{MW}]{voit19};
{  g1.12e12 has been retained for this comparison because its mass is closer to  that of the Milky Way, but the results for g7.55e11 are remarkably similar.}

SN feedback expels baryons and lowers the mass of hot CGM from $M_{\rm hot}\sim 6\times 10^{10}{\rm\,M}_\odot$ to $M_{\rm hot}\sim 3\times 10^{10}{\rm\,M}_\odot$.
 Hence, the electron densities at $20{\rm\,kpc}\lsim r\lsim 200{\rm\,kpc}$ are lower in the simulation with feedback than in the simulation without it (in Fig.~\ref{MW}, the red circles are
 below the black circles). The circles stop at $r_{\rm vir}\simeq 210\,$kpc, but a simple extrapolation shows that, at $r>r_{\rm vir}$, the situation is reversed.
 The electron densities with feedback move above those without it, as expected if some of the baryons at  $r<r_{\rm vir}$ have been moved to $r>r_{\rm vir}$.
 
At $r\lsim 10\,$kpc, $n_{\rm e}$ is higher in the simulation with feedback because, {  even though there are outflows, the hot CGM is constantly replenished by SN-driven winds}.
Fig.~\ref{MW} shows that a lot of the gas reheated by SNe remains in the central region, especially in massive galaxies, where the hot CGM confines the winds
(in dwarf galaxies, the situation is quite different; \citealp{tollet_etal19}).

The density of the hot gas in the halo of the Milky Way has been studied with many different methods.
The thermal pressure of the ISM in the Solar neighbourhood can be robustly measured from ultraviolet absorption lines \citep{jenkins_shaya79}.
The green rectangle in Fig.~\ref{MW} corresponds to \citet{jenkins_tripp11}'s comprehensive analysis of the available observational data.
The hot CGM beyond the Solar neighbourhood can be probed by the emission and absorption of X-ray lines.
The region bordered in cyan and labelled X-ray emission corresponds to \citet{henley_shelton13}'s emission-line analysis.
The blue dotted-dashed, blue dashed and red dashed lines are the results of \citet{miller_bregman13,miller_bregman15}, who focussed on oxygen lines.
The region bordered in brown and labelled X-ray absorption is \citet{voit19}'s reassessment of \citet{miller_bregman13}'s O VII absorption measurements (the thin red dashed line).
Measurements based on emission and absorption lines depend on the metallicity of the CGM.
The data shown in Fig.~\ref{MW} are corrected for the metallicity gradient in \citet{voit19}.

Another method uses dispersion. Radio waves of longer wavelengths travel through the CGM faster than those of shorter wavelengths.
\citet{anderson_bregman10} used dispersion measurements on pulsars in the Large Magellanic Cloud to derive an upper limit for $n_{\rm e}$.
Their results are shown by the green line in Fig.~\ref{MW}.

Additional constraints come from ram-pressure stripping models of dwarf galaxies that orbit the Milky Way, i.e. the Large Magellanic Cloud (\citealp{salem_etal15}, magenta polygon),
Carina (\citealp{gatto_etal13}, red polygon), Sextans  (\citealp{gatto_etal13}, blue polygon), Fornax  (\citealp{grcevich_putman09}, yellow polygon) and Sculptor (\citealp{grcevich_putman09}, black polygon).
Ram-pressure constraints have the advantage of being independent of metallicity.
Circumgalactic pressures can also be derived from $21\,$cm measurements of H I in high-velocity clouds (\citealp{putman_etal12}, vertical black error bar) and similar observations of clouds in the Magellanic Stream
(\citealp{stanimirovic_etal02}, inverted orange open triangle).

The final set of constraints comes from galaxies more massive that the Milky Way.
\citet{singh_etal18} used a power-law profile to model stacked X-ray and cosmic-microwave-background data for galaxies with $10^{12.6}{\rm\,M}_\odot<M_{\rm vir}<10^{13.0}{\rm\,M}_\odot$.
Their results are shown by the region bordered in gray and labelled ``CMBXstacks".

The CGM densities measured in NIHAO for g1.12e12 are broadly consistent with the observations in Fig.~\ref{MW}, even though they lie on the high side of what is observationally permissible.
The inefficient cooling of the hot CGM in NIHAO cannot be due to densities that are too low.

The greatest difference between NIHAO and the observations is the central $n_{\rm e}$ excess in the simulation with feedback. This finding has no observational counterpart and appears in conflict with the density of the ISM in the Solar neighbourhood 
(\citealp{jenkins_tripp11}; Fig.~\ref{MW}, green rectangle).

{  The comparison is based on $n_{\rm e}$ measured assuming a spherical distribution, while the hot ISM is mainly confined to the plain of the disc, both in  NIHAO (Fig.~3 of \citealp{tollet_etal19}) and the observations.
However, in NIHAO the hot CGM as a whole is remarkably spherical
 \citep{gutcke_etal17}. It is even more so with our definition that excludes
the hot ISM. While this questions the meaning of comparing the red circles to the green rectangle, the hot CGM should be less dense, not denser than the hot ISM.
This makes the discrepancy even more acute.}

This finding is remarkable because the simulation with feedback is the more physical one. The NIHAO simulations with feedback form realistic central galaxies \citep{wang_etal15,tollet_etal19}.
 Those without it do not.
A possible interpretation is that the hot winds in the simulations with feedback are not venting out or cooling as rapidly as in real galaxies, but this problem is beyond the scope of the current article.

\begin{figure}
\includegraphics[width=1.1\hsize]{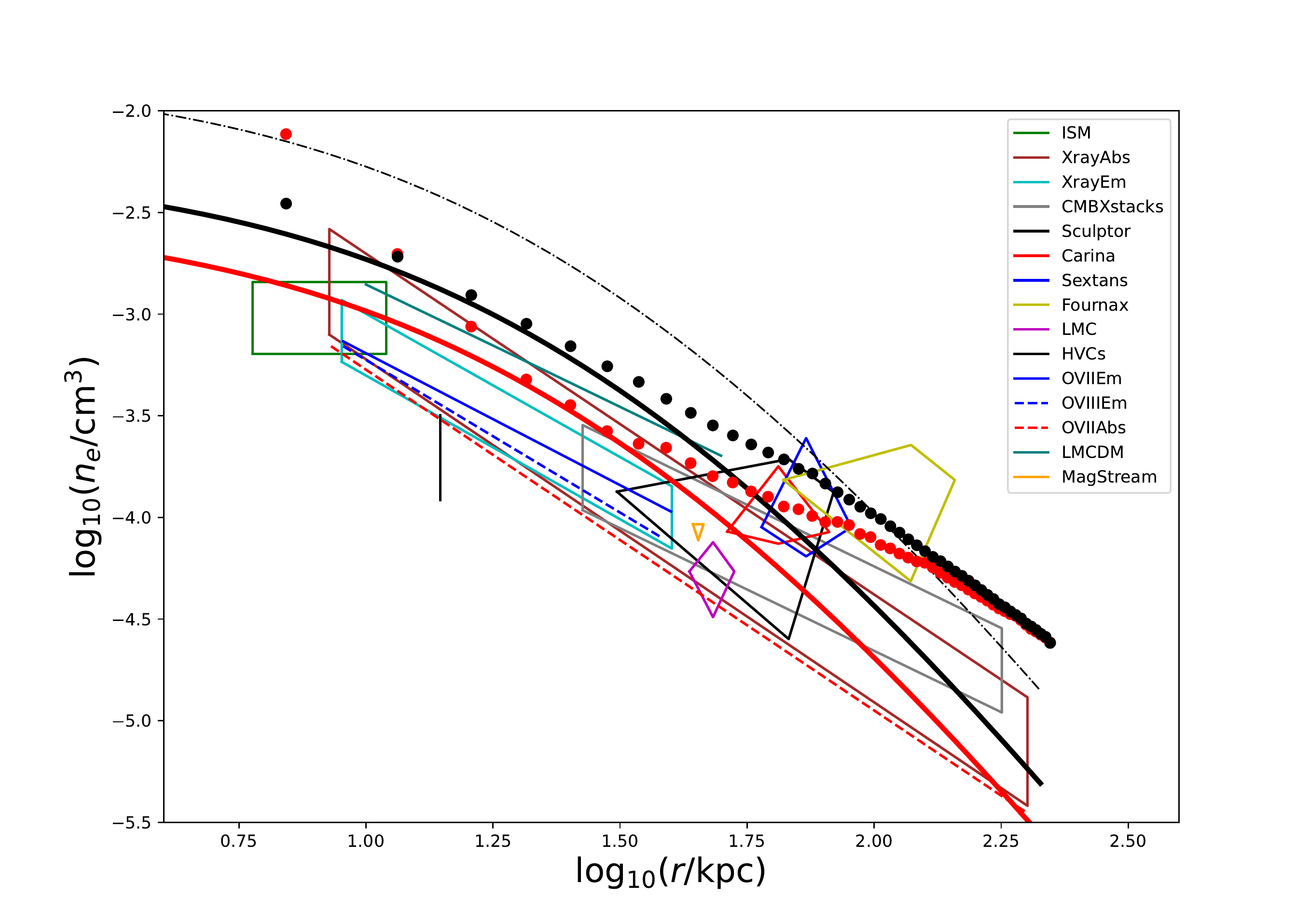} 
\caption{
The electron-density profile of the hot CGM {  in g1.12e12} (circles; red: with feedback, black: without  it)
and the {  Milky Way} (polygons, straight lines). The legend shows the observations  that correspond to each colour/line-style. The green rectangle is for the local ISM and comes from ultraviolet absorption lines \citep{jenkins_tripp11}.
The brown parallelogram (X-ray absorption) is \citet{voit19}'s reassessment of \citet{miller_bregman13}'s O VII absorption-line measurements (red dashes).
The blue solid line and the blue dashes are \citet{miller_bregman13}'s O VII and O VIII emission-line measurements, respectively.
The cyan parallelogram  shows \citet{henley_shelton13}'s X-ray emission-line measurements.
The black, red, blue, yellow and magenta polygons are the constraints from ram-pressure-stripping models for Sculptor \citep{grcevich_putman09}, Carina \citep{gatto_etal13}, Sextans \citep{gatto_etal13}, Fornax  \citep{grcevich_putman09}
and the Large Magellanic Cloud \citep{salem_etal15}, respectively. The black vertical error bar is the constraint from high-velocity clouds (21\,cm data; \citealp{putman_etal12}). The inverted orange triangle is the result of similar observations
for the Magellanic Stream \citep{stanimirovic_etal02}. The green line comes from Large-Magellanic-Cloud dispersion measurements \citep{anderson_bregman10}.
The gray parallelogram comes from stacked cosmic-microwave-background and X-ray data \citep{singh_etal18}; these are the only data  in this figure for galaxies other than the Milky Way. {  The curves show the KS solution for the DM parameters measured in NIHAO for g1.12e12 and a normalisation equal to $100\%$ (thin black-dotted dashed curve), $35\%$ (thick black curve) and $20\%$ (thick red curve) of the one expected if all the baryons were in the hot CGM.}
}
\label{MW}
\end{figure} 

It is interesting to compare the density profiles in Fig.~\ref{MW} to the model by
\citet[KS]{komatsu_seljak01}, which is the exact solution for a polytropic gas in hydrostatic equilibrium with the gravitational potential of halo described
by the \citet*[NFW]{navarro_etal97} profile. 
Even if these assumptions   are correct, the  solution  still contains three free parameters:
the central density, the central temperature and the polytropic index.
KS  constrained them by requiring that the density $\rho$ of the {  hot gas} follows the density $\rho_{\rm  NFW}$ of the DM halo ($\rho/\rho_{\rm NFW}\simeq \Omega_{\rm b}/\Omega_{\rm m}$)
not only at $r=r_{\rm vir}$, where the equality is required to hold exactly, but also over the entire radial range $r_{\rm vir}/2<r<2r_{\rm vir}$.
This approach gives results in good agreement with X-ray observations of galaxy clusters, where most of the baryons are in the intracluster medium and where there is direct observational evidence that the intracluster
medium is approximately polytropic \citep{ghirardini_etal19}.

The difficulty is how to transport the model to lower masses, where $M_{\rm hot}/M_{\rm vir}\ll  \Omega_{\rm b}/\Omega_{\rm m}$.
{  The thin black dotted-dashed curve in Fig.~\ref{MW} shows the KS solution using the virial radius and the halo concentration for g1.12e12 at $z=0$
($r_{\rm vir}=229\,$kpc and $c=13.3$ in the simulation without feedback; with SNe, the concentration grows to $c=14.1$ but the curves are almost indistinguishable)
and the universal baryon fraction in NIHAO ($\Omega_{\rm b}/\Omega_{\rm m}\simeq 0.15$).
Normalising $\rho(r)$ so that $\rho/\rho_{\rm NFW}= \Omega_{\rm b}/\Omega_{\rm m}$ at $r_{\rm vir}$ overestimates the density at $r<100\,$kpc with respect to both the observations and NIHAO.

The thick curves show  the KS solution rescaled by $M_{\rm hot}/(f_{\rm b}M_{\rm vir})=0.35$ (as in the simulation without feedback; black curve)
and $M_{\rm hot}/(f_{\rm b}M_{\rm vir})=0.2$ (as in the simulation with feedback; red curve).
The rescaled KS model overlaps the observations better than the simulations do, especially with the lower normalisation,
 but the decrease of $n_{\rm e}(r)$ is much steeper than the observations suggest.
 In contrast,  NIHAO tends to overpredict $n_{\rm e}$ at all radii but the slope is about right (except  for the central cusp in the simulation with feedback).
A cautionary note is that the difference in slope between {\sc GalICS~2.1} and the observations may partly derive from the single power-law models $n_{\rm e}\propto r^\alpha$ used by the observers to fit their data.
This choice precludes them from finding profiles with a central core and a steeper outer region but is an assumption and not an observational result.

Comparing the curves with the filled circles shows that, in NIHAO, the hot gas does not trace the density distribution of the DM at large radii but spreads further out.
This occurs even without feedback and is additional proof of the spillage phenomenon discussed earlier in the article.
Even without rescaling, the black circles still lie above the thin black dotted-dashed curve at large radii.}

\section{Conclusion}

This article contains two sets of conclusions: one that derives purely from the NIHAO simulations, the other that derives from comparing the results of the {\sc GalICS~2.1} SAM to those of the simulations.
We shall start from the former and move to the latter in the second part of this section.

The hot and the cold CGM are different thermodynamic phases (Fig.~\ref{T_rho_diagram}) with different geometries (Figs.~\ref{poststamp_images} and~\ref{poststamp_images_S2})
and  kinematics (Figs.~\ref{characteristic_speeds} and \ref{u1}).
{ The transition from the cold mode to the hot mode is a real physical phenomenon} visible through all  these different aspects. It does not derive from separating the cold phase and the hot phase at an arbitrary temperature, which appears nowhere in our article.

{The most physical criterion to separate the hot phase from the cold phase is to use the equation of state of the IGM}, which is close to being adiabatic.
The cold mode is characterised by entropies that fall below the entropy the IGM due to radiative cooling.
The hot mode is characterised by entropies that rise above it due to shock-heating.
The quantitative results obtained with this criterion (Fig.~\ref{fhot}) are consistent with the qualitative picture from temperature maps (Figs.~\ref{poststamp_images}--\ref{poststamp_images_S2}).

The distribution of the maximum entropy $K_{\rm max}$ of baryonic SPH particles reveals four types of particles:
\begin{enumerate}
\item{Low and lower-intermediate entropy baryons ($K_{\rm max}\lsim 1{\rm\,keV\,cm^2}$) that have never been in the hot phase. }
\item{Upper-intermediate entropy baryons ($1{\rm\,keV\,cm^2}< K_{\rm max}< 10{\rm\,keV\,cm^2}$) that were shock-heated but have cooled or are in a cooling flow.}
\item{High entropy baryons ($10{\rm\,keV\,cm^2}\lsim K_{\rm max}< 100{\rm\,keV\,cm^2}$) that were shock-heated and are still in the hot phase at $z=0$.}
\item{Very high entropy baryons ($K_{\rm max}\gsim 100{\rm\,keV\,cm^2}$) that spilled out of the virial radius after shock-heating.}
\end{enumerate}
The first type corresponds to the cold mode, the other three to the hot mode.

Without feedback, the four groups of particles account for about 50, 6, 25 and 19 per cent of the total mass 
of all baryons that have been {  within the virial radius of g7.55e11} {\it at some point}.
{  The values for g1.12e12 are quite similar: 40, 4, 28 and 28 per cent, respectively.}

SN feedback lowers the halo baryon fraction by $\sim 25$ per cent and increases 
 $K_{\rm max}$ for the  baryons within $r_{\rm vir}$.
Most of the low-entropy baryons heated by SNe move to an intermediate-entropy warm-hot phase without crossing the line that separates the hot phase from the cold phase. 
Those that do are rarely reheated to $K> 10{\rm\,keV\,cm^2}$.
Hence, they have short cooling times compared to the bulk of the hot CGM.
{  The cooling of gas heated by SNe raises the contribution of the hot mode to the stellar mass at $z=0$ from 7--13 per cent to 20--40 per cent 
and its contribution to the SFR at $z=0$ from  10--20 per cent to 30--70 per cent (Fig.~\ref{hot_mode_sf}; the lower values are for g7.55e11, the upper values for g1.12e12).

Our quantitative results should be taken with caution for three reasons. They are based on only two objects. The differences from one object to the another can be significant.
They are also likely to depend on how  feedback is implemented in the simulations. 

 In relation to the first point, one may be concerned that g7.55e11 and g1.12e12 yield similar results because we have selected similar objects.
Both were selected  discarding the 10 per cent of the NIHAO objects with a  history of major merging and both have a final virial mass comparable to that of the Milky Way,
although $M_{\rm vir}(z=0)$  is 40 per cent larger for g1.12e12 than it is 
for g7.55e11. This is where the similarity stops, however.
At $t=2\,$Gyr, the virial mass of g1.12e12 is almost double than that of g7.55e11.
A hot quasi-static atmosphere emerges at $z\sim 1$ in g7.55e11  but at $z\sim 3$ in g1.12e12.
The baryons that spill out of the virial radius even without feedback are about 20 per cent for g7.55e11 but almost 40 per cent for g1.12e12.
The DM halo of g1.12e12 is significantly more concentrated ($c=13$--$14$) than the halo of g7.55e11 ($c=9$--$11$).
Hence, g7.55e11 and g1.12e12 have no particular similarity beyond being both massive spirals.

In relation to third concern, one may also wonder to what extent our results depend on the feedback model in NIHAO (which \citealp{wang_etal15} had carefully calibrated to reproduce the observations).
Any kind of feedback will reduce or delay the formation of stars to some extent, but feedback must heat the gas above a critical entropy to affect our analysis.
Weaker feedback will simply make our results resemble more closely the case without it.
Stronger feedback can have two opposite effects.
If more baryons are heated to upper-intermediate entropies, then star formation will be delayed but there will be more cooling later on; hence, cooling will become {\it more} important (unless an AGN switches on in the meantime and cooling is shut down altogether). 
An extremely violent feedback that heats the gas to $K>10{\rm\,keV\,cm}^2$ will suppress cooling rather than promoting it, but that is hardly conceivable with SNe\footnote{An AGN can heat the gas to very high entropy and quench star formation, but this article is concerned with massive spirals like the Milky Way and not with the formation of early-type galaxies.}.
Finally, although the model for SN feedback can change the contribution of cooling to the accretion rate onto galaxies, the cooling of SN-heated gas does not alter our fundamental conclusion
that the gas heated by accretion shocks cools very inefficiently.

Despite these uncertainties, the qualitative picture that emerges from the NIHAO simulations is robust and very much} the same with or without feedback:
\begin{itemize}
\item{Cold accretion is the main mode of galaxy formation even though the halo has been in the hot mode throughout most of the cosmic lifetime.
The cooling of hot gas contributes to $\lsim 20$ per cent of the final stellar mass.}
\item{The hot CGM has such long cooling times that it is approximately adiabatic
($K\simeq K_{\rm max}$; Figs.~\ref{ent_dist2} and \ref{ent_dist3}).}
\end{itemize}
The first point means that our results are in substantial agreement with \citet{dekel_birnboim06} and \citet{dekel_etal09}.
The second point gives us a hint of how we can model the entropy distribution $K(M)$ of the hot CGM.
Let $K_{\rm ad}(M)$ be the entropy distribution of the baryons in an adiabatic simulation (i.e. the baseline entropy profile of \citealp{voit_etal05}), and let $M_{\rm cold}$ be the total mass of the halo baryons that have never been hot or have cooled
(i.e. the particles of types i and ii).
{\it The quasi-adiabatic behaviour of the hot CGM implies that $K(M)\simeq K_{\rm ad}(M)$ for $M>M_{\rm cold}$}.

Fig.~\ref{MW} shows that the inefficient cooling of the hot CGM in the NIHAO simulations is not due to densities that are too low. 
The electron densities in NIHAO are at the upper boundary of what is observationally permissible if we neglect the central $\sim 10\,$kpc where $n_{\rm e}$ is much higher in the simulation with feedback than it is in the observations.

After summarising the results of the simulations themselves, 
we shall now discuss the ability of {\sc GalICS~2.1} to predict their results. {\sc GalICS~2.1} uses the shock-stability argument of \citet{birnboim_dekel03} and \citet{dekel_birnboim06}
to decide to whether a halo is accreting in the cold or the hot mode.
When $t_{\rm cool}<\Gamma_{\rm c}t_{\rm comp}$ ($\Gamma_{\rm c}=5/7$), the gas cools faster than it is heated by compression. Hence, the shock is isothermal (it does not increase the temperature of the gas).
When $t_{\rm cool}\ge \Gamma_{\rm c} t_{\rm comp}$, a stable shock propagates through the gas and heats it to temperatures of the order of the virial temperature; 
the cold filaments are destroyed and replaced by a quasi-static hot spherical atmosphere.
The cooling time $t_{\rm cool}$ and the compression time $t_{\rm comp}$ are computed assuming that the filaments are in free fall all the way down to the shock radius $r_{\rm s}$  and that they are
accreted from a constant solid angle $\Omega=4\pi/C$ with $C=9$.
 
The NIHAO simulations (Figs.~\ref{poststamp_images} and \ref{fhot}) show that this picture is an approximation:
$f_{\rm hot}\sim 0.2$--$0.3$ even at the highest redshifts, where shock fronts are visible alongside the filaments.
{  The gas is not exactly in free fall.}
There are times when the infall speed $u_1$ at the virial radius deviates from $v_{\rm vir}$ (Fig.~\ref{u1}).
Even small deviations may not be negligible because $u_1$ enters Eq.~(\ref{tcomp_over_tcool2}) at the fifth power.
{  Our estimate of $\Omega$ is based on purely cosmological considerations and neglects hydrodynamical effects.
The normalisation of $\Omega$ (set by $C$) may compensate other errors and not reflect the true solid angle of accretion.}

Despite these limitations, which are the inevitable consequence of applying a highly idealised geometry (purely radial conical inflows) to complex hydrodynamics, the NIHAO simulations confirm
that the assumptions and predictions of {\sc GalICS~2.1} are fundamentally correct:
\begin{itemize}
\item{The infall speed is normally consistent with $v_{\rm vir}$ at $r_{\rm vir}$.
It increases at small radii following a freefall law at high $z$ when $r_{\rm s}\ll r_{\rm vir}$.
The trend is reversed once the shock radius reaches the virial radius. }
\item{When the filaments are still visible,
the $\Omega$ computed by {\sc GalICS~2.1} is consistent with the findings of NIHAO within a factor $<2$.}
\item{$f_{\rm hot}$ strongly correlates with $\Gamma=t_{\rm comp}/t_{\rm cool}$.}
\item{{\sc GalICS~2.1} correctly predicts the epoch at which the transition from the cold mode to the hot mode occurs.}
\end{itemize}

\section*{Data availability statement}

The results of these articles do not depend on any non-public data.

\section*{Acknowledgements}

 AC thanks  Louis Carton for compiling the data used to plot Fig.~\ref{MW} and making them available to us.
 
\bibliographystyle{mn2e} 

\bibliography{ref_av}

\renewcommand{\thefigure}{S1}

\begin{figure*}
\begin{center}$
\begin{array}{cccc}
  \includegraphics[width=0.25\hsize]{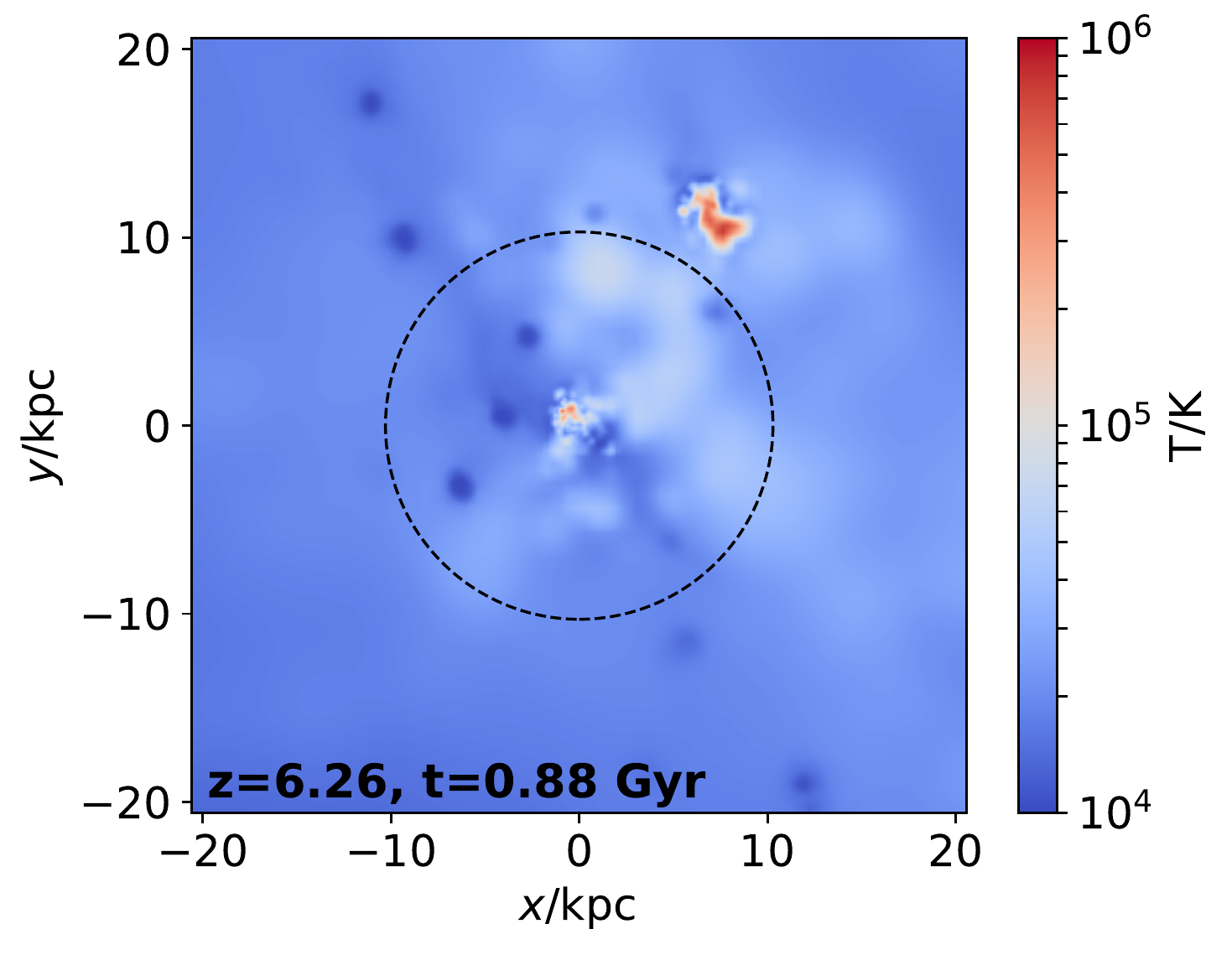} 
  \includegraphics[width=0.25\hsize]{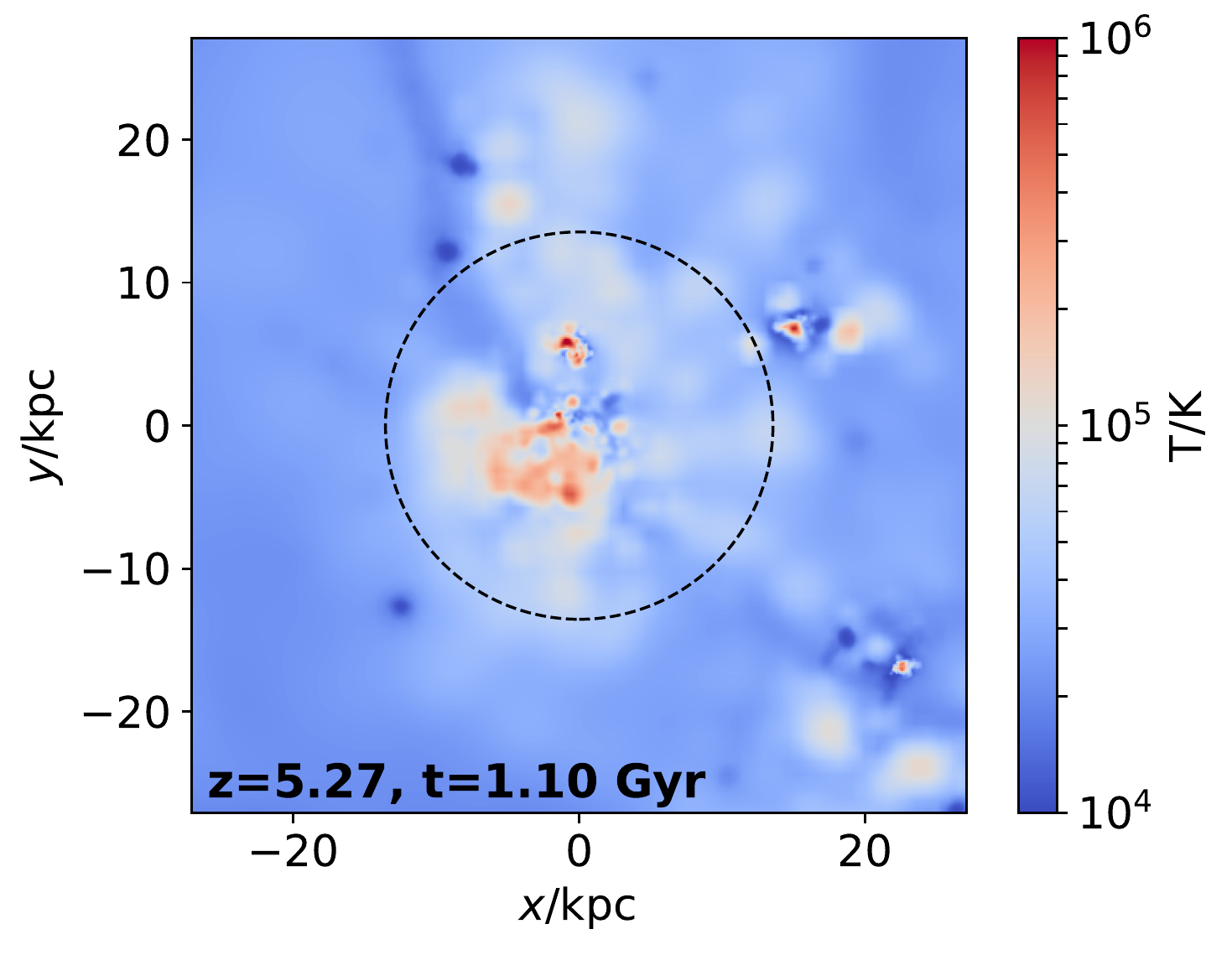}
   \includegraphics[width=0.25\hsize]{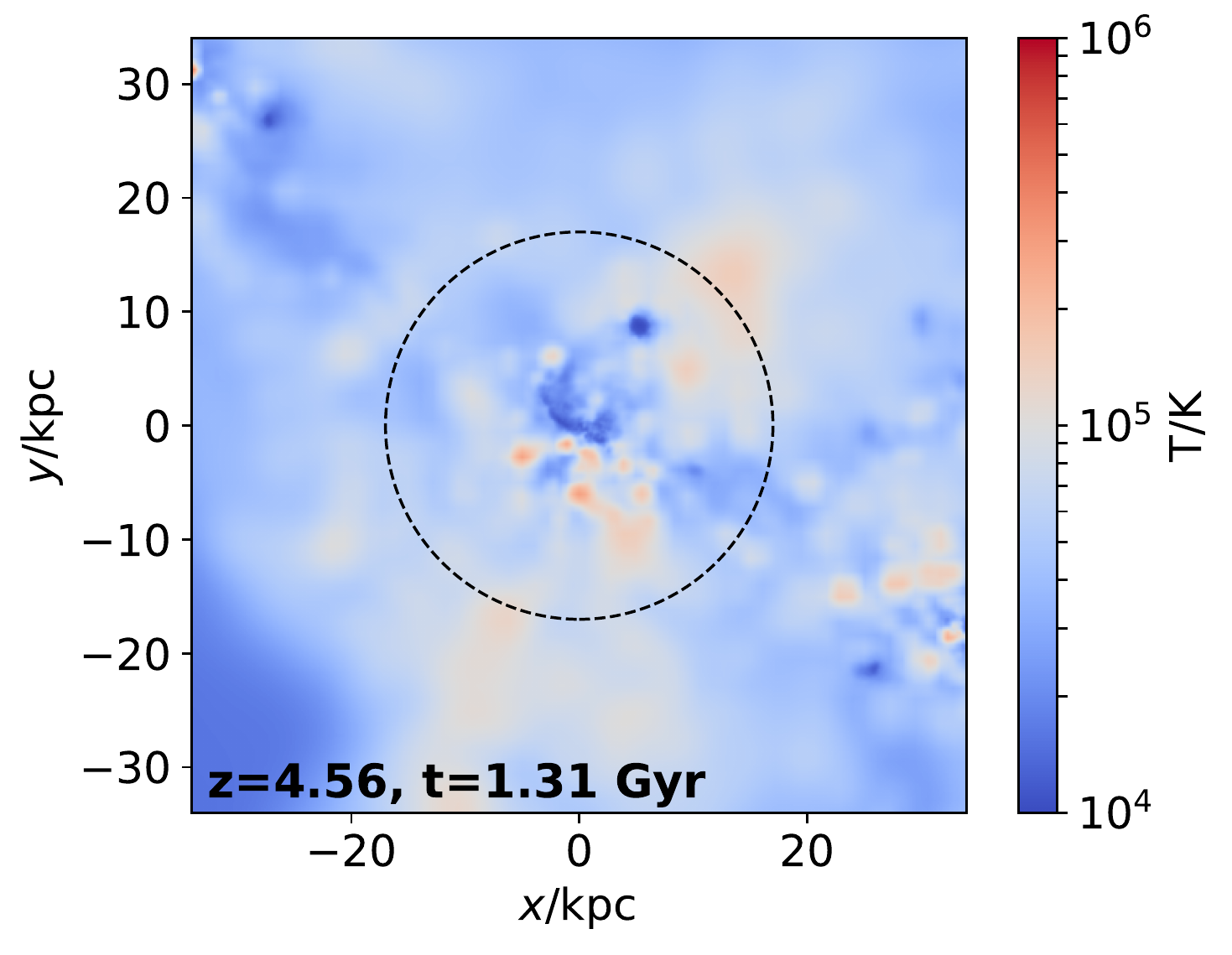}
  \includegraphics[width=0.25\hsize]{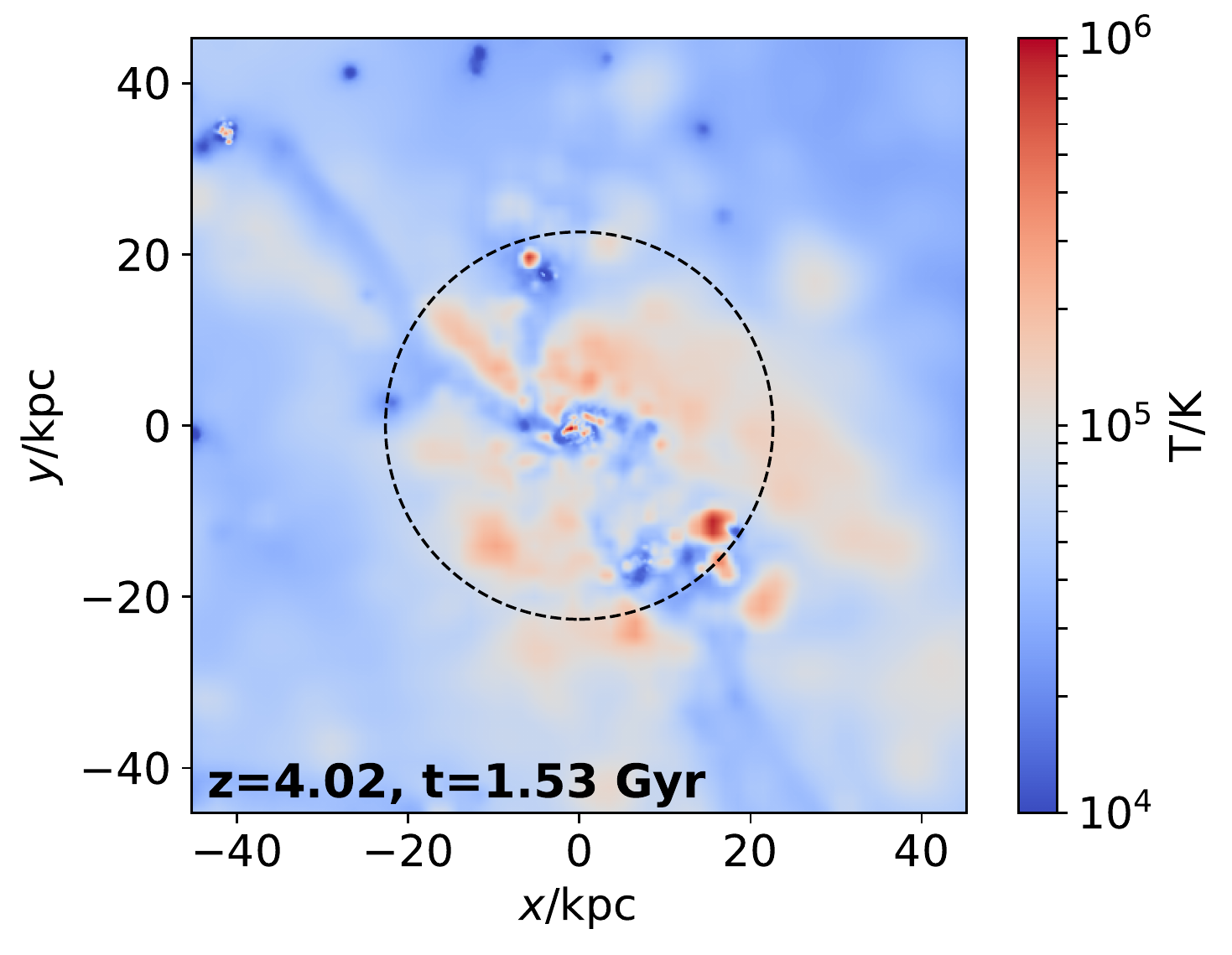}\\
  \includegraphics[width=0.25\hsize]{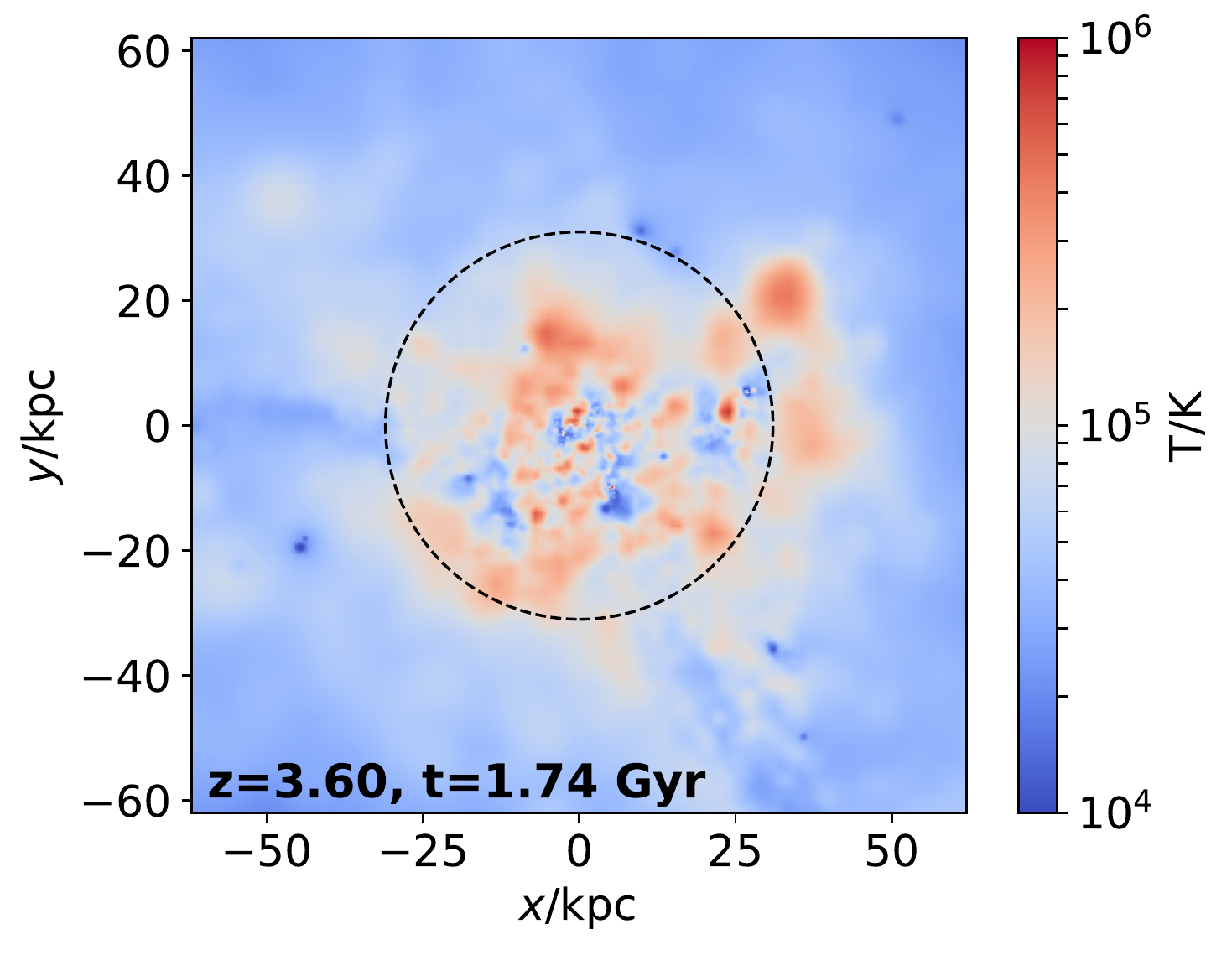}
  \includegraphics[width=0.25\hsize]{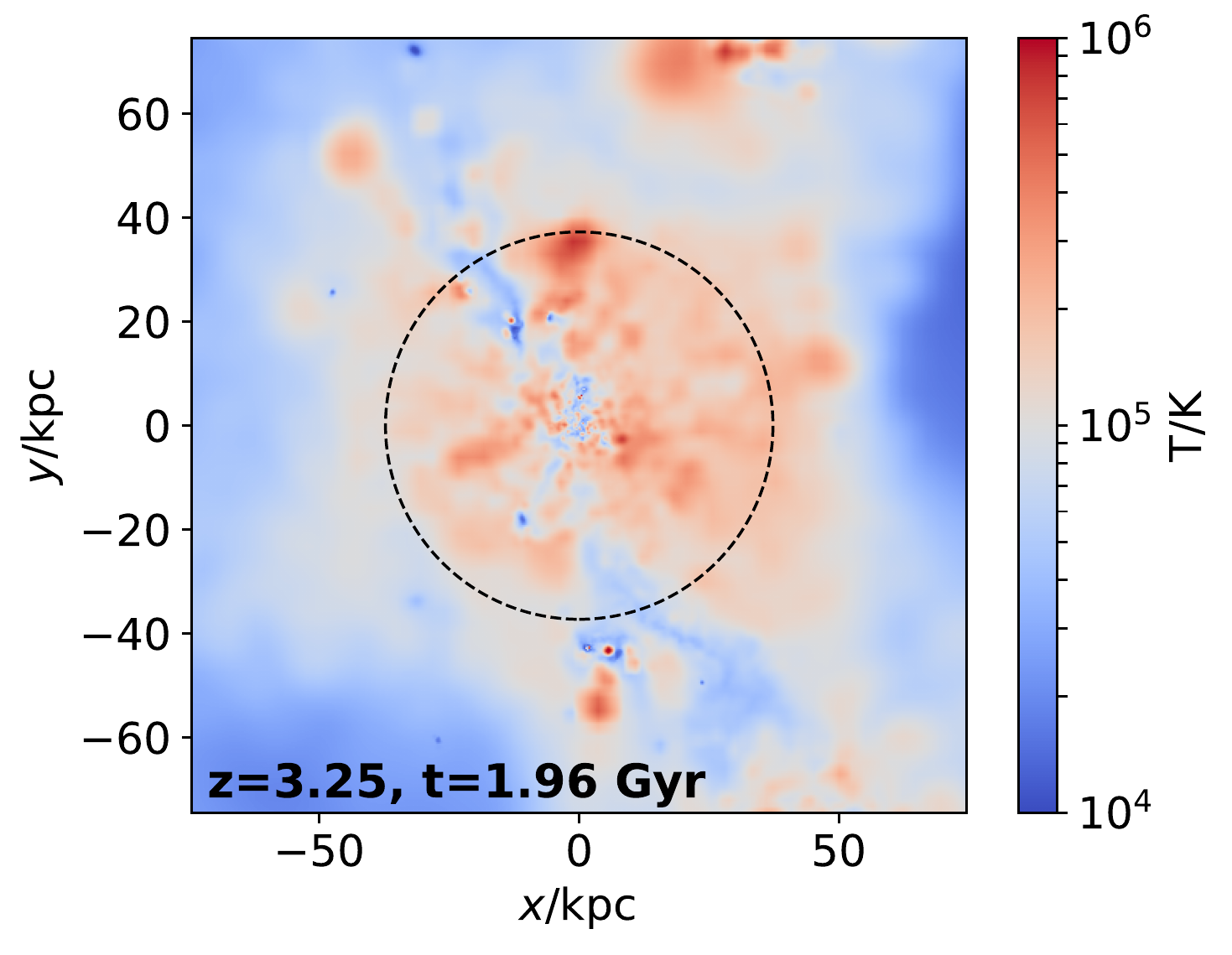}
    \includegraphics[width=0.25\hsize]{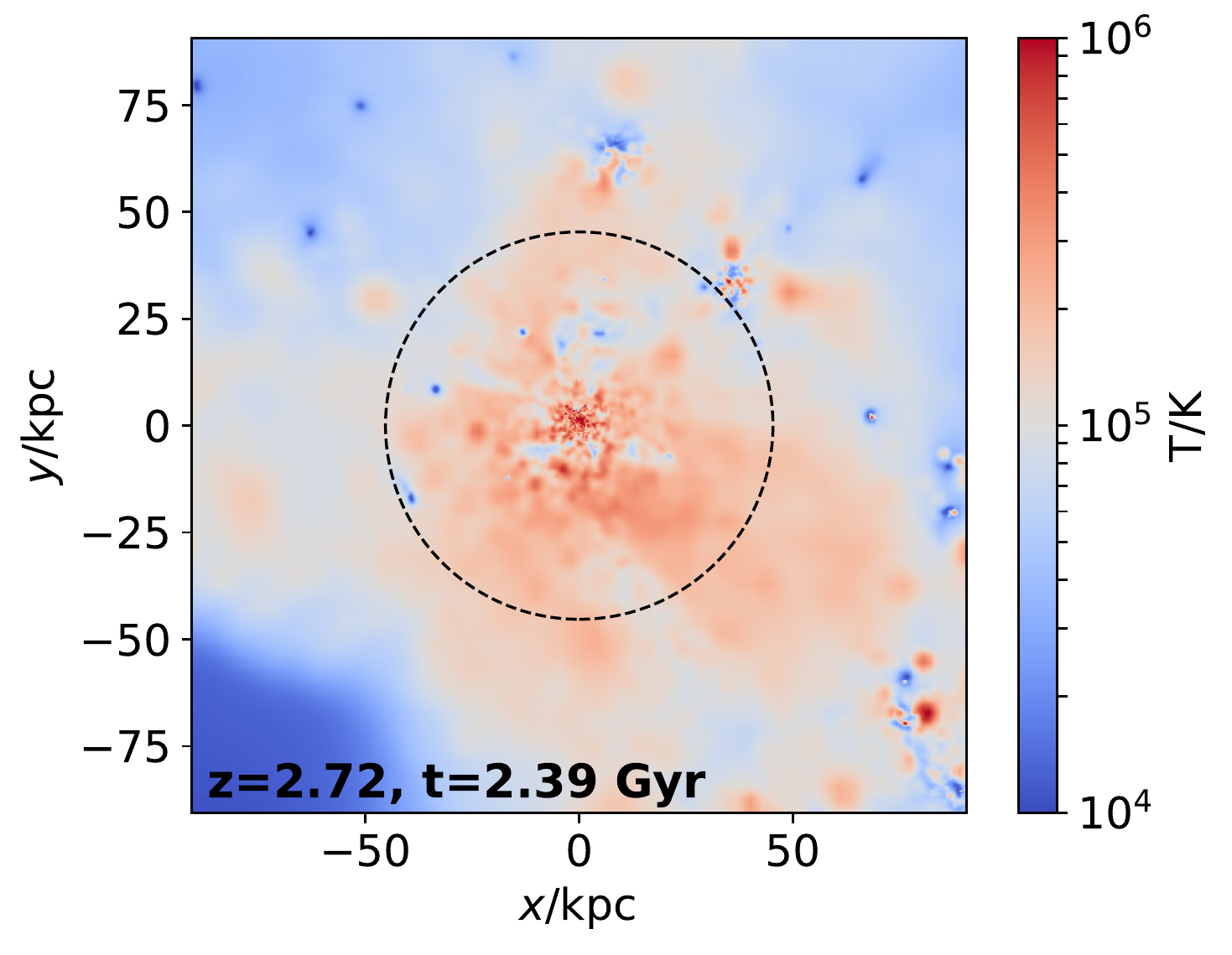}
  \includegraphics[width=0.25\hsize]{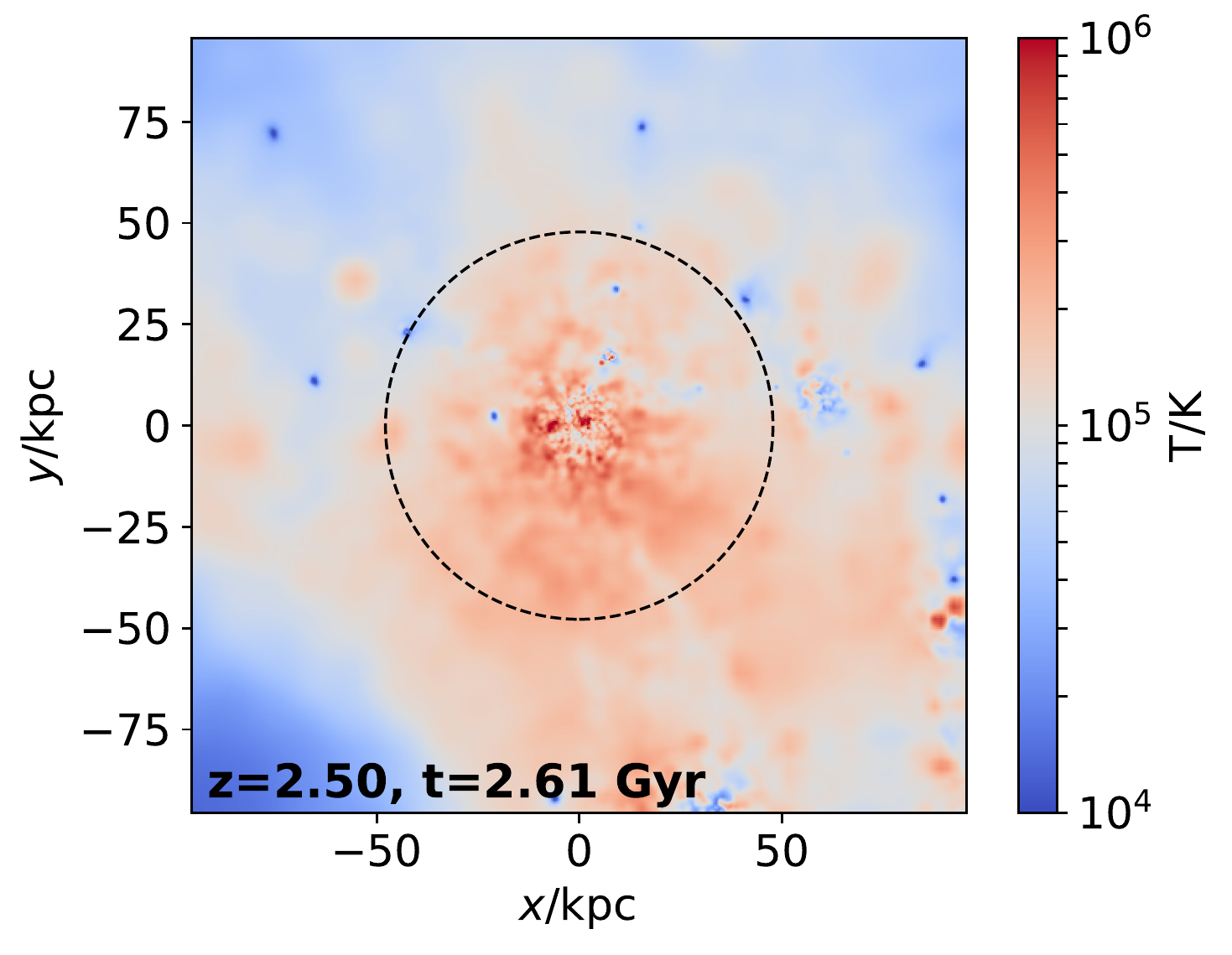}\\
  \includegraphics[width=0.25\hsize]{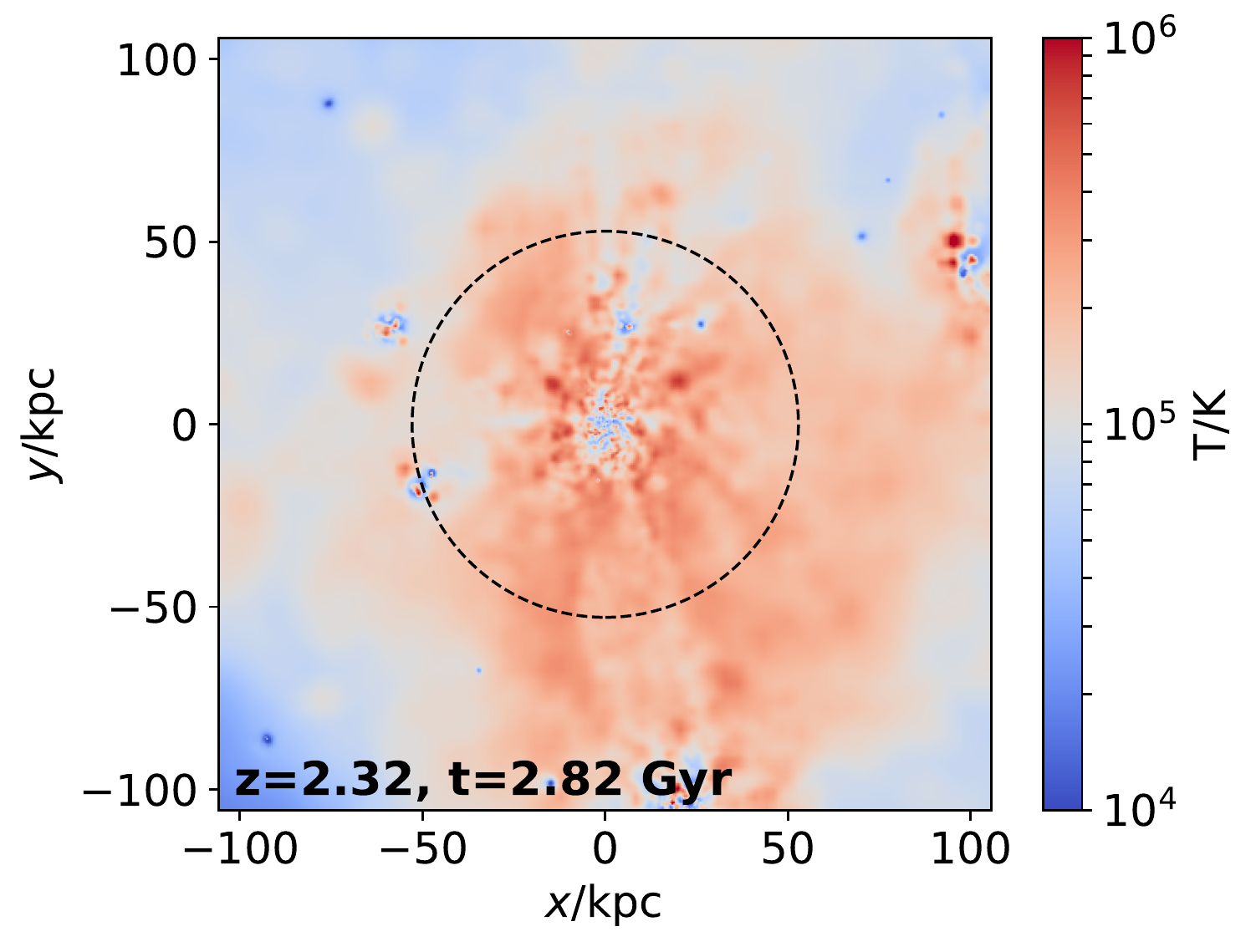}
  \includegraphics[width=0.25\hsize]{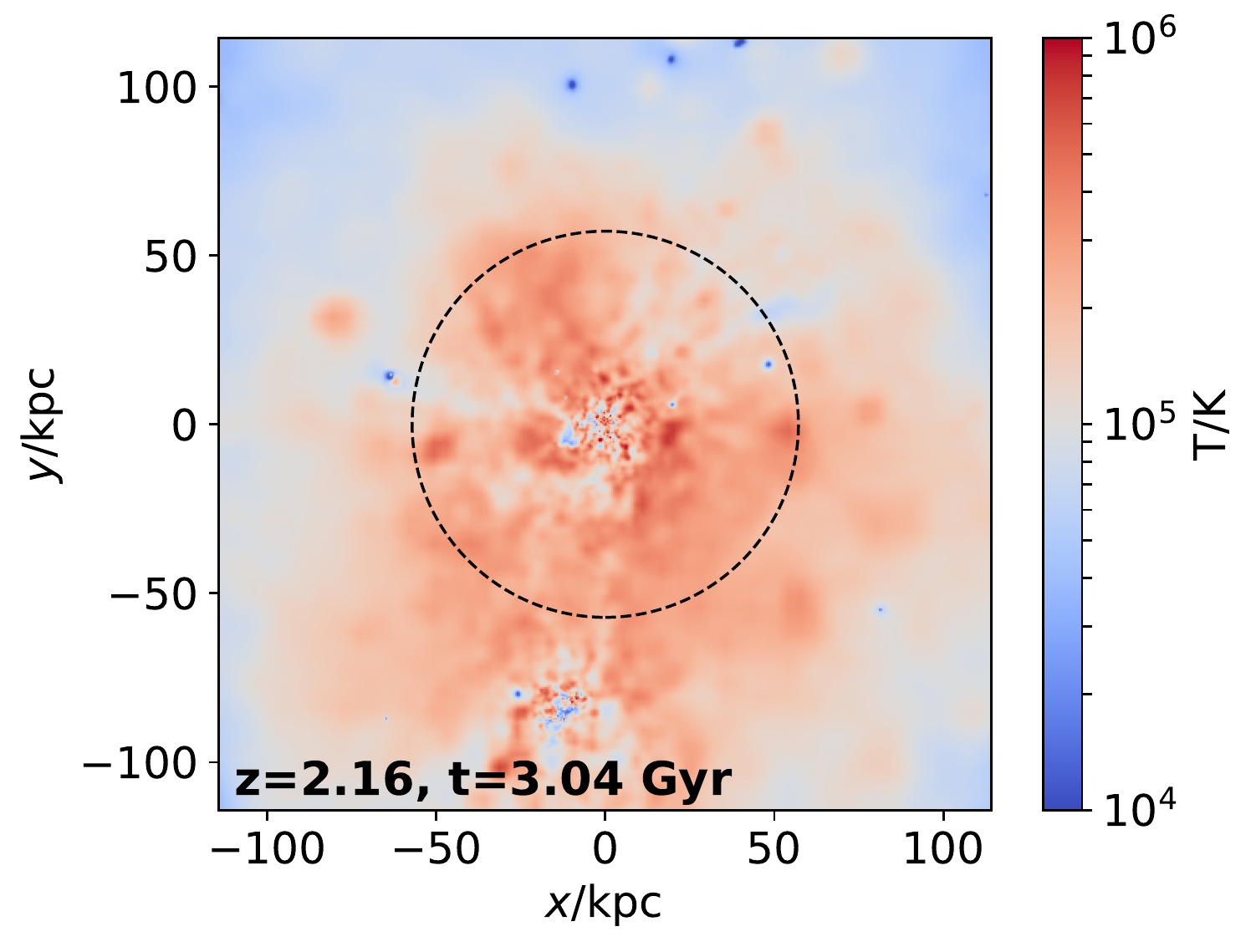}
  \includegraphics[width=0.25\hsize]{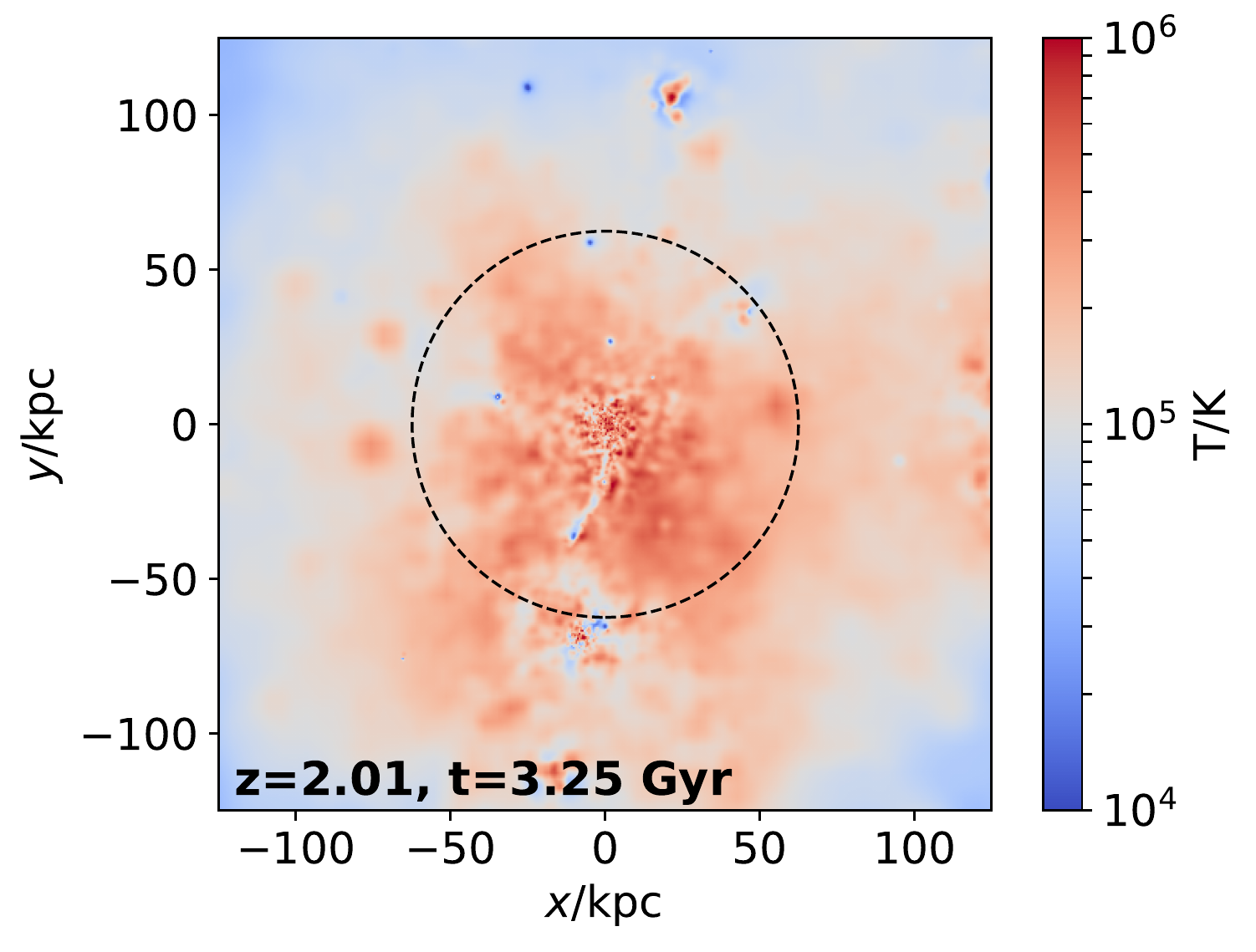}
  \includegraphics[width=0.25\hsize]{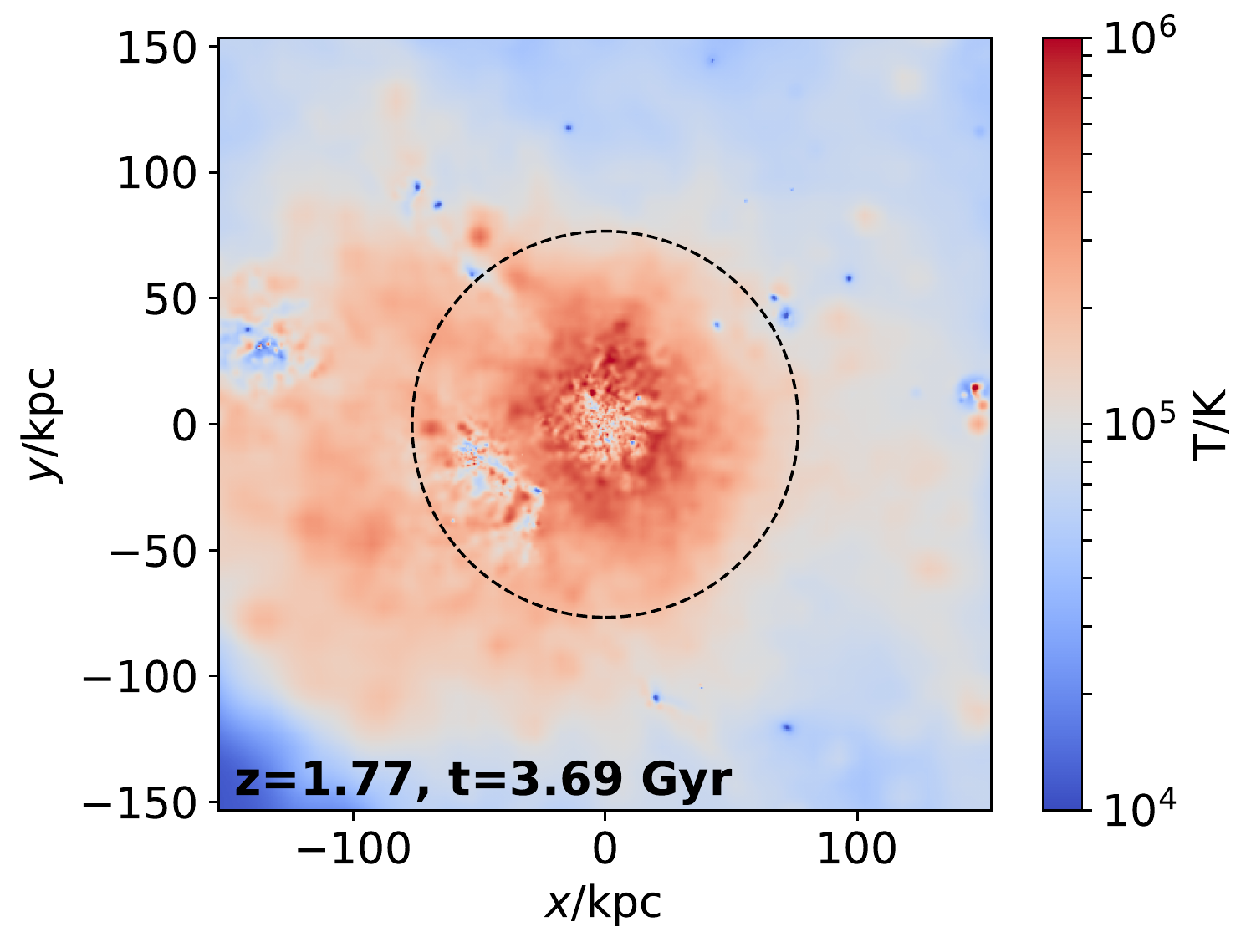}\\
  \includegraphics[width=0.25\hsize]{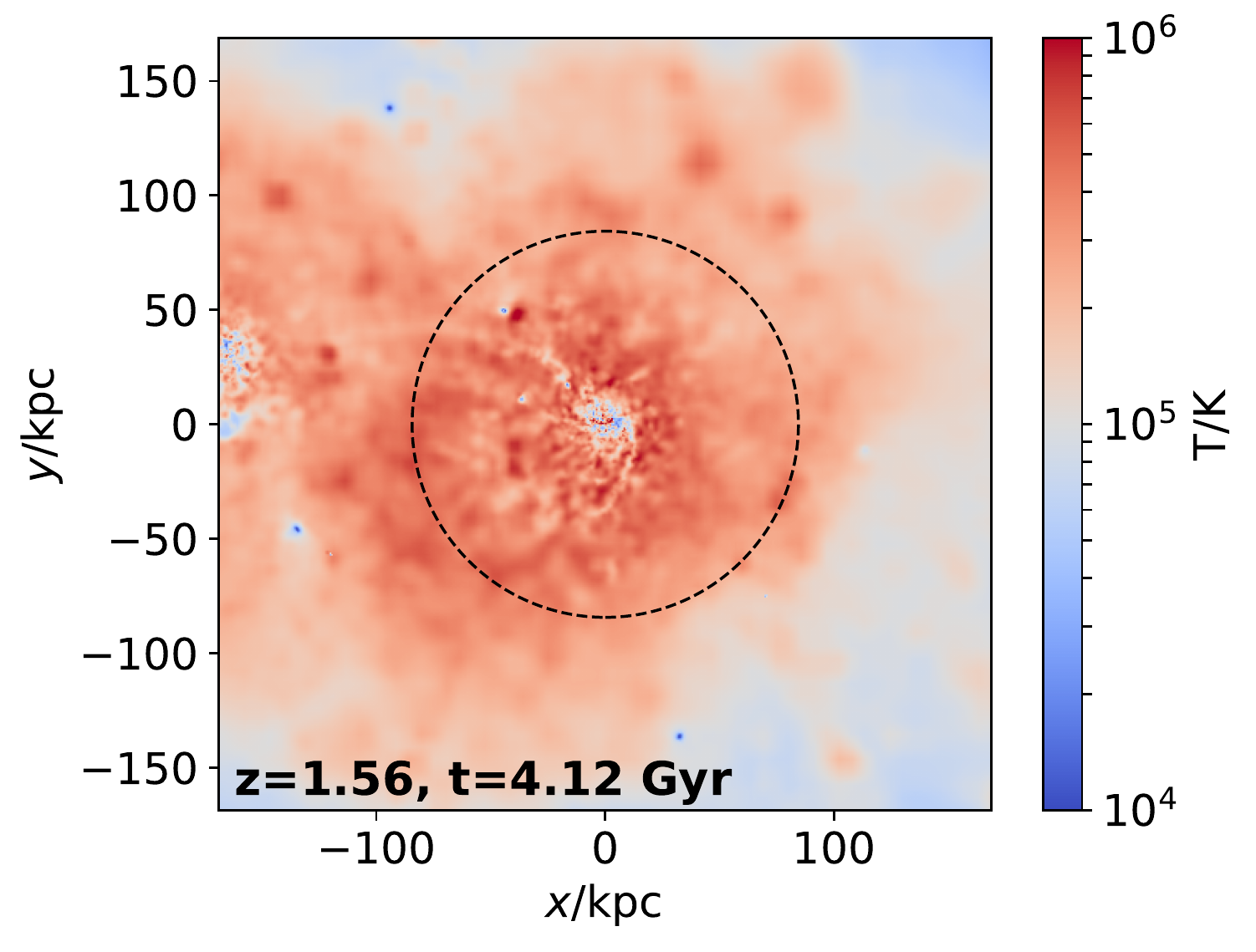}
  \includegraphics[width=0.25\hsize]{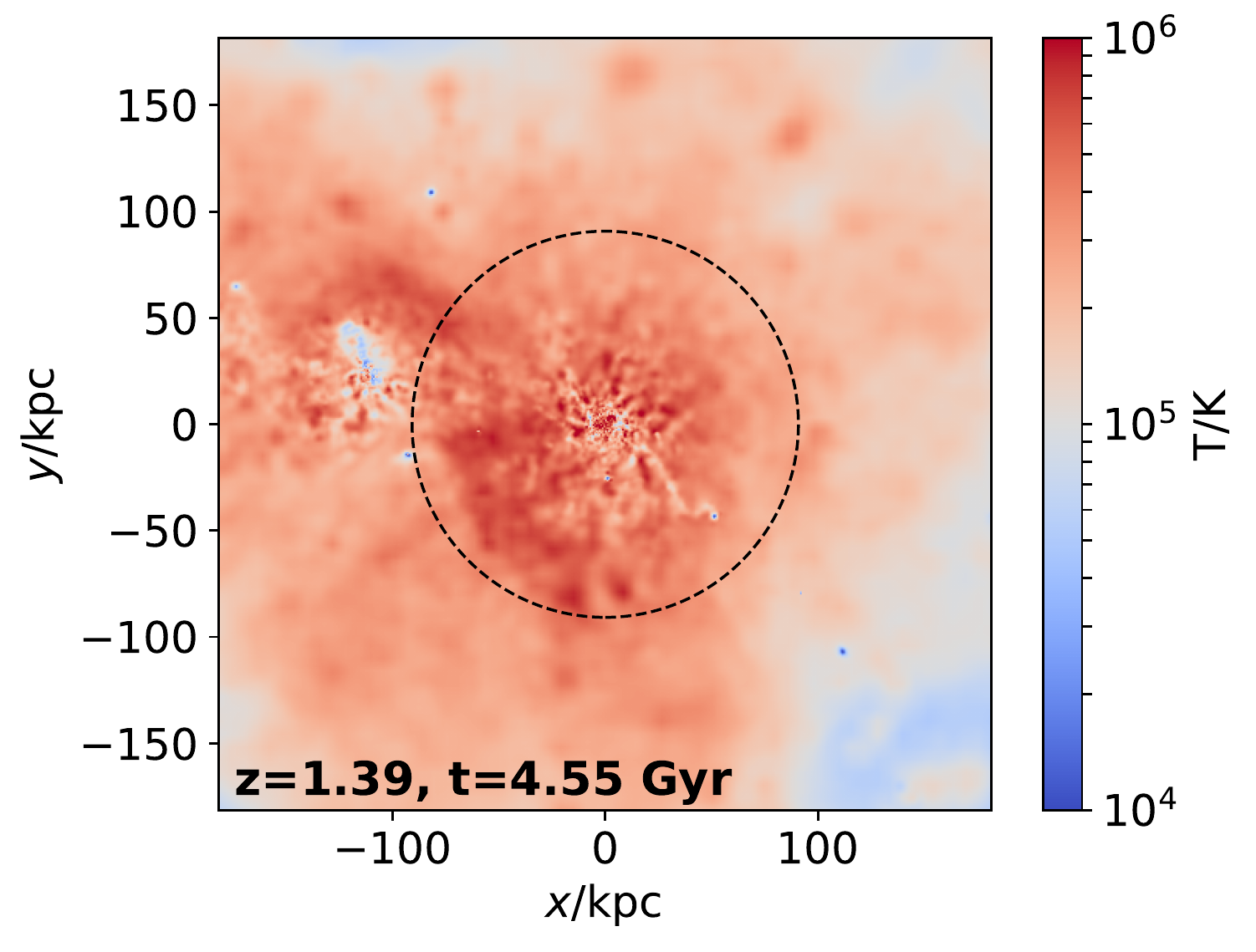}
  \includegraphics[width=0.25\hsize]{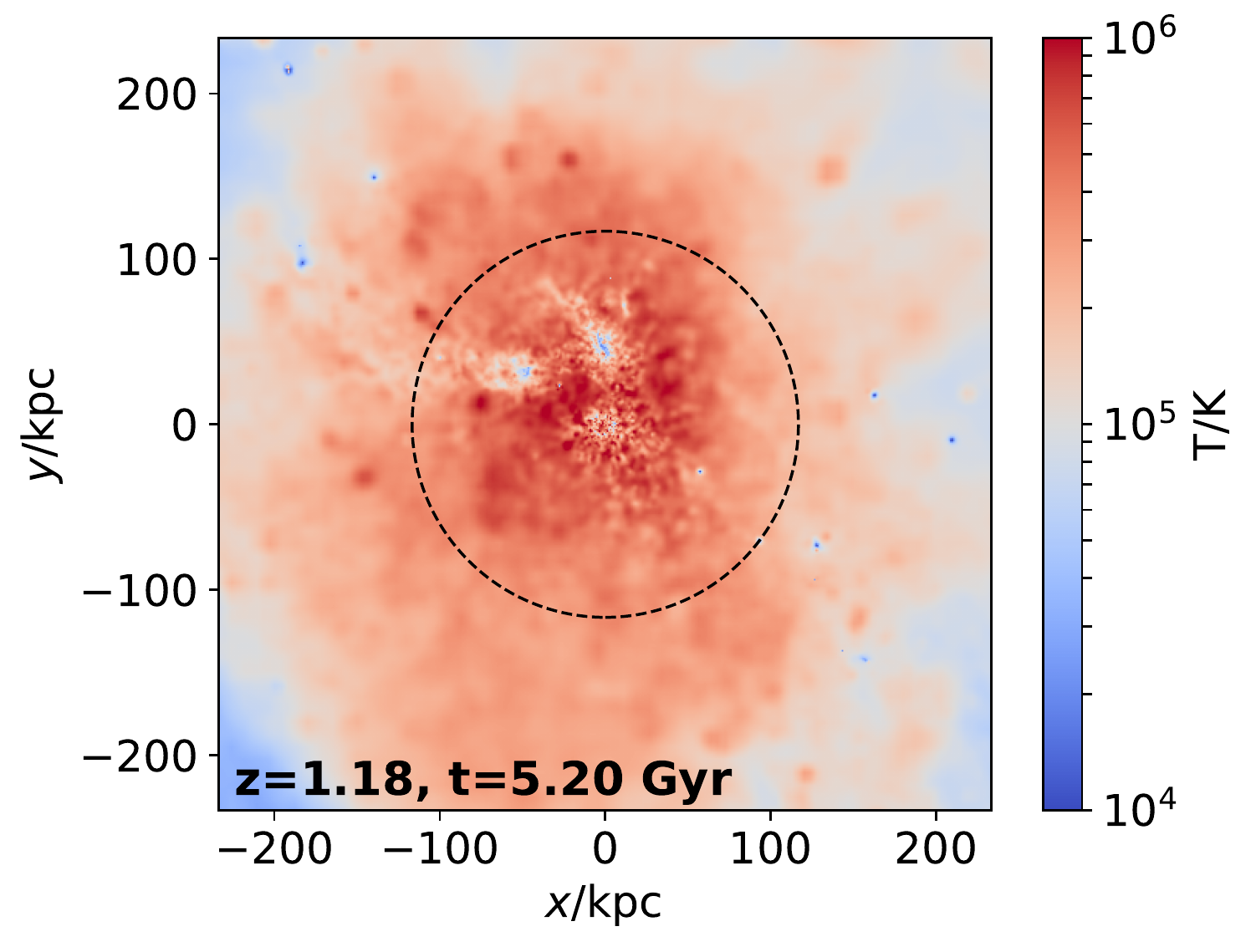}
  \includegraphics[width=0.25\hsize]{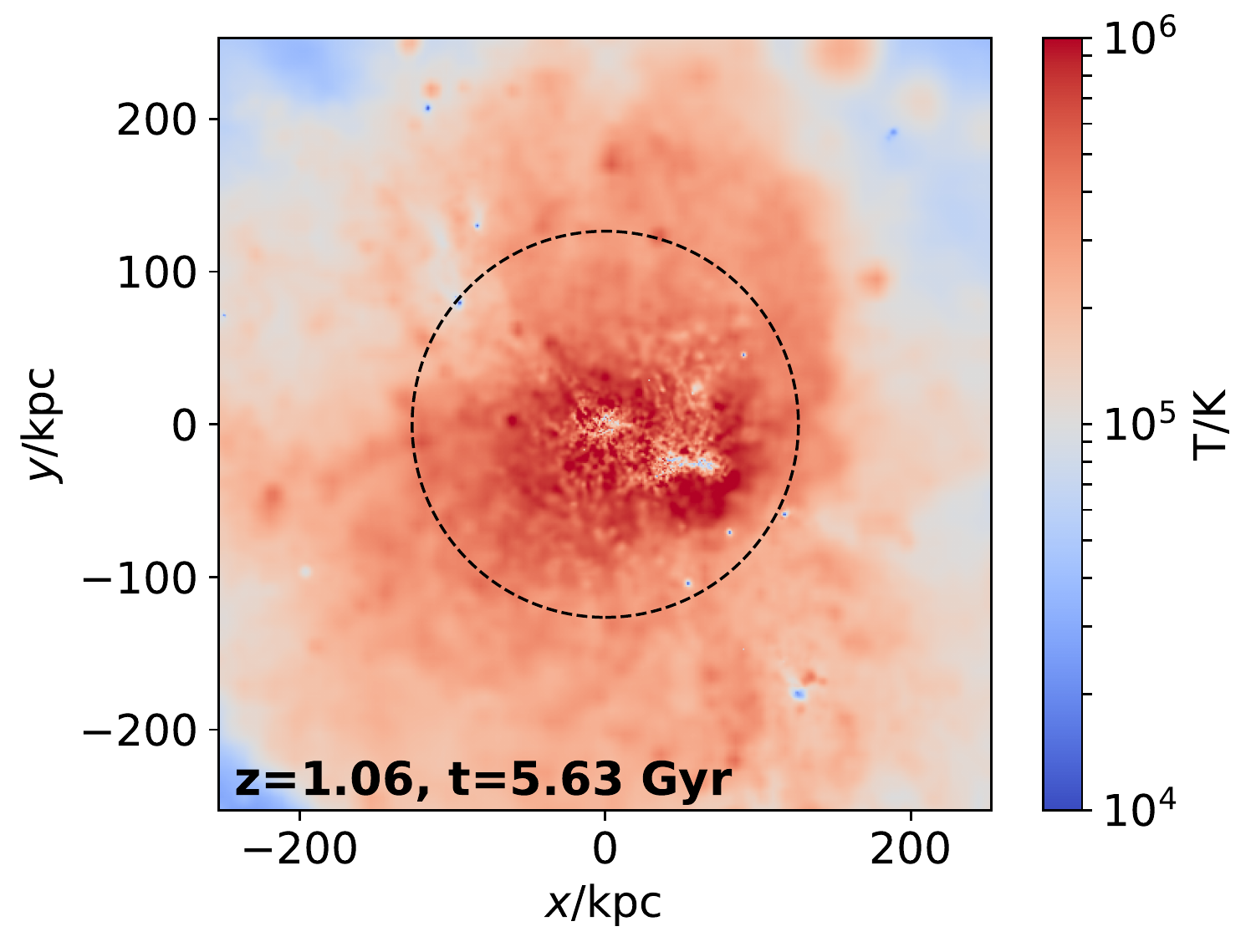}\\
  \includegraphics[width=0.25\hsize]{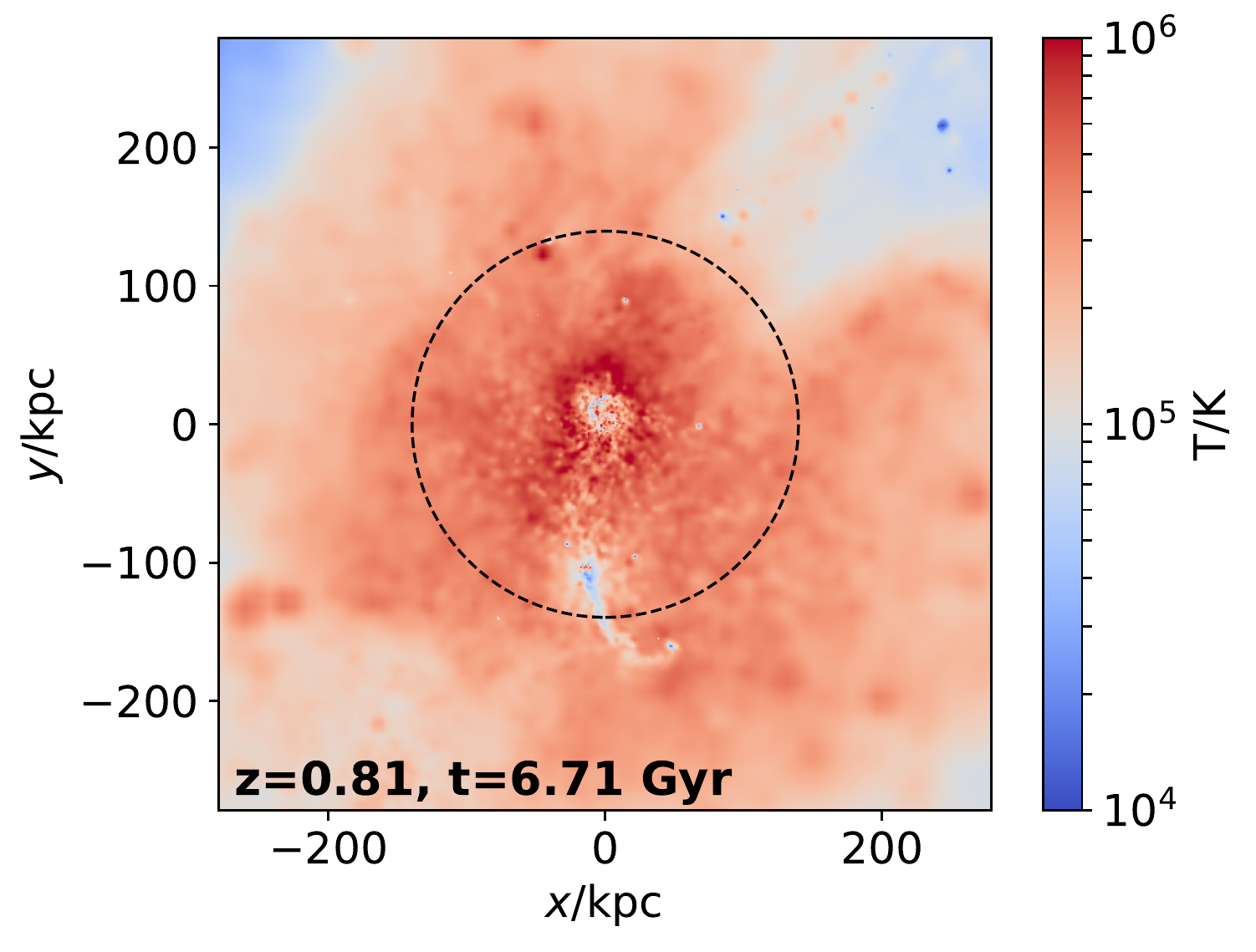}
  \includegraphics[width=0.25\hsize]{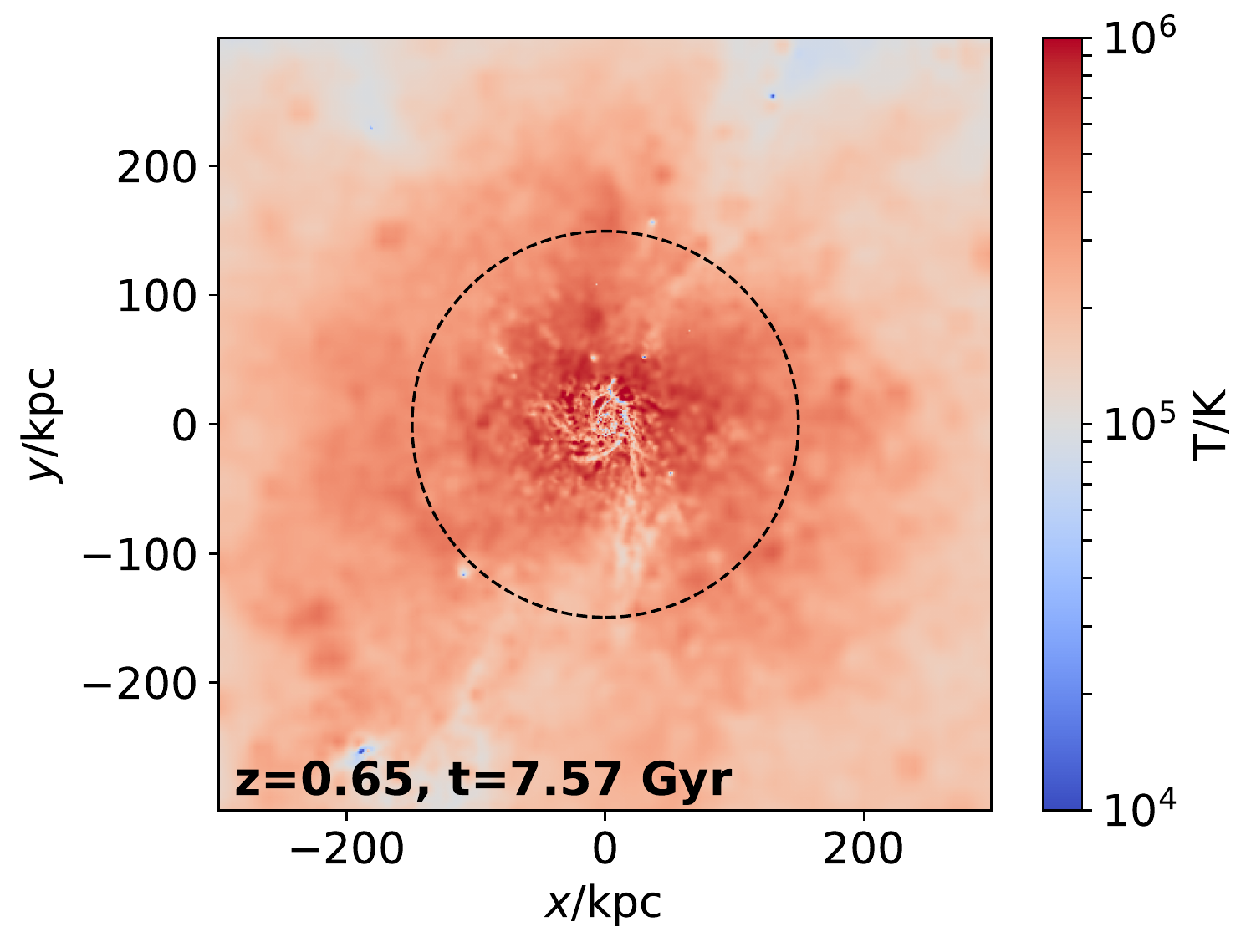}
  \includegraphics[width=0.25\hsize]{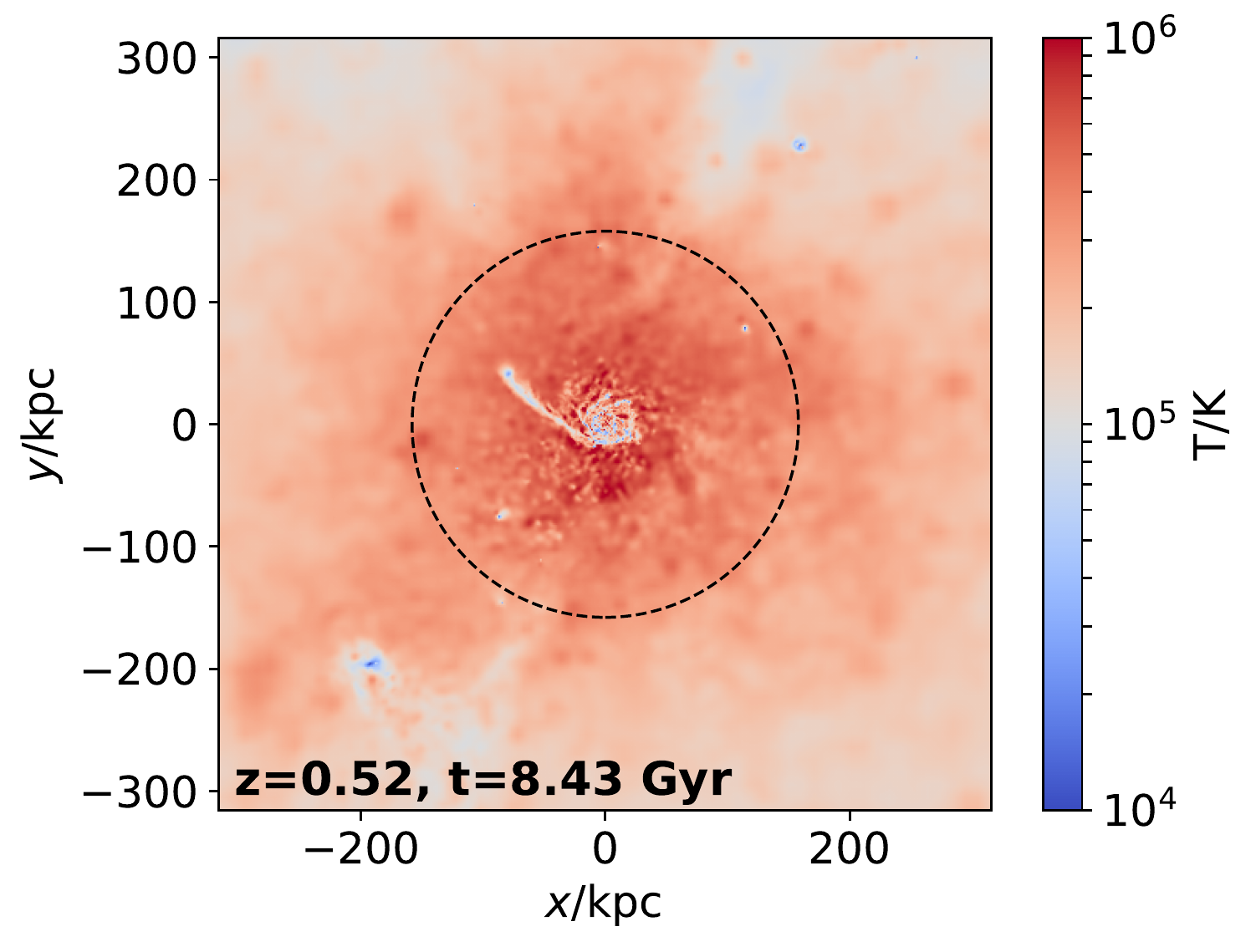}
    \includegraphics[width=0.25\hsize]{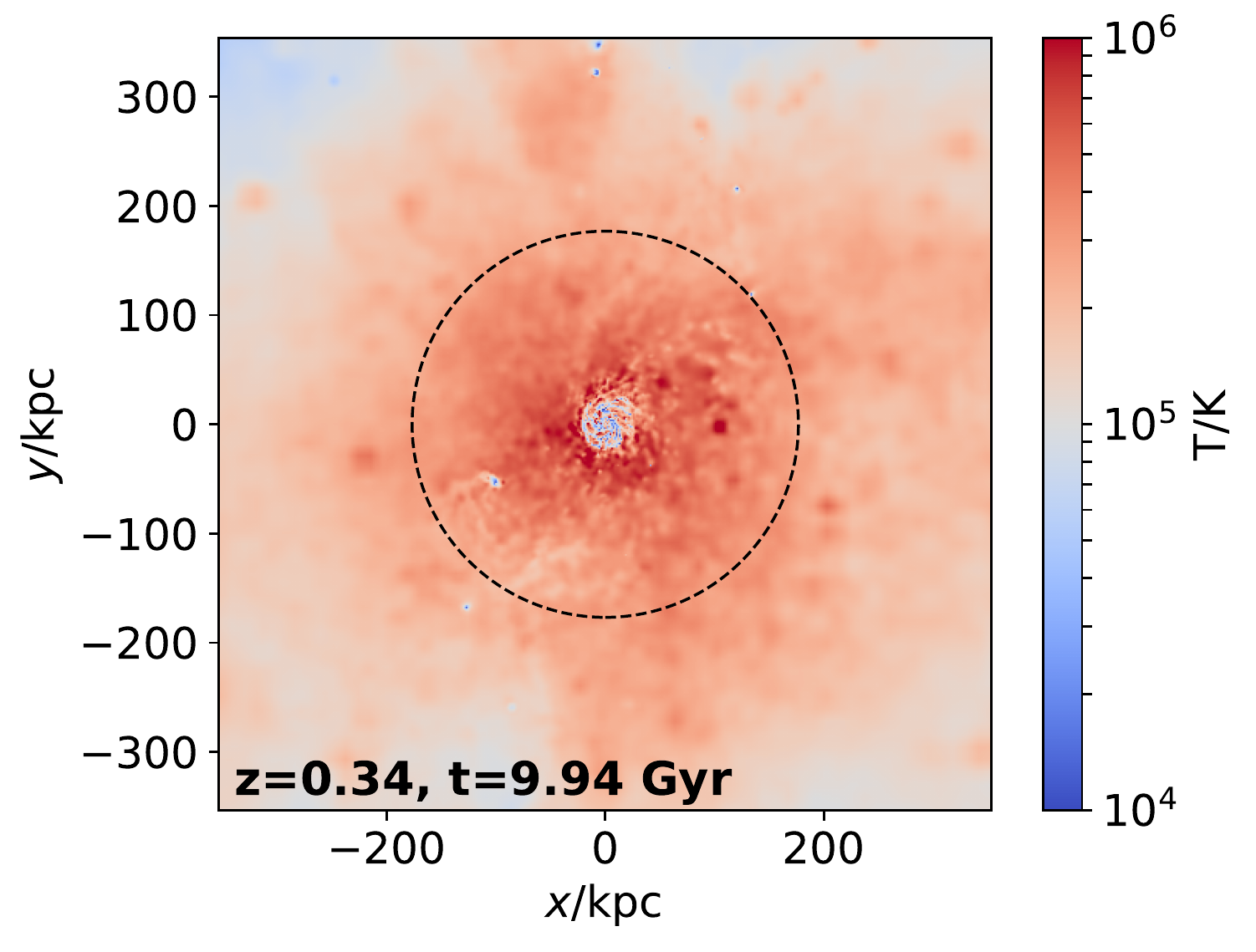}\\
  \includegraphics[width=0.25\hsize]{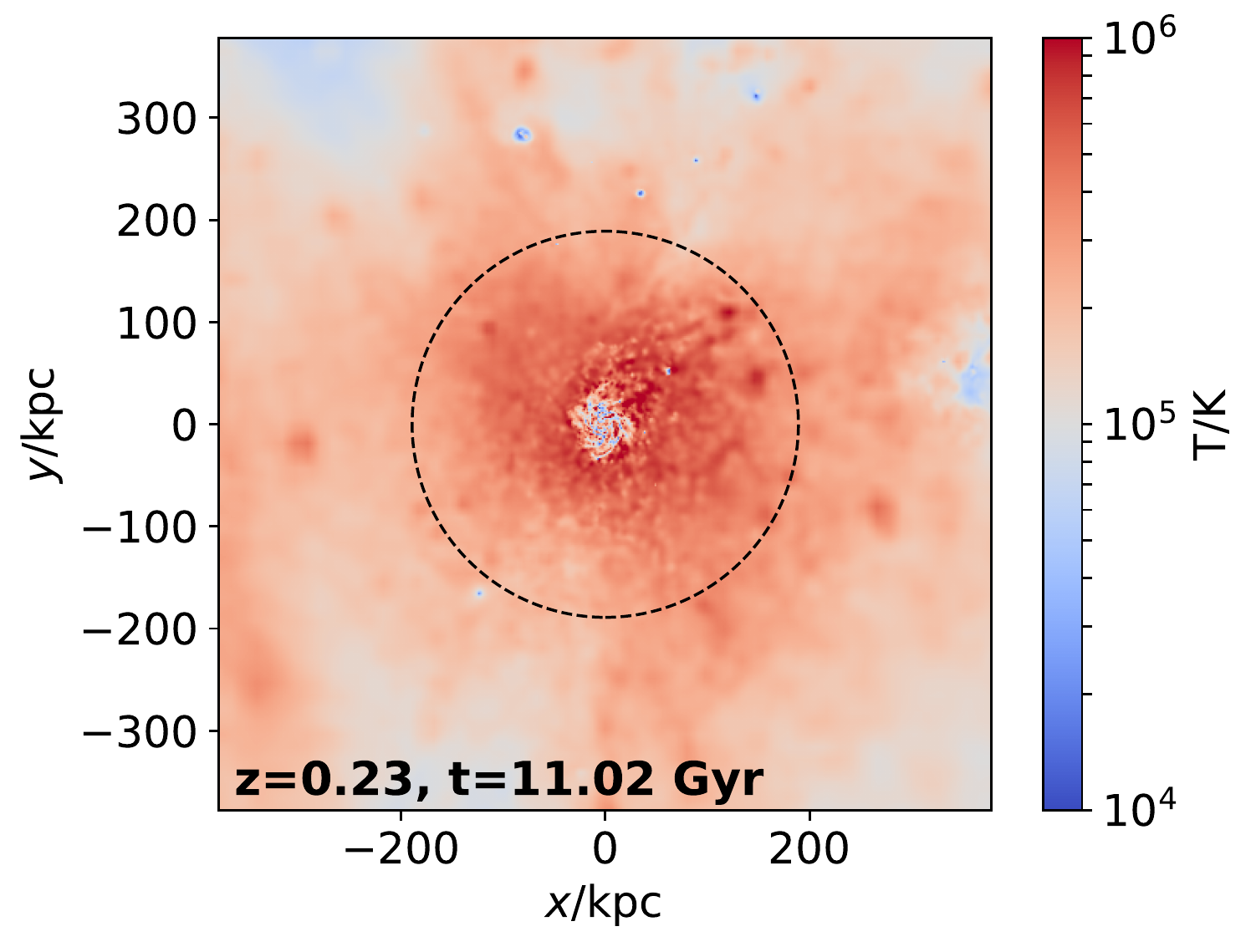}
  \includegraphics[width=0.25\hsize]{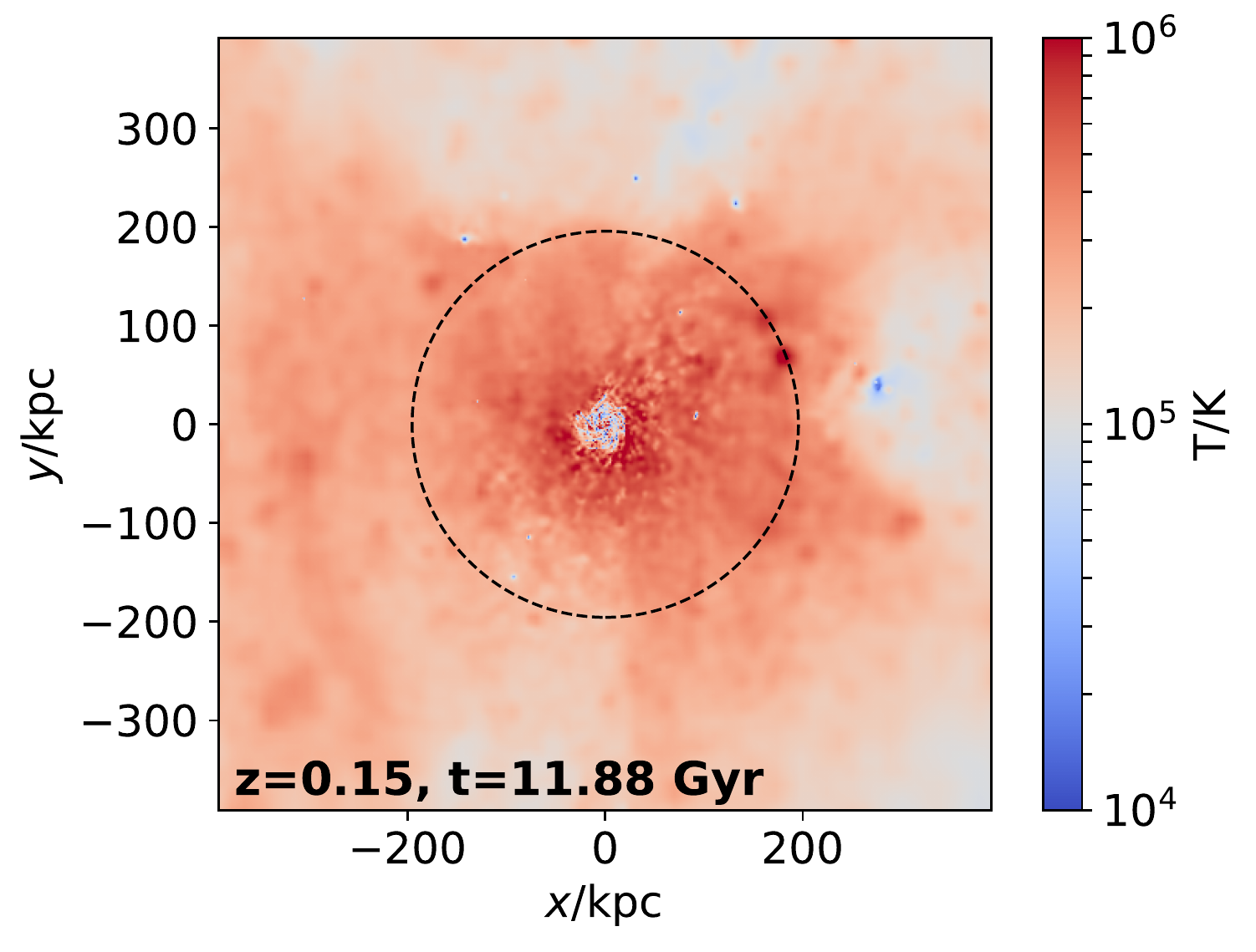}
   \includegraphics[width=0.25\hsize]{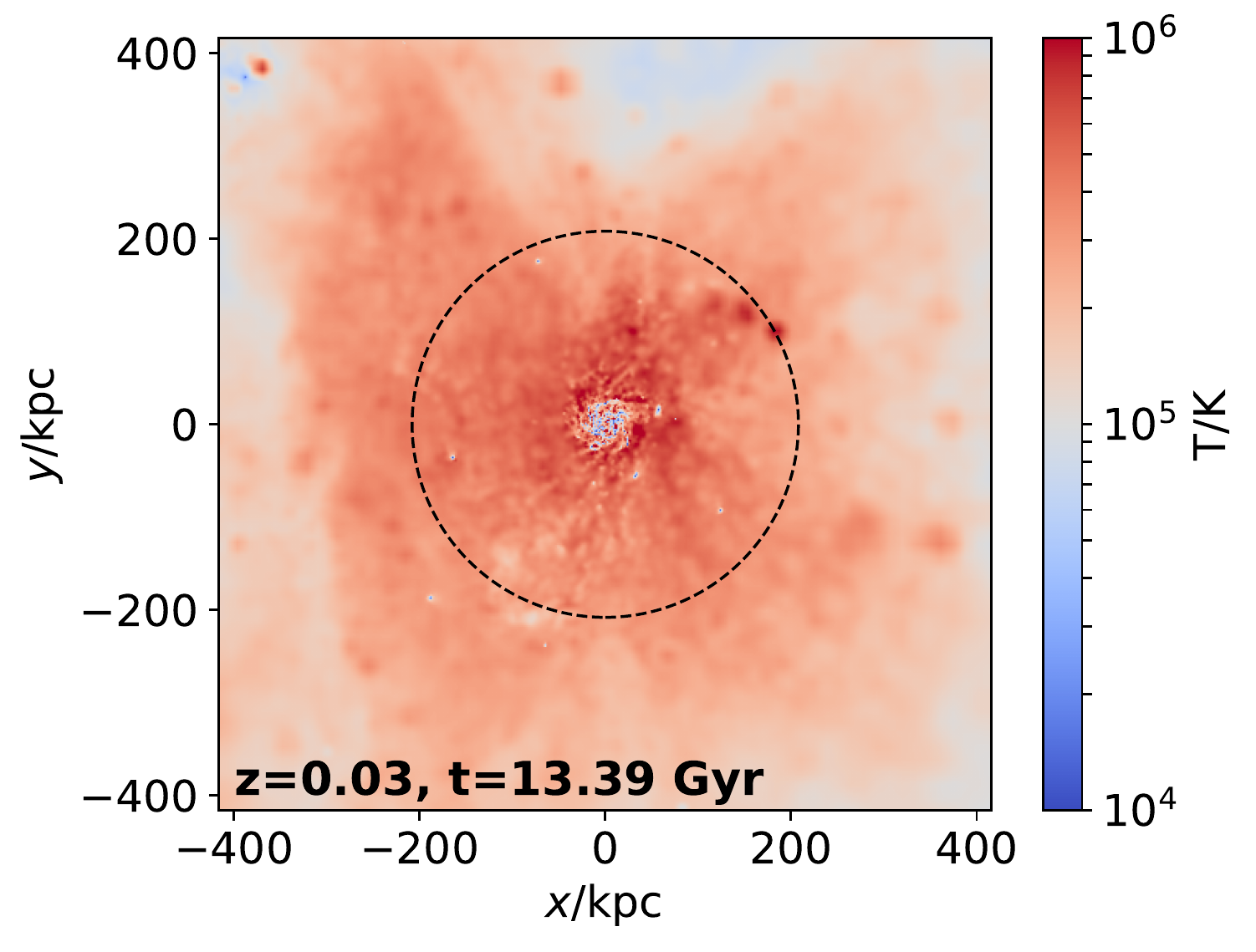}
   \includegraphics[width=0.25\hsize]{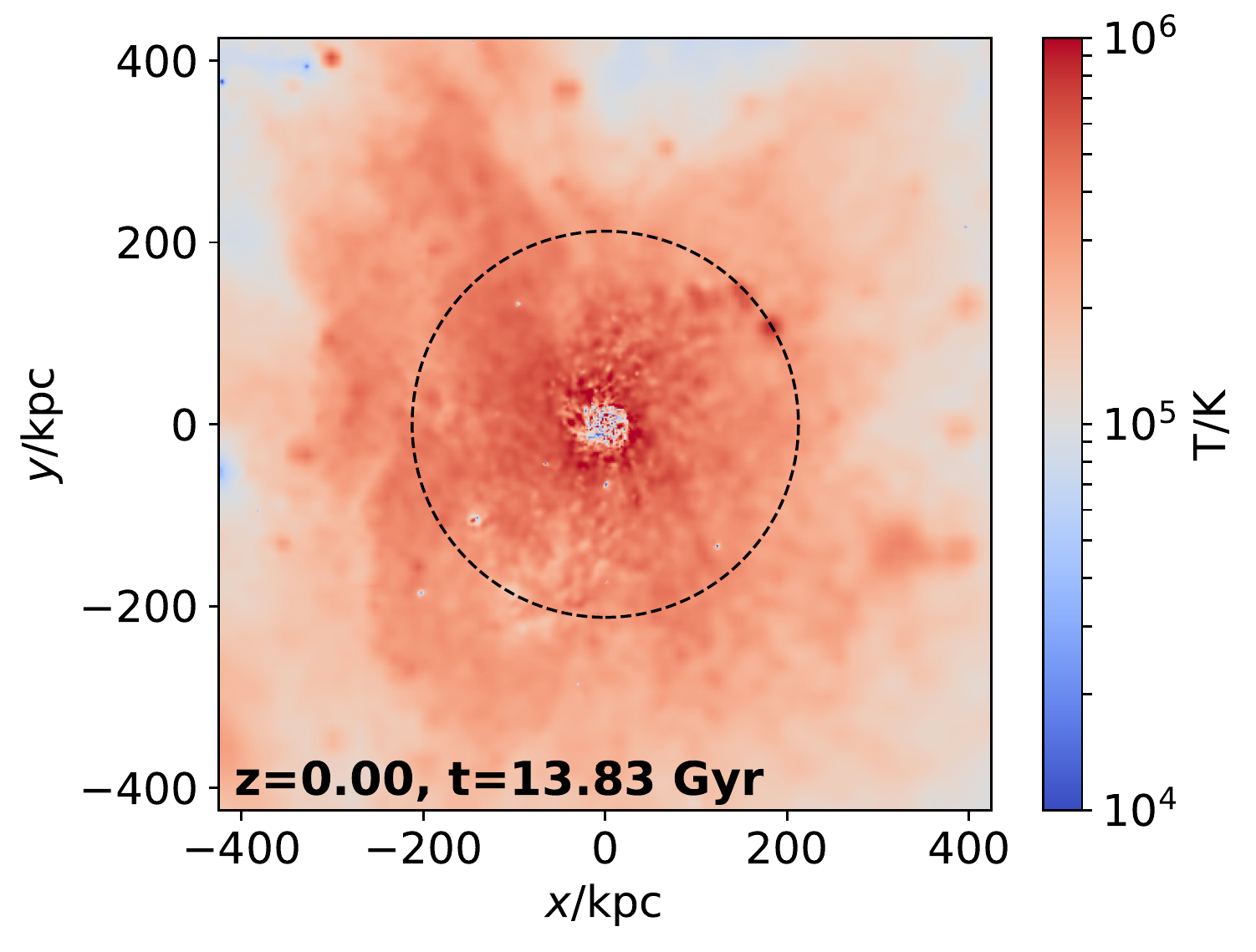}
\end{array}$
\end{center}
\caption{{\it Online-only supplementary material.} Temperature maps for the NIHAO simulation g7.55e11 with feedback (face-on view).}
\label{Supplementary1}
\end{figure*}

\renewcommand{\thefigure}{S2}

\begin{figure*}
\begin{center}$
\begin{array}{cccc}
  \includegraphics[width=0.25\hsize]{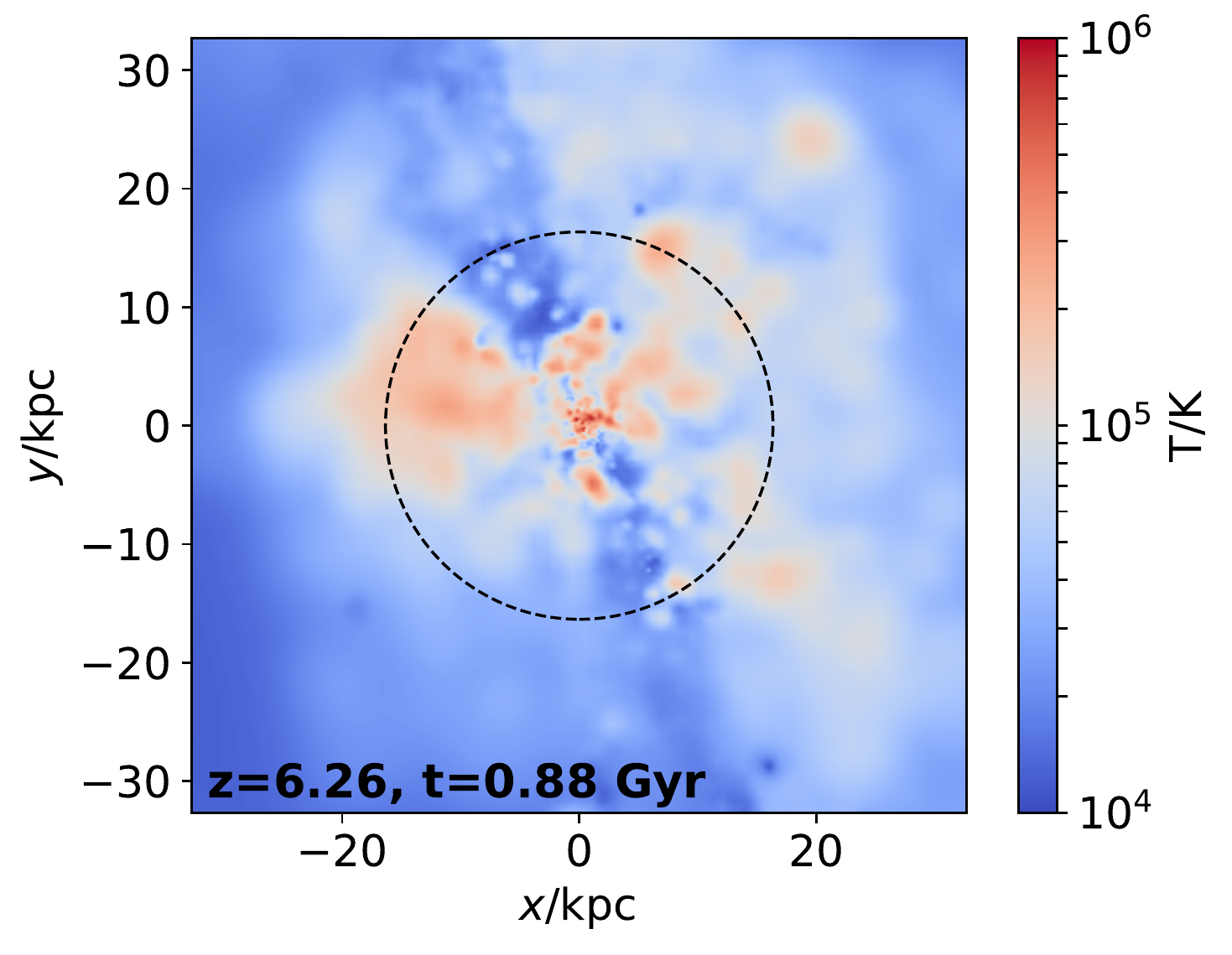} 
  \includegraphics[width=0.25\hsize]{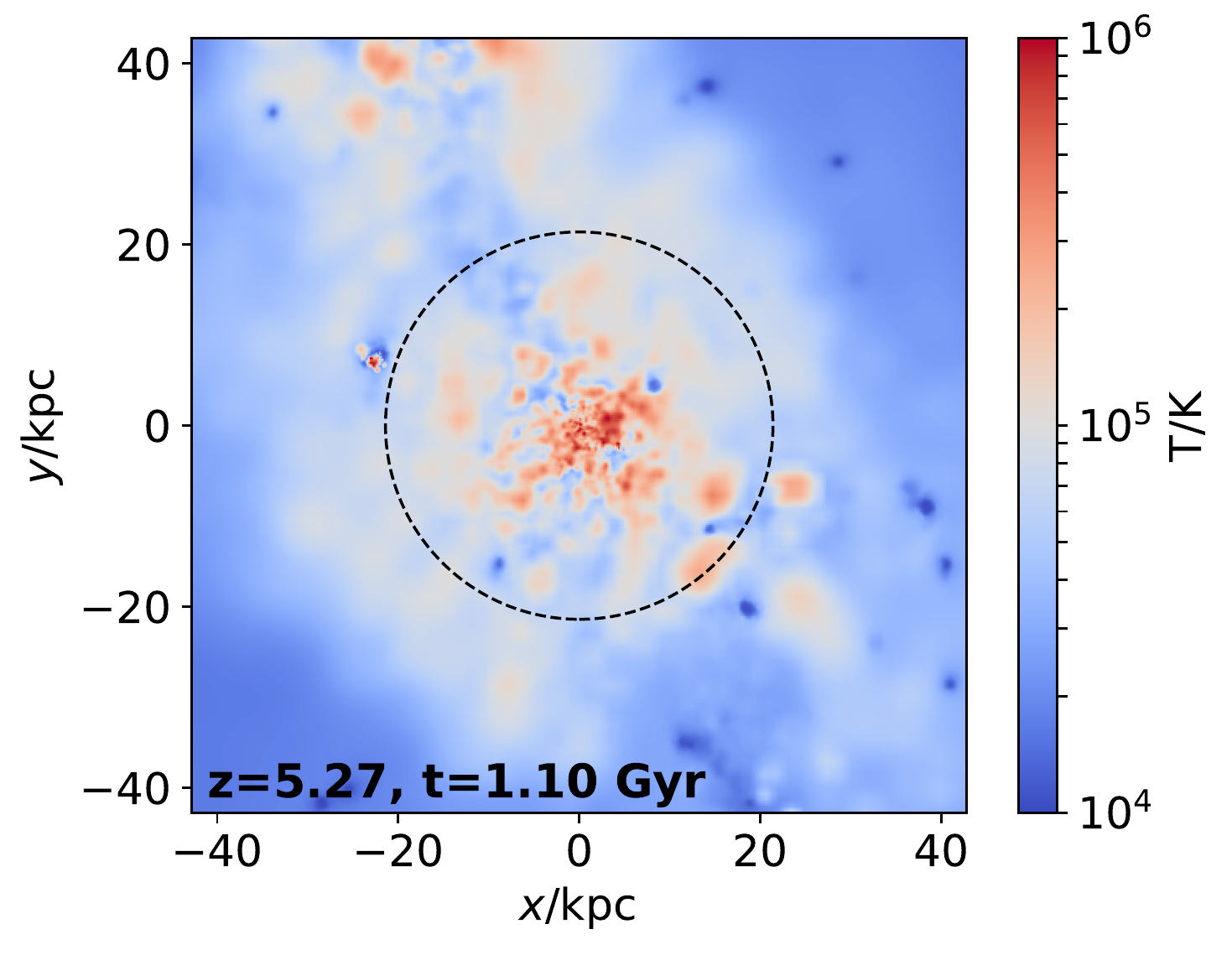}
   \includegraphics[width=0.25\hsize]{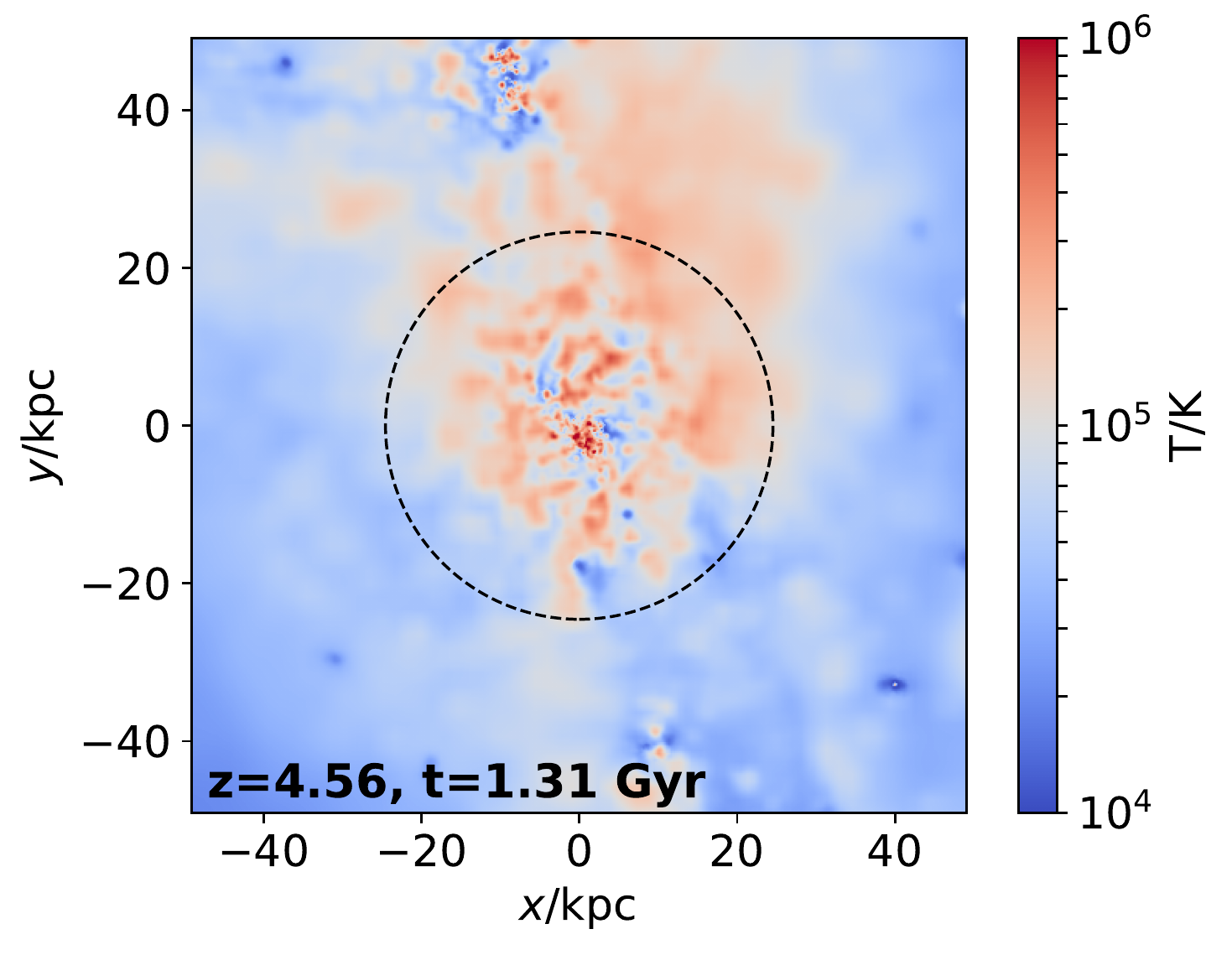}
  \includegraphics[width=0.25\hsize]{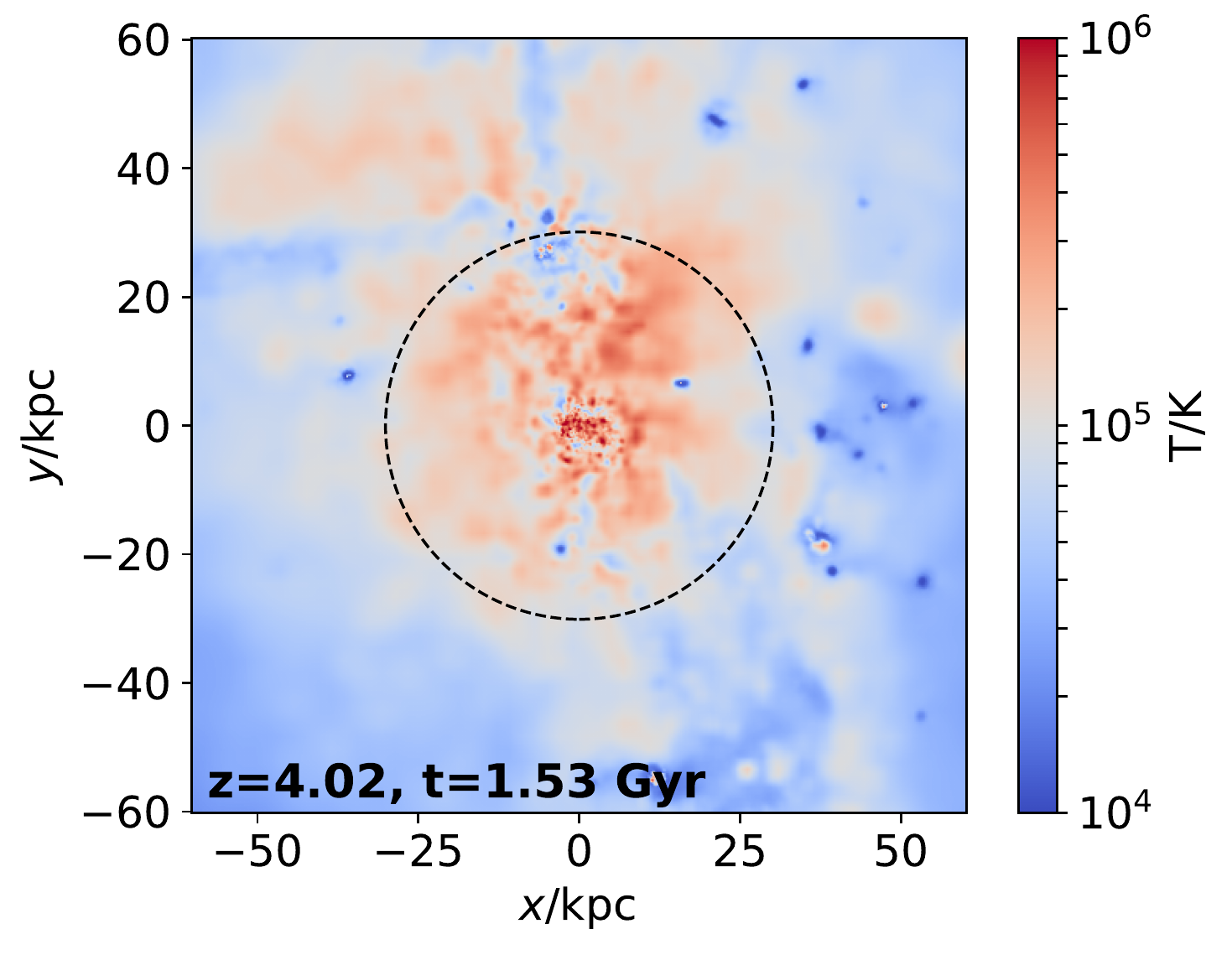}\\
  \includegraphics[width=0.25\hsize]{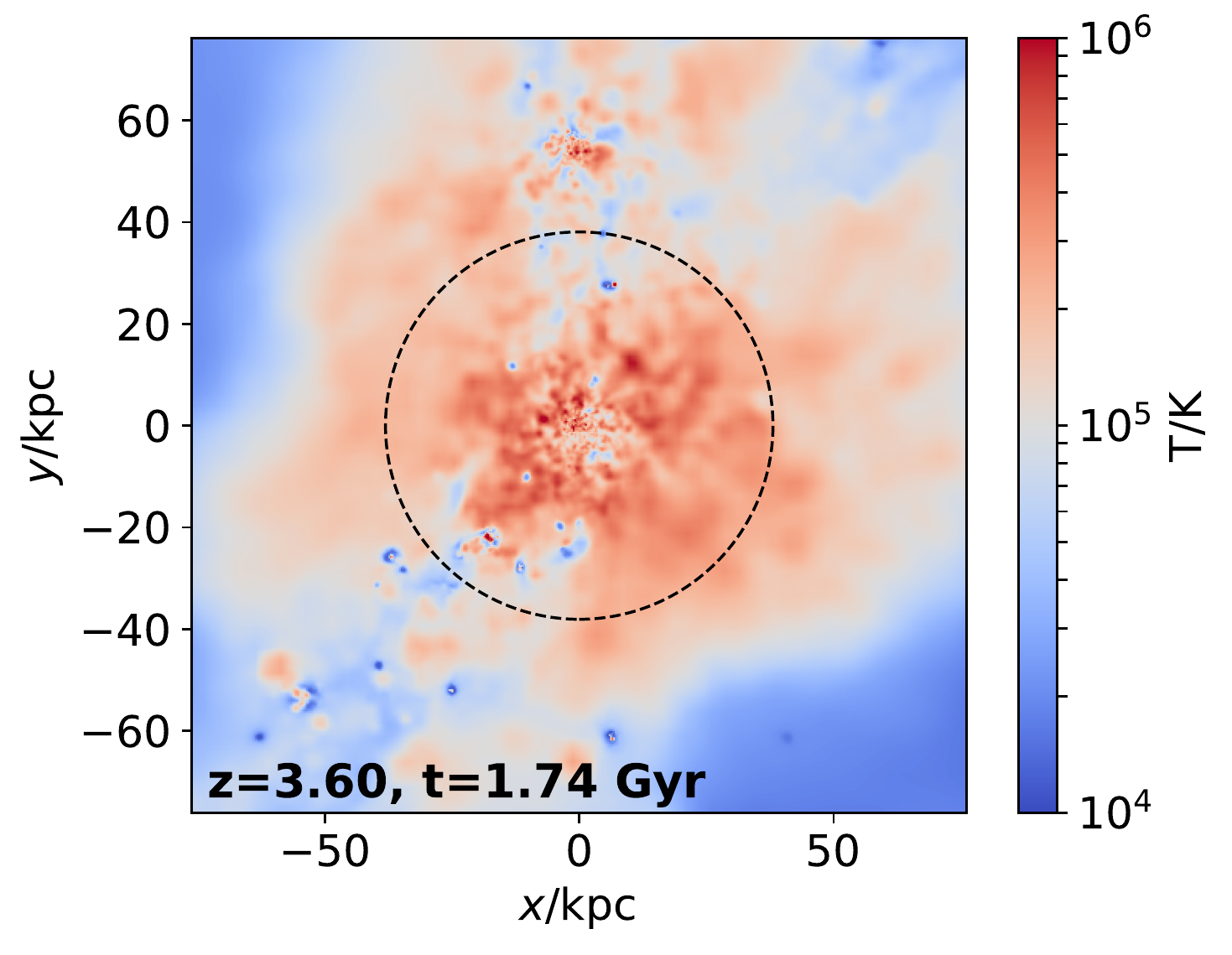}
  \includegraphics[width=0.25\hsize]{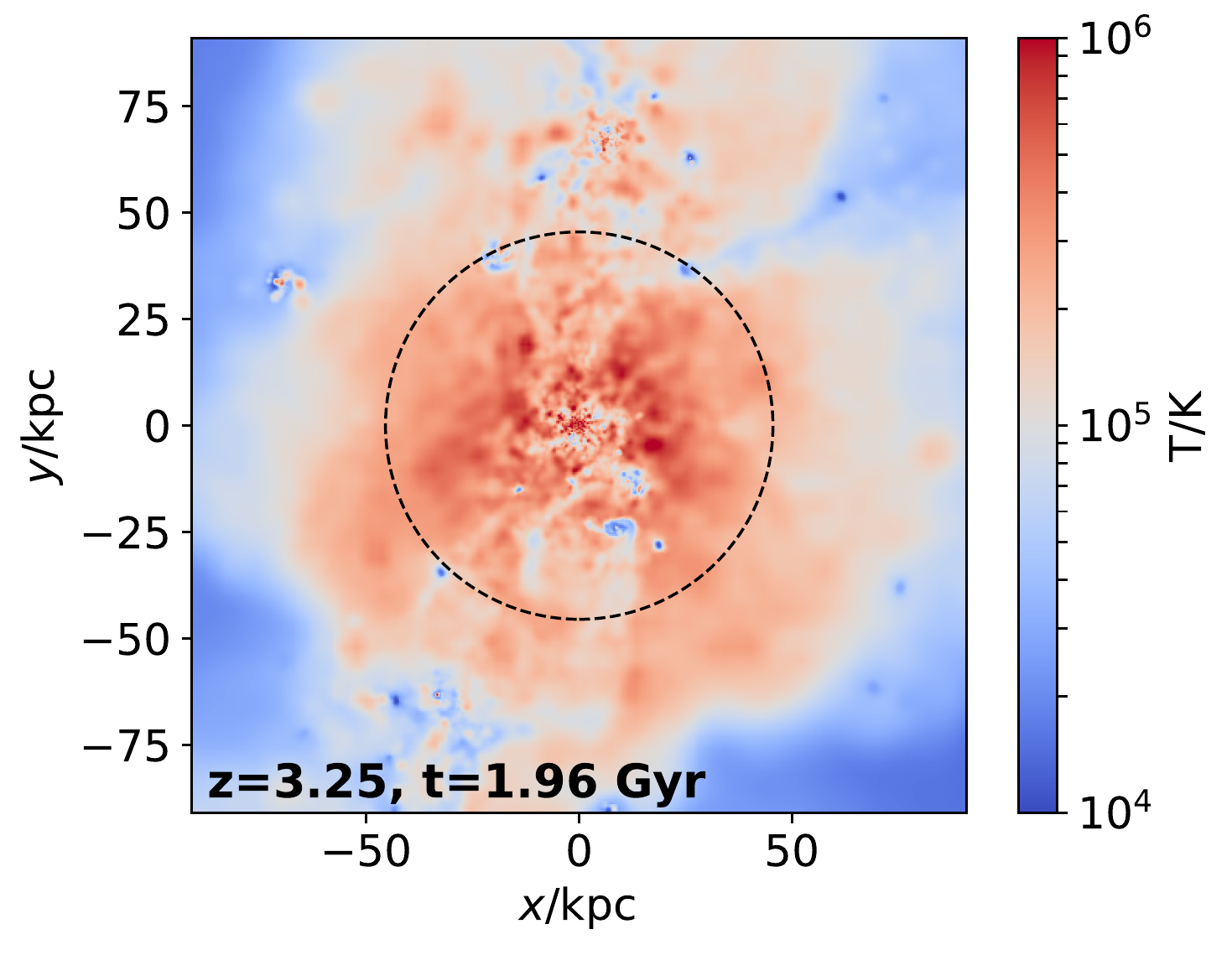}
    \includegraphics[width=0.25\hsize]{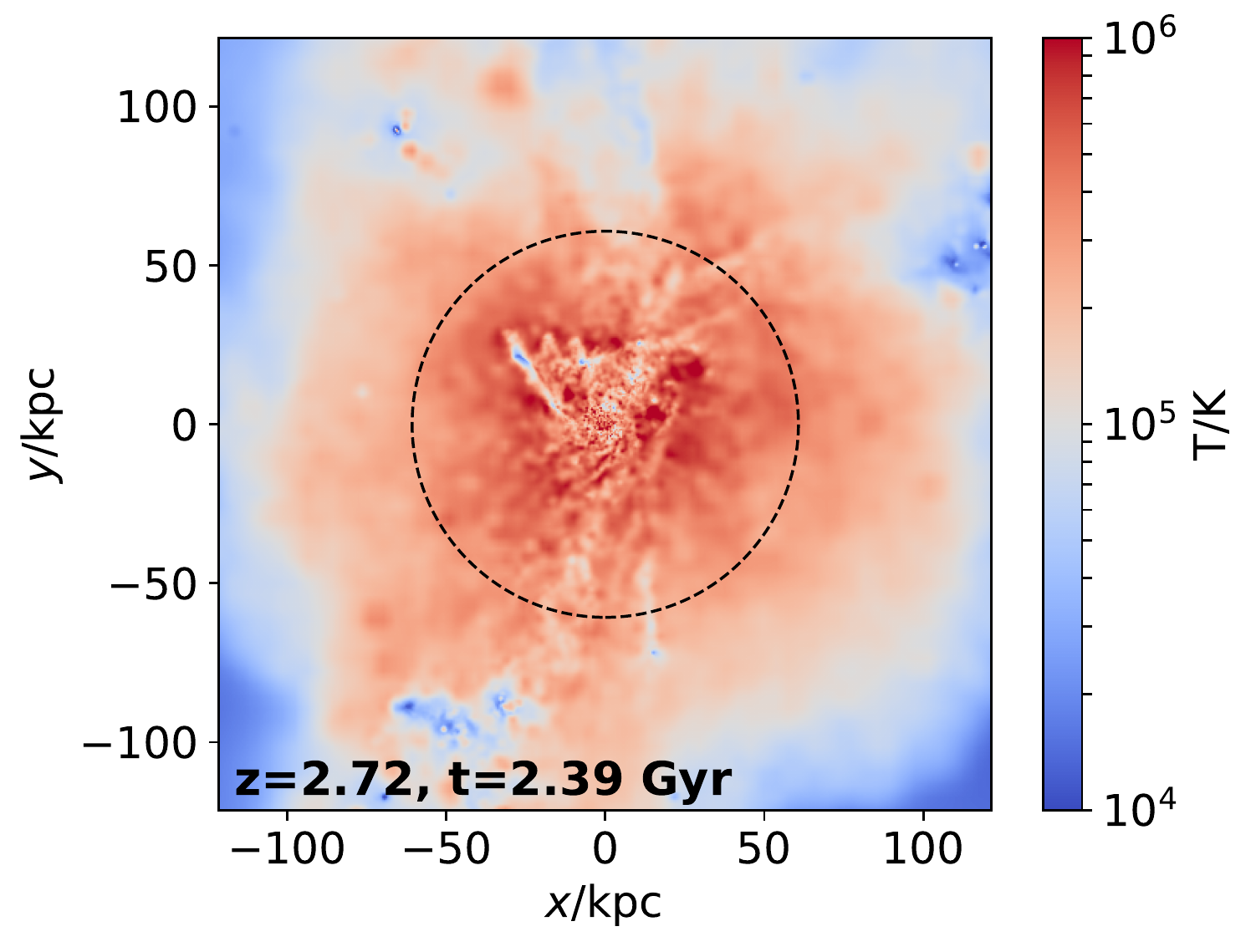}
  \includegraphics[width=0.25\hsize]{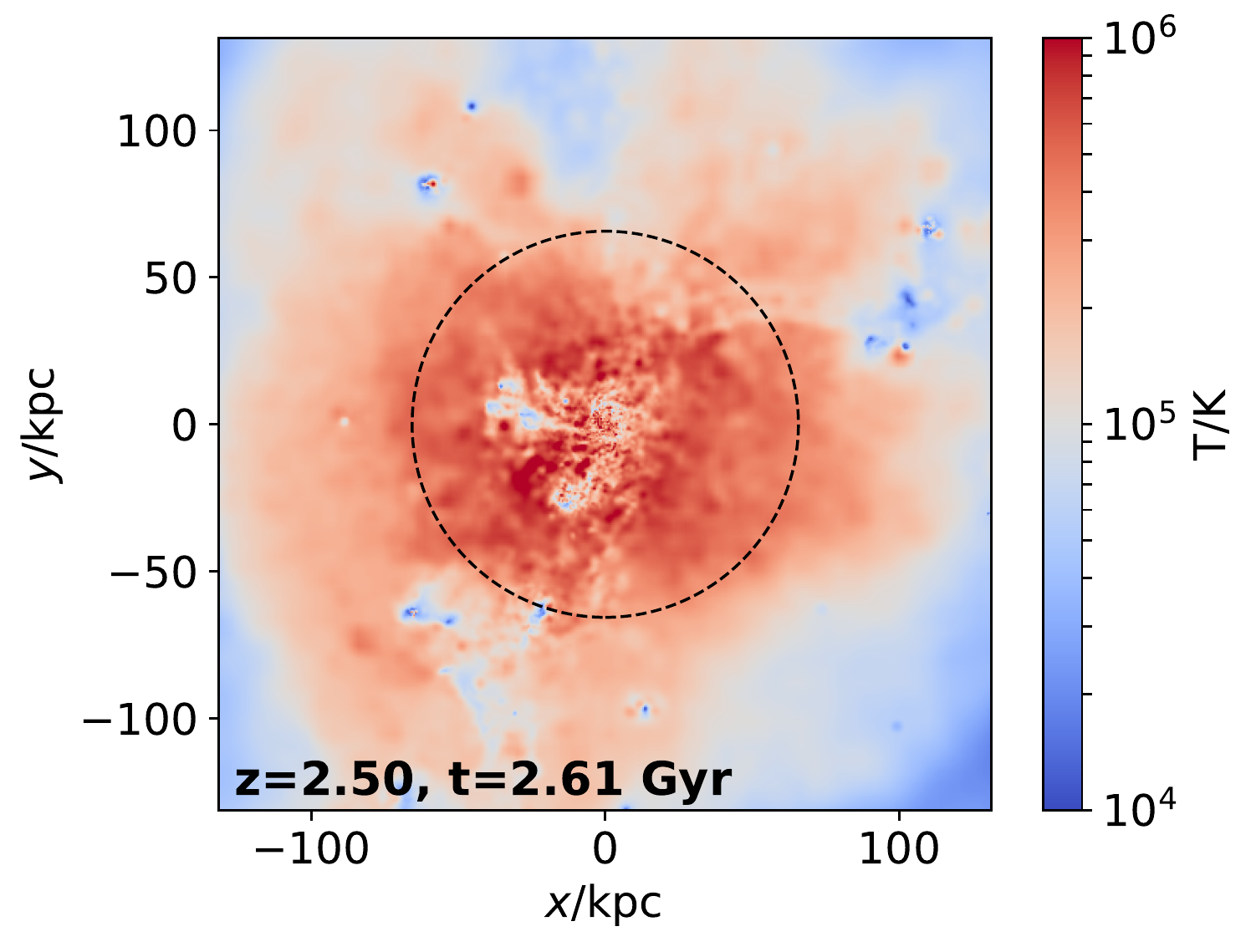}\\
  \includegraphics[width=0.25\hsize]{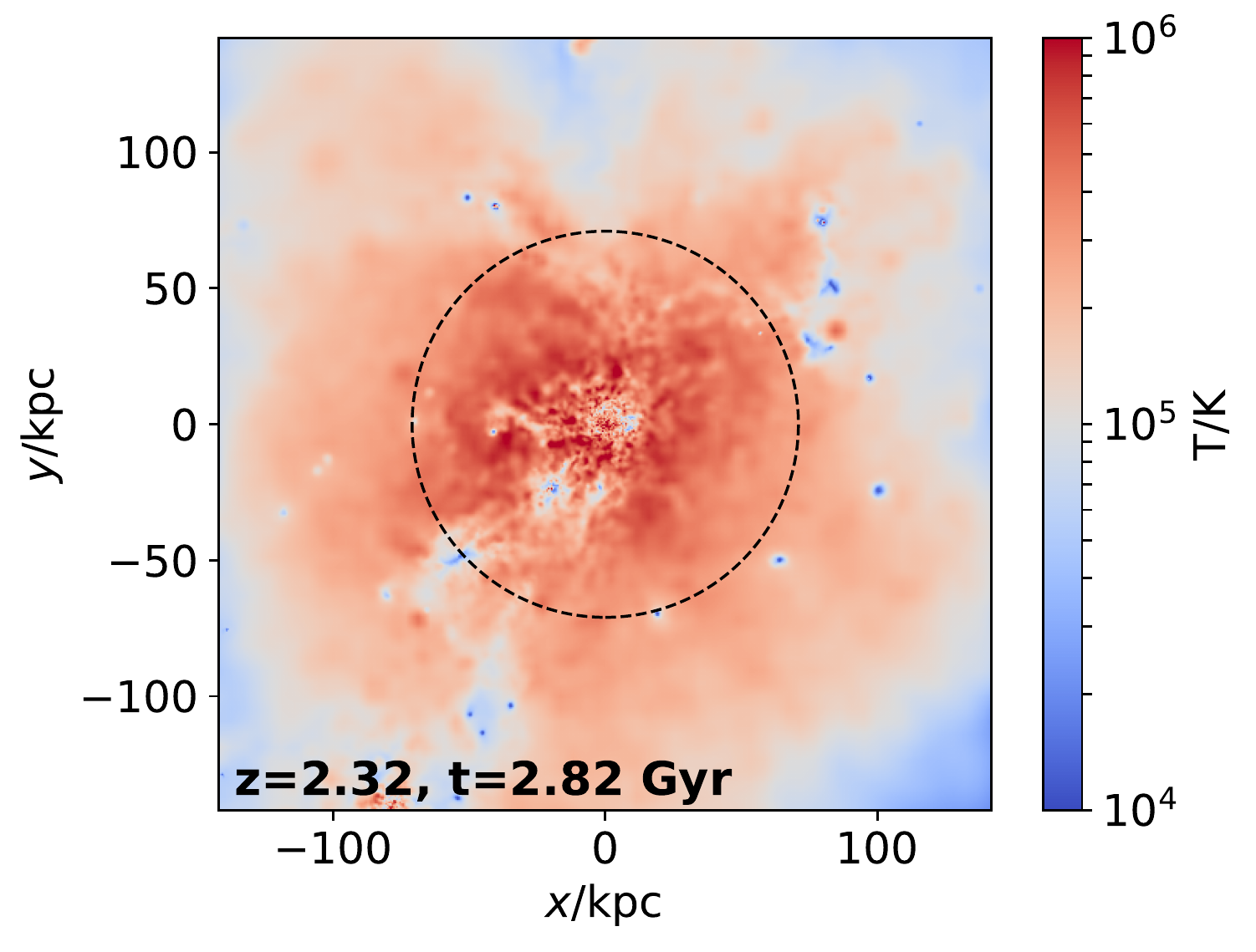}
  \includegraphics[width=0.25\hsize]{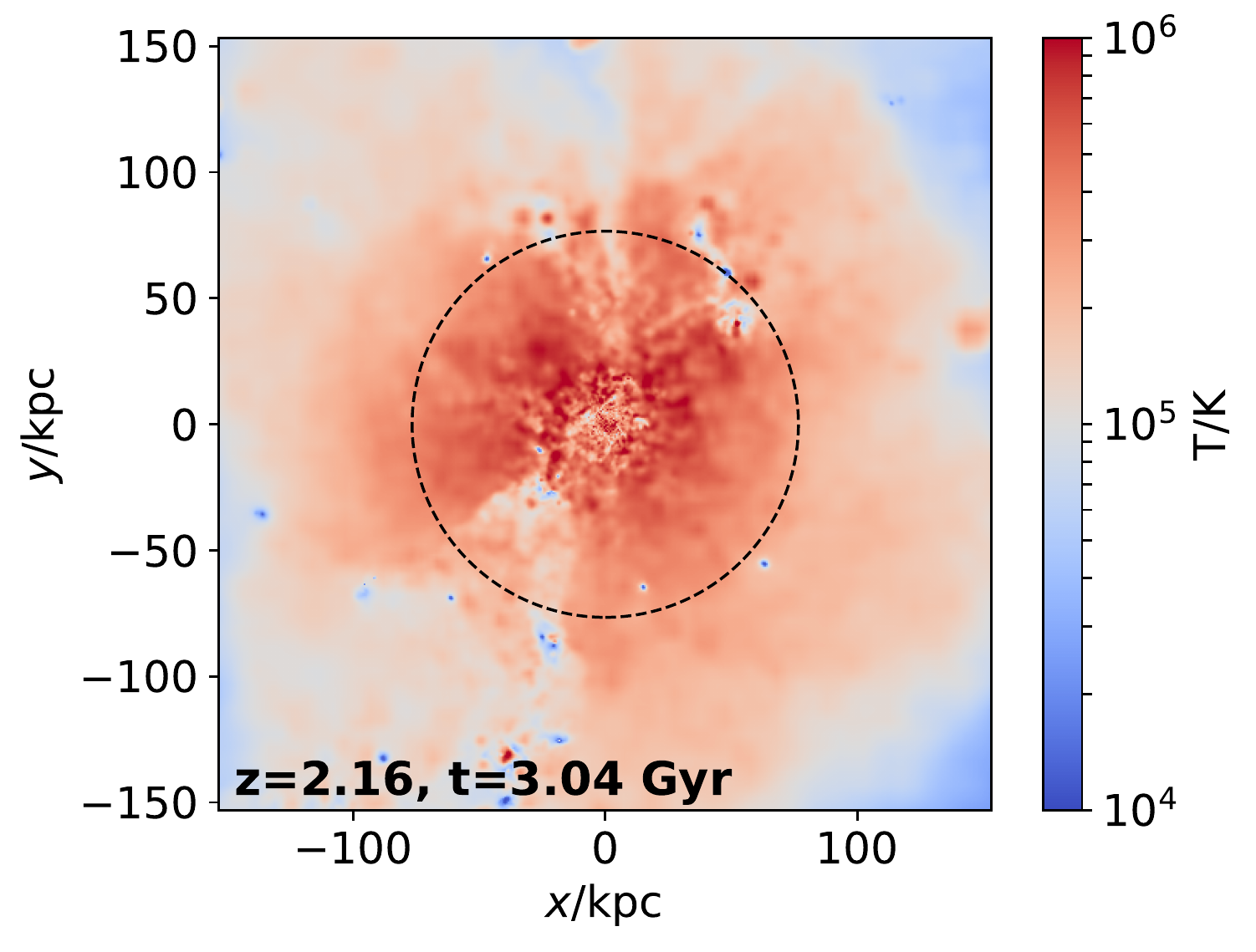}
  \includegraphics[width=0.25\hsize]{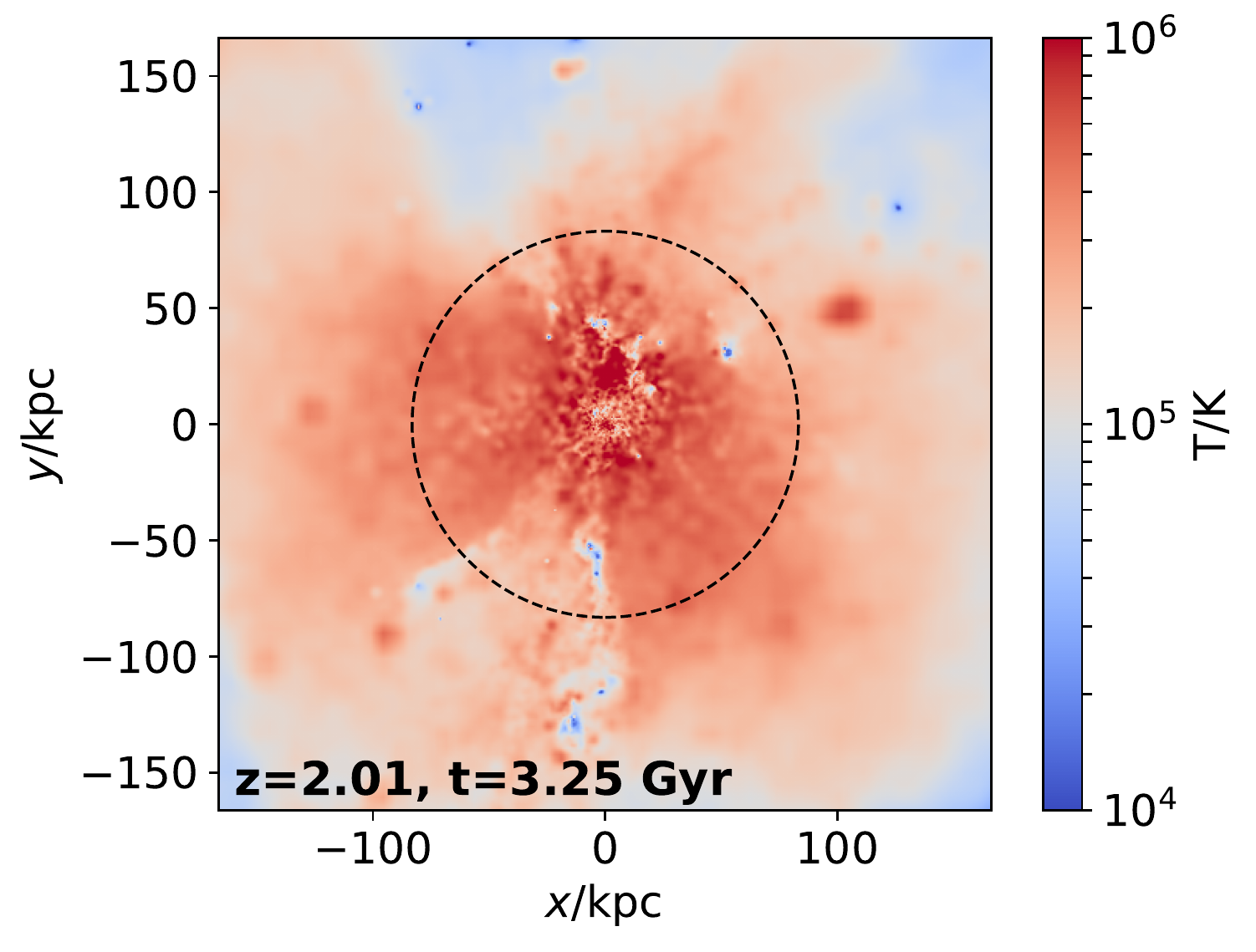}
  \includegraphics[width=0.25\hsize]{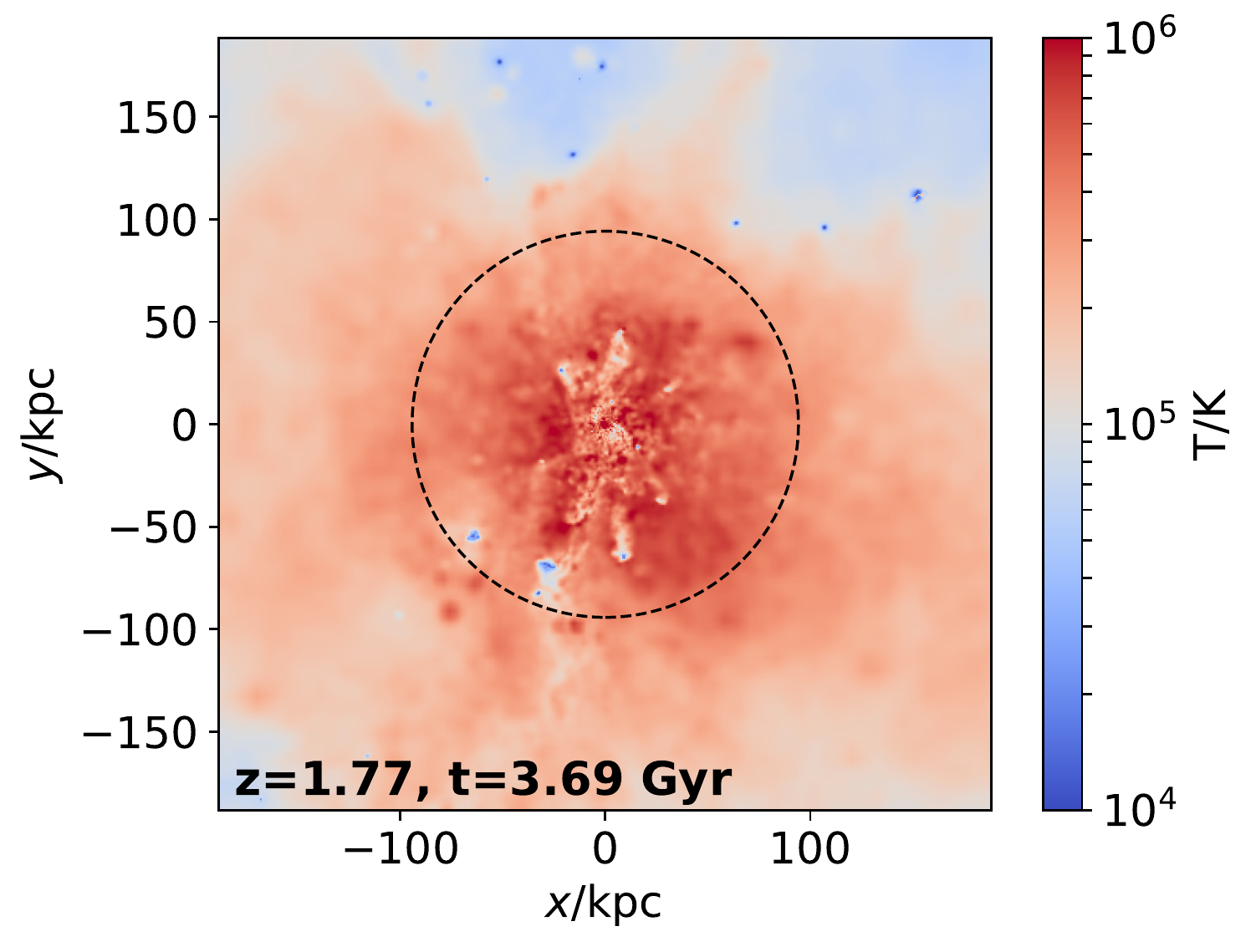}\\
  \includegraphics[width=0.25\hsize]{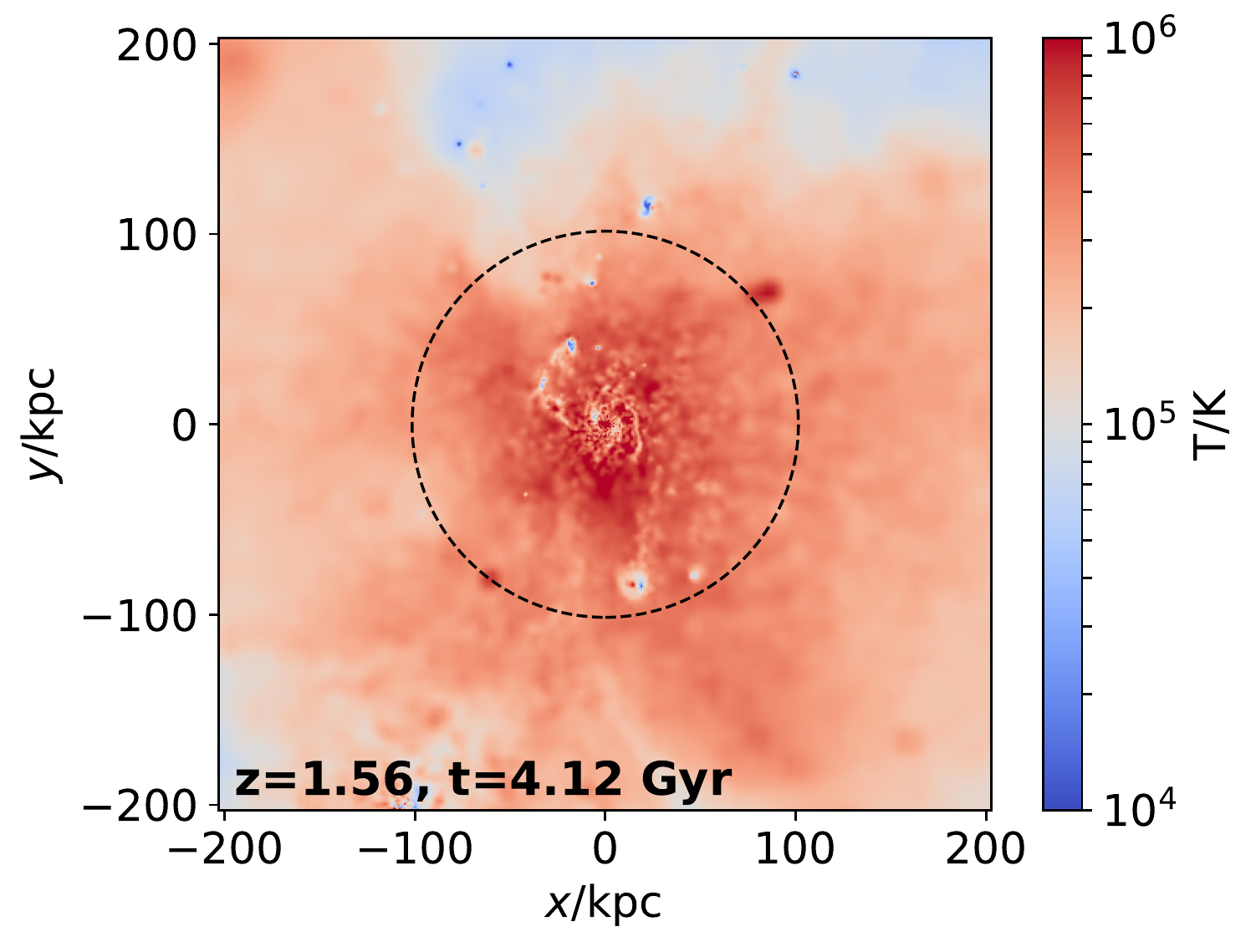}
  \includegraphics[width=0.25\hsize]{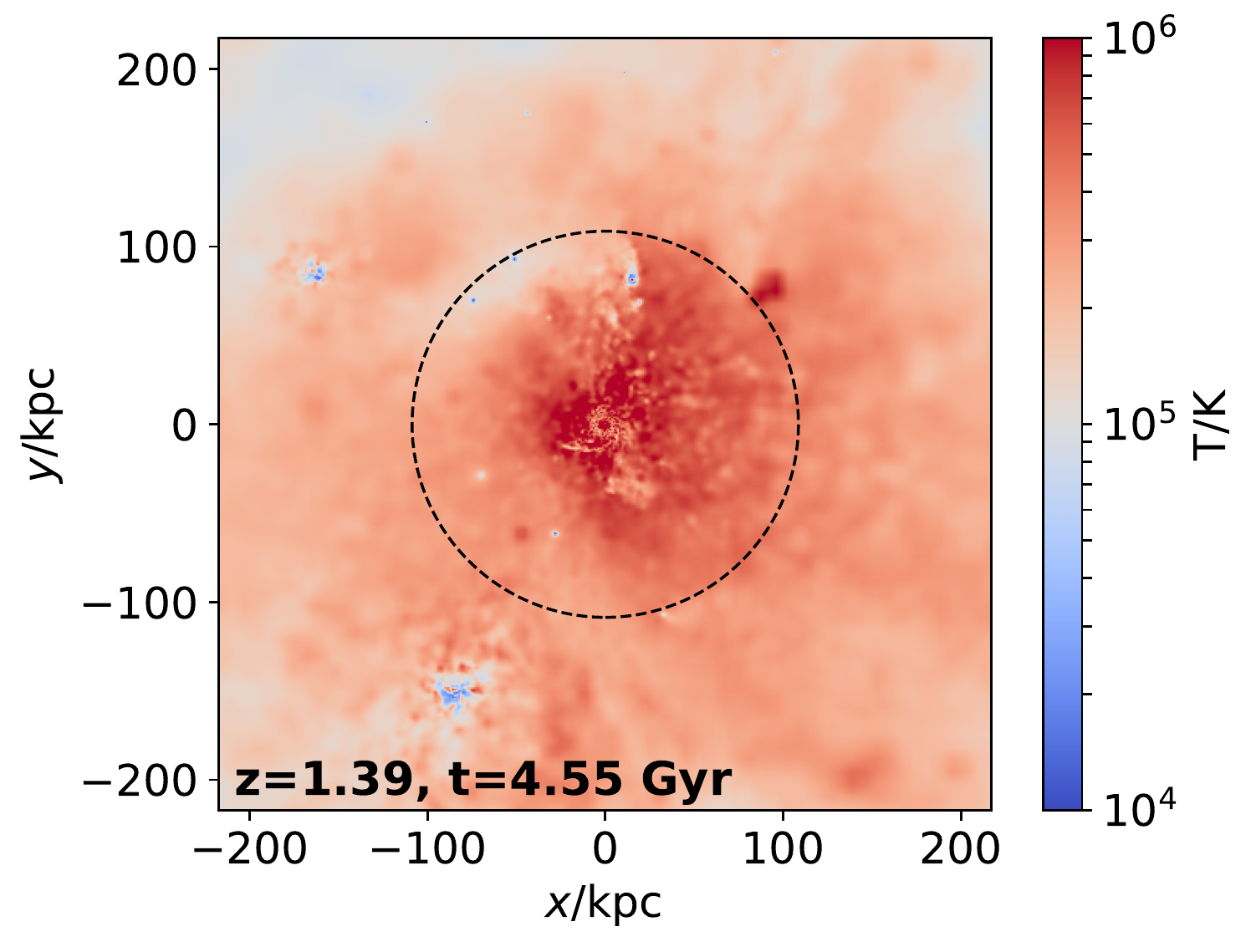}
  \includegraphics[width=0.25\hsize]{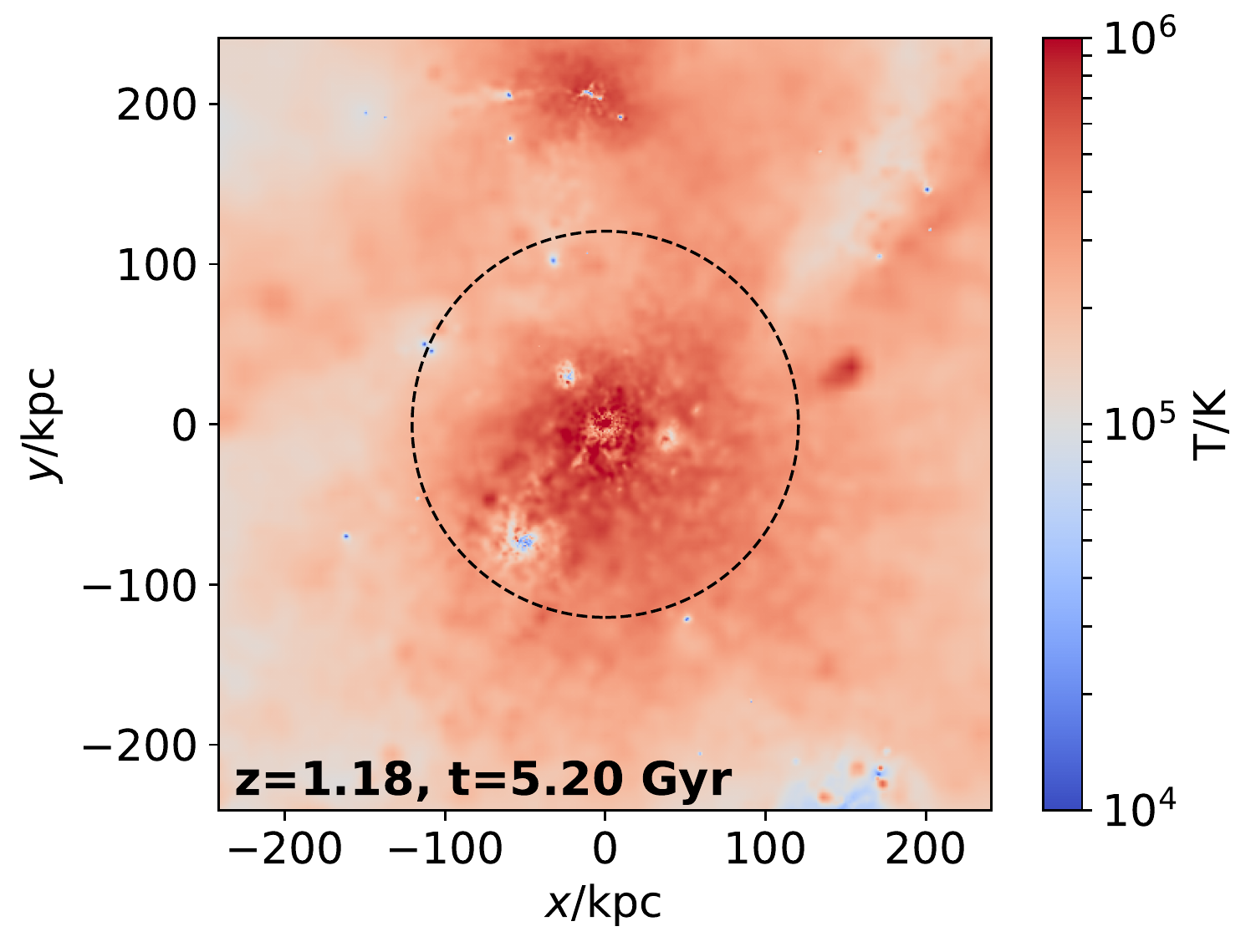}
  \includegraphics[width=0.25\hsize]{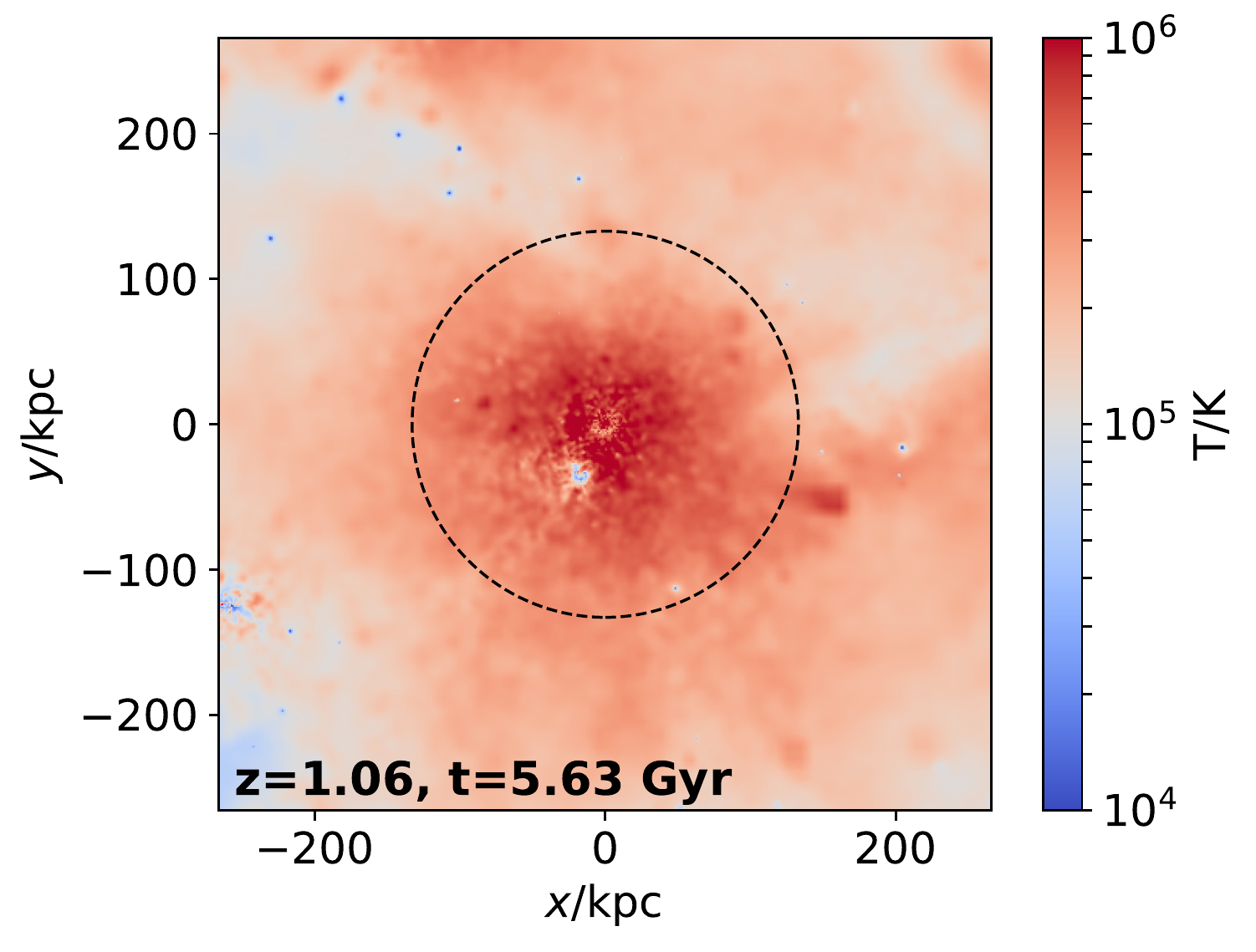}\\
  \includegraphics[width=0.25\hsize]{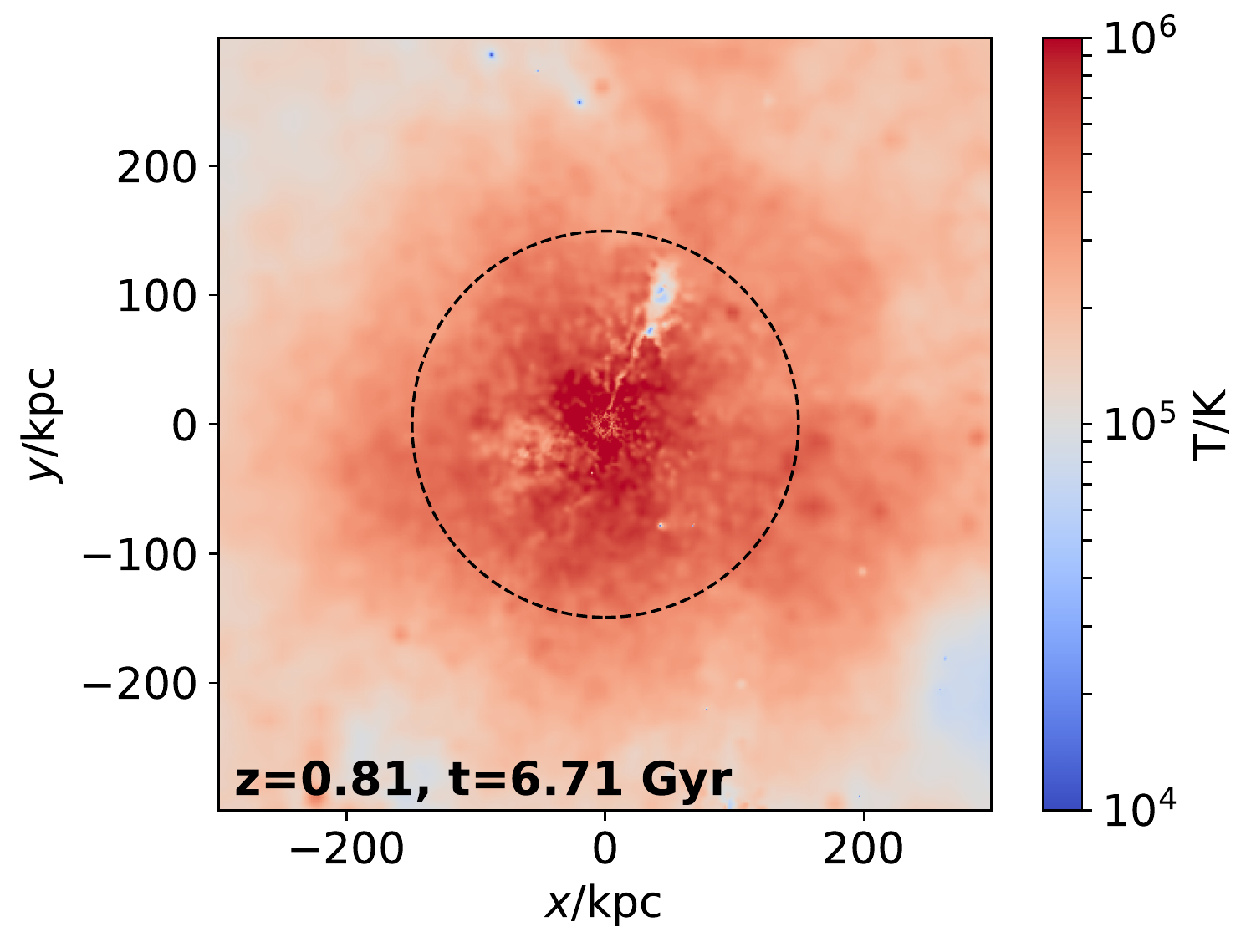}
  \includegraphics[width=0.25\hsize]{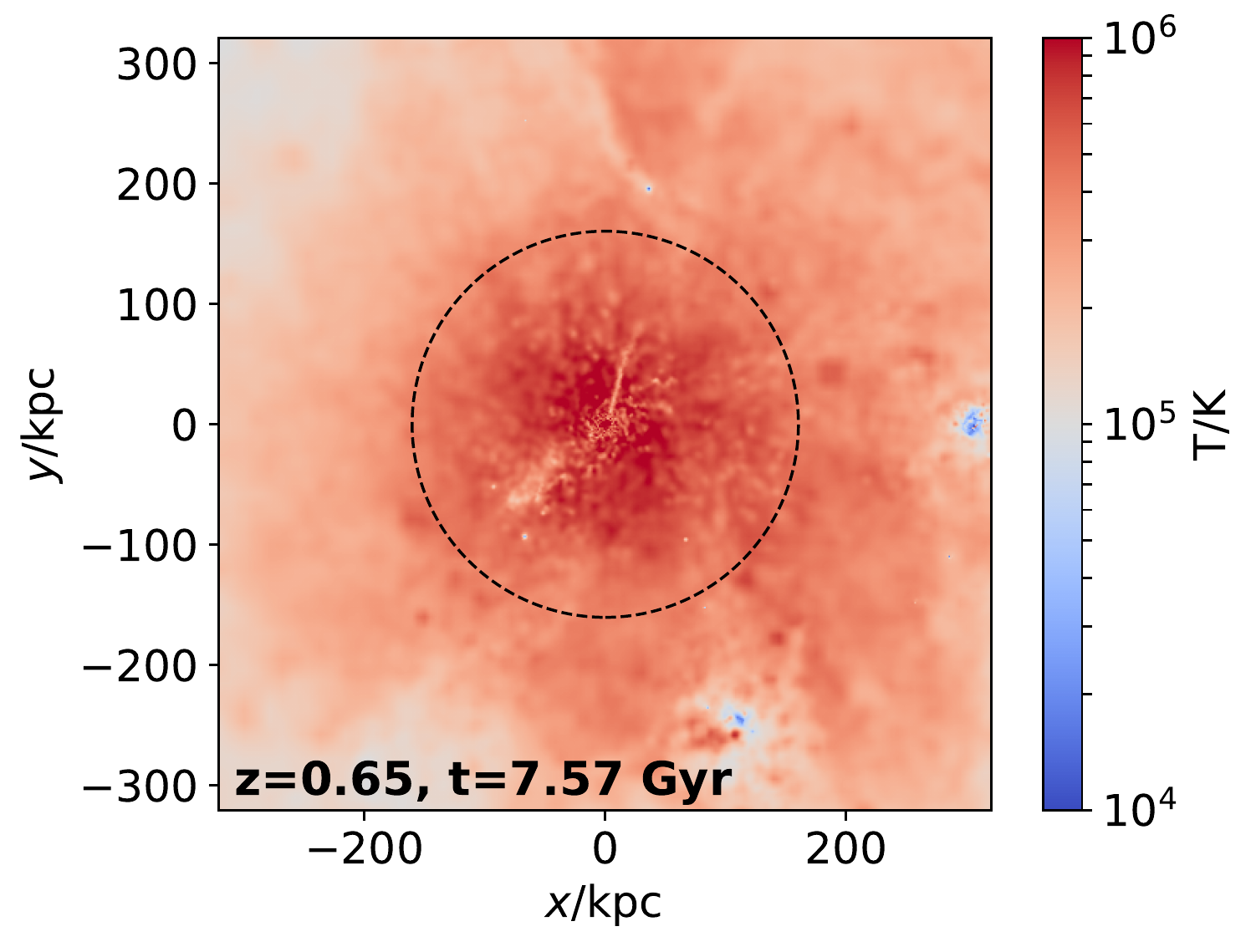}
  \includegraphics[width=0.25\hsize]{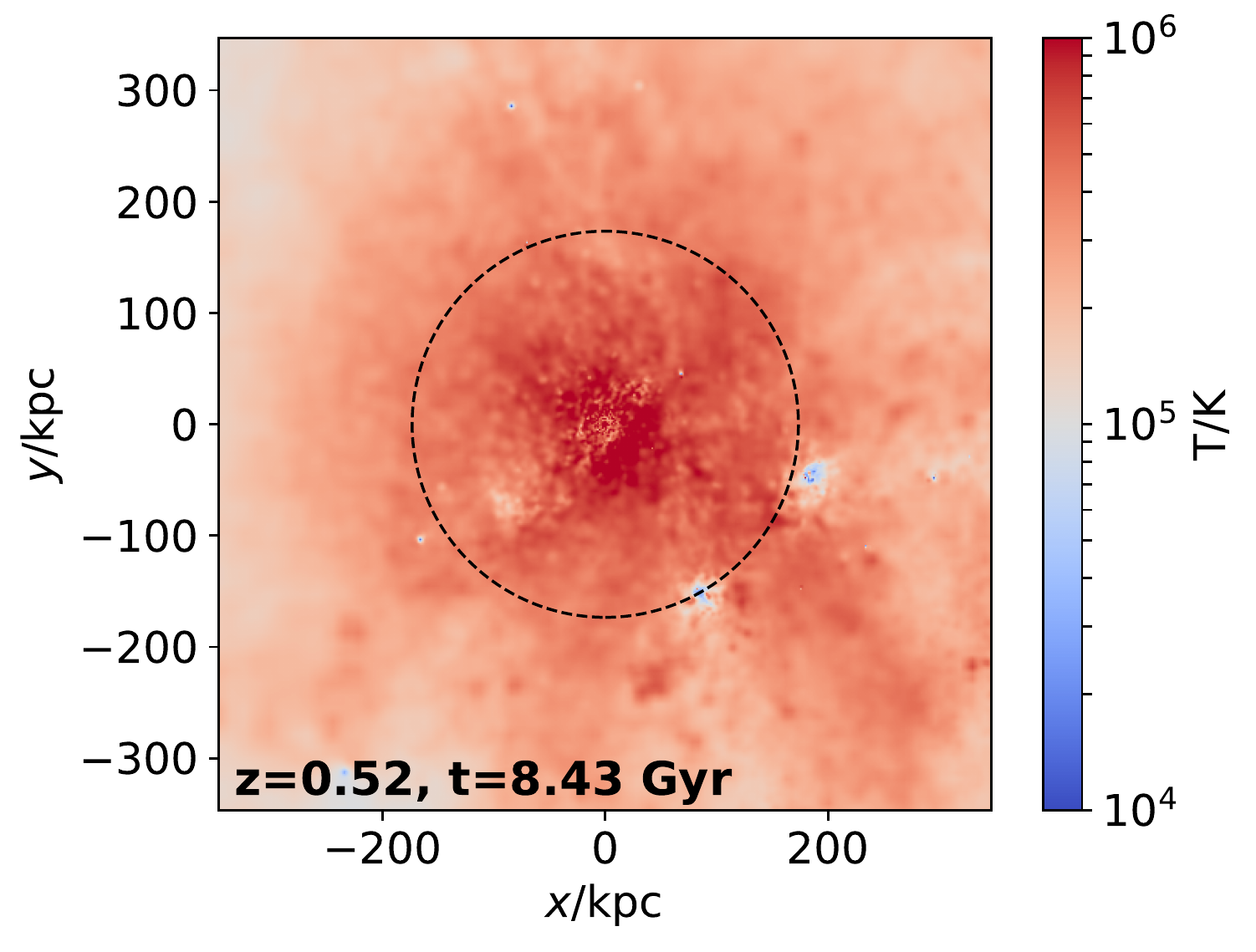}
    \includegraphics[width=0.25\hsize]{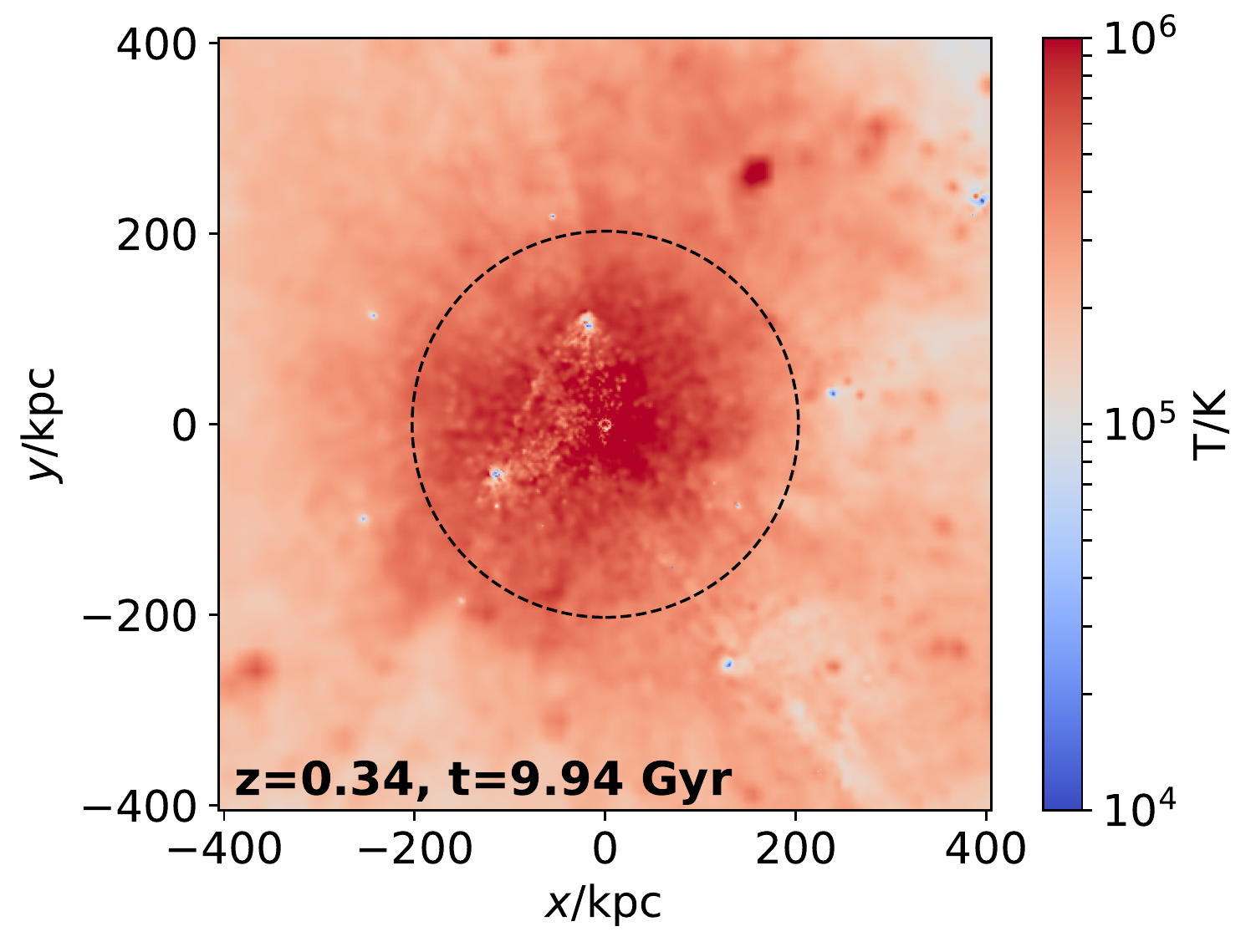}\\
  \includegraphics[width=0.25\hsize]{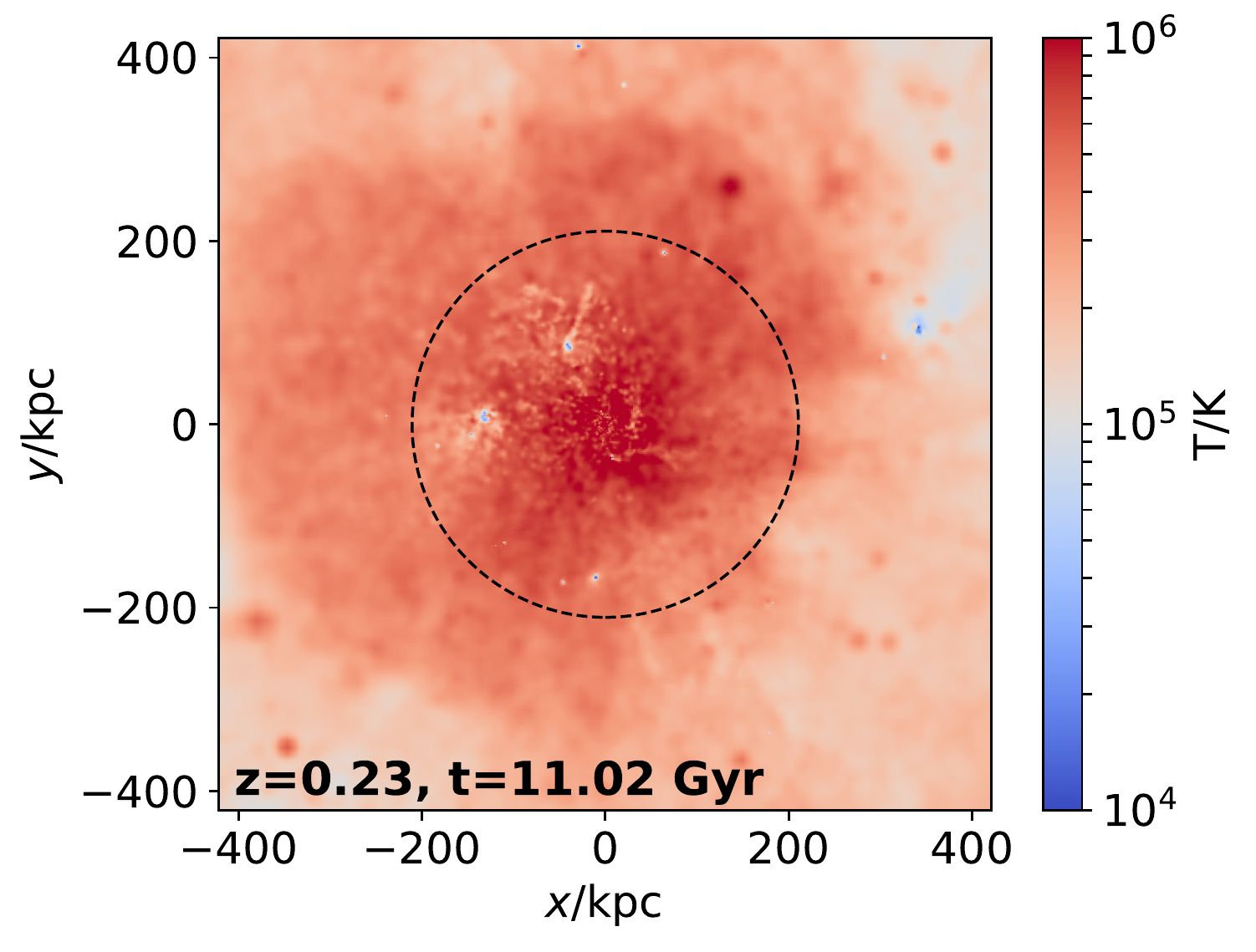}
  \includegraphics[width=0.25\hsize]{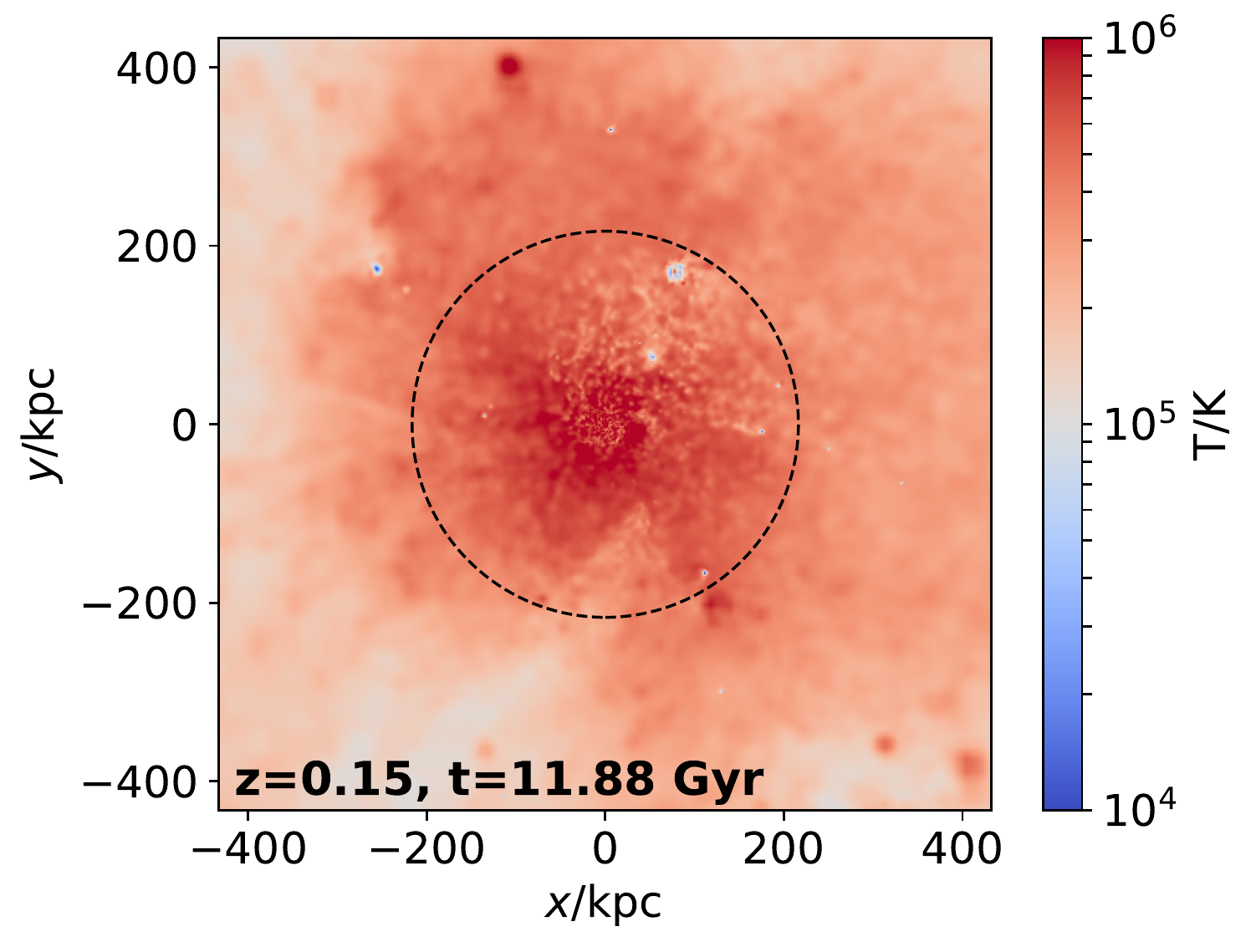}
   \includegraphics[width=0.25\hsize]{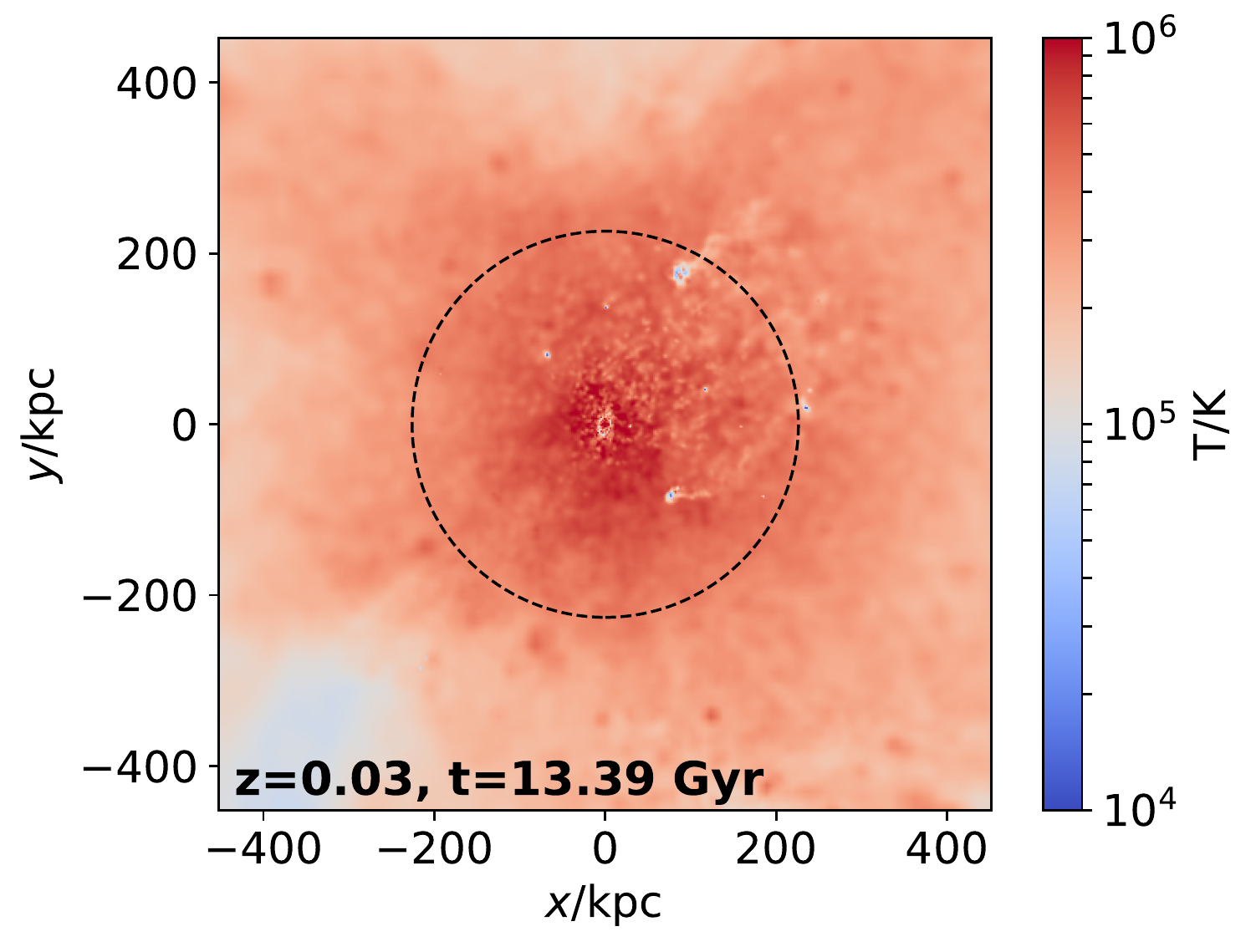}
   \includegraphics[width=0.25\hsize]{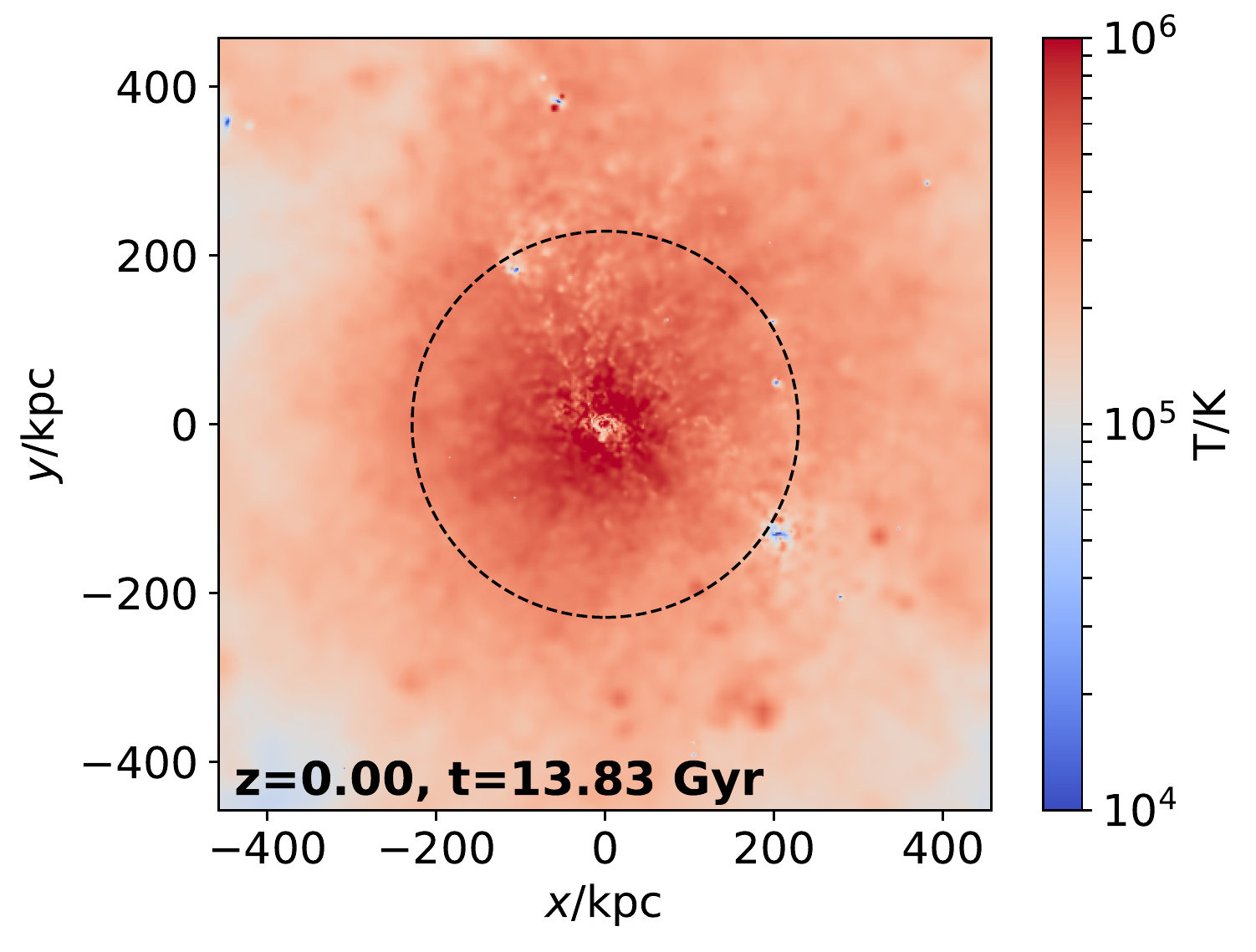}
\end{array}$
\end{center}
\caption{{\it Online-only supplementary material.} Temperature maps for the NIHAO simulation g1.12e12 with feedback (face-on view).}
\label{Supplementary2}
\end{figure*} 

\label{lastpage}
\end{document}